\documentclass{uzhthesis}
\usepackage[svgnames,x11names]{xcolor}
\definecolor{col}{rgb}{.4,.4,1}
\definecolor{col1}{rgb}{.4,.4,1}
\definecolor{col2}{rgb}{.4,.5,1}
\definecolor{col3}{rgb}{.4,.6,1}
\definecolor{col4}{rgb}{.4,.7,1}
\definecolor{hard}{rgb}{0.56,0.69,0.19}
\definecolor{stuff}{rgb}{0.88,0.88,0.88}
\definecolor{soft}{rgb}{0.88,0.61,0.14}
\definecolor{math1}{rgb}{0.37,0.51,0.71}
\definecolor{math2}{rgb}{0.88,0.61,0.14}
\definecolor{math3}{rgb}{0.56,0.69,0.19}
\definecolor{math4}{rgb}{0.92,0.39,0.21}
\definecolor{math5}{rgb}{0.53,0.47,0.70}
\definecolor{charge}{rgb}{0.8 0.15 0.15}
\definecolor{lightgray}{HTML}{D5D5D5}
\definecolor{lightergray}{HTML}{EEEEEE}

\usepackage{xspace}
\def\CDR{\text{\scshape cdr}\xspace}
\def\HV{\text{\scshape hv}\xspace}
\def\FDH{\text{\scshape fdh}\xspace}

\usepackage[numbers,sort&compress]{natbib}
\usepackage{amsmath}
\usepackage{amsthm}
\theoremstyle{definition} 
\newtheorem{exmp}{Example}[chapter]
\usepackage{amsfonts}
\usepackage{amssymb}
\usepackage{upgreek}
\usepackage{bbm}
\usepackage{slashed}
\usepackage[T1]{fontenc}
\usepackage{multirow}
\usepackage{multibib}\newcites{my}{Mine}\newcites{mules}{Mules}
\usepackage[caption=false]{subfig}

\usepackage{multicol}
\usepackage[printonlyused,nohyperlinks]{acronym}
\makeatletter
    \renewcommand{\ac}[1]{\AC@placelabel{#1}\acs{#1}}
\makeatother

\usepackage{listings}
\lstdefinelanguage{vim}{
  morekeywords={
  set, let
  map, nmap,
  filetype,
  on, off,
  autocmd,
  Plugin,
  call,
   },
morecomment=[l]{"}, % l is for line comment
morestring=[b]' % defines that strings are enclosed in double quotes
}

\usepackage[colorlinks=true
,urlcolor=blue
,anchorcolor=blue
,citecolor=blue
,filecolor=blue
,linkcolor=blue
,menucolor=blue
,linktocpage=true
,pdfproducer=medialab
]{hyperref}

\usepackage{tikz}
\usetikzlibrary{decorations.pathreplacing,calligraphy}
\usepackage{tikz-feynman}
\usetikzlibrary{decorations.pathmorphing,positioning,decorations.pathreplacing,shapes,calc}
\tikzset{
    photon/.style={decorate, decoration={snake,amplitude=1pt,segment length=6pt}},
    tightphoton/.style={decorate, decoration={snake,amplitude=1pt,segment length=3pt}},
    zigzag it/.style={decorate, decoration={zigzag,amplitude = .7mm,segment length=2mm}},
    gluon/.style={decorate, draw=black,decoration={coil,amplitude=4pt, segment length=5pt}},
    tightgluon/.style={decorate, draw=black,decoration={coil,amplitude=2pt, segment length=3pt}}
}
\def\centerarc[#1](#2,#3)(#4:#5:#6)% Syntax: [draw options] (center) (initial angle:final angle:radius)
    { \draw[#1] ({#2+#6*cos(#4)},{#3+#6*sin(#4)}) arc (#4:#5:#6); }

\usepackage{enumitem}
\makeatletter
    \newlist{treelist}{itemize}{5}
    \setlist[treelist]{label=\treelist@label}

    \tikzset{treelist line/.style={thick, line cap=round, rounded corners}}
    \def\treelist@label{%
        \begin{tikzpicture}[remember picture, baseline={([yshift=-.6ex] treelist-bullet-\the\enit@depth.center)}]
            \draw [treelist line] (0, 0) -- node (treelist-bullet-\the\enit@depth) {} ++(.5em, 0);
        \end{tikzpicture}%
        \ifnum\enit@depth>1
            \tikz[remember picture, overlay] \draw [treelist line] (treelist-bullet-\the\numexpr\enit@depth-1\relax.center) |- (treelist-bullet-\the\enit@depth.center);%
        \fi
    }
\makeatother

\newcommand{\mtextsc}[1]{\relax\ifmmode\textsc{#1}\else\ac{#1}\fi}

\newcommand\mcmule{{\sc McMule}}

\renewcommand{\d}[1]{\ensuremath{\operatorname{d}\!{#1}}}

\def\D{\mathrm{d}}

\newcommand{\wideeq}[2][1.5]{
  \mathrel{\overset{#2}{\scalebox{#1}[1]{$=$}}}
}
\newcommand{\widesim}[2][1.5]{
  \mathrel{\overset{#2}{\scalebox{#1}[1]{$\sim$}}}
}
\newcommand{\pder}[2][]{\frac{\partial#1}{\partial#2}}
\catcode`,\active
\newcommand*\pFq{\begingroup
        \catcode`\,\active
        \def ,{\mskip\pFqskip\relax}%
        \dopFq
}
\catcode`\,12
\def\dopFq#1#2#3#4#5{%
        {}_{#1}F_{#2}\biggl[\genfrac..{0pt}{}{#3}{#4};#5\biggr]%
        \endgroup
}

\newcommand{\M}[2]{\mathcal{M}_{#1}^{(#2)}}
\newcommand{\fM}[2]{\mathcal{M}_{#1}^{(#2)f}}

\newcommand\eik{\mathcal{E}}
\newcommand\ieik{\hat{\mathcal{E}}}
\def\xc{\xi_{c}}
\newcommand{\cdis}[2][c]{\left(\frac{1}{#2}\right)_{\hspace*{-3pt}#1}}

\def\pref#1{%
 \ifnum0<0#1\relax
   \newcount\foo%
   \foo=0%
   \loop
     \advance\foo +1
     \D\Upsilon_{\the\foo}
   \ifnum\foo<#1
   \repeat
 \else
   \prod_{i=1}^{#1}\D\Upsilon_i
 \fi
  \D \Phi_{n,#1}
}

\makeatletter
\def\thickhline{%
  \noalign{\ifnum0=`}\fi\hrule \@height \thickarrayrulewidth \futurelet
   \reserved@a\@xthickhline}
\def\@xthickhline{\ifx\reserved@a\thickhline
               \vskip\doublerulesep
               \vskip-\thickarrayrulewidth
             \fi
      \ifnum0=`{\fi}}
\makeatother

\newlength{\thickarrayrulewidth}
\usepackage{chngcntr}
\counterwithout{footnote}{chapter}

\usepackage{ifoddpage}
\usepackage{marginnote}

\makeatletter
  \def\ps@headings{%
      \let\@oddfoot\@empty\let\@evenfoot\@empty
      \def\@evenhead{\thepage\hfil\slshape\leftmark}%
      \def\@oddhead{{\slshape\rightmark}\hfil\thepage}%
      \let\@mkboth\markboth
    \def\chaptermark##1{%
      \markboth {{%
        \ifnum \c@secnumdepth >\m@ne
            \@chapapp\ \thechapter. \ %
        \fi
        ##1}}{}}%
    \def\sectionmark##1{%
      \markright {{%
        \ifnum \c@secnumdepth >\z@
          \thesection. \ %
        \fi
        ##1}}}}
\makeatother

\title{Muon-Electron Scattering at NNLO}
\author{Tim Engel}
\from{Zollikon ZH}
\committee{
    \member{Prof. Dr. Adrian Signer (Vorsitz)}
    \member{Prof. Dr. Thomas Gehrmann}
    \member{PD Dr. Michael Spira}
}
\abstractE{
Precision experiments with leptons are an essential part of the search of physics beyond the Standard Model. In order to meet the experimental precision the calculation of next-to-leading order (NLO) or even next-to-next-to-leading order (NNLO) QED corrections has become mandatory. This has triggered the development of \mcmule{}, a Monte Carlo integrator for processes with muons and other leptons. The main difference of higher-order calculations in QED compared to QCD is that fermion masses are typically not neglected. Collinear divergences are thus regularised by finite fermion masses which greatly simplifies the infrared structure. At the same time, it also results in an increased complexity in the calculation of loop integrals. Furthermore, the small electron mass typically gives rise to strong scale hierarchies complicating a numerically stable implementation of amplitudes.
\newline \indent
This thesis provides a pedagogical overview of the theoretical foundations of the \mcmule{} framework. Among other things, we show how the simple infrared structure in QED can be exploited to construct FKS$^\ell$, a subtraction scheme for soft singularities to all orders in perturbation theory. Furthermore, we present the method of massification as a solution to the problem of multi-scale integrals in the presence of large scale hierarchies. Finally, we introduce next-to-soft stabilisation as an elegant tool to stabilise the numerically delicate real-virtual contribution. To this end, we generalise the Low-Burnett-Kroll theorem for massive fermions to one loop. This allows for a straightforward application of the method without the need of explicit calculations.
\newline \indent
We have developed all of these techniques with fully differential NNLO QED calculations in mind and have successfully applied them to many processes such as the muon decay as well as Bhabha and M{\o}ller scattering. One of the main drivers of these developments has been the MUonE experiment requiring a high-precision theory prediction for muon-electron ($\mu$-$e$) scattering at the level of $10\, \text{ppm}$. The multi-scale nature of $\mu$-$e$ scattering makes this process particularly challenging from a technical point of view. Only the combined application of FKS$^\ell$, massification, and next-to-soft stabilisation makes the corresponding calculation possible. This thesis therefore presents for the first time the fully differential calculation of the complete set of NNLO corrections to $\mu$-$e$ scattering. This represents a major step towards the ambitious $10\, \text{ppm}$ target precision of the MUonE experiment.
}

\acknowledgement{
\textit{
I would like to start by thanking my supervisor Adrian Signer. I feel extremely fortunate for having been able to work with and learn from him during these past years. His door has always been open for any kind of problem and his advice helpful. He has been able to create a working environment that is both challenging and pleasant. His playful approach to physics has conveyed the perfect balance of ambition and fun. All of this has made this period particularly instructive and enjoyable for me.
\newline \indent
Next, I would like to thank Yannick Ulrich. It has been a privilege
to work with him so closely for the past years. His guidance and support has been incredibly useful during my master thesis and continued to be so during my PhD. His helpful attitude combined with a broad knowledge in physics as well as in computer science has been incredibly valuable. Without him the work presented here would not have been possible.
\newline \indent
I would like to express my appreciation also for all other members of the \mcmule{} team who I had the pleasure of working with during this time. In particular, I would like to thank Pulak Banerjee, Marco Rocco, Nicolas Schalch, Luca Naterop, Andrea Gurgone, Sophie Kollatzsch, Franziska Hagelstein, and Vladyslava Sharkovska.
\newline \indent
I would like to extend these compliments to the entire PSI theory group. It has been a great experience to be part of such a diverse and lively group of people. I am particularly thankful for numerous interesting lunch discussions covering a wide range of topics from jacobian peaks to ambergris. I am grateful to the two group leaders, Adrian Signer and Michael Spira, for making this possible. 
\newline \indent
Finally, I would like to thank the Bachelor and Master students that I had the pleasure of supervising. This includes David Urwyler, Guglielmo Coloretti, Nicolas Schalch, and Luca Naterop. I am grateful for this experience and honored for having been trusted with this responsibility.
}
}

\declaration{This thesis is based on the following works to which the author contributed to directly: \nocitemy{Engel:2018,Engel:2018fsb,Engel:2018fsb,Engel:2019nfw,Banerjee:2020tdt,Banerjee:2020rww,Banerjee:2021mty,Banerjee:2021qvi,Engel:2021ccn,Frixione:2022ofv} \bibliographymy{thesis}{} \bibliographystylemy{JHEP}
    \noindent
    Hence, text may be copied in verbatim without direct reference.
\nocite{Engel:2018,Engel:2018fsb,Engel:2018fsb,Engel:2019nfw,Banerjee:2020tdt,Banerjee:2020rww,Banerjee:2021mty,Banerjee:2021qvi,Engel:2021ccn,Frixione:2022ofv}
The work presented in the thesis has been supported by the Swiss National Science Foundation (SNF) under contract 178967 and 207386.
 }

\begin{document}

\maketitle

\chapter{Introduction}\label{chap:introduction}

The Standard Model (\ac{SM}) of particle physics has been verified to an astonishing precision in a variety of experiments that span a wide range of different observables and energies. Nevertheless, it is an inherently incomplete theory since it is unable to explain numerous fundamental physical phenomena such as gravity, dark matter, dark energy, and the matter-antimatter asymmetry. Furthermore, there are also a number of unresolved theoretical issues within the SM, two prominent examples of which are the hierarchy and the strong CP problems. As a consequence, the search for physics beyond the SM (\ac{BSM}) is the main quest of particle physics today.

Two complementary experimental approaches are pursued in this endeavour. For sufficiently high energies the direct production of unknown heavy particles is possible. The \textit{high-energy frontier} can thus provide clear and concrete evidence for BSM physics along the same lines as the discovery of the Higgs boson at the \ac{LHC}. This strategy, however, relies on a BSM energy scale that lies within reach of collider experiments. The \textit{high-intensity frontier}, on the other hand, aims at an indirect measurement of BSM physics via their quantum effects. This requires intense sources as well as ultra-sensitive investigations in order to reach sufficiently high precision. Such experiments often have the potential to investigate higher energies and weaker interactions than those directly accessible at high-energy particle colliders. However, if a BSM signal is observed, constraints on the precise form of new physics are more indirect.

Precision experiments with leptons are an essential part of the high-intensity program. Often these measurements rely on highly accurate SM calculations either for direct comparisons or background predictions. At low energies these processes are dominated by quantum electrodynamics (\ac{QED}).  As a consequence, high-precision QED calculations have become ever more important in recent years. This has triggered the development of \mcmule{}~\cite{Banerjee:2020rww}, a Monte Carlo integrator for processes with muons and other leptons. 

Higher-order calculations in QED are in many ways simpler than in quantum chromodynamics (\ac{QCD}). Its abelian nature results in significantly fewer Feynman diagrams and thus in simpler expressions from an algebraic point of view. The main difference lies, however, in the treatment of fermion masses. QCD observables are defined such that final-state collinear singularities cancel. The small quark masses can thus be neglected in the calculation. Precision QED observables, on the other hand, are typically not collinear finite in the massless limit and the corresponding mass logarithms are physical. Finite mass effects therefore have to be taken into account in QED calculations.

This results in a significant simplification of the infrared structure compared to QCD. At the same time, however, there are also serious challenges related to finite fermion masses. They result in an increased complexity in the calculation of loop integrals. Furthermore, the small electron mass typically gives rise to strong scale hierarchies complicating a numerically stable implementation of amplitudes as well as hampering the numerical phase-space integration. The \mcmule{} framework is based on the philosophy of exploiting the simplicities  of QED while at the same time solving the corresponding ubiquitous problems in a process independent way. This approach has allowed us to perform many fully differential higher-order QED calculations relevant for low-energy precision experiments. A list of processes currently implemented in \mcmule{} is shown in Table~\ref{tab:mcmule}.

The objective of this thesis is to provide an overview of these methods by example of muon-electron ($\mu$-$e$) scattering. This process is of high phenomenological relevance because of the MUonE experiment requiring a high-precision theory prediction at the level of $10\,\text{ppm}$. This has triggered a large theory effort~\cite{Banerjee:2020tdt} with the goal of developing two completely independent Monte Carlo generators that ensure this level of precision. In addition to its phenomenological relevance, the multi-scale nature of $\mu$-$e$ scattering is also interesting from a technical point of view. Promoting the corresponding calculation to the main theme of the thesis therefore allows us to cover many interesting aspects of fully differential QED calculations.

The following sections provide a gentle introduction to the topic. While it is assumed that the reader is familiar with quantum field theory (\ac{QFT}), some basic notions needed for later discussion are explained in Section~\ref{sec:pert_theory}. Section~\ref{sec:muone} presents the physics case of the MUonE experiment and explains why a high-precision calculation of $\mu$-$e$ scattering at the level of $10\,\text{ppm}$ is required. The following Section~\ref{sec:10ppm} then assesses the feasibility of this ambitious target precision and discusses recent developments in this endeavour. We conclude this introductory chapter with an overview of the thesis. A compilation of the conventions and notation used throughout the thesis can be found in Appendix~\ref{sec:notation}.

\begin{table}
\centering
\begin{tabular}{ |c|c| } 
 \hline
 $\ell\to \ell'\bar{\nu}\nu$ & NNLO + resumm.  \\ 
 $\ell\to \ell'\bar{\nu}\nu \gamma$ & NLO  \\ 
 $\ell\to \ell'\bar{\nu}\nu \ell' \ell'$ & NLO  \\ 
 $\ell \ell'\to \ell \ell'$ & NNLO  \\ 
 $\ell p\to \ell p$ & dominant NNLO  \\ 
 $\ell\ell\to \ell\ell$ & NNLO  \\ 
 $\ell\ell \to \gamma \gamma$ & NNLO  \\ 
 \hline
\end{tabular}
	\caption{Processes currently implemented in \mcmule. These include charged leptons ($\ell$/$\ell'$), neutrinos ($\nu$), protons ($p$), and photons $\gamma$. In the case of $\ell p\to \ell p$ only the dominant lepton line corrections are included.}
\label{tab:mcmule}
\end{table}

\section{Perturbation theory}\label{sec:pert_theory}

This section introduces some basic notions of QFT perturbation theory. Even though this is done using QED, most concepts are also relevant for QCD where fermions are typically considered to be massless. An emphasis is therefore put on the ramifications connected to the different treatment of fermion masses. 

The QED Lagrangian is given by
\begin{align}\label{eq:qed_lagrangian}
	\mathcal{L}_\text{QED}
	=\sum_{i=e,\mu,\tau} \bar{\psi}_i(i\slashed{\partial}-m_i)\psi_i
	-\frac{1}{4}F_{\mu\nu}F^{\mu\nu}
	- e \sum_{i=e,\mu,\tau} \bar{\psi}_i\slashed{A}\psi_i
\end{align}
where $\psi_i$ are the spinor fields of the leptons, $A^\mu$ the photon field, $F^{\mu\nu}$ the electromagnetic field tensor, $m_i$ the lepton masses, and $e$ the electromagnetic coupling. We have used the standard Feynman slash notation $\slashed{a}\equiv \gamma_\mu a^\mu$. At low energies the quark fields become non-perturbative and are therefore left out in~\eqref{eq:qed_lagrangian}. The treatment of these non-perturbative contributions is closely related to the MUonE experiment and is therefore a recurring topic in the thesis. In what follows we only consider electrons and muons as external states. We thus use the simplified notation  
\begin{align}
	m_e \to m, \quad\quad m_\mu \to M
\end{align}
for the corresponding masses.

The calculation of a $2\to n$ cross section consists of two parts, the computation of the scattering amplitude $\mathcal{A}_n$ based on the Lagrangian $\mathcal{L}_\text{QED}$ and its subsequent integration over the $n$-particle phase space $\D\Phi_n$
\begin{align}\label{eq:xsection_unphysical}
	\sigma_n 
	= \int\D\Phi_n\, |\mathcal{A}_n|^2  \Theta_\text{cuts} \,.
\end{align}
The measurement function $\Theta_\text{cuts}$ has two roles. It enters the binning procedure for histograms in the calculation of fully differential observables. Furthermore, it can be used to approximate the detector geometry of the experiment. As a consequence, it is important to be able to modify this function independently from the rest of the calculation. A numerical approach for the phase-space integration in~\eqref{eq:xsection_unphysical} is therefore clearly advantageous.

It is not possible to compute $\mathcal{A}_n$ exactly due to the photon-fermion interaction term in~\eqref{eq:qed_lagrangian}. However, the coupling $e$ is sufficiently small to allow for a perturbative expansion
\begin{align}\label{eq:amp_pertexp}
	\mathcal{A}_n
	= \mathcal{A}_n^{(0)} + e^2 \mathcal{A}_n^{(1)} + e^4 \mathcal{A}_n^{(2)} + \mathcal{O}(e^6)
\end{align}
with $\mathcal{A}_n^{(0)}$ the leading order (\ac{LO}), $\mathcal{A}_n^{(1)}$ the next-to-leading order (\ac{NLO}), and $\mathcal{A}_n^{(2)}$ the next-to-next-to-leading order (\ac{NNLO}) contribution. To make the power counting more transparent, all couplings appearing already at LO are absorbed in $\mathcal{A}_n^{(l)}$. Each term in this expansion permits a representation in terms of Feynman diagrams. For $\mu$-$e$ scattering, for example, we have
\begin{align}\label{eq:pert_muone}
	\mathcal{A}_2
	=
	\begin{tikzpicture}[scale=0.3,baseline={(0,.3)}]
	    
    \draw[line width=.3mm]  (-1.7,.1) -- (1.7,.1);
    \draw[line width=.3mm]  (-1.7,-.1) -- (1.7,-.1);
    \draw[line width=0.15mm, tightphoton] (0,2.5)--(0,.1);
    \draw[line width=.3mm]  (-1.7,2.5) -- (1.7,2.5);

    %\draw[line width=.3mm]  [charge,fill=charge] (0,2.5) circle (0.1);
    %\draw[line width=.3mm]  [charge,fill=charge] (0,0) circle (0.1);

	\end{tikzpicture}
	+ e^{\color{charge}2}\,
	\Bigg(\quad
	\begin{tikzpicture}[scale=0.3,baseline={(0,.3)}]
	\input{tikz/oneloop_pert}
	\end{tikzpicture}
	+ …
	\quad \Bigg)
	+ e^{\color{charge}4}\,
	\Bigg(\quad
	\begin{tikzpicture}[scale=0.3,baseline={(0,.3)}]
	\input{tikz/twoloop_pert}
	\end{tikzpicture}
	+ …
	\quad \Bigg)
	+ \mathcal{O}(e^{\color{charge}6})
\end{align}
where each additional vertex contributes one power of the coupling. The fact that the muon is much heavier than the electron is emphasised with a double line.

Higher-order corrections involve loops of virtual particles. The corresponding mathematical expression requires the integration over the unconstrained loop momenta. It is the calculation of these loop integrals that represents one of the main bottlenecks in the calculation of the amplitudes. Both analytic and numerical methods are used to calculate these integrals. For phenomenological applications an analytic approach is often advantageous since the fast evaluation of the amplitude facilitates the phase-space integration.\footnote{To avoid this problem, numerical computations typically rely on an interpolation grid. In this case, it is, however, challenging to rigorously quantify the error introduced by the interpolation.} However, for multi-loop multi-scale integrals one often has to resort to numerical methods because even state-of-the-art analytic loop integral techniques are insufficient. On the other hand, if these scales follow a strong hierarchical structure, an alternative approach is the expansion in small parameters. This reduces the number of active scales and therefore facilitates the analytic evaluation of the integrals. This is the strategy employed throughout this thesis.

The evaluation of loop integrals is complicated by the occurrence of divergences originating in the ultraviolet (\ac{UV}) and the infrared (\ac{IR}) region of the loop momenta. Both types of singularities can be efficiently regularised with \textit{dimensional regularisation} which shifts the number of space-time dimensions to $d=4-2\epsilon$. In particular, this preserves all symmetries of the QED Lagrangian~\eqref{eq:qed_lagrangian}.\footnote{This is not the case for Lagrangians that include $\gamma_5$.} The UV singularities can then be consistently absorbed via \textit{renormalisation}, a redefinition of the parameters $\psi_i$, $A_\mu$, $m_i$, and $e$.\footnote{The QED renormalisation constants up to two loops and in all relevant schemes of dimensional regularisation can be found in Appendix~B of \cite{Ulrich:2020frs}.} However, even after UV renormalisation the loop integrals and therefore also the cross section $\sigma_n$ are still singular in the IR region. The observable as defined in~\eqref{eq:xsection_unphysical} is in fact unphysical.

This is explained by the observation that the final state of any scattering process is only meaningfully defined up to the emission of any number of soft photons with vanishing momentum $k$. Since any physically realisable detector has a finite energy resolution, photons with an energy below this threshold, i.e. $E_k < \Delta$, will escape the measurement. Thus, only the combination
\begin{equation}\label{eq:xsection_phys}
	\sigma = \sigma_n + \sigma_{n+1} + …
\end{equation}
with
\begin{equation}\label{eq:xsection_real}
	\sigma_{n+1} = \int_{E_k < \Delta} \d\Phi_{n+1} |\mathcal{A}_{n+1}|^2 \Theta_\text{cuts}
\end{equation}
gives an observable quantity. Even though an infinite number of soft photon emissions contributes in~\eqref{eq:xsection_phys}, only finitely many have to be taken into account at a fixed order in perturbation theory. At NLO, for example, only $\sigma_{n+1}$ has to be added to the virtual one-loop correction. In the case of $\mu$-$e$ scattering, we have the perturbative expansion
\begin{align}\label{eq:pert_muone_radiative}
	\mathcal{A}_3
	=
	e^{\color{charge}1}\,
	\Bigg(\quad
	\begin{tikzpicture}[scale=0.3,baseline={(0,.3)}]
	\input{tikz/born_rad_pert}
	\end{tikzpicture}
	+ …
	\quad \Bigg)
	+ e^{\color{charge}3}\,
	\Bigg(\quad
	\begin{tikzpicture}[scale=0.3,baseline={(0,.3)}]
	\input{tikz/oneloop_rad_pert}
	\end{tikzpicture}
	+ …
	\quad \Bigg)
	+ \mathcal{O}(e^{\color{charge}5})
\end{align}
in analogy to~\eqref{eq:pert_muone}. Here, we display all additional couplings relative to the non-radiative tree-level amplitude $\mathcal{A}_2^{(0)}$. The powers of $e$ thus match up with the ones from~\eqref{eq:pert_muone} at the level of the squared amplitude.

Radiative amplitudes have a singular dependence on the energy of the emitted photon. For instance, we have
\begin{equation}\label{eq:matel_real}
	\mathcal{A}_{n+1}
	\sim \frac{1}{p_i \cdot k}
	= \frac{1}{E_i E_k(1-\beta_i  \cos \sphericalangle(p_i,k))}
\end{equation}
where $\beta_i=(1-m_i^2/E_i^2)^{1/2}$ is the velocity of the emitting fermion with momentum $p_i$, energy $E_i$, and mass $m_i$. As a consequence, the amplitude is singular in the soft region. Due to the corresponding double pole in the squared amplitude, the combination with $\d\Phi_{n+1}\sim E_k$ in the complete integrand~\eqref{eq:xsection_real} still has a non-integrable $E_k^{-1}$ singularity giving rise to a logarithmically divergent integral.  These soft singularities then cancel the ones from the virtual corrections rendering~\eqref{eq:xsection_phys} finite. This is the powerful statement of the \textit{Bloch-Nordsieck theorem}~\cite{Bloch:1937pw}. In this context, it is important that the measurement function $\Theta_\text{cuts}$ in \eqref{eq:xsection_unphysical} and \eqref{eq:xsection_real} does not spoil this cancellation. In particular, the measurement function is only \textit{infrared safe} if its value does not change under infinitely soft emission~\cite{Kunszt:1992tn}
\begin{align}\label{eq:soft_safety}
	\Theta_\text{cuts}(p_1,…p_n,k=0)
	= \Theta_\text{cuts}(p_1,…,p_n)\, .
\end{align}

Despite this IR cancellation, the occurrence of soft singularities in intermediate contributions has important implications. Due to the $E_k^{-1}$ behaviour of the integrand in~\eqref{eq:xsection_real}, each soft photon emission introduces logarithms of the form $L_\Delta=\log(\Delta^2/S)$ with $S$ the energy scale of the considered process. For $\Delta^2 \ll S$ these logarithms become large and can hamper the reliability of the perturbative expansion. Furthermore, the cancellation is non-trivial to obtain in practice since the phase-space integration is typically performed numerically. As a consequence, dedicated techniques such as subtraction methods have to be developed to cope with divergences in the numerical integration.

In the case of massless fermions, we have $\beta_i=1$ and the amplitude~\eqref{eq:matel_real} develops an additional singularity for collinear photon emission where $\sphericalangle(p_i,k)=0$. Similar to the soft photon case, this is related to the kinematic indistinguishability of a single massless fermion and a collinear fermion-photon pair (\textit{jet}). Both objects have the same invariant mass $p_i^2=(p_i+k)^2=0$. As a consequence, the real-emission contribution~\eqref{eq:xsection_real} has to include also the phase-space region where $\sphericalangle(p_i,k)$ is below a certain threshold. The final-state collinear singularities then cancel in the sum of virtual and real contributions according to the \textit{\ac{KLN} theorem}~\cite{Kinoshita:1962ur,Lee:1964is}. This is completely analogous to the soft photon case. The infrared safety condition \eqref{eq:soft_safety} now has to be modified to also include collinear splittings
\begin{align}\label{eq:coll_safety}
	\Theta_\text{cuts}(p_1,…(1-\kappa)p_i,…p_n,k=\kappa p_i)
	= \Theta_\text{cuts}(p_1,…,p_i,…,p_n)
\end{align}
with $0\leq\kappa<1$. Initial-state collinear divergences, on the other hand, are different in nature. They remain present in the cross section and need to be absorbed in a redefinition of the initial particle states in terms of so-called \textit{structure functions}.

Furthermore,  also the collinear splitting of a photon into a massless fermion-antifermion pair, $k \to p_i + p_j$, gives an infrared divergence. This has important ramifications and is a crucial difference between massive and massless fermions. The additional collinear divergence can again be understood from the indistinguishability relation $k^2=(p_i+p_j)^2=0$. A physical observable in the $m_i=0$ case thus has to be defined such that these additional final states are included via the definition of a photon jet. In the massive case, on the other hand, the production of an additional fermion pair is a physically distinguishable process. In this case, it does not have to be included in order to obtain IR finite results.

In reality, all fermions in the SM have non-zero masses. In particular in QED, finite lepton masses have to be taken into account because their effects are measured in experiments. Nevertheless, even at the high-intensity frontier, the electron mass is small compared to the energy scale $S$ of the experiments. The fermion velocity satisfies $\beta_i \sim 1-m^2/S \lesssim 1$ in this case and the amplitude in~\eqref{eq:matel_real} exhibits a collinear pseudo-singularity (\ac{CPS}) for $\sphericalangle(p_i,k)=0$. At the cross-section level this introduces additional large logarithms of the form $L_m=\log(m^2/S)$ that have to be included in the power counting of the perturbative expansion. Taking into account also the soft logarithms as well as the $1/(4\pi)^2$ from the loop measure we have
\begin{align}
	\sigma = \sigma^{(0)}+\sigma^{(1)}+\sigma^{(2)}+...
\end{align}
where
\begin{align}\label{eq:pertexp_pc}
	\sigma^{(n)}
	= \Big(\frac{\alpha}{4\pi}\Big)^n 
	\sum_{n_1,n_2=0}^{n} c_{n_1,n_2} L_\Delta^{n_1} L_m^{n_2}
\end{align}
with $L_\Delta = \log(\Delta^2/S)$, $L_m=\log(m^2/S)$, and $\alpha=e^2/(4\pi)$. Since now all the logarithmically enhanced terms are given separately, it can be naively assumed that $c_{n_1,n_2}\sim 1$. The above power counting can thus be used as a rough estimate of the $n$-th order contribution to the cross section. 

For sufficiently large scale hierarchies $m^2, \Delta^2 \ll S$ the logarithmic enhancement in~\eqref{eq:pertexp_pc} results in a breakdown of the naive perturbative expansion. In this case, the large logarithms have to be resummed. The leading logarithmic (\ac{LL}) contribution $n_1=n_2=n$ can be calculated by a \textit{parton shower}. Some effects at next-to-leading-order logarithmic (\ac{NLL}) accuracy, where $n_1+n_2=2n-1$, can also be captured in this way. The feasibility of a strict resummation of these logarithms, however, depends on the precise definition of the observable and has to be performed case by case. In QED, a strict resummation is often not required since the fine structure constant $\alpha \sim 1/137$ is sufficiently small. Instead, the large logarithms only have to be taken into account up to a certain order in $\alpha$.

\section{The MUonE experiment and the Muon $g-2$}\label{sec:muone}

\begin{figure}
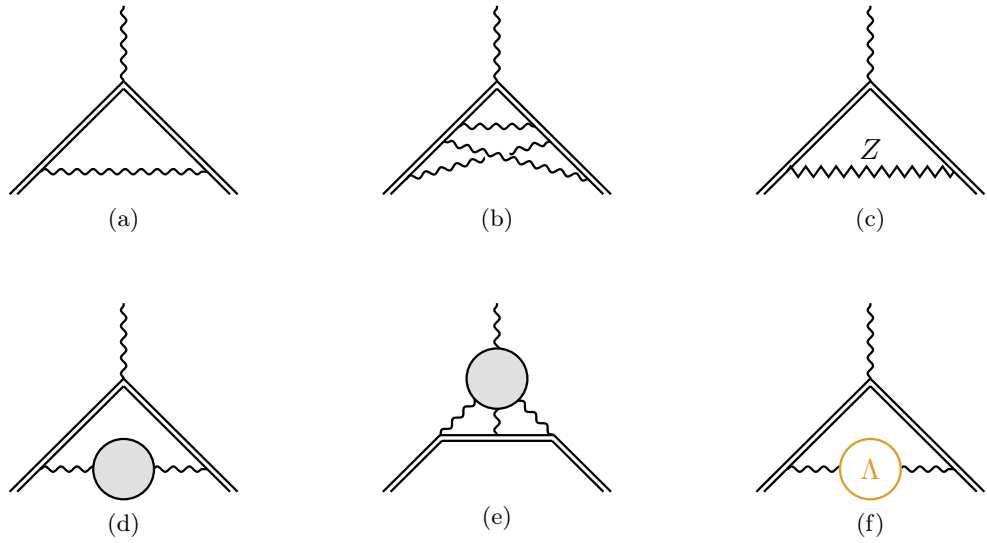

    \centering
    \subfloat[]{
        \begin{tikzpicture}[scale=1,baseline={(0,0)}]
        \input{tikz/g2_lo}
    \end{tikzpicture}
    \label{fig:g2_lo}
    }
    \hspace{1.5cm}
    \subfloat[]{
        \begin{tikzpicture}[scale=1,baseline={(0,0)}]
        \input{tikz/g2_qed}
    \end{tikzpicture}
    \label{fig:g2_qed}
    }
    \hspace{1.5cm}
    \subfloat[]{
        \begin{tikzpicture}[scale=1,baseline={(0,0)}]
        \input{tikz/g2_ew}
    \end{tikzpicture}
    \label{fig:g2_ew}
    } \\
    \vspace{.5cm}
    \subfloat[]{
        \begin{tikzpicture}[scale=1,baseline={(0,0)}]
        \input{tikz/g2_hvp}
    \end{tikzpicture}
    \label{fig:g2_hvp}
    }
    \hspace{1.5cm}
    \subfloat[]{
        \begin{tikzpicture}[scale=1,baseline={(0,0)}]
        \input{tikz/g2_hlbl}
    \end{tikzpicture}
    \label{fig:g2_hlbl}
    }
    \hspace{1.5cm}
    \subfloat[]{
        \begin{tikzpicture}[scale=1,baseline={(0,0)}]
        \input{tikz/g2_bsm}
    \end{tikzpicture}
    \label{fig:g2_bsm}
    }
\caption{Figures (a)-(e) show various examples for SM contributions to the anomalous magnetic moment of the muon, while possible BSM effects are illustrated in Figure (f). The LO SM diagram is shown in (a). A three-loop QED diagram and a leading EW contribution are given in (b) and (c), respectively. Figure (d) depicts the leading non-perturbative HVP diagram while the HLbL contribution is shown in (e). The grey blobs represent loops of hadrons.}
\label{fig:g2_diags}
\end{figure}

A prime observable for high-intensity BSM searches is the anomalous magnetic moment of the muon $a_\mu = (g-2)_\mu$. This quantity vanishes at tree level in the SM since the $g$-factor is exactly $2$. The leading contribution is thus given by the one-loop diagram of Figure~\ref{fig:g2_lo}. A selection of additional contributing SM diagrams is depicted in Figures~\ref{fig:g2_qed}-\ref{fig:g2_hlbl}. The muon life time is sufficiently long as to allow for a high-statistics test of these effects. Moreover, possible BSM contributions, such as the one illustrated in Figure~\ref{fig:g2_bsm}, enter as
\begin{align}
	\frac{\delta a_\mu^\text{BSM}}{a_\mu}
	\sim \frac{M^2}{\Lambda^2}
\end{align}
with $\Lambda$ the scale of BSM physics. Compared to the anomalous magnetic moment of the electron, $a_e$, the sensitivity to BSM effects is therefore significantly boosted for $a_\mu$ due to the much larger muon mass $M \sim 200 m$. Taking into account both the BNL E821~\cite{Muong-2:2006rrc} and the recent Fermilab measurement~\cite{Muong-2:2021ojo}, the current experimental average is
\begin{align}
	a_\mu^\text{exp} = 116592061(41)\times 10^{-11}\, .
\end{align}
This result deviates from the SM theory prediction~\cite{Aoyama:2020ynm}
\begin{align}\label{eq:g2_sm}
	a_\mu^\text{SM} = 116591810(43)\times 10^{-11}
\end{align}
by $4.2\sigma$. This anomaly observed in the anomalous magnetic moment of the muon is one of the most intriguing hints for BSM physics today. It is therefore of utmost importance both to ensure the validity of these values and to further improve their precision. On the experimental side, the error will soon be reduced by up to a factor 4 due to the ongoing experiment at Fermilab. Furthermore, the planned experiment at J-PARC~\cite{Saito:2012zz} will serve as a completely independent validation of these results. Significant progress on the theoretical side is therefore necessary to match the future experimental precision.

\begin{figure}
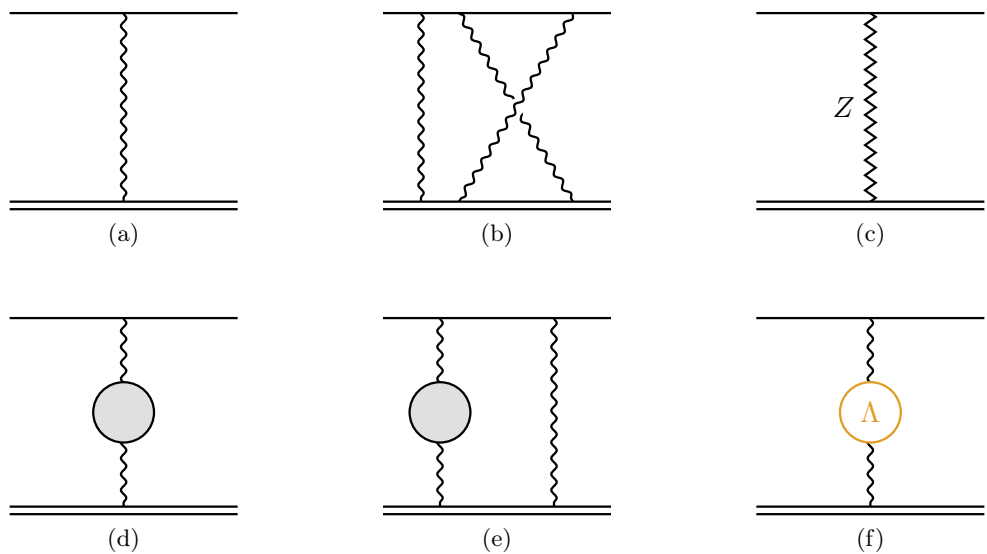

    \centering
    \subfloat[]{
        \begin{tikzpicture}[scale=1,baseline={(0,0)}]
        \input{tikz/muone_tree}
    \end{tikzpicture}
    \label{fig:muone_tree}
    }
    \hspace{1.5cm}
    \subfloat[]{
        \begin{tikzpicture}[scale=1,baseline={(0,0)}]
        \input{tikz/muone_qed}
    \end{tikzpicture}
    \label{fig:muone_qed}
    }
    \hspace{1.5cm}
    \subfloat[]{
        \begin{tikzpicture}[scale=1,baseline={(0,0)}]
        \input{tikz/muone_ew}
    \end{tikzpicture}
    \label{fig:muone_ew}
    } \\
    \vspace{.5cm}
    \subfloat[]{
        \begin{tikzpicture}[scale=1,baseline={(0,0)}]
        \input{tikz/muone_hvp}
    \end{tikzpicture}
    \label{fig:muone_hvp}
    }
    \hspace{1.5cm}
    \subfloat[]{
        \begin{tikzpicture}[scale=1,baseline={(0,0)}]
        \input{tikz/muone_hvp_nnlo}
    \end{tikzpicture}
    \label{fig:muone_hvp_nnlo}
    }
    \hspace{1.5cm}
    \subfloat[]{
        \begin{tikzpicture}[scale=1,baseline={(0,0)}]
        \input{tikz/muone_bsm}
    \end{tikzpicture}
    \label{fig:muone_bsm}
    }
\caption{Various SM contributions to $\mu$-$e$ scattering are shown in (a)-(e) and possible BSM contaminations are illustrated in (f). The tree-level SM diagram is shown in (a). A two-loop QED diagram and the leading EW contribution are given in (b) and (c), respectively. Figures~(d) and (e) depict leading and subleading hadronic contributions. The grey blobs represent loops of hadrons. Contrary to the anomalous magnetic moment, these corrections are entirely due to HVP with HLbL scattering only appearing at subsubleading order.
}
\label{fig:muone_diags}
\end{figure}

The SM prediction obtains contributions from pure QED as well as electroweak (\ac{EW}) and hadronic corrections and can thus be written as
\begin{align}\label{eq:g2_split}
	a_\mu^\text{SM} = a_\mu^\text{QED}+a_\mu^\text{EW}+a_\mu^\text{had}\, .
\end{align}
Mixed QED-EW and QED-hadronic corrections are included in $a_\mu^\text{EW}$ and $a_\mu^\text{had}$, respectively. 
An example of a three-loop QED and a LO EW contribution is shown in Figures~\ref{fig:g2_qed} and \ref{fig:g2_ew}, respectively. The number given in~\eqref{eq:g2_sm} includes QED corrections up to five loops~\cite{Aoyama:2012wk,Aoyama:2019ryr} and EW contributions up to two loops~\cite{Czarnecki:2002nt,Gnendiger:2013pva}. The theory error for $a_\mu^\text{QED}$ and $a_\mu^\text{EW}$ is therefore under good control. While these perturbative calculations are highly challenging and the achieved precision astonishing, the theory prediction for $a_\mu^\text{had}$ is more delicate. This is due to the low-energy nature of the observable where the non-perturbative regime of QCD becomes relevant. For an ab initio determination of the hadronic contribution one therefore has to resort to lattice QCD simulations~\cite{Aoki:2021kgd}. Alternatively, one can use a data-driven approach that relies on experimental input to capture the non-perturbative effects.

Figure~\ref{fig:g2_hvp} shows the leading hadronic contribution to $a_\mu$ which is due to the hadronic vacuum polarisation (\ac{HVP})
\begin{align}\label{eq:hvp_def}
	\begin{tikzpicture}[scale=1,baseline={(0,-.1)}]

  \draw[line width=.3mm,photon] (-1.1,0) -- (1.1,0);
  \draw[line width=.3mm,fill=stuff] (0,0) circle (0.4);
  \draw (-.85,.3) node[] {$q$};

	\end{tikzpicture}
	=i \Pi_\text{had}(q^2)(g^{\mu\nu}q^2-q^\mu q^\nu).
\end{align}
At subleading order, a new topology arises given by the hadronic light-by-light (\ac{HLbL}) diagram depicted in Figure~\ref{fig:g2_hlbl}. In the case of HLbL, data-driven and lattice QCD calculations are in perfect agreement averaging to~\cite{Aoyama:2020ynm}
\begin{align}\label{eq:g2_hlbl}
	a_\mu^\text{HLbL} = 92(19)\times10^{-11}\, .
\end{align}
Since a significantly higher relative precision is needed for the larger HVP contribution, most lattice predictions are not yet competitive with the data-driven calculations. They are therefore not included in the world average~\cite{Davier:2017zfy,Keshavarzi:2018mgv,Colangelo:2018mtw,Hoferichter:2019mqg,Davier:2019can,Keshavarzi:2019abf}
\begin{align}\label{eq:g2_hvp}
	a_\mu^\text{HVP,LO} = 6931(40)\times10^{-11}\, .
\end{align}
There is, however, one recent lattice simulation that has been able to reach a competitive precision for this contribution~\cite{Borsanyi:2020mff}. Interestingly, its prediction significantly deviates from the above data-driven value. A resolution of this discrepancy is therefore of high priority in order to arrive at a completely robust theoretical prediction for the HVP contribution. This is particularly pressing since the HVP prediction~\eqref{eq:g2_hvp} dominates the theory error in~\eqref{eq:g2_sm}.

The data-driven calculations that enter in~\eqref{eq:g2_hvp} follow a dispersive approach. Based on unitarity (optical theorem) and causality (analyticity) of the SM, the HVP function, $\Pi_\text{had}$, can be related by means of the dispersion relation~\cite{Jegerlehner:2017gek}
\begin{align}\label{eq:hvp_dispersive}
	\frac{\Pi_\text{had}(q^2)}{q^2} =
	 \frac{\alpha}{3 \pi} \int_{4m_\pi^2}^{\infty}
	\frac{\D z}{z} \frac{R_\gamma^\text{had}(z)}{q^2-z+i \delta}
\end{align}
to the hadronic $R$-ratio
\begin{align}\label{eq:rratio_had}
	R_\gamma^\text{had}(s) = \sigma(e^+e^- \to \gamma^* \to \text{had})\Big/ \frac{4\pi\alpha^2}{3s}
\end{align}
with $s$ the centre-of-mass energy of the scattering process, $m_\pi$ the pion mass, and $\delta>0$. This formula differs from the one in~\cite{Jegerlehner:2017gek} by a sign due to the different sign convention in the definition of $\Pi_\text{had}$. The precise definition used in this thesis is given in Appendix~\ref{sec:vacuum_polarisation}. The normalisation factor in~\eqref{eq:rratio_had}  corresponds to the tree-level cross section for $\sigma(e^+e^- \to \gamma^* \to \mu^+\mu^-)$ in the high-energy limit $s\gg M^2$. Based on~\eqref{eq:hvp_dispersive} it is possible to determine the HVP contribution to $a_\mu$ from the experimental measurement of hadron production.

\begin{figure}
    \centering
    \includegraphics[scale=1]{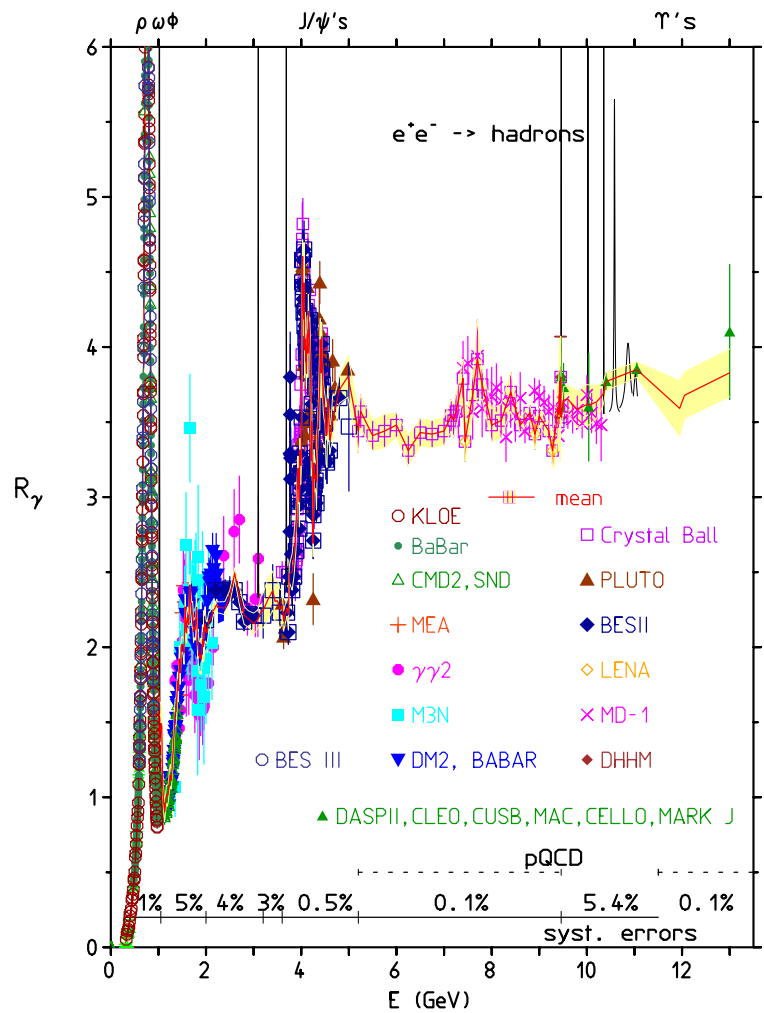}
\caption{This figure is taken from~\cite{refId0} and illustrates the hadronic resonances in the $R$-ratio that enter the time-like dispersive integral~\eqref{eq:hvp_dispersive}. Many dedicated experiments are needed that operate at different production thresholds to cover these non-perturbative regions.}
\label{fig:rratio}
\end{figure}

There are, however, a number of problems related to this approach. As already mentioned, it currently contributes the largest error to the SM prediction for $a_\mu$. At the same time, the achieved precision is limited by the available data. Further improvements therefore crucially depend on more precise measurements of the $R$-ratio as well as a better understanding of the corresponding background based on theory predictions. A large effort is therefore underway to improve upon this situation~\cite{Abbiendi:2022liz}. On a more fundamental note, however, the $R$-ratio is a challenging object to accurately measure to begin with. This is due to narrow hadronic resonances that contribute to the dispersion integral~\eqref{eq:hvp_dispersive} and that are non-perturbative in nature. Multiple experiments are therefore required that operate at different production thresholds. This situation is illustrated in Figure~\ref{fig:rratio}. It is therefore a challenging task to ensure that all systematic uncertainties are properly accounted for when aggregating all of these results. In light of the aforementioned discrepancy between the recent lattice simulation and dispersive results, an additional independent determination of the HVP correction is therefore highly desirable.

This is the main motivation of the MUonE experiment~\cite{Abbiendi:2016xup}, a recent proposal to perform a very precise measurement of $\mu$-$e$ scattering. A comparison of the experimental data with perturbative calculations can be used to extract the HVP as depicted in Figure~\ref{fig:muone_hvp}. Contrary to the $R$-ratio in the dispersive approach, this measures the HVP function in the space-like region ($q^2<0$) where it is \textit{smooth and free of hadronic resonances}. Following~\cite{CarloniCalame:2015obs} this measurement can then directly be used to obtain a completely independent prediction for $a_\mu^\text{HVP}$ given by the integral
\begin{align}
	a_\mu^\text{HVP} 
	= \frac{\alpha}{\pi} \int_0^1 \D x (1-x) \Delta \alpha_\text{had}\big(t(x)\big)
\end{align}
with the hadronic contribution to the running of the effective fine-structure constant
\begin{align}
	\Delta \alpha_\text{had}\big(q^2\big)
	= \frac{\alpha}{1-\text{Re}\, \Pi_\text{had}(q^2)}
\end{align}
and
\begin{align}
	t(x) = \frac{x^2 M^2}{x-1} < 0 \, .
\end{align}
This, in turn, would yield valuable information to resolve the aforementioned discrepancy between data-driven and lattice results and would contribute to the further improvement of the theory precision.

The MUonE approach suffers from the disadvantage of measuring a subleading effect. The contribution of the leading HVP changes the differential cross section by up to $\mathcal{O}(10^{-3})$, depending on the scattering angle of the outgoing electron. In order to match the error of the current value~\eqref{eq:g2_hvp}, the HVP needs to be extracted from $\mu$-$e$ data with a precision below one percent. Hence, the accuracy of the total experimental and theoretical error should not exceed $10\,\text{ppm}$. This is only feasible since the most important experimental systematic effects can be measured in the so-called \textit{normalisation region}, where hadronic effects are negligible~\cite{Abbiendi:2022oks}. Nevertheless, this level of precision still represents a daunting task both on the experimental and the theoretical side. Figure~\ref{fig:muone_diags} shows a selection of Feynman diagrams that contribute at this level of precision. The precise theory requirement for the SM prediction is investigated in the following Section~\ref{sec:10ppm}.

A contamination of the signal with BSM physics, as illustrated in Figure~\ref{fig:muone_bsm}, can a priori not be excluded. However, the dedicated studies performed in \cite{Masiero:2020vxk,Dev:2020drf} have shown that these effects are expected to lie below MUonE's sensitivity. A crucial point in these analyses is the aforementioned normalisation of the cross section using the low-signal region to cancel larger BSM effects. This explains the different conclusion reached in~\cite{Schubert:2019nwm} where this cancellation was not taken into account. While this additional study is therefore not particularly relevant regarding the determination of the HVP, it raises the question of MUonE's potential to search for BSM physics. This idea was investigated in~\cite{diCortona:2022zjy,Galon:2022xcl} with the conclusion that new parameter space for light new physics can indeed be explored.

The proposal of MUonE is to scatter the $150\, \mathrm{GeV}$ muon beam currently available at the CERN North Area on a Beryllium fixed target. The scattering angles of the electron and the muon, $\theta_e$ and $\theta_\mu$, are measured very precisely, but no further kinematic information is assumed to be available. From an idealised point of view we thus consider
\begin{align}\label{eq:muone_process}
	e^-(p_1)\mu^{\pm}(p_2) \to e^-(p_3)\mu^{\pm}(p_4) + X(p_5)
\end{align}
where the initial-state electron is at rest and $X$ stands for any further radiation. The incoming muon beam consists of either positively or negatively charged muons and is about 80\% polarised. With the energy of the incoming muon set to $E_2=150\, \mathrm{GeV}$, the centre-of-mass energy is fixed as $s=m^2+M^2+2mE_2\approx(400\,\mathrm{MeV})^2$. The HVP, $\Pi_\text{had}(q^2)$, is then extracted from the differential cross section measured as a function of the squared momentum transfer in the space-like domain $q^2<0$. This is, however, only a well-defined quantity in the case of elastic events where $p_5=0$ and thus $q^2=t=(p_1-p_3)^2=(p_2-p_4)^2$. It is therefore important to define the observable such that inelastic events are rejected. This can be done, for example, by using the fact that the electron and muon scattering angles are not independent in the elastic case. While in the centre-of-mass system (\ac{CMS}) we trivially have ${\theta_\mu^\text{el}}^*=\pi-{\theta_e^\text{el}}^*$, in the laboratory (\ac{LAB}) frame the two angles are correlated by the elasticity condition~\cite{Banerjee:2020tdt}
\begin{align}\label{eq:elasticity_band}
	\tan \theta_\mu^\text{el}
	= \frac{2 \tan \theta_e^\text{el}}{(1+\gamma^2\tan^2\theta_e^\text{el})(1+g_\mu^*)-2}
\end{align}
where
\begin{align}
	g^*_\mu = \frac{E_2 m + M^2}{E_2 m + m^2}, \quad\quad \gamma=\frac{E_2+m}{\sqrt{s}}\, .
\end{align}
Inelastic contributions can thus be suppressed by rejecting events that do not lie in the vicinity of this elasticity curve. 

Since the suppression is experimentally limited by the finite angular resolution of the detector, inelastic contributions still need to be studied as a background. Without any elasticity requirement, the following processes are kinematically allowed: photon radiation, the emission of a neutral pion ($X=\pi^0$), and the production of an electron, muon, and pion pair ($X=e^+e^-,\mu^+\mu^-,\pi^+\pi^-,\pi^0\pi^0$). It was shown in~\cite{Budassi:2022kqs} that when realistic elastic event selections are taken into account the contribution from $X=\pi^0$ is well below $10^{-5}$ and that pion pair production is kinematically forbidden. Furthermore, in~\cite{Budassi:2021twh} also the effect due to $X=\mu^+\mu^-$ was calculated to be negligible due to the tiny allowed phase space. The production of an electron pair, on the other hand, was shown to be highly relevant due to the presence of the large logarithms $L_m$. The corresponding process thus needs to be incorporated in any Monte Carlo generator for MUonE. This is obviously also true for photon radiation that obtain additional soft enhancements due to the elasticity cut restricting hard emission.

\section{Muon-electron scattering at 10 ppm}\label{sec:10ppm}

This section investigates the theory requirements needed for MUonE's $10\,\text{ppm}$ goal and presents an overview of the results that have already been obtained in this endeavour. Analogously to the anomalous magnetic moment, the SM theory prediction for elastic $\mu$-$e$ scattering can be split into the contributions
\begin{align}\label{eq:muone_split}
	\sigma = \sigma_\text{QED} + \sigma_\text{EW} + \sigma_\text{had}\, .
\end{align}

The LO HVP contribution, contributing to $\sigma_\text{had}$, is shown in Figure~\ref{fig:muone_hvp} and corresponds to the signal of the experiment. It is a crucial property of $\mu$-$e$ scattering, in this context, that also at subleading order all hadronic corrections are due to HVP. The more difficult HLbL diagrams only occur at subsubleading order and are therefore below the sensitivity of the MUonE experiment. The subleading HVP corrections, such as the one shown in Figure~\ref{fig:muone_hvp_nnlo}, have been calculated in~\cite{Fael:2019nsf} with a dispersive approach based on~\eqref{eq:hvp_dispersive}. This prediction depends on time-like data which spoils the independence of MUonE's HVP measurement. To avoid this problem, it was shown in~\cite{Fael:2018dmz} that the hyperspherical method~\cite{Levine:1974xh,Levine:1975jz} can be used instead which only integrates over space-like data. This allows for an iterative fit where in a first step only the LO HVP is extracted with the subleading hadronic corrections switched off. Next, the corresponding data can be used as an input to the hyperspherical method to predict these missing contributions. 

The EW corrections in~\eqref{eq:muone_split} are suppressed by 
\begin{align}
	\frac{\sigma_\text{EW}}{\sigma_\text{QED}}
	\sim \Big(\frac{s}{M_Z^2}\Big)^2 \sim 10^{-5}
\end{align}
with $M_Z$ the mass of the $Z$ boson. For the purpose of the MUonE experiment, it is therefore sufficient to only include the tree-level $Z$-exchange. This statement is corroborated by the dedicated studies of the NLO EW corrections performed in~\cite{Alacevich:2018vez}.

The main challenge in reaching the $10\,\text{ppm}$ precision goal is therefore the evaluation of the perturbative expansion of $\sigma_\text{QED}$ to a sufficiently high order. To estimate the size of the $n$-th order correction we can rely on the power counting given in~\eqref{eq:pertexp_pc}. We identify the scale $S$ with the CMS energy $s$. As mentioned above, the soft logarithms originate from the elasticity cut that restricts hard photon radiation above $\Delta$. 
In order to get a rough estimate for $\Delta$ we determine the energy resolution of the scattered electron from the expected angular resolution of $0.02\, \text{mrad}$~\cite{Abbiendi:2016xup}. Assuming non-radiative kinematics we find $\Delta \sim 10\, \text{MeV}$, which then yields for the N$^{4}$LO correction the estimate 
\begin{align}
	\sigma_\text{QED}^{(4)}
	\sim \Big( \frac{\alpha}{4\pi} \Big)^4 L_\Delta^4 L_m^4
	 \sim  10^{-5}\, .
\end{align}
This lies within the sensitivity of the MUonE experiment due to strong soft and collinear logarithmic enhancements. A parton shower approach to resum the LL contributions is therefore unavoidable. The remaining NLL corrections then amount to
\begin{subequations}
\begin{align}
	\sigma_\text{QED}^{(2),\text{NLL}} &\sim \Big( \frac{\alpha}{4\pi} \Big)^2 L_\Delta L_m^2 \sim  10^{-4}\, ,\\
	\sigma_\text{QED}^{(3),\text{NLL}} &\sim \Big( \frac{\alpha}{4\pi} \Big)^3 L_\Delta^2 L_m^3 \sim  10^{-5}\, , \\
	\sigma_\text{QED}^{(4),\text{NLL}} &\sim \Big( \frac{\alpha}{4\pi} \Big)^3 L_\Delta^3 L_m^4 \sim  10^{-6}\, .
\end{align}
\end{subequations}
We can thus tentatively assume that a fixed-order calculation at N$^{3}$LO is sufficient for the $10\, \text{ppm}$ target precision if properly matched to a parton shower. Unfortunately, this still seems to be an impossible endeavour even with the current state-of-the-art. However, it is only the contributions that correct the electron line that obtain the full collinear enhancement. All other corrections, on the other hand, are naively expected to be of less relevance since they include logarithms of the much larger muon mass. It is therefore reasonable to assume that it is sufficient to only calculate the dominant electronic effects at N$^{3}$LO. As discussed in the outlook at the end of the thesis, such a calculation does indeed seem feasible in the near-term future.

The LO unpolarised differential cross section for $\mu$-$e$ scattering is given by
\begin{align}\label{eq:mue_born}
	\frac{\d\sigma^{(0)}_\text{QED}}{\d t}
	= 4\pi\alpha \frac{(m^2+M^2-s)^2+s t+t^2/2}{t^2 \lambda(s,m^2,M^2)}
\end{align}
with
\begin{align}
t=(p_1-p_3)^2=(p_2-p_4)^2=2m^2-2m E_3
\end{align}
and the Källén function
\begin{align}
	\lambda(x,y,z) = x^2 + y^2 + z^2-2xy-2yz-2zx\, .
\end{align}
This result remains unchanged even if the muon polarisation, $n_\mu$, is taken into account. This is a consequence of the parity invariance of the QED Lagrangian~\eqref{eq:qed_lagrangian} which implies for the amplitude
\begin{align}
	\mathcal{A}(n_\mu) = \mathcal{A}(-n_\mu)\, .
\end{align}
If the target electrons were also polarised this conclusion would not hold anymore due to parity invariant combinations such as $n_e \cdot n_\mu$. The only effect of the polarisation is thus due to the EW contributions coming from the $Z$-boson exchange. Furthermore, because $\mathcal{A}_n\sim 1/t$ and, hence, $\d\sigma_\text{QED}^{(0)}/\d t \sim 1/t^2$ the total cross section is not well-defined. In~\cite{Melnikov:1996na} it was shown that this singularity is naturally regularised by finite beam size effects. These considerations are in this case, however, purely academic in nature. The problematic region $t\sim 0$ is automatically avoided with a cut on the minimal electron energy or equivalently on the maximal electron scattering angle. 

Beyond LO it is convenient to split the amplitude into separately gauge invariant subsets by introducing formally different charges, $Q_e$ and $Q_\mu$, for the electron and the muon, respectively. For the virtual one-loop amplitude the corresponding decomposition reads
\begin{align}\label{eq:oneloop_split}
	\mathcal{A}_n^{(1)}
	= Q_e^2 \mathcal{A}^{(1)}_{2,0}
	+ Q_e Q_\mu \mathcal{A}^{(1)}_{1,1}
	+ Q_\mu^2 \mathcal{A}^{(1)}_{0,2}\, ,
\end{align}
where the LO charges are omitted in the counting. These contributions are expected to satisfy the hierarchy
\begin{align}\label{eq:charge_hierarchy}
	\mathcal{A}^{(1)}_{0,2} \ll  \mathcal{A}^{(1)}_{1,1} \ll \mathcal{A}^{(1)}_{2,0}
\end{align}
due to different powers of the large collinear logarithm $L_m$. As already mentioned in the context of the electronic N$^3$LO correction, this separation is useful in order to disentangle dominant from subleading effects. Furthermore, it makes the dependence on the muon charge in~\eqref{eq:muone_process} explicit. The mixed subset, $Q_e Q_\mu$, therefore changes sign when switching from negatively to positively charged muons. The purely electronic and muonic contributions, on the other hand, remain unchanged. 

The gauge-invariant split for the tree-level real-emission amplitude is given by
\begin{align}\label{eq:oneloop_rad_split}
	\mathcal{A}_{n+1}^{(0)}
	= Q_e \mathcal{A}_{1,0}^{(0)}
	+ Q_\mu \mathcal{A}_{0,1}^{(0)}\, .
\end{align}
For the squared amplitude this yields again the three subsets $\{Q_e^2,Q_e Q_\mu,Q_\mu^2\}$ and combines with~\eqref{eq:oneloop_split} to the physical NLO correction. These effects have been extensively studied with the fully-differential Monte Carlo code \textsc{MESMER} in~\cite{Alacevich:2018vez}. The corresponding analysis shows that the hierarchy~\eqref{eq:charge_hierarchy} is satisfied for most observables. Nevertheless, an enhancement of the subleading corrections is observed in the phase-space region of small scattering angles.

At NNLO the gauge invariant split for the virtual-virtual, real-virtual, and real-real amplitudes reads
\begin{subequations}\label{eq:twoloop_split}
\begin{align}
	\mathcal{A}_{n}^{(2)}
	&= Q_e^4 \mathcal{A}^{(2)}_{4,0}
	+ Q_e^3 Q_\mu \mathcal{A}^{(2)}_{3,1}
	+ Q_e^2 Q_\mu^2 \mathcal{A}^{(2)}_{2,2}
	+ Q_e^3 Q_\mu \mathcal{A}^{(2)}_{1,3}
	+ Q_\mu^4 \mathcal{A}^{(2)}_{0,4}\, , \\
	\mathcal{A}_{n+1}^{(1)}
	&= Q_e^3 \mathcal{A}_{3,0}^{(1)}
	+ Q_e^2 Q_\mu \mathcal{A}_{2,1}^{(1)}
	+ Q_e Q_\mu^2 \mathcal{A}_{1,2}^{(1)}
	+ Q_\mu^3 \mathcal{A}_{0,3}^{(1)}\, , \\
	\mathcal{A}_{n+2}^{(0)}
	&= Q_e^2 \mathcal{A}_{2,0}^{(0)}
	+ Q_e Q_\mu \mathcal{A}_{1,1}^{(0)}
	+ Q_\mu^2 \mathcal{A}_{0,2}^{(0)}\, .
\end{align}
\end{subequations}
The main bottleneck at this order is the calculation of the mixed two-loop amplitudes $\mathcal{A}^{(2)}_{3,1}$, $\mathcal{A}^{(2)}_{2,2}$, and $\mathcal{A}^{(2)}_{1,3}$. The purely electronic and muonic contributions, $\mathcal{A}^{(2)}_{4,0}$ and $\mathcal{A}^{(2)}_{0,4}$, on the other hand, can be computed based on the heavy quark form factor~\cite{Mastrolia:2003yz,Bonciani:2003ai,Bernreuther:2004ih,Gluza:2009yy} and combined with the corresponding subsets of the real-virtual and real-real contributions. 
We have implemented these effects in \mcmule{}~\cite{Banerjee:2020rww} and found perfect agreement in a comparison with the \textsc{MESMER} collaboration~\cite{CarloniCalame:2020yoz}. This check is particularly strong since completely different techniques have been used to cope with IR divergences. While the \mcmule{} framework is based on dimensional regularisation and a subtraction method, \textsc{MESMER} uses a photon-mass regulator combined with a slicing approach.

An exact calculation is currently not feasible for the mixed contributions. Instead, a considerable theory effort has been put into the calculation of the massless ($m=0, M\neq 0$) two-loop integrals~\cite{Mastrolia:2017pfy,DiVita:2018nnh} culminating in the recent completion of the full massless amplitude~\cite{Bonciani:2021okt}. This result can, however, only be used directly in the calculation of collinear safe observables as defined in~\eqref{eq:coll_safety}. Observables in lepton experiments and in particular also in MUonE typically only satisfy the soft safety condition~\eqref{eq:soft_safety}. The mass logarithms $L_m$ are measured in these experiments and thus have to be predicted by the calculation. Nevertheless, there is a close connection between these physical mass logarithms and the collinear $1/\epsilon$ poles in the massless calculation. In fact, following~\cite{Becher:2007cu,Penin:2005eh} it is possible to predict the leading mass effects in a process-independent way based on the massless amplitude. This `massification' approach therefore allows us to obtain an approximation of the massive amplitude based on the massless result of~\cite{Bonciani:2021okt}. 

Since the dominant electronic corrections are known with full mass dependence, only the mixed amplitude has to be massified. It can therefore be expected that the error of the massified approximation lies well below the sensitivity of the MUonE experiment. This is supported by the dedicated study of Section~\ref{sec:massification_error} where a massification error of $\mathcal{O}(1\%)$ in the NNLO coefficient is observed in the case of the muon decay. A fully differential calculation of the full set of NNLO QED corrections to $\mu$-$e$ scattering is therefore possible. This represents a major step towards the  ambitious $10\,\text{ppm}$ goal and is the main topic of the thesis.

\section{Overview of the thesis}\label{sec:overview}

The main objective of the thesis is to give a pedagogical overview of the theoretical foundations of the \mcmule{} framework. We have developed the corresponding methods with fully differential NNLO QED calculations in mind and have successfully applied them to many processes (see Table~\ref{tab:mcmule}). Muon-electron scattering is particularly interesting in this regard due to its multi-scale nature. This process therefore allows us to cover many interesting aspects of higher-order QED calculations. Most importantly, this includes IR subtraction, massification and next-to-soft stabilisation.

Using the simple exponentiating structure of soft singularities in QED we have developed FKS$^\ell$~\cite{Engel:2019nfw}, a subtraction scheme for soft singularities to all orders in perturbation theory. Due to the difficulties related to massive loop integrals, many two-loop amplitudes in QED are only known in the massless limit and can not be directly used in \mcmule{}. For this reason we have extended the method of massification to heavy external states in~\cite{Engel:2018fsb}. Exploiting the universality of collinear degrees of freedom, massification determines all mass effects that are not polynomially suppressed based on the massless amplitude. These developments leave a numerically stable implementation of the real-virtual contribution as the remaining bottleneck. In the case of Bhabha and M{\o}ller scattering~\cite{Banerjee:2021mty, Banerjee:2021qvi} we have found an elegant solution to this problem by expanding the amplitude in the soft photon momentum up to and including subleading power. In order to facilitate the application of the method to other processes such as $\mu$-$e$ scattering, we have systematised the corresponding calculation by extending the Low-Burnett-Kroll (\ac{LBK}) theorem to one loop~\cite{Engel:2021ccn}. This makes it trivial to determine the subleading term in the soft expansion from the non-radiative amplitude.

Chapter~\ref{chap:foundations} starts with a concise and basic introduction to loop integrals and the method of regions (\ac{MoR}). The second part of the chapter is then devoted to a review of massification and the FKS$^\ell$ subtraction scheme. An emphasis is put on the occurrence of the factorisation anomaly which significantly complicates the application of massification to closed fermion loops. The semi-numerical hyperspherical method is therefore presented in Chapter~\ref{chap:fermionic} as an alternative approach for these fermionic contributions. The subsequent Chapter~\ref{chap:cps} then presents multi-channeling combined with a dedicated tuning of the phase space as a means to cope with the problem of CPS. The method of next-to-soft stabilisation is introduced in Chapter~\ref{chap:nts}. This includes a detailed derivation of the LBK theorem at one loop. At this point of the thesis all methodology relevant for $\mu$-$e$ scattering at NNLO has been covered. Before presenting the corresponding results in Chapter~\ref{chap:results}, a novel collinear factorisation formula for radiative QED amplitudes at one loop is derived in Chapter~\ref{chap:coll}. While this formula has not yet been used for practical purposes, there are many possible applications such as leading-collinear stabilisation and the subtraction of CPS. Furthermore, it represents an important building block for the massification of radiative amplitudes. This approach is discussed in more detail in the concluding Chapter~\ref{chap:outlook} where possible strategies for the electronic \ac{N$^3$LO} corrections are proposed.

\chapter{Theoretical foundations}\label{chap:foundations}

This chapter presents a selection of topics relevant for higher-order calculations in QED. The topics are chosen with the aim of providing the reader with an overview of relevant techniques and to develop the technical basis for the main part of the thesis. When considered helpful the reader is pointed to literature where more detailed information can be found. Sections~\ref{sec:loop} and~\ref{sec:mor} present a short introduction to loop integrals and the MoR. A strong emphasis is put on the difficulties related to the multi-scale nature of QED, which is particularly relevant for $\mu$-e scattering. Sections~\ref{sec:massification} and~\ref{sec:ir_divergences} then discuss the method of massification and the FKS$^\ell$ subtraction scheme. Both techniques have been developed specifically for higher-order QED calculations and represent essential building blocks of the \mcmule{} framework. All topics in this chapter are presented in a way to avoid unnecessary overlap with an earlier thesis~\cite{Ulrich:2020frs} and to provide complimentary information. 

\section{Loop integrals}\label{sec:loop}

One of the main challenges in perturbative calculations in QFT is the computation of loop integrals. A huge effort has been put into the development of corresponding analytic and numerical methods. Detailed reviews of analytic techniques can be found in~\cite{Weinzierl:2006qs,Weinzierl:2022eaz,Smirnov:2012gma}. Here, we limit ourselves to a very basic introduction to the topic in order to highlight the main difficulties related to the presence of multiple scales.

Loop integrals are of the general form
\begin{align}\label{eq:loop_integral}
	I = \int \prod_{l=1}^{L}\left[\mathrm{d}\ell_l\right] 
	\frac{1}{\mathcal{P}_1^{n_1} \ ...\ \mathcal{P}_p^{n_p}}\, ,
\end{align}
where we conventionally define the loop measure in $d=4-2\epsilon$ dimensions as
\begin{align}
	\left[\mathrm{d}\ell_l\right]
	=C(\epsilon)\frac{d^d \ell_l}{i\pi^{d/2}}, \quad\quad
	C(\epsilon) = \mu^{2\epsilon}\Gamma(1-\epsilon)
\end{align}
with the scale of dimensional regularisation denoted by $\mu$. The measure is chosen such that the Gaussian integral\footnote{The gaussian integral in non-integer dimensions $d$ is defined by promoting the integer dimension $n$ in the standard solution of the integral to $d$.}
\begin{align}\label{eq:gaussian}
	\int \left[\D\ell_l\right] e^{i \ell_l^2} 
	= C(\epsilon) i^{-d/2}
	\equiv \mathcal{G}
\end{align}
is normalised in $d=4$ dimensions. The propagators $\mathcal{P}_i$ are at most quadratic in the loop momenta $\ell=\{\ell_1,...,\ell_L\}$ and can therefore be generically written as
\begin{align}
	\mathcal{P}_j = \ell^\top M_j \ell - 2 Q_j^\top(r_i) \ell+J_j(s_i)+i \delta\, ,
\end{align}
with $M_j$ a symmetric $L\times L$-matrix, $Q_j$ a $L$-vector, and the external momenta and the kinematical invariants denoted by $r_i$ and $s_i$, respectively. The causal structure of the propagators is implemented with the small positive imaginary part $+i \delta$.

An equivalent representation of the loop integral $I$ more suitable for analytic calculation can be derived using the Schwinger-parameter identity
\begin{align}\label{eq:schwinger_param}
	\frac{1}{\mathcal{P}_j^{n_j}}
	= \frac{(-i)^{n_j}}{\Gamma(n_j)}
	   \int_0^\infty \mathrm{d}\alpha_j \alpha_j^{n_j-1} e^{i \alpha_j \mathcal{P}_j}\, .
\end{align}
This integral is only well-defined because of the positive imaginary part $+i\delta$ in the propagators $\mathcal{P}_j$ which ensures an exponential damping for $\alpha_j\to\infty$. We can then write the loop integral~\eqref{eq:loop_integral} as
\begin{align}
	I = \frac{(-i)^N}{\prod_{j=1}^p \Gamma(n_j)}
	  \int \prod_{l=1}^L [\D\ell_l]
	  \int_0^\infty \Big[\prod_{j=1}^p \D\alpha_j\, \alpha_j^{n_j-1}\Big] 
	  e^{i E}\, ,
\end{align}
where $N=\sum_j n_j$ and all propagators are now combined in the exponent
\begin{align}
	 E = \sum_{j=1}^p \alpha_j \mathcal{P}_j
	  = \ell^\top M \ell-2Q^\top \ell + J + i\delta
\end{align}
with $M=\sum_j M_j \alpha_j$, $Q=\sum_j Q_j \alpha_j$, and $J=\sum_j J_j \alpha_j$. This makes the integration over the loop momenta in terms of the standard Gaussian integral~\eqref{eq:gaussian} possible. To this end, we perform the shift $\ell\rightarrow \ell+M^{-1}Q$ to cancel the $\ell$-linear term in $E$ and diagonalise $M$ via a rotation $\ell \rightarrow R \ell$. This yields
\begin{align}\label{eq:exp_trsf}
	E \to \ell^\top M_{\mathrm{diag}} \ell - \Delta + i\delta
	  = \sum_l \lambda_l \ell_l^2-\Delta + i\delta
\end{align}
with
\begin{align}
	\Delta=QM^{-1}Q-J
\end{align}
and $\lambda_l$ the eigenvalues of $M$. As a last transformation, we rescale $\ell_l\rightarrow \lambda_l^{-1/2}\ell_l$ and find
\begin{subequations}
\begin{align}
	I &= \frac{(-i)^N}{\prod_{j=1}^p \Gamma(n_j)}
	  \int \prod_{l=1}^L [\D\ell_l] e^{i \ell^2}
	  \int_0^\infty \Big[\prod_{j=1}^p \D\alpha_j\, \alpha_j^{n_j-1}
	  \Big] (\mathrm{det}M)^{-d/2} e^{-i(\Delta-i\delta)}
	  \\ 
	  &= \frac{(-i)^N \mathcal{G}^L }{\prod_{j=1}^p \Gamma(n_j)}
	  \int_0^\infty \Big[\prod_{j=1}^p \D\alpha_j\, \alpha_j^{n_j-1} \Big]
	  \label{eq:schwinger_rep}
	  \mathcal{U}^{-d/2} e^{-i\mathcal{F}/\mathcal{U}}\, ,
\end{align}
\end{subequations}
where we have used the eigenvalue property $\prod_l\lambda_l=\mathrm{det}M$ and introduced the \textit{Symanzik polynomials}
\begin{align}\label{eq:symanzik}
	\mathcal{U}(\alpha_j)= \mathrm{det}M\, , 
	\quad\quad 
	\mathcal{F}(\alpha_j,s_i)= \Delta\, \mathrm{det}M -i\delta\, .
\end{align}
Both polynomials are homogeneous in the Schwinger parameters $\alpha_j$ with $L$ and $(L+1)$ the degrees of $\mathcal{U}$ and $\mathcal{F}$, respectively. The loop integral in the form~\eqref{eq:schwinger_rep} is called the \textit{Schwinger-parameter representation}.

More often, however, one relies on a different parametric representation. This alternative version can be derived from~\eqref{eq:schwinger_rep} using the identity
\begin{align}
	1 = \int_{-\infty}^{\infty} \D t\, \delta\Big(t-\sum_{j\in\nu} \alpha_j \Big)
	  = \int_{0}^{\infty} \D t\, \delta\Big(t-\sum_{j\in\nu} \alpha_j \Big)\, ,
\end{align}
where $\emptyset\neq\nu\subset\{1,...,p\}$ and we have used that $\alpha_j \geq 1$. Combined with the rescaling
\begin{align}\label{eq:schwinger_feynamn_param}
x_j = \alpha_j/t
\end{align}
this gives the identity
\begin{align}
	\int_0^\infty \Big[\prod_{j=1}^p\D\alpha_j\Big] f(\alpha_j)
	= \int_0^\infty \Big[\prod_{j=1}^p\D x_j\Big]
          \delta\Big(1-\sum_{j\in\nu} x_j\Big)
	  \int_0^\infty \D t\, t^{p-1} f(t x_j)
\end{align}
which we can apply to~\eqref{eq:schwinger_rep}. Because of the homogeneity of the Symanzik polynomials the integral over the auxiliary $t$ variable factorises and can thus be solved generically as
\begin{align}
	\int_0^\infty \D t\, t^{N-Ld/2-1} e^{-i t \mathcal{F}/\mathcal{U}}
	= \Big(i\frac{\mathcal{F}}{\mathcal{U}}\Big)^{Ld/2-N}\Gamma(N-Ld/2)\, .
\end{align}
This directly yields the \textit{Feynman-parameter representation}
\begin{align}\label{eq:feynman_rep}
	I= C(\epsilon)^L (-1)^N
	\frac{\Gamma(N-L d/2)}{\prod_{j=1}^p \Gamma(n_j)}
	\int_0^\infty \Big[\prod_{j=1}^p \mathrm{d}x_j  x_j^{n_j-1}\Big]
	\  \delta\Big( \sum_{j\in\nu} x_j-1 \Big)
	\frac{\mathcal{U}^{N-(L+1)d/2}}{\mathcal{F}^{N-Ld/2}}\, .
\end{align}
The freedom to restrict the sum of the delta function to any non-empty subset of $\{1,...,p\}$ is very useful. It allows us, for example, to set a suitably chosen $x_j=1$ and to integrate the remaining parameters over the full domain without any constraints. 

An alternative way to derive~\eqref{eq:feynman_rep} is based on the Feynman-parameter identity
\begin{align}
	\frac{1}{\mathcal{P}_1^{n_1}...\mathcal{P}_p^{n_p}}
	=\frac{\Gamma(N)}{\prod_{j=1}^p \Gamma(n_j)} \int_0^\infty
	\Big[\prod_{j=1}^p\mathrm{d}x_j x_j^{n_j-1}\Big]
	\ \delta\Big( \sum_{j=1}^p x_j -1 \Big)
	\frac{1}{(\sum_j x_j \mathcal{P}_j)^N}\, .
\end{align}
The transformation~\eqref{eq:exp_trsf} can then be used to map the loop integrand to the standard form
\begin{align}\label{eq:loop_standard}
	\int \prod_{l=1}^L \left[\mathrm{d}\ell_l\right]
	\frac{1}{\left(\ell^2-\Delta+i\delta \right)^N}
	= C(\epsilon)^L (-1)^{N}
	\frac{\Gamma(N-dL/2)}{\Gamma(N)}\left(\frac{1}{\Delta-i\delta}\right)^{N-dL/2}\, .
\end{align}
This evaluation of the loop integral is based on a Wick rotation that turns the loop momenta into Euclidean vectors. More details about this procedure can be found in Section~\ref{sec:loop_integration} where the more delicate case of non-shift invariant integrands is studied.

The following example illustrates the usefulness of the representation~\eqref{eq:feynman_rep} in the analytic calculation of loop integrals. Furthermore, it introduces the Mellin-Barnes representation as a convenient tool to solve the integral over the Feynman paramaters.

\vspace{.1cm}
\begin{exmp}\label{ex:loop_integral}
We consider the simple one-loop triangle integral
\begin{align}\label{eq:triangle}
	I_\triangle 
	&= \int [d\ell] 
	\frac{1}{[\ell^2+i\delta][\ell^2+2\ell\cdot p_1+i\delta] [\ell^2 + 2\ell\cdot
	p_3+i\delta]}
\end{align}
from the calculation of the vertex correction for $\mu$-$e$ scattering (see Figure~\ref{fig:muone_triangle}). Using the on-shell conditions $p_1^2=p_3^2=m^2$ and $t=(p_1-p_3)^2$, the Feynman-parameter representation~\eqref{eq:feynman_rep} takes the form
\begin{align}\label{eq:alpha_rep_ex}
	I_\triangle =
	-\frac{C(\epsilon)\Gamma(-2\epsilon)\Gamma(1+\epsilon)}{\Gamma(1-2\epsilon)}
	\int_0^{\infty} \d x_2  (1+x_2)^{2\epsilon}
	\big(-(t+i\delta) x_2 + m^2 (1 + x_2)^2\big)^{-1-\epsilon}\, ,
\end{align}
where we have set $x_3=1$ and performed the trivial integration over $x_1$ in terms of Gamma functions. To solve the remaining integral we need to disentangle the two scales $t$ and $m^2$ in the integrand. The substitution that achieves this is hard to find (if it exists at all).\footnote{In this simple case,~\eqref{eq:alpha_rep_ex} can actually be mapped directly to the integral representation of a hypergeometric function. The Mellin-Barnes representation is only applied to introduce the method.}  It turns out that the disentanglement can be accomplished with the Mellin-Barnes representation~\cite{Smirnov:1999gc,Tausk:1999vh}
\begin{equation}\label{eq:mellin_barnes}
	\frac{1}{(A+B)^\lambda}=\frac{1}{\Gamma(\lambda)}\frac{1}{2\pi i}
	\int_{-i\infty}^{+i\infty}\mathrm
{d}\sigma \ \Gamma(\lambda+\sigma)\Gamma(-\sigma)A^\sigma B^{-\sigma-\lambda}\, ,
\end{equation} 
where the contour is chosen such that it separates the left poles from gamma functions of the form $\Gamma(...+\sigma)$ from the right ones associated with $\Gamma(...-\sigma)$. After applying~\eqref{eq:mellin_barnes} to~\eqref{eq:alpha_rep_ex} the integral over $x_2$ can be trivially performed yielding
\begin{equation}
	I_\triangle 
	= -\Big(\frac{\mu^2}{m^2}\Big)^\epsilon
	  \frac{\Gamma(1-\epsilon) \Gamma(-2\epsilon)}{m^2 \Gamma(1-2\epsilon)}
	  \int_{-i\infty}^{+i \infty} \mathrm{d}\sigma
	  \Big(\frac{t+i\delta}{m^2}\Big)^\sigma
	  \frac{\Gamma(-\sigma) \Gamma(1+\sigma)^2
	  \Gamma(1+\epsilon+\sigma)}{\Gamma(2+2\sigma)}\, .
\end{equation}

\begin{figure}
	\captionsetup{justification=raggedright,singlelinecheck=false}
	\centering
	\input{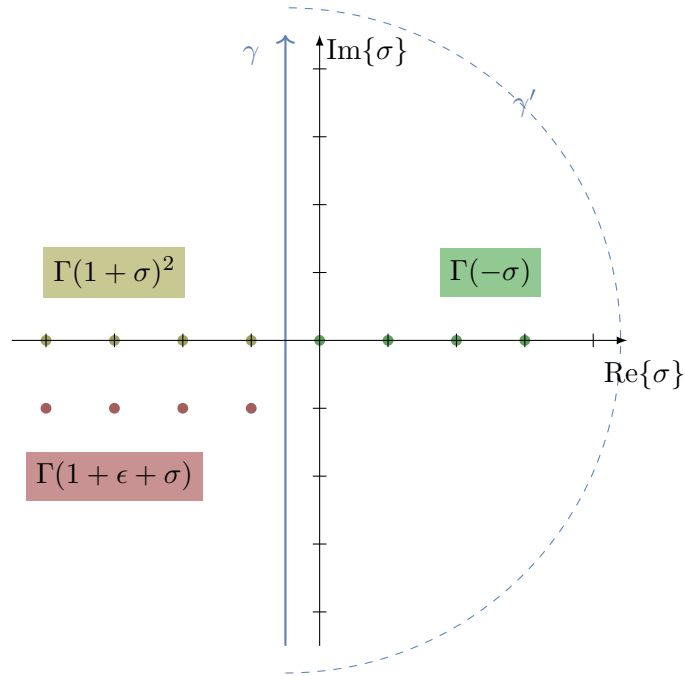}
\caption{The Mellin-Barnes integration contour $\gamma$ is defined such that it separates left poles of the form $\Gamma(...+\sigma)$ from right ones originating from $\Gamma(...-\sigma)$. The integral over $\gamma$ is calculated by closing the contour at infinity with the half circle $\gamma^\prime$. Afterwards, the residues of the infinite number of enclosed poles are summed up.} 
	\label{fig:mb_contour}
\end{figure}

\noindent
	The Mellin-Barnes representation has thus replaced a parameter integral with a contour integral. The advantage of the latter is that it can be solved using the residue theorem. We can close the contour with a half circle at infinity to the right as depicted in Figure~\ref{fig:mb_contour}. The resulting sum over the residues of $\Gamma(-\sigma)$ can then be re-written to match the series representation of the hypergeometric function $_2F_1(1,1+\epsilon,3/2,t/4m^2)$. Finally, we expand our expression in $\epsilon$ with the \texttt{Mathematica} package \texttt{HypExp}~\cite{Huber:2005yg,Huber:2007dx} that yields a result in terms of harmonic polylogarithms (\ac{HPL})~\cite{Remiddi:1999ew} that can in turn be evaluated with the code \texttt{HPL}~\cite{Maitre:2005uu,Maitre:2007kp}. These functions extend the notion of polylogarithms by generalising the order $o$ to a weight vector $\vec{w}$. In particular, introducing $\vec{w}=\{0\}$, $\vec{w}=\{0,0\}$, and $\vec{w}=\{-1,0\}$ we have
\begin{align}
	H_0(x) = \log(x)\, , \quad\quad
	H_{0,0}(x) = \frac{\log^2(x)}{2}\, , \quad\quad
	H_{-1,0}(x) = \log(x)\log(1+x)+\text{Li}_2(-x)\, .
\end{align}
Our final result then reads
\begin{equation}\label{eq:triangle_sol}
	I_\triangle
	= \Big(\frac{\mu^2}{m^2}\Big)^\epsilon \frac{x}{m^2 (1-x^2)} 
	  \Big(
	  -\frac{H_0(x)}{\epsilon} 
	  -H_{0,0}(x)+2 H_{-1,0}(x)+\zeta_2 
	  + \mathcal{O}(\epsilon)
	  \Big)
\end{equation}
with 
\begin{align}
	x
	= \frac{\sqrt{4m^2-t-i\delta}-\sqrt{-t-i\delta}}
	  {\sqrt{4m^2-t-i\delta}+\sqrt{-t-i\delta}}\, .
\end{align}
This result can be checked with existing one-loop libraries such as \texttt{Package-X}~\cite{Patel:2015tea}. In the physical region where $t<0$ and thus $0<x<1$ we can drop the $i\delta$ prescription since all HPLs are real. However, this picture changes if we want to connect this result to the process $e^- e^+ \to \mu^-\mu^+$ via the amplitude crossing relation
\begin{equation}
	\mathcal{A}\Big\{e^-(p_1)e^+(p_2)\to \mu^-(p_3)\mu^+(p_4)\Big\}
	= \mathcal{A}\Big\{e^-(p_1)\mu^-(-p_4)\to e^-(-p_2)\mu^-(p_3)\Big\}\, .
\end{equation}
The two processes are thus related through $s \leftrightarrow t$. As a consequence, we need to analytically continue~\eqref{eq:triangle_sol} to the region $t \to s>4m^2>0$. Because of the $i\delta$ prescription this can be done unambiguously via
\begin{equation}
	x\to \tilde{x} + i\delta\, , 
	\quad
	\tilde{x} = \frac{\sqrt{s-4m^2}-\sqrt{s}}
	{\sqrt{s-4m^2}+\sqrt{s}}\, .
\end{equation}
\end{exmp}
\vspace{.1cm}

The above example illustrates how the Mellin-Barnes representation can be used to calculate simple loop integrals. For more complicated cases, however, multiple contour integrals have to be introduced. For an introduction on more elaborate Mellin-Barnes applications we refer the reader to~\cite{Engel:2018}. Furthermore, we have produced a \texttt{Mathematica} code to facilitate the workflow for this type of calculation~\cite{Urwyler:2019}. However, for higher loop orders and multiple scales the Mellin-Barnes approach turns out to be rather limited. State-of-the-art loop calculations in QCD and QED therefore use more sophisticated techniques such as differential equations~\cite{Henn:2014qga}. Nevertheless, for all calculations encountered in this thesis Mellin-Barnes techniques have turned out to be sufficient. This is thanks to the MoR which can be used to exploit scale hierarchies to simplify loop calculations.

\section{Method of regions}\label{sec:mor}

Example~\ref{ex:loop_integral} has demonstrated that the degree of complexity in the calculation of loop integrals increases with the number of scales. In many applications these scales exhibit a large hierarchy. In QED, for example, the mass of the electron is typically much smaller than the energy scale of the process. In this case it is advantageous to expand the loop integral at the integrand level in order to reduce the number of active scales. The MoR allows us to do this in a systematic and convenient manner. We first introduce the standard version of this technique formulated in momentum space~\cite{Beneke:1997zp} and afterwards describe the analogous method in parameter space~\cite{Smirnov:1999bza}. While the former is closely connected to physical properties of the considered process, the latter is more convenient from a computational point of view. A more extensive introduction to the MoR from the perspective of effective field theory (\ac{EFT}) can be found in~\cite{Becher:2014oda}.

\subsection{Momentum-space formulation}\label{sec:mor_mom}

For a generic loop integral the integration and the expansion in a small parameter do not commute. This is the case whenever there are regions in the integration domain where additional parameters in the integrand become small and the original scale hierarchy breaks down. The simplest example is the soft region where the loop momentum itself becomes small, i.e. $\ell \ll 1$. The MoR yields an elegant solution to this problem given by the  following prescription:
\begin{enumerate}
 \item Introduce intermediate factorisation scales to divide the loop integral into regions where the loop momenta have a definite scaling with respect to the small scale.
 \item Perform a Taylor expansion in parameters that are small in a given region.
 \item \textit{Ignore the factorisation scales} and integrate over the entire loop integration domain in each region.
\end{enumerate}
The third step is only allowed if the calculation is performed in dimensional regularisation where the additional contributions vanish due to the absence of a scale in the loop integral. It is precisely this step that makes the MoR so powerful. It avoids unphysical factorisation scales in intermediate steps and renders each region separately Lorentz invariant. Furthermore, even though there is in principle an infinite number of regions, only a finite number gives contributions that are not scaleless. In the following, we apply the MoR to the simple one loop integral of Example~\ref{ex:loop_integral}.

\vspace{.1cm}
\begin{exmp}\label{ex:mor_mom}
We consider the integral~\eqref{eq:triangle} and assume the scale hierarchy
\begin{equation}\label{eq:triangle_scaling}
  p_1^2=p_3^2=m^2 \sim \lambda^2 \ll t = (p_1-p_3)^2 \sim \lambda^0\, .
\end{equation}
The book-keeping parameter $\lambda$ is introduced to facilitate the power counting. It is straightforward to perform the small-mass expansion of the exact result~\eqref{eq:triangle_sol} yielding
\begin{equation}\label{eq:triangle_exp}
  I_\triangle 
  = \frac{1}{t} \Big(\frac{\mu^2}{m^2}\Big)^\epsilon
  \Big(\frac{\log{y}}{\epsilon}+\frac{1}{2}\log^2{y}-\zeta_2 + \mathcal{O}(\epsilon)
  \Big) + \mathcal{O}(\lambda)
\end{equation}
with $y=-m^2/t$. This result can also be obtained with the MoR.

To identify the contributing momentum regions the large and small components of the momenta $p_i=E_i(1,\vec{e}_i\beta_i)$ have to be disentangled. The velocities $\beta_i = (1-m^2/E_i^2)^{1/2} = 1 - \mathcal{O}(\lambda^2)$ are close to 1 as expected for highly energetic particles. The disentanglement can be achieved via a decomposition into the light-cone basis vectors
\begin{align}
\{ e_i = (1, \vec{e}_i)/\sqrt{2} , \bar{e}_i = (1, -\vec{e}_i)/\sqrt{2} \}
\end{align}
as
\begin{align}\label{eq:lightcone_decomposition}
	p_j = (e_i\cdot p_j) \bar{e}_i+(\bar{e}_i\cdot p_j)e_i+ p_j^{(\perp,i)}  \, .
\end{align}
In the following we use the more compact notation
\begin{subequations}
\begin{align}
    p_j
    &= p_j^{(+,i)} + p_j^{(-,i)} + p_j^{(\perp,i)} \\
    &= (e_i\cdot p_j, \bar{e}_i\cdot p_j, p_{j,\perp})_i\, ,
\end{align}
\end{subequations}
with $p_j^{(+,i)}$ and $p_j^{(-,i)}$ the component of $p_j$ in the direction of $ \bar{e}_i$ and $e_i$, respectively.

In the case of the energetic particle $p_i$ this takes the form of the
desired decomposition
\begin{equation}
    p_i = \big(E_i(1-\beta_i)/\sqrt{2},E_i(1+\beta_i)/\sqrt{2},p_{i,\perp}\big)_i \sim (\lambda^2,1,\lambda)_i
\end{equation}
where the scaling of the perpendicular component can be deduced from
\begin{equation}\label{eq:lightcone_square}
    p_i^2 = 2 p_i^{(+,i)}\cdot p_i^{(-,i)}+\big(p_i^{(\perp,i)}\big)^2=m^2\sim\lambda^2\, .
\end{equation}
Applying the light-cone decomposition to the external momenta in our
integral we find
\begin{subequations}
\begin{align}
    &p_1 \sim (\lambda^2,1,\lambda)_1\sim (1,1,1)_3\, , \\
    &p_3 \sim (1,1,1)_1\sim (\lambda^2,1,\lambda)_3\, .
\end{align}
\end{subequations}
Based on the light-cone decomposition, a region as described in Step~$1$ of the MoR recipe can be defined as a specific scaling of the loop momentum $\ell \sim (\lambda^a, \lambda^b, \lambda^c)_i$. It turns out that it is permissible to restrict to on-shell scalings only where $c=(a+b)/2$. This ensures a homogeneous scaling in~\eqref{eq:lightcone_square}. All other choices for $a$, $b$, and $c$ result in scaleless integrals. A corresponding proof can be found in Appendix~A of~\cite{Engel:2018} in the specific case of the muon decay.

The following three momentum regions can now be identified that do not yield scaleless integrals:
\begin{subequations}
\begin{alignat}{3}
    &\mbox{hard:}                  \quad &\ell& \sim (1,1,1)_1               \ \ &\sim&\ \  (1,1,1)_3 \\
    &\mbox{$p_1$-collinear:}       \quad &\ell& \sim (\lambda^2,1,\lambda)_1 \ \ &\sim&\ \  (1,1,1)_3 \\
    &\mbox{$p_3$-collinear:}       \quad &\ell& \sim (1,1,1)_1               \ \ &\sim&\ \  (\lambda^2,1,\lambda)_3
\end{alignat}
\end{subequations}
An example of a scaleless region is discussed at the end of the example. The hard momentum scaling projects onto the large components of the external momenta, i.e. $\ell \cdot p_i = \ell \cdot p_i^{(-,i)}+\mathcal{O}(\lambda)$. Because $(p_i^{(-,i)})^2=0$ the leading term in the expansion of the hard region corresponds to the massless version of the integral. We find
\begin{subequations}\label{eq:triangle_hard}
\begin{align}
  I_\triangle^\text{hard} 
  &= 
  \int [d\ell] \frac{1}{[\ell^2][\ell^2+2\ell\cdot p_1^{(-,1)}] [\ell^2 + 2\ell\cdot p_3^{(-,3)}]}
  + \mathcal{O}(\lambda) \\
  &= 
  \frac{1}{t} \Big(-\frac{\mu^2}{t}\Big)^\epsilon \Big(
  \frac{1}{\epsilon^2} + \mathcal{O}(\epsilon)
  \Big) + \mathcal{O}(\lambda)\, .
\end{align}
\end{subequations}
As expected, this result only depends on the hard scale $t$. Furthermore, since collinear divergences are not regulated anymore with finite masses as in the original integral we obtain  a $1/\epsilon^2$ pole from the overlap of soft and collinear singularities. This pole has to cancel against the remaining regions to combine to the result~\eqref{eq:triangle_exp}. This feature can be very helpful to check whether all regions have been correctly identified. Similarly, we find for the $p_1$-collinear region
\begin{subequations}\label{eq:triangle_coll}
\begin{align}
	I_\triangle^{p_1\text{-coll}}
 	&= 
  	\int [d\ell] \frac{1}{[\ell^2][\ell^2+2\ell\cdot p_1] [2\ell\cdot p_3^{(+,1)}]}
	+ \mathcal{O}(\lambda) \\
	&= 
	-\frac{1}{2t}\Big(\frac{\mu^2}{m^2}\Big)^\epsilon \Big( 
	\frac{1}{\epsilon^2} + \zeta_2 + \mathcal{O}(\epsilon)
	\Big) + \mathcal{O}(\lambda)
\end{align}
\end{subequations}
where only a non-trivial dependence on the collinear scale $m$ remains. The Feynman-parameter representation~\eqref{eq:feynman_rep} can also be used in the case of linear propagators encountered here. Furthermore, since the integral $I_\triangle$ is symmetric under $p_1 \leftrightarrow p_3$ we have $I_\triangle^{p_3\text{-coll}}=I_\triangle^{p_1\text{-coll}}$. As a consequence, the soft-collinear $1/\epsilon^2$ poles indeed cancel in the sum of the three regions. Furthermore, they correctly combine to the expanded result of~\eqref{eq:triangle_exp}, i.e.
\begin{equation}
	 I_\triangle^\text{hard}+I_\triangle^{p_1\text{-coll}}+I_\triangle^{p_3\text{-coll}}
	 = I_\triangle + \mathcal{O}(\lambda)\, .
\end{equation}

We have thus identified correctly all contributing momentum regions. All other loop momentum scalings lead to scaleless integrals. As an example we consider the soft region $\ell \sim (\lambda, \lambda, \lambda)$ that yields the integral
\begin{align}\label{eq:triangle_soft}
	I_\triangle^\text{soft}
 	=  \int [d\ell] \frac{1}{[\ell^2][2\ell\cdot p_1^{(-,1)}] [2\ell\cdot p_3^{(-,3)}]}
	 + \mathcal{O}(\lambda)\, .
\end{align}
To proof that this integral is scaleless we apply the transformation
\begin{align}
	\ell \to \kappa \ell
\end{align}
which leaves the integral invariant but changes the integrand. This gives the relation
\begin{align}
	I_\triangle^\text{soft}
 	= \kappa^{d-4} I_\triangle^\text{soft}
\end{align}
that is only satisfied if $ I_\triangle^\text{soft}$ is either zero or infinite. In dimensional regularisation the integral is finite and thus $I_\triangle^\text{soft}=0$.
\end{exmp}
\vspace{.1cm}

In this example it is straightforward to verify that all contributing regions have been taken into account since the exact result is known. Furthermore, the contributing regions have a clear physical interpretation which facilitates their identification. In many applications this is, however, not the case. The formulation of the MoR in the parametric representation is immensely helpful in such a situation.

\subsection{Parameter-space formulation}\label{sec:mor_param}

Following~\cite{Smirnov:1999bza} it is possible to formulate the MoR in the Feynman-parameter representation~\eqref{eq:feynman_rep}. In this case, a region is defined as a scaling of the Feynman parameters $(x_1, ... , x_n) \sim (\lambda^{a_1}, ... , \lambda^{a_n})$. This is only uniquely defined up to a shift $a_i \to a_i + r$ due to the homogeneity of the Symanzik polynomials~\eqref{eq:symanzik}. To indicate this, we write $(x_1, ... , x_n) \stackrel{r}{\sim} (\lambda^{a_1}, ... , \lambda^{a_n})$. In all other aspects the method follows the same steps as in momentum space. There is, however, one crucial advantage. It yields an algorithmic means to find the contributing regions. To see this it is instructive to consider the scaleless soft integral~\eqref{eq:triangle_soft} in the parameter representation
\begin{subequations}
\begin{align}
	I_\triangle^\text{soft}
	&\sim \int_0^\infty \d x_1\d x_2\d x_3\, 
	x_3^{-1+2\epsilon} (-t x_1 x_2)^{-1-\epsilon} \delta\Big(1-x_3\Big) \\
	&= \int_0^\infty  \d x_1\d x_2\, (-t x_1 x_2)^{-1-\epsilon}\, .
\end{align}
\end{subequations}
The absence of scales can easily be seen from the transformation $x_1 \to \kappa x_1$ that gives the identity $I_\triangle^\text{soft} = \kappa^{-\epsilon} I_\triangle^\text{soft}$. This, in turn, is an immediate consequence of the homogeneity of the graph polynomials w.r.t. a strict subset of Feynman parameters (in this case $\{x_1\}$). This represents a simple sufficient condition for an integral to be scaleless. Reinterpreting this condition in terms of geometrical properties of the Symanzik polynomials then yields an efficient way to find all contributing regions. The corresponding algorithm is implemented in the \texttt{Mathematica} code \texttt{asy.m}~\cite{Jantzen:2012mw} and was recently included in \texttt{pySecDec}~\cite{Heinrich:2021dbf}.

The two formulations of the MoR are related in that there exists  a one-to-one correspondence between the contributing regions. This can be seen from the identity~\eqref{eq:schwinger_param} that serves as a basis for the Feynman-parameter representation.  The Feynman parameters $x_j$ correspond to the Schwinger parameters $\alpha_j$ up to the rescaling~\eqref{eq:schwinger_feynamn_param}. From this we can deduce a relationship between the regions in the Feynman-parameter space and the scaling of the propagators in momentum space. In particular we find
\begin{align}\label{eq:mom_param_rel}
	(x_1, ... , x_n) \stackrel{r}{\sim} (\mathcal{P}_1^{-1}, ... , \mathcal{P}_n^{-1})\, .
\end{align}
Let us illustrate this with the previous example.

\vspace{.1cm}
\begin{exmp}\label{ex:mor_param}

With the code \texttt{asy.m} it is straightforward to identify the three contributing regions in Feynman-parameter space for the integral~\eqref{eq:triangle} and the scale hierarchy~\eqref{eq:triangle_scaling}. We obtain the following output:
\begin{subequations}
\begin{align}
	r_1: \quad\quad (x_1, x_2, x_3)  &\sim \{1,1,1\} \\
	r_2: \quad\quad (x_1, x_2, x_3) &\sim \{1,1,\lambda^2\} \\
	r_3: \quad\quad (x_1, x_2, x_3)  &\sim \{1,\lambda^2,1\}
\end{align}
\end{subequations}
Going back to the momentum regions of Example~\ref{ex:mor_mom} we find the following scalings of the propagators:
\begin{subequations}
\begin{alignat}{2}
    &\mbox{hard:}                  \quad & \{\mathcal{P}_1^{-1},\mathcal{P}_2^{-1},\mathcal{P}_3^{-1}\} & \sim (1,1,1)   \\
    &\mbox{$p_1$-collinear:} \quad & \{\mathcal{P}_1^{-1},\mathcal{P}_2^{-1},\mathcal{P}_3^{-1}\} & \sim (\lambda^{-2},\lambda^{-2},1)  \\
    &\mbox{$p_3$-collinear:} \quad & \{\mathcal{P}_1^{-1},\mathcal{P}_2^{-1},\mathcal{P}_3^{-1}\} & \sim (\lambda^{-2},1,\lambda^{-2})
\end{alignat}
\end{subequations}
With~\eqref{eq:mom_param_rel} we can identify $r_1$ with the hard, $r_2$ with the $p_1$-collinear, and $r_3$ with the $p_3$-collinear region. An explicit calculation of these contributions indeed reproduces the results of the individual regions given in~\eqref{eq:triangle_hard} and~\eqref{eq:triangle_coll}.

\end{exmp}
\vspace{.1cm}

\section{Massification}\label{sec:massification}

The previous section has illustrated how the MoR can be used to alleviate the problem of multi-scale integrals in the case of small masses. Nevertheless, the corresponding calculation can become difficult and lengthy. This section presents the method of \textit{massification} that allows us to determine these mass effects based on the massless amplitude without the need of explicit computations. Fermion masses act as regulators of collinear divergences. Massification is therefore closely related to the well-known fact~\cite{Frenkel:1976bj} that the collinear divergences factorise into the wave function renormalisation constant. The universality of these mass effects can best be studied in the modern language of soft-collinear effective theory (\ac{SCET})~\cite{Bauer:2001yt,Bauer:2000yr,Beneke:2002ph}. SCET is an EFT for processes with highly-energetic particles. A didactic introduction to the topic can be found in~\cite{Becher:2014oda}. Here, we touch the topic only briefly in order to motivate the factorisation formula for massification.

\subsection{Factorisation formula}\label{sec:factorisation}

SCET describes soft and collinear degrees of freedom in the presence of a hard interaction. A prime example for a SCET application is jet production at hadron colliders. But it is also the appropriate EFT for the small-mass expansion discussed in Example~\ref{ex:mor_mom}. To be specific, we consider a scattering process $|i\rangle\to| f \rangle$ with $|i\rangle =|0\rangle$ and $n$ highly-energetic fermions in the final state $|f\rangle = | \psi_{e_1}…\psi_{e_n}\rangle$. For simplicity we do not consider hard photons in the final state. The directions of the outgoing fermions are labelled by $e_j$ in reference to the light-cone decomposition introduced in Section~\ref{sec:mor_mom}. Crossing symmetry can be used to reshuffle the final-state particles to the initial state. 
In analogy to~\eqref{eq:triangle_scaling} we have the scale hierarchy
\begin{align}\label{eq:massification_scaling}
	p_j^2=m_j^2 \sim \lambda^2 \ll S\, ,
\end{align}
with $S$ the scale of the hard scattering. This scenario is the generalisation of Example~\ref{ex:mor_mom} to an arbitrary number of fermions with small but non-vanishing masses.

SCET can be used to describe this process at \textit{leading power} (\ac{LP}) in the small-mass expansion where power-suppressed terms of $\mathcal{O}(\lambda)$ are neglected. The corresponding Lagrangian has the structure\footnote{There are no soft fermion fields since we take them to have small but non-zero masses.}
\begin{align}\label{eq:scet_lagrangian}
	\mathcal{L}_n 
	= \sum_{j=1}^n \mathcal{L}_\text{SCET}(\psi_{e_j},A_{e_j}^\mu,A_s^\mu)
	+ C_n \mathcal{O}_n\big(\{\psi_{e_j}\},\{A_{e_j}^\mu\}\big)\, .
\end{align}
The hard degrees of freedom of the fermion and photon fields, $\psi$ and $A^\mu$, have been integrated out leaving their inprint only in the \textit{Wilson coefficient} $C_n$ of the hard scattering operator $\mathcal{O}_n$. For simplicity and without loss of generality, we assume that there is only one operator that induces the hard scattering. The remaining dynamical degrees of freedom are captured by the soft photon field $A_s^\mu$ and the high-energy collinear modes $\psi_{e_j}$ and $A_{e_j}^\mu$. The interaction among these fields is described by the SCET Lagrangian $\mathcal{L}_\text{SCET}$. The dependence of the hard scattering operator, $\mathcal{O}_n$, on the collinear photon fields, $A_{e_j}^\mu$, is introduced to ensure gauge invariance.

One of the most important properties of SCET is that soft and collinear interactions can be decoupled via a field redefinition of the collinear modes
\begin{align}\label{eq:decoupling_transformation}
	\psi_{e_j} \to T(A_s^\mu) \tilde{\psi}_{e_j}\, , \quad\quad\quad
	A_{e_j}^\mu \to T(A_s^\mu) \tilde{A}_{e_j}^\mu T^\dagger(A_s^\mu)
\end{align}
known as \textit{decoupling transformation}~\cite{Bauer:2001yt}. The Lagrangian~\eqref{eq:scet_lagrangian} now reads
\begin{align}\label{eq:scet_lagrangian_decoupled}
	\mathcal{L}_n
	= \sum_{j=1}^n \mathcal{L}_\text{coll}(\tilde{\psi}_{e_j},\tilde{A}_{e_j}^\mu)
	+  \mathcal{L}_\text{soft}(A_s^\mu)
	+ C_n \tilde{\mathcal{O}}_n\big(\{\tilde{\psi}_{e_j}\},\{\tilde{A}_{e_j}^\mu\},A_s^\mu\big)
\end{align}
with the soft modes completely removed from the interaction Lagrangian and instead entering in the hard scattering term.
Based on this decoupled version of the Lagrangian we can now show that the SCET prediction for the on-shell scattering amplitude, $\mathcal{A}_n$, factorises into a soft part and a contribution for each collinear direction $e_j$.

The \ac{LSZ} formula relates scattering amplitudes to correlation functions. In this case we have
\begin{align}\label{eq:massification_correlator}
	\mathcal{A}_n(S,\{m_j\})\sim \langle 0 | \psi_{e_1}…\psi_{e_n} e^{i S_n} |0\rangle
\end{align}
with the action
\begin{align}
	S_n 
	= \int \D^4 x\, \mathcal{L}_n
	=\sum_{j=1}^n S_\text{coll}^j+S_\text{soft}+\int \D^4 x\, C_n \tilde{\mathcal{O}}_n\, .
\end{align}
At LP in the EFT we then find
\begin{align}
	\mathcal{A}_n
	 \sim   C_n \langle 0 | \tilde{\psi}_{e_1}…\tilde{\psi}_{e_n}
	  \bar{\mathcal{O}}_n 
	  e^{i \sum_{j=1}^n S^j_\text{coll}+S_\text{soft}} |0\rangle
	 + \mathcal{O}(\lambda)
\end{align}
where we have absorbed the additional contributions from the decoupling transformation~\eqref{eq:decoupling_transformation} of the external fields $\psi_{e_j}$ into the redefinition $\tilde{\mathcal{O}}_n \to \bar{\mathcal{O}}_n$. Assuming a standard operator form that consists only of fields and their derivatives, we can perform the split
\begin{align}
	\bar{\mathcal{O}}_n\big(\{\tilde{\psi}_{e_j}\},\{\tilde{A}_{e_j}^\mu\},A_s^\mu\big)
	= \Big(\prod_{j=1}^n \bar{\mathcal{O}}_n^j(\tilde{\psi}_{e_j},\tilde{A}_{e_j} ) \Big)
	 \bar{\mathcal{O}}_n^s(A_s^\mu)
\end{align}
into a soft part, $\bar{\mathcal{O}}_n^s$, and a collinear contribution, $\bar{\mathcal{O}}_n^j$, for each direction $e_j$. Next, we insert a complete set of states
\begin{align}
	1 = \sum_m | m \rangle  \langle m |
\end{align}
and use the orthogonality relation
\begin{align}
 	\langle 0 | \tilde{\psi}_{e_j} \bar{\mathcal{O}}_n^j e^{i S^j_\text{coll}} |m\rangle
	\sim \delta_{0 m}
\end{align}
to arrive at the aforementioned factorising structure
\begin{align}\label{eq:massification_amp}
	\mathcal{A}_n \sim 
	\Big( \prod_{j=1}^n \langle 0 | \tilde{\psi}_{e_j} \bar{\mathcal{O}}^j_n e^{i S^j_\text{coll}} |0\rangle \Big)
	\, \langle 0 | \bar{\mathcal{O}}_n^s e^{i S_\text{soft}} | 0 \rangle 
	\, C_n 
	+ \mathcal{O}(\lambda)\, .
\end{align}
The collinear matrix elements
\begin{align}\label{eq:jet_matel}
\langle 0 | \tilde{\psi}_{e_j} \bar{\mathcal{O}}^j_n e^{i S^j_\text{coll}} |0\rangle
\end{align}
only depend on one direction $e_j$ at a time and are thus \textit{process independent}. This is not the case for the soft contribution
\begin{align}\label{eq:soft_matel}
	 \langle 0 | \bar{\mathcal{O}}_n^s e^{i S_\text{soft}} | 0 \rangle 
\end{align}
that connects different directions. Also the Wilson coefficient $C_n$ is process dependent and has to be determined through a matching calculation. This can be done assuming $m_j=0$ since $C_n$ does not depend on the collinear scales $m_j$.  All loop integrals in SCET are scaleless in this case and thus
\begin{align}
	C_n \sim \mathcal{A}_n(S,\{m_j=0\})\, .
\end{align}

\begin{figure}
    \centering
    \begin{tikzpicture}[scale=1,baseline={(1,0)}]
    	\input{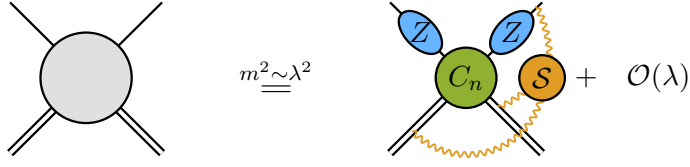}
    \end{tikzpicture}
    \caption{Schematic illustration of the small-mass factorisation formula~\eqref{eq:massification} for $\mu$-$e$ scattering. Each light electron contributes one power of the universal collinear massification constant $Z$. The process dependent soft function $\mathcal{S}$ connects the different external directions.}
    \label{fig:massification_fac}
\end{figure}

Putting all of this together, we find at the level of the squared amplitude the small-mass factorisation formula
\begin{equation}\label{eq:massification}
	\mathcal{M}_n(S,\{m_j\}) 
	= \Big( \prod_j Z(m_j) \Big) \,   \mathcal{S}(S,\{m_j\}) \,  \mathcal{M}_n(S,\{m_j=0\}) 
	+ \mathcal{O}(\lambda) \, .
\end{equation}
The \textit{universal massification constant} $Z$ corresponds to~\eqref{eq:jet_matel} and only depends on the collinear scales $m_j$.\footnote{As we will see in Section~\ref{sec:massification_twoloop}, an additional scale dependence is introduced in the presence of the factorisation anomaly.} The \textit{soft function} $\mathcal{S}$, originating from~\eqref{eq:soft_matel}, also depends on the hard scale $S$ and is process dependent. In the case of QED, it can be shown to all orders in perturbation theory that the soft function only receives contributions from diagrams with closed fermion loops~\cite{Becher:2007cu}. The jet and soft function combined turn the collinear $1/\epsilon$ poles in the massless amplitude into logarithms $\log(m_j^2/S)$. A schematic illustration of this factorisation formula is shown in Figure~\ref{fig:massification_fac}.

For many QED processes the massless two-loop amplitudes can be extracted from known QCD results~\cite{Bern:2000ie,Anastasiou:2002zn}. Once the process-independent massification constant $Z$ is known, the factorisation formula~\eqref{eq:massification} can be used to efficiently determine the leading mass effects. The only explicit calculation that remains to be done is the typically rather simple soft function. We emphasise that this massification procedure not only correctly determines the logarithmically enhanced but also the constant terms, only neglecting polynomially suppressed contributions.

Computing the massification constant based on the SCET Lagrangian~\eqref{eq:scet_lagrangian_decoupled} is impractical. Instead, one can resort to the MoR for this calculation. In fact, there is a one-to-one correspondence between the momentum regions in the MoR and the contributions from soft and collinear fields in SCET. This, in turn, implies that for physical quantities, only hard, soft, and collinear momentum scalings are expected to contribute. Other regions can exist at the level of individual diagrams but must drop out in the end. In the following Section~\ref{sec:massification_oneloop} we show how $Z$ can be determined at NLO. We then remark on the corresponding NNLO calculation that we have performed in~\cite{Engel:2018fsb} in the subsequent Section~\ref{sec:massification_twoloop}.

\subsection{Massification at one loop}\label{sec:massification_oneloop}

\begin{figure}
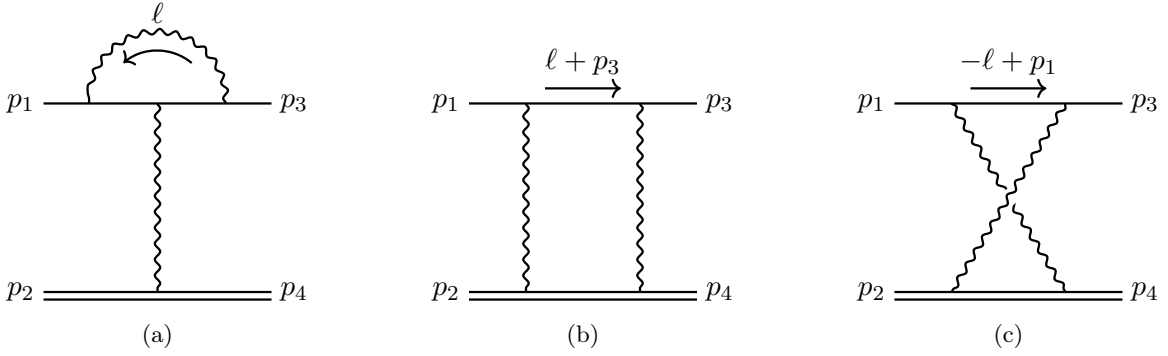

    \centering
	\subfloat[]{
        \begin{tikzpicture}[scale=1,baseline={(0,0)}]
        \input{tikz/muone_triangle}
    \end{tikzpicture}
    \label{fig:muone_triangle}
    }
    \hspace{1cm}
    \subfloat[]{
        \begin{tikzpicture}[scale=1,baseline={(0,0)}]
        \input{tikz/muone_box}
    \end{tikzpicture}
    \label{fig:muone_box}
    }
    \hspace{1cm}
    \subfloat[]{
        \begin{tikzpicture}[scale=1,baseline={(0,0)}]
        \input{tikz/muone_xbox}
    \end{tikzpicture}
    \label{fig:muone_xbox}
    }
\caption{The one-loop vertex, box, and crossed-box corrections to $\mu$-$e$ scattering. The arrows together with the corresponding labels fix the momentum routing.
}
\end{figure}

In the following, we discuss in detail how the collinear massification constant $Z$ of~\eqref{eq:massification} can be determined at one loop. The same methods turn out to be relevant also for the collinear limit discussed in Chapter~\ref{chap:coll}. Since the result is universal we are free to choose any process for the computation. We pick $\mu$-$e$ scattering
\begin{align}
	e(p_1) \mu(p_2) \to e(p_3) \mu(p_4)
\end{align}
and consider the one-loop vertex correction to the electron line shown in Figure~\ref{fig:muone_triangle}. It is permissible to restrict to this subset of corrections since it is gauge invariant. In fact, it exactly corresponds to $\mathcal{A}_{2,0}^{(1)}$ of~\eqref{eq:oneloop_split}. The scale hierarchy \eqref{eq:massification_scaling} reads
\begin{equation}
	m^2 \sim \lambda^2 \ll S
\end{equation}
with $S$ corresponding to any of the Mandelstam invariants $s=(p_1+p_2)^2$, $t=(p_1-p_3)^2$, and $u=(p_1-p_4)^2$.

Since the topology of the diagram exactly corresponds to the integral discussed in Example~\ref{ex:mor_mom} we can directly deduce that there are only three regions that contribute to the squared amplitude at one-loop
\begin{equation}
	\mathcal{M}_n^{(1)}(m)
	= 2 \text{Re} \big\{ A_n^{(1)} {A_n^{(0)}}^\dagger \big\}
	=\mathcal{M}_n^{(1),\text{hard}}+\mathcal{M}_n^{(1),p_1\text{-coll}}+\mathcal{M}_n^{(1),p_3\text{-coll}}
	  + \mathcal{O}(\lambda)\, .
\end{equation}
In addition to the diagram of Figure~\ref{fig:muone_triangle} we also need to take into account UV renormalisation. While the wave function renormalisation of the incoming electron enters $\mathcal{M}^{(1),p_1\text{-coll}}$, the one for the outgoing electron needs to be taken into account in $\mathcal{M}^{(1),p_3\text{-coll}}$. 

There are two possible strategies for this calculation. One could either perform the tensor reduction first and subsequently expand the resulting scalar integrals as well as the corresponding coefficients. This approach, however, has several disadvantages. First of all, the scale hierarchy is not exploited in regards of the complexity of the tensor reduction. In more involved applications this can be a significant disadvantage since the tensor reduction can be a formidable task. Secondly, the tensor reduction can produce artificial poles in $\lambda$ in the coefficients of the scalar integrals that will later cancel requiring the higher-order expansion of the integrals. Lastly and most importantly, the tensor reduction can mix various momentum regions preventing the correct identification of the corresponding contributions. For our specific example, this can be seen easily by observing that the full tensor reduction yields the scalar integral
\begin{align}
A_0 = \int [\d \ell] \frac{1}{[\ell^2-m^2]}
\end{align}
where it is not possible to determine whether this contributes to the $p_1$- or the $p_3$-collinear region. All of this can be avoided if the amplitude is expanded in the momentum regions before applying the tensor reduction. The main disadvantage of this approach is that the tensor reduction has to be performed with linear propagators which is not always supported in commonly available tools such as \texttt{Package-X}.\footnote{This issue can be avoided by reintroducing a subleading $\ell^2$ term in the linear propagator and re-expanding again after tensor reduction.}

With the second approach and based on the discussion in Example~\ref{ex:mor_mom} regarding the relationship of the hard region with the massless theory, it is easy to see that
\begin{equation}\label{eq:massification_oneloop_hard}
	\mathcal{M}_n^{(1),\text{hard}} = \mathcal{M}_n^{(1)}(m=0)\, .
\end{equation}
Furthermore, an explicit calculation yields for the collinear contributions the simple result
\begin{equation}\label{eq:massification_oneloop_coll}
	\mathcal{M}^{(1),p_1\text{-coll}}
	= \mathcal{M}^{(1),p_3\text{-coll}}
	= Z^{(1)} \mathcal{M}_n^{(0)}(m=0)
\end{equation}
with
\begin{equation}\label{eq:massification_constant}
  	Z^{(1)} =
    	\frac{\alpha(4\pi)^\epsilon}{4\pi\Gamma(1-\epsilon)}
    	\Big(\frac{\mu^2}{m^2}\Big)^\epsilon
    	\Big(
    	\frac{2}{\epsilon^2}
    	+\frac{1}{\epsilon}
    	+3+2\zeta_2
    	\Big)
    	+\mathcal{O}(\epsilon)\, .
\end{equation}
This is the result in the four-dimensional helicity scheme (\ac{FDH})~\cite{Gnendiger:2017pys}. A more general form that can also be used for other major flavours of dimensional regularisation is given in~\eqref{eq:jetfunction_general}. Taking~\eqref{eq:massification_oneloop_hard} and~\eqref{eq:massification_oneloop_coll} together we indeed observe that the factorisation formula~\eqref{eq:massification} is satisfied with $Z = 1 + Z^{(1)} + \mathcal{O}(\alpha^2)$. In particular, there is no soft contribution and thus $\mathcal{S}=1+\mathcal{O}(\alpha^2)$.

It is useful to investigate what happens if also box contributions are taken into account in addition to the vertex diagram of Figure~\ref{fig:muone_triangle}. They correspond to a different gauge invariant subset and are therefore not needed to determine the process independent massification constant. As a consequence, these additional contributions are not expected to contribute to the collinear region.  It turns out that this is indeed the case due to a cancellation between the box and the crossed box diagrams depicted in Figures~\ref{fig:muone_box} and \ref{fig:muone_xbox}. This cancellation can be understood without the need of explicit computations. As indicated in the figures the only difference between the two diagrams is the momentum flowing through the electron line propagator that changes from $\ell+p_3$ to $-\ell+p_1$. Since we neglect polynomially suppressed electron mass effects, the two contributions are therefore related via $p_3 \leftrightarrow -p_1$ combined with an overall change of sign.\footnote{The transformation applied to the spinors has no effect at the level of the squared amplitude since it changes the sign twice.} Because the collinear contributions only depend on the collinear scale $m^2$ this transformation only changes the sign and thus results in the aforementioned cancellation. As a consequence, these box diagrams only modify the process-dependent hard part of the factorisation formula~\eqref{eq:massification}. The same cancellation is again encountered in Chapter~\ref{chap:coll} when expanding radiative amplitudes in the limit of collinear photon emission.

\subsection{Massification at two loop}\label{sec:massification_twoloop}

\begin{figure}
        \centering
        \input{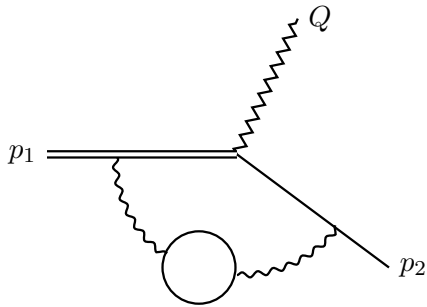}
        \caption{Closed fermion loop correction to the muon decay at two loop. It is the only diagram that gives rise to a soft contribution as well as the factorisation anomaly.}
        \label{fig:muondecay_fermionloop}
\end{figure}

The two-loop contribution to the massification constant, $Z^{(2)}$, was first calculated based on the heavy quark form factor with the aim of massifying the massless two-loop amplitude for Bhabha scattering~\cite{Becher:2007cu,Penin:2005eh}. In order to be able to use the same approach for $\mu$-$e$ scattering we have extended this formalism to the case of heavy external states in \cite{Engel:2018fsb} based on the muon decay
\begin{equation}
	\mu(p_1) \to e(p_2)\bar{\nu}\nu\, .
\end{equation}
In what follows we summarise the most important aspects of this calculation. The explicit result for $Z^{(2)}$ can be found in~(3.12) of \cite{Engel:2018fsb}.

At two loop, a momentum region is given by a pair of on-shell scalings
\begin{align}\label{eq:2loop_scaling}
	\ell_1 \sim (\lambda^{a_1},\lambda^{b_1},\lambda^{(a_1+b_1)/2})_i\, , \quad
	\ell_2 \sim (\lambda^{a_2},\lambda^{b_2},\lambda^{(a_2+b_2)/2})_i
\end{align}
with $\ell_1$ and $\ell_2$ the two loop momenta and $i$ indicating the direction of the light-cone decomposition. Beyond one loop there are therefore mixed regions where $\ell_1$ and $\ell_2$ scale differently. In this context it is helpful to consider the massification formula~\eqref{eq:massification} expanded at two loop. In the case of the muon decay, there is only one collinear direction given by the electron momentum $p_2$ and thus $i=2$ in~\eqref{eq:2loop_scaling}. Furthermore, there is no one-loop soft contribution, $\mathcal{S}^{(1)}$, due to the absence of closed fermion loops at this order. We then find 
\begin{align}\label{eq:muondecay_massification}
	\mathcal{M}_n^{(2)}(m)
	=\mathcal{M}_n^{(2)}(0)
	+Z^{(2)}\mathcal{M}_n^{(0)}(0)
	+S^{(2)}\mathcal{M}_n^{(0)}(0)
	+Z^{(1)}\mathcal{M}_n^{(1)}(0)
	+\mathcal{O}(\lambda)\, .
\end{align}
Each individual contribution can now be interpreted in the context of the MoR. The first three terms can be identified with the following pure momentum scalings:
\begin{subequations}
\begin{alignat}{3}
    &\mbox{hard-hard:} 		\quad &\ell_1& \sim (1,1,1)_2,  \quad &\ell_2& \sim (1,1,1)_2 \\
    &\mbox{coll-coll:}          		\quad &\ell_1& \sim (\lambda^2,1,\lambda)_2,  \quad  &\ell_2& \sim (\lambda^2,1,\lambda)_2 \\
    &\mbox{soft-soft}           	\quad &\ell_1& \sim (\lambda,\lambda,\lambda)_2,  \quad  &\ell_2& \sim (\lambda,\lambda,\lambda)_2
\end{alignat}
\end{subequations}
The contribution of the pure collinear region determines the two-loop massification constant $Z^{(2)}$. The last term in~\eqref{eq:muondecay_massification}, on the other hand, corresponds to the following mixed regions:
\begin{subequations}
\begin{alignat}{3}
    &\mbox{coll-hard:}          	\quad &\ell_1& \sim (\lambda^2,1,\lambda)_2,  \quad  &\ell_2& \sim (1,1,1)_2 \\
    &\mbox{hard-coll}           	\quad &\ell_1& \sim (1,1,1)_2,  \quad  &\ell_2& \sim (\lambda^2,1,\lambda)_2
\end{alignat}
\end{subequations}
These regions are thus entirely due to one-loop quantities and can be predicted based on the results from the previous section. This represents a strong consistency check of the calculation. Finally, we remark that the above identified regions are the only ones that are expected to contribute to the amplitude. At the level of individual diagrams, however, additional contributions can emerge that cancel in the full result. In this specific calculation, it turns out that the ultra-soft scaling  $\ell_i \sim(\lambda^2,\lambda^2,\lambda^2)$ gives rise to such an unphysical momentum region.

Many interesting aspects of the calculation occur in the soft-soft region. Since the unrenormalised soft function receives only contributions from the closed fermion loop diagram of Figure~\ref{fig:muondecay_fermionloop}, its projection onto the vector form factor can be compactly written as
\begin{equation}\label{eq:muondecay_soft}
	\mathcal{S}^{(2)} =
	\frac{\alpha^2}{4\pi} \int [\d\ell] 
	\frac{(-2 p_{1,\mu}) (-2 p_{2,\nu}^{(-,2)})}{[\ell^2]^2[2\ell \cdot p_1][2\ell \cdot p_2^{(-,2)}]} \Pi^{\mu\nu}(\ell^2,m^2)
\end{equation}
with the tensorial VP, $\Pi^{\mu\nu}$, defined in~\eqref{eq:hvp_def}. Even though this integral looks very similar to the scaleless integral~\eqref{eq:triangle_soft} it does not vanish. This is due to the occurrence of the VP which depends on the scale $m$. Applying the transformation $\ell \to \kappa \ell$ that we have used to show that~\eqref{eq:hvp_def} is scaleless introduces a non-factorisable dependence on $\kappa$ in $\Pi^{\mu\nu}$.

The integral in~\eqref{eq:muondecay_soft} turns out to be divergent even in dimensional regularisation. The occurrence of such unregulated singularities is a well-known feature of the MoR~\cite{Smirnov:1997gx}. It is most conveniently treated with analytic regularisation by shifting the power of a suitable propagator. In our case we perform the replacement
\begin{equation}\label{eq:analytic_regulatarisation}
	\frac{1}{[2 \ell \cdot p_1 ]} \to  \frac{(-\nu^2)^\eta}{[2 \ell \cdot p_1 ]^{1+\eta}}
\end{equation}
with $\nu$ the scale of the analytic regularisation. At the level of the QED Lagrangian~\eqref{eq:qed_lagrangian} this regularisation entails modifications of the form
\begin{align}
	(i\slashed{\partial}-m_i) \to (i\slashed{\partial}-m_i)^{1+\eta}\, .
\end{align}
While obviously preserving Lorentz invariance this breaks the QED gauge symmetry since the power of the photon field $A_\mu$ in~\eqref{eq:qed_lagrangian} remains unchanged. In order to avoid gauge dependent results it is therefore mandatory to first expand in $\eta$ and only afterwards in $\epsilon$. The soft contribution now contains poles both in $\eta$ and in $\epsilon$. It turns out that similar unregulated integrals also arise in the calculation of the collinear region. These additional $1/\eta$ poles exactly cancel the ones from the soft region and render the combination finite. This is not particularly surprising since the small-mass expansion of a finite result has to be finite.
Despite their cancellation, these divergences leave a significant imprint on the structure of the result. They introduce new logarithms that break the clean factorisation of scales as well as the equivalence of different collinear regions.

This can be understood from a field theoretic perspective in the context of SCET. After the decoupling transformation each sector of the SCET Lagrangian becomes individually equivalent to QCD and therefore invariant under separate Lorentz boosts. This enhanced classical symmetry can, however, be broken by loop corrections. It is precisely this quantum anomaly that enforces the breaking of naive factorisation realised as unregulated divergences in individual regions. In~\cite{Becher:2010tm} this phenomenon is referred to as collinear or factorisation anomaly while~\cite{Chiu:2011qc} calls it the rapidity divergence.

At this point it is interesting to study the case where no heavy external state is present. Taking the muon to be light as well, we replace  $p_1$ with $p_1^{(-,1)}$ in~\eqref{eq:muondecay_soft}. Since $(p_i^{(-,i)})^2=0$ the soft function now only depends on the invariant $p_1^{(-,1)}\cdot p_2^{(-,2)}$. As a consequence, the integral remains invariant under the simultaneous rescaling $p_1^{(-,1)}\to\kappa p_1^{(-,1)}$ and $p_2^{(-,2)}\to\kappa^{-1} p_2^{(-,2)}$. However, after introducing the analytic regulator~\eqref{eq:analytic_regulatarisation} the integrand itself does transform and we find the identity
\begin{align}
	\mathcal{S}^{(2)} = \kappa^{-\eta} \mathcal{S}^{(2)}\, .
\end{align}
The soft contribution is thus scaleless and vanishes in the absence of a heavy external particle.

In summary, the factorisation anomaly in the fermion loops introduces an additional scale dependence in the massification constant and therefore spoils the naive process independence that is expected from the factorisation formula~\eqref{eq:massification_amp}. From the perspective of massification, these fermionic corrections are thus conceptually very difficult to cope with. In addition, the factorisation breaking logarithms induce a larger massification error and thus make the approximation less reliable. On the other hand, fermionic contributions are typically the easiest ones to calculate from the point of view of standard loop integral computations. It is therefore advisable to calculate these diagrams with the full dependence on the masses based on other techniques. This has the added advantage of removing any soft contribution from the factorisation formula~\eqref{eq:massification} rendering massification completely process independent. As discussed in detail in Section~\ref{chap:fermionic}, a semi-numerical approach is well-suited for this kind of calculation.

\subsection{Massification error}\label{sec:massification_error}

Massification can be used to obtain the leading mass effects based on the massless amplitude without the need of explicit computations. In particular, it determines both the mass logarithms as well as the constant terms and only neglects polynomially suppressed contributions. The reliability of this approximation obviously depends on the precise form of the scale hierarchy~\eqref{eq:massification_scaling}. It is therefore advisable to study the massification error case by case. Nevertheless, we can consider the muon decay as a first reference for the reliability of the approximation.

\begin{figure}
    \centering
    \includegraphics[width=.9\textwidth]{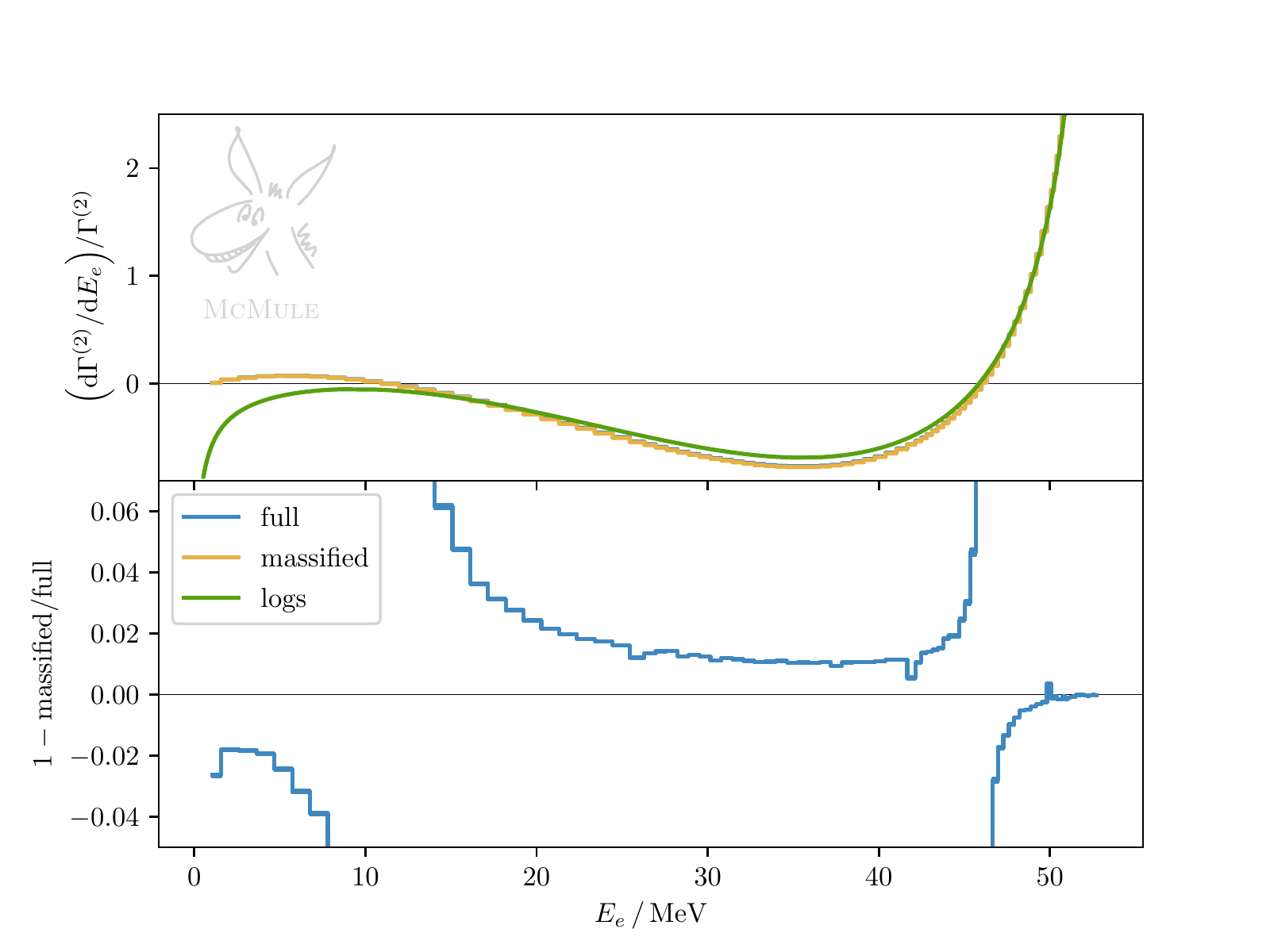}
    \caption{The photonic NNLO correction to the muon decay electron energy spectrum in the logarithmic and massified approximation as well as with full mass dependence. The massification error is shown in the lower panel.}
\label{fig:massification_error}
\end{figure}

The muon decay allows for an exact determination of the massification error since the NNLO corrections are known with full mass dependence. The corresponding calculation for the electron energy spectrum has been performed numerically some time ago~\cite{Anastasiou:2005pn}. More recently, the exact master integrals have been computed analytically in~\cite{Chen:2018dpt}. This has allowed us to perform the corresponding calculation in~\cite{Engel:2019nfw} for the first time fully differentially. Figure~\ref{fig:massification_error} shows the comparison of the exact with the massified result for the energy spectrum. We only include photonic corrections here since the fermionic contributions can be calculated semi-numerically with full mass dependence. We also show the purely logarithmic approximation from~\cite{Arbuzov:2002pp,Arbuzov:2002cn} in order to assess the impact of the constant terms. The upper panel depicts the NNLO correction to the differential decay width normalised to the corresponding integrated quantity. The effect of the constant terms is clearly visible. The lower panel shows the relative difference between the massified and the exact result. The divergent behaviour at $10\,\text{MeV}$ and $45\, \text{MeV}$ is due to zero-crossings.

Massification thus provides a significantly better approximation than the purely logarithmic result.  In this case, the massification error amounts to $\sim1\%$ in the NNLO coefficient. It is useful to compare this to the expected size of the missing higher order corrections. In~\cite{Engel:2019nfw} we have found the NLO and NNLO correction to be of $\sim 10^{-2}$ and $\sim 10^{-4}$, respectively, in the bulk of the spectrum. A naive extrapolation therefore yields an estimate that is of the same size as the massification error. We can thus conclude for the muon decay that massification gives a highly robust approximation. An even smaller massification error can be expected in processes where the scale hierarchy~\eqref{eq:massification_scaling} is more pronounced.

\section{Infrared subtraction}\label{sec:ir_divergences}

The previous sections were devoted to the calculation of amplitudes and in particular of virtual loop corrections. As discussed in Section~\ref{sec:pert_theory}, this is only the first step in the prediction of measurable cross sections. For this the squared amplitude has to be integrated over the phase space according to~\eqref{eq:xsection_unphysical}. However, the virtual corrections still contain $1/\epsilon$ poles -- originating from soft photons -- even after UV renormalisation. Following~\eqref{eq:xsection_phys}, an IR finite and thus measurable observable is obtained when the emission of an arbitrary number of soft photons is taken into account. The soft singularities from virtual loops then cancel the soft divergences in the phase-space integral of real-emission contributions.

In practice, it is non-trivial to obtain this cancellation. The measurement function in the phase-space integral~\eqref{eq:xsection_real} makes an analytic integration impractical and in most cases impossible. We are therefore left with the challenging task of performing a divergent integration numerically. Subtraction schemes are an effective means to cope with this issue. The basic idea is to construct a counterterm that has the same singular behaviour in the soft limit as the amplitude but at the same time is sufficiently simple to be integrated analytically in $d=4-2\epsilon$ dimensions. Schematically, we then write the integral over the one-photon phase space $\D\Phi_\gamma$ as
 \begin{equation}\label{eq:subtr_schematic}
 	\int \d\Phi_\gamma \mathcal{M}_{n+1} \Theta_\text{cuts} =
	\int \d\Phi_\gamma \Big( \mathcal{M}_{n+1}-\mathcal{M}_\text{CT} \Big) \Theta_\text{cuts}
	+ \hat{\mathcal{M}}_\text{CT}\,  \Theta_\text{cuts}
 \end{equation}
 with
 \begin{align}
 	\hat{\mathcal{M}}_\text{CT} = \int \d\Phi_\gamma \mathcal{M}_\text{CT}\, .
 \end{align}
The subtracted integrand is now finite and can be integrated numerically in $4$ dimensions. The soft singularity is explicit as $1/\epsilon$ poles in the integrated counterterm $\hat{\mathcal{M}}_\text{CT}$ which can be combined with the virtual contribution to form a second finite quantity. A crucial point in this procedure is the factorisation of the measurement function from the integrated counterterm which follows from the soft safety condition~\eqref{eq:soft_safety}.
 
In order to construct suitable counterterms the infrared behaviour of both real and virtual amplitudes has to be studied. In the case of soft singularities in QED, the seminal work of Yennie, Frautschi, and Suura (\ac{YFS})~\cite{Yennie:1961ad} provides all the necessary ingredients. They have shown that the soft limit of the $\ell$-loop correction to any radiative squared amplitude simplifies to
\begin{align}\label{eq:soft_limit}
	\lim_{\xi \to 0} \xi^2 \mathcal{M}_{n+1}^{(\ell)}
	= \eik \mathcal{M}_{n}^{(\ell)}
\end{align}
with
\begin{align}
	\xi = \frac{2 E_k}{\sqrt{s}}
\end{align}
and $s$ the CMS energy. The \textit{eikonal factor}
\begin{align}\label{eq:eikonal}
	\eik = - \sum_{ij} Q_i Q_j \frac{p_i \cdot p_j}{(n \cdot p_i)(n \cdot p_j)}\, ,
	\quad\quad 
	\text{with}
	\quad\quad
	k = n\, \xi\, ,
\end{align}
sums over all pairs of external legs with charges $Q_i$ and momenta $p_i$. Furthermore, they have proven that the virtual soft singularities exponentiate as
\begin{equation}\label{eq:soft_exp}
	e^{\ieik} \sum_{\ell=0}^\infty \mathcal{M}_n^{(\ell)}  
	= \mathcal{M}_n^{(0)} + \Big( \mathcal{M}_n^{(1)} + \ieik \mathcal{M}_n^{(0)} \Big) + \mathcal{O}(\alpha^2)
	= \sum_{\ell=0}^\infty \mathcal{M}_n^{(\ell)f}
	= \text{finite}\, .
\end{equation}
The exponential subtracts all soft singularities in $\mathcal{M}_n^{(\ell)}$ rendering $\mathcal{M}_n^{(\ell)f}$ finite. The integrated eikonal $\ieik$ is related to the eikonal factor $\eik$ through integration over the photon phase space analogous to the definition of $\hat{\mathcal{M}}_\text{CT}$ in~\eqref{eq:subtr_schematic}. Since this integration is done in $d$ dimensions the soft singularity is transformed to a $1/\epsilon$ pole. The explicit form of the integrated eikonal $\ieik$ can be found in~\cite{Frederix:2009yq} or in Appendix~A of~\cite{Engel:2019nfw}.

In \cite{Engel:2019nfw} we have exploited this simple IR structure to extend the \ac{FKS} subtraction scheme~\cite{Frixione:1995ms,Frederix:2009yq} initially
developed for NLO QCD calculations to FKS$^\ell$, a subtraction scheme for QED calculations at all orders in perturbation theory. The starting point for this construction is the LP soft limit~\eqref{eq:soft_limit}. The following section presents a modern proof of this all-order statement based on the MoR. This will serve as a basis for the study of the next-to-leading power (\ac{NLP}) soft behaviour in Chapter~\ref{chap:nts}. The FKS$^\ell$ subtraction scheme is presented in Section~\ref{sec:fks}.

\subsection{Soft limit}\label{sec:yfs}

In the following, we consider the behaviour of generic radiative QED amplitudes in the soft limit. We assume the energy of the emitted photon, $E_k$, to be small compared to all other scales $S$ in the process
\begin{align}\label{eq:hierarchy_soft}
	E_k \sim \lambda_s \ll S
\end{align}
with $\lambda_s$ the soft power-counting parameter. At tree level, it is straightforward to derive the soft limit in a process independent way and the corresponding proof can be found in any standard QFT textbook. When loop corrections are taken into account, the derivation is complicated by loop momentum regions that disturb the naive scale hierarchy. We can use the MoR to disentangle these contributions. Because there is no collinear scale present in~\eqref{eq:hierarchy_soft}, only hard and soft modes are expected to contribute.\footnote{From a formal EFT point of view the physics in this limit is governed by heavy quark effective theory (HQET).} In the following, we calculate these regions at LP in a generic way and show that this reduces to the YFS formula~\eqref{eq:soft_limit}.

\paragraph{Hard}
Let us start with the calculation of the purely hard region where all loop momenta scale as
\begin{align}
	\ell_i \sim \lambda_s^0\, .
\end{align}
We split the hard contribution up into external and internal emission according to
\begin{align}\label{eq:hard_split}
    \mathcal{A}_{n+1}^\text{hard} =
    \sum_i \Bigg(
    \begin{tikzpicture}[scale=.8,baseline={(1,-.1)}]
    	\input{tikz/yfs_external}
    \end{tikzpicture}
    \hspace{0.185cm} \Bigg)
    +
    \begin{tikzpicture}[scale=.8,baseline={(1,-.1)}]
    	\input{tikz/yfs_internal}
    \end{tikzpicture}
    =
    \mathcal{A}_{n+1}^\text{ext} + \mathcal{A}_{n+1}^\text{int}.
\end{align}
The grey blobs for $\Lambda$, $\Gamma^\text{ext}$, and $\Gamma^\text{int}$ symbolise all possible one-particle irreducible (\ac{1PI}) insertions in the diagram. In the case of $S_F$, it denotes the mass renormalised and resummed fermion propagator. Since all loop momenta are hard, there are no $1/k$ poles in $\Gamma_\text{int}$ and thus $\mathcal{A}_{n+1}^\text{int}\sim \lambda_s^0$. In case of the hard region we can therefore ignore internal emission at LP. The contribution from external emission is given by
\begin{subequations}\label{eq:yfs_hard}
\begin{align}	
	\mathcal{A}_{n+1}^\text{ext}
	&= \sum_i \Gamma^\text{ext}(p_i-k)S_F(p_i-k)\Lambda^\mu(p_i,k) u(p_i) \epsilon_\mu \\
	&= \frac{1}{\lambda_s}\sum_i \Gamma^\text{ext}(p_i) \frac{i Z_2}{\slashed{p_i}-\slashed{k}-m_i}
	   Z_1^{-1} (i Q_i \gamma^\mu) u(p_i)\epsilon_\mu + \mathcal{O}(\lambda_s^0) \\
	&= \frac{1}{\lambda_s}\sum_i \Gamma^\text{ext}(p_i) \frac{i}{\slashed{p_i}-\slashed{k}-m_i}
	   (i Q_i \gamma^\mu) u(p_i)\epsilon_\mu + \mathcal{O}(\lambda_s^0) \\
	&=  \frac{1}{\lambda_s}\sum_i Q_i \frac{p_i\cdot\epsilon}{k\cdot p_i}\mathcal{A}_n + \mathcal{O}(\lambda_s^0)\, . \label{eq:yfs_amplitude}
\end{align}
\end{subequations}
Going from the first to the second line, we have inserted the vertex and fermion field strength renormalisation constants $Z_1$ and $Z_2$. In particular, we have used the on-shell renormalisation conditions
\begin{subequations}
\begin{align}
	\Lambda^\mu(p_i,k)&= Z_1^{-1}(iQ_i\gamma^\mu) +\mathcal{O}(\lambda_s), \\
	S_F(p_i-k) &= \frac{1}{\lambda_s}\frac{i Z_2}{\slashed{p_i}-\slashed{k}-m_i}+\mathcal{O}(\lambda_s^0)\, ,
\end{align}
\end{subequations}
where the scaling in $\lambda_s$ of the fermion propagator $S_F$ follows from
\begin{align}
	\frac{1}{\slashed{p_i}-\slashed{k}-m_i}
	= \frac{\slashed{p_i}-\slashed{k}+m_i}{k^2-2k\cdot p_i}\, .
\end{align}

Higher-order corrections to these renormalisation factors then cancel as a consequence of the Ward identity $Z_1=Z_2$. Squaring this result and summing over spins and polarisations already yields the full YFS formula~\eqref{eq:soft_limit}. We are therefore left to show that at LP the soft contribution vanishes to all orders in perturbation theory.

\paragraph{Soft}
We consider the soft regions where one or more loop momenta satisfy the soft scaling relation
\begin{align}
	\ell_i \sim \lambda_s\, .
\end{align}
In this case, internal emission is not necessarily subleading due to the presence of the additional soft scales. This is different from the hard momentum region. However, the loop propagator structure is significantly simpler here. Only the propagator
\begin{align}\label{eq:prop_scale}
	\Big(\sum_{j} \ell_j + p_i - k\Big)^2-m_i^2
	= \lambda_s \Big( \sum_{j} 2\ell_j\cdot p_i - 2k\cdot p_i \Big)
	+ \mathcal{O}(\lambda_s^2)
\end{align}
where $p_i^2=m_i^2$ is not homogeneous in the loop momenta. Consequently, the soft loop integral is scaleless if not at least one propagator of this form is present. This, in turn, is only the case if the following conditions are satisfied. At least one soft virtual photon has to attach to the emitting leg. In addition to that, hard loops are only allowed to connect to the fermion line after this soft attachment as well as after the emission of the on-shell photon. In all other cases, a hard scale is introduced that prevents the occurrence of the propagator~\eqref{eq:prop_scale}. Diagrammatically, every non-zero soft contribution can therefore be represented by
\begin{align}\label{eq:soft_rep_yfs}
    \mathcal{R}_i^{m,r} =
    \begin{tikzpicture}[scale=.8,baseline={(1,0)}]
    	
    \draw[line width=.3mm] (-2.1,0) node[left]{$p_i$} -- (1,0);
    \draw[line width=.3mm]  [tightphoton] (-.9,0) -- (-.9,1.2) node[right]{$k$};
    \draw[line width=.3mm] (1,0)  -- (2,1);
    \draw[line width=.3mm] (1,0)  -- (2,-1);

    \centerarc [line width=0.3mm, tightphoton](.1,0)(0:180:.7);
    \centerarc [line width=0.3mm, tightphoton](.25,0)(0:180:.55);
    
    \centerarc [line width=0.3mm, tightphoton](-.5,0)(0:-180:1.4);
    \centerarc [line width=0.3mm, tightphoton](-.35,0)(0:-180:1.25);
    \centerarc [line width=0.3mm, tightphoton](-.2,0)(0:-180:1.1);

    \draw[line width=.3mm]  [fill=stuff] (1,0) circle (0.5) node[]{$\Gamma$};

    \draw [decorate, decoration = {brace},line width = 0.3mm] (-2,.2) --  (-1.3,.2);
    \draw (-1.65,.6) node[] {$m\times$};

    \draw [decorate, decoration = {brace},line width = 0.3mm]  (-.1,-.2) -- (-.7,-.2);
    \draw (-.4,-.6) node[] {$r\times$};

    \end{tikzpicture}
\end{align}
where $m$ and $r$ indicate the number of soft attachments before and after the emission of the external soft photon. We denote the total number of attachements by $n=m+r>0$. Note that there can be an arbitrary number of soft and hard loops in $\Gamma$. In principle, one also has to take into account hard corrections to the vertices and the propagators of the emitting line. However, they cancel at LP by virtue of the Ward identity in analogy to~\eqref{eq:yfs_hard}. The full soft contribution can then be written as
\begin{align}\label{eq:full_soft}
	\mathcal{S} = \sum_i  \sum_n \sum_{m+r=n} \mathcal{R}_i^{m,r}\, .
\end{align}
In the following, we show that a cancellation occurs among all $\mathcal{R}_i^{m,r}$ with $i$ and $n$ fixed. In particular, this results in
\begin{align}
	\sum_{m+r=n} \mathcal{R}_i^{m,r} = 0
\end{align}
and thus in the vanishing of the soft contribution
\begin{align}
	\mathcal{S} = 0\, .
\end{align}

The numerator of $\mathcal{R}_i^{m,r}$ simplifies at LP to
\begin{align}
	\mathcal{N}_i^{m,r}
	= 2^{m+r+1}\Gamma^{\mu_1...\mu_{m+r}} u(p_i) 
	  p^{\mu_1}...p_i^{\mu_{m+r}} p_i \cdot \epsilon
	\equiv
	2^{n+1} \mathcal{N}_i^{n=m+r}\, ,
\end{align}
where we have indicated the independence on the individual values for $m$ and $r$. The LP denominator reads
\begin{align}
	 \mathcal{D}_i^{m,r} 
	 = 2^{n+1}
	   \underbrace{
	   \Big[ \sum_j^{m+r} \ell_j\cdot p_i - k \cdot p_i \Big] …}_{r}
	   \Big[ \sum_j^{m} \ell_j\cdot p_i - k\cdot p_i \Big]
	   \underbrace{
	   \Big[ \sum_j^{m} \ell_j\cdot p_i \Big] ...
	   \Big[ \ell_1\cdot p_i \Big]}_{m}\, .
\end{align}
The partial fraction identity
\begin{align}\label{eq:partialfraction_yfs}
	\Big[ \sum_j^{m} \ell_j\cdot p_i - k \cdot p_i \Big]^{-1}
	\Big[ \sum_j^{m} \ell_j\cdot p_i \Big]^{-1}
	=
	\frac{1}{k\cdot p_i} \Bigg(
	\Big[ \sum_j^{m} \ell_j\cdot p_i - k \cdot p_i \Big]^{-1}
	-
	\Big[ \sum_j^{m} \ell_j\cdot p_i \Big]^{-1}
	\Bigg)
\end{align}
then allows us to write for $m\neq0$ and $r\neq 0$
\begin{align}\label{eq:softrep_split}
	\mathcal{R}_i^{m,r} 
	= \frac{\mathcal{N}_i^n}{k\cdot p_i}
	  \Big(\frac{1}{\tilde{\mathcal{D}}^{m,r}}
	  -\frac{1}{\tilde{\mathcal{D}}^{m+1,r-1}}\Big)
	\equiv \tilde{\mathcal{R}}_i^{m,r} - \tilde{\mathcal{R}}_i^{m+1,r-1} 
\end{align}
with
\begin{align}
	\tilde{\mathcal{D}}^{m,r}
	&=  \Big[ \sum_j^{m+r} \ell_j\cdot p_i - k \cdot p_i \Big] ... \,
	   \Big[ \sum_j^{m} \ell_j\cdot p_i - k\cdot p_i \Big] \,
	   \Big[ \sum_j^{m-1} \ell_j\cdot p_i \Big] ...
	   \Big[ \ell_1\cdot p_i \Big]\, .
\end{align}
Furthermore, we have
\begin{align}\label{eq:softrep_special}
	\mathcal{R}_i^{m,0} 
	= \tilde{\mathcal{R}}_i^{m,0} + \text{scaleless}, \quad\quad
	\mathcal{R}_i^{0,r} 
	= - \tilde{\mathcal{R}}_i^{1,r-1}\, ,
\end{align}
where the former only holds up to a contribution that is scaleless.

Inserting~\eqref{eq:softrep_split} and~\eqref{eq:softrep_special} in~\eqref{eq:full_soft} we find
\begin{subequations}
\begin{align}
	\sum_{m+r=n} \mathcal{R}_i^{m,r}
	&=  \mathcal{R}_i^{0,n} +  \mathcal{R}_i^{1,n-1} + … +  \mathcal{R}_i^{n-1,1} +  \mathcal{R}_i^{n,0} \\
	&= -\tilde{\mathcal{R}}_i^{1,n-1} + (\tilde{\mathcal{R}}_i^{1,n-1} - \tilde{\mathcal{R}}_i^{2,n-2})
	+ … + (\tilde{\mathcal{R}}_i^{n-1,1} - \tilde{\mathcal{R}}_i^{n,0}) + \tilde{\mathcal{R}}_i^{n,0} \\
	&= 0\, .
\end{align}
\end{subequations}
The total soft contribution thus vanishes due to various cancellations among neighbouring summands. This shows that the hard contribution~\eqref{eq:yfs_amplitude} is not modified by soft corrections. This concludes the proof for the YFS soft limit~\eqref{eq:soft_limit}.

\subsection{The FKS$^\ell$ subtraction scheme}\label{sec:fks}

This section presents FKS$^\ell$, a QED subtraction scheme for soft divergences to all orders in perturbation theory. While we restrict to the main concepts here, a much more detailed discussion can be found in~\cite{Engel:2019nfw}. The basic idea of FKS$^\ell$ is to construct the counterterms based on the YFS formula~\eqref{eq:soft_limit} that we have proven in the previous section. This is done by means of the identity
\begin{equation} \label{eq:fks_split}
	\xi^{-1-2\epsilon}
	= -\frac{\xi_c^{-2\epsilon}}{2\epsilon} \delta(\xi) + \Big( \frac{1}{\xi} \Big)_c + \mathcal{O}(\epsilon)
\end{equation}
where the distribution $(1/\xi)_c$ acts on a test function $f(\xi)$ as 
\begin{equation}
	\int_0^{\xi_\text{max}} \Big(\frac{1}{\xi} \Big)_c f(\xi)
	= \int_0^{\xi_\text{max}} \d\xi \frac{f(\xi)-f(0)\theta(\xi_c-\xi)}{\xi}\, .
\end{equation}
The upper bound on the photon energy is given by 
\begin{align}
	\xi_\text{max} = 1-\frac{(\sum_i m_i)^2}{s}
\end{align}
where the sum is over all masses in the final state. The dependence on the parameter $\xi_c$ then cancels between the two terms on the r.h.s of~\eqref{eq:fks_split} as long as $0<\xi_c\leq\xi_\text{max}$. Within this constraint, the value for the unphysical $\xi_c$ parameter can be chosen arbitrarily. Making explicit the energy dependence of the photon phase space
\begin{align}
	\d\Phi_\gamma^{d=4-2\epsilon} = \xi^{1-2\epsilon}\D\Upsilon^{d=4-2\epsilon}\, ,
\end{align}
we can use \eqref{eq:fks_split} to write
\begin{subequations}
\begin{align}
	\int \d\Phi_\gamma^{d=4-2\epsilon} \mathcal{M}_{n+1}
	&=  \int \d\Upsilon^{d=4-2\epsilon} \d\xi\, \xi^{-1-2\epsilon} \Big( \xi^2 \mathcal{M}_{n+1} \Big) \\
	&=  -\frac{\xi_c^{-2\epsilon}}{2\epsilon}\int \d\Upsilon^{d=4-2\epsilon}  \eik \mathcal{M}_n
	    + \int  \d\Upsilon^{d=4-2\epsilon} \d\xi\, \Big( \frac{1}{\xi} \Big)_c \Big(\xi^2 \mathcal{M}_{n+1}\Big) \\
	 \label{eq:fks_schematic}
	 &= \ieik(\xi_c) \mathcal{M}_n
	    + \int  \d\Phi_\gamma^{d=4} \Big( \frac{1}{\xi} \Big)_c \Big( \xi \mathcal{M}_{n+1}\Big) \, .
\end{align}
\end{subequations}
For simplicity, we have omitted the measurement function, $\Theta_\text{cuts}$, in these expressions. In the second line we have used both the FKS split~\eqref{eq:fks_split} as well as the YFS soft limit formula~\eqref{eq:soft_limit}. As a consequence, the corresponding term becomes sufficiently simple to allow for an analytic integration over the photon phase space, effectively replacing the eikonal factor $\eik$ with the integrated eikonal $\ieik$. The counterterm is now in a form where it can be combined with the virtual contribution cancelling the soft $1/\epsilon$ poles analytically. Furthermore, the distributional subtraction renders the second term in~\eqref{eq:fks_schematic} finite\footnote{The integrand still has an integrable $1/\sqrt{\xi}$ divergence.} and therefore suitable for numerical integration in $4$ dimensions.

This leads us directly to the formulation of the original FKS scheme (restricted to soft singularities) where the physical cross section is divided into the two separately finite contributions
\vspace{-.4cm}
\begin{subequations}
\label{eq:fks_nlo}
\begin{align}
\sigma^{(1)} 
&= \sigma^{(1)}_n(\xc)  + \sigma^{(1)}_{n+1}(\xc)\, , \\
\sigma^{(1)}_n(\xc) &= \int\!
   \D\Phi_n^{d=4}\,\bigg(
    \M n1
   +\ieik(\xc)\,\M n0
\bigg) =
\int\! \D\Phi_n^{d=4}\, \fM n1(\xc)\, ,
\\
\sigma^{(1)}_{n+1}(\xc) &= \int\!
 \D\Phi^{d=4}_{n+1}
  \cdis{\xi} \Big(\xi \, \fM{n+1}0 \Big)\, .
\end{align}
\end{subequations}
The dependence on the unphysical parameter $\xi_c$ cancels exactly between the two pieces. This property of the subtraction scheme is immensely useful to verify both the correctness of the implementation as well as the reliability of the numerical integration. We will perform this check explicitly in Chaper~\ref{chap:cps} when studying the numerical problems arising from CPS.

Because of the simple exponentiating structure of soft singularities~\eqref{eq:soft_exp}, it is possible to extend this scheme to any order in perturbation theory. For example, at NNLO we find
\begin{subequations}
\label{eq:fks_nnlo}
\begin{align}
\begin{split}
\sigma^{(2)}_n(\xc) &= \int\!
   \D\Phi_n^{d=4}\,\bigg(
    \M n2
   +\ieik(\xc)\,\M n1
   +\frac1{2!}\M n0 \ieik(\xc)^2
\bigg) =
\int\! \D\Phi_n^{d=4}\, \fM n2(\xc)
\, ,
\end{split}\label{eq:fks_nnlo:n}
\\
\sigma^{(2)}_{n+1}(\xc) &= \int\!
 \D\Phi^{d=4}_{n+1}
  \cdis{\xi_1} \Big(\xi_1\, \fM{n+1}1(\xc)\Big)\label{eq:fks_nnlo:n1}
\, ,\\
\sigma^{(2)}_{n+2}(\xc) &= \int\!
  \D\Phi_{n+2}^{d=4}
   \cdis{\xi_1}\,
   \cdis{\xi_2}\,
     \Big(\xi_1\xi_2\, \fM{n+2}0\Big) \label{eq:fks_nnlo:n2}\, .
\end{align}
\end{subequations}
The twofold distributional subtraction in~\eqref{eq:fks_nnlo:n2} comes from applying the FKS split~\eqref{eq:fks_split} to $\xi_1^{-1-2\epsilon} \xi_2^{-1-2\epsilon}$ simultaneously. This introduces two a priori different $\xi_c$ parameters and generates three additional terms. The product of the two delta functions gives the double soft limit resulting after integration over both photon phase spaces in the third term in~\eqref{eq:fks_nnlo:n}. The factor $1/2!$ is due to the indistinguishability of the two photons and prevents a double counting in the integration over the phase space. The two other contributions yield complicated process-dependent integrals. However, if all the $\xi_c$ parameters are chosen equal, the contributions can be combined with the subtracted real-virtual contribution $(1/\xi_1)_c \mathcal{M}_{n+1}^{(1)}$ which still contains explicit $1/\epsilon$ poles. This combination cancels the pole analytically and replaces $\mathcal{M}_{n+1}^{(1)}$ with its finite version $\mathcal{M}_{n+1}^{(1)f}$. No process dependent integrals are therefore left in~\eqref{eq:fks_nnlo}. This can be viewed as a direct consequence of the YFS exponentiation whose build up can be nicely observed in~\eqref{eq:fks_nnlo:n}. Furthermore, it is the main reason for the striking simplicity of the FKS$^\ell$ subtraction scheme.

\chapter{Hyperspherical method}\label{chap:fermionic}

For many processes it is helpful to split the radiative corrections into \textit{fermionic} and \textit{photonic} contributions. The former are corrections that include at least one closed fermion loop while the latter take into account all other contributions. 
As discussed in Section~\ref{sec:pert_theory}, the production of an additional fermion-antifermion pair corresponds to a physically distinguishable final state and does not have to be considered here. In particular, it does not have to be taken into account to define IR safe quantities since fermions are considered to be massive. The reason for this split is that fermionic contributions have many special features that make a separate treatment useful and often unavoidable.

The fermionic corrections can be further subdivided into a leptonic and a hadronic part that are conceptually very different. Contrary to the leptons, the hadronic degrees of freedom become non-perturbative at low energies. These non-perturbative hadronic effects are thus important for low-energy experiments and have to be properly accounted for. A fully analytic calculation is therefore not possible in this case since one has to rely on experimental input that captures the non-perturbative contribution. This is of course closely related to the discussion of Section~\ref{sec:muone} about the non-perturbative hadronic corrections to the anomalous magnetic moment of the muon and the MUonE experiment.

The leptonic corrections, on the other hand, can be calculated from first principle with perturbation theory. While this calculation is often simpler than for photonic corrections the determination of the full-mass dependence at two loop is still a formidable task. In many cases, one would therefore have to rely on massification. As explained in Section~\ref{sec:massification_twoloop} it is exactly these contributions that give rise to the factorisation anomaly. This makes the corresponding massification conceptually difficult. Even more importantly, however, the anomaly results in additional large logarithms that hamper the reliability of the approximation. It is therefore not advisable to use massification for leptonic corrections.

The best approach is therefore a simultaneous treatment of both leptonic and hadronic contributions. The simplest fermionic corrections are due to insertions of the QED vacuum polarisation (\ac{VP}) tensor~\eqref{eq:hvp_def} defined as
\begin{align}\label{eq:vp_definition}
	i\Pi^{\mu\nu}(q)
	=i\Pi(q^2)(g^{\mu\nu}q^2-q^\mu q^\nu)
	=i\Pi(q^2) t^{\mu\nu}
	= -\int \D^4x e^{i q x} \langle 0 | T\{ j_\text{em}^\mu(x) j_\text{em}^\nu(0) | 0 \rangle
\end{align}
where $j_\text{em}(x)=\sum_i Q_i \bar{\psi}_i(x) \gamma^\mu \psi_i(x)$ is the electromagnetic current and the sum runs over fermions with charge $Q_i$. This definition varies in the literature up to an overall sign. We have chosen our convention such that the resummed photon propagator is given by
\begin{align}
	\frac{-i t^{\mu\nu}}{q^2 \big( 1-\Pi(q^2)\big)}
	= \frac{-i t^{\mu\nu}}{q^2}\Big( 1+\Pi(q^2)+\Pi(q^2)^2+… \Big)
\end{align}
We can take all fermions into account by including them in the renormalised VP
\begin{align}
	\Pi = \Pi_e + \Pi_\mu + \Pi_\tau + \Pi_\text{had}\, .
\end{align}
For many processes, and in particular for $\mu$-$e$ scattering, VP corrections are the only fermionic contributions at NNLO. We will therefore restrict the following discussion to this case. Nonetheless, we keep in mind that for other processes such as $e^+ e^- \to \gamma \gamma$ and for higher orders more complicated corrections occur such as light-by-light scattering (\ac{LbL}).

The one-loop contribution to the leptonic VP can be calculated straightforwardly and the two-loop result can be extracted from~\cite{Djouadi:1993ss}. The corresponding formulas are given in Appendix~\ref{sec:vacuum_polarisation}. For the non-perturbative hadronic VP (HVP), one has to rely on experimental data that capture the non-perturbative degrees of freedom. As already discussed in Section~\ref{sec:muone}, it is possible to relate the HVP to the hadronic $R$-ratio by means of the dispersive integral~\eqref{eq:hvp_dispersive}. In order to cover the full non-perturbative regime multiple experiments are needed that run at individual hadron production thresholds. This, in turn, requires a careful analysis and combination of these experiments. Specialised tools exists in this case such as \texttt{alphaQED}~\cite{Jegerlehner:2001ca,Jegerlehner:2006ju,Jegerlehner:2011mw} and \texttt{HVPTools}~\cite{Davier:2010rnx}.

\begin{figure}
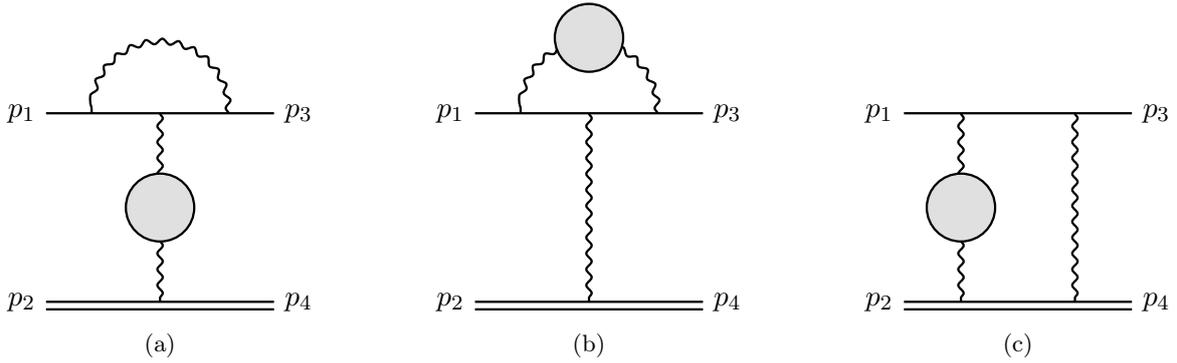

    \centering
    \subfloat[]{
    \begin{tikzpicture}[scale=1,baseline={(0,0)}]
        \input{tikz/muone_fermionic_factorisable}
    \end{tikzpicture}
    \label{fig:muone_fermionic_factorisable}
    }
    \hspace{1cm}
    \subfloat[]{
    \begin{tikzpicture}[scale=1,baseline={(0,0)}]
        \input{tikz/muone_fermionic_vertex}
    \end{tikzpicture}
    \label{fig:muone_fermionic_vertex}
    }
    \hspace{1cm}
    \subfloat[]{
    \begin{tikzpicture}[scale=1,baseline={(0,0)}]
        \input{tikz/muone_fermionic_box}
    \end{tikzpicture}
    \label{fig:muone_fermionic_box}
    }
\caption{Three sample diagrams of VP corrections to $\mu$-$e$ scattering at NNLO. In the case of the diagram (a), the VP factorises from the one-loop vertex correction and can therefore easily be calculated. The non-factorisable diagrams shown in (b) and (c) are more difficult. They can be calculated either dispersively or with the hyperspherical method.}
\label{fig:muone_fermionic}
\end{figure}

For many contributions, the VP factorises from the rest of the diagram. Such a case is shown in Figure~\ref{fig:muone_fermionic_factorisable} for $\mu$-$e$ scattering. In this case the correction reduces to quantities that have already been computed at NLO. The diagrams depicted in Figures~\ref{fig:muone_fermionic_vertex} and~\ref{fig:muone_fermionic_box} are much more complicated to calculate. These non-factorisable VP contributions have the form
\begin{equation}\label{eq:fermionic_nonfactorisable}
	\mathcal{M}_\text{NF}
	= \int [\D \ell]\, \frac{\Pi(\ell^2)}{\ell^2} \tilde{\mathcal{M}}_\text{NF}(\ell,p_i)
\end{equation}
where $\tilde{\mathcal{M}}_\text{NF}$ denotes the rest of the diagram and the loop momentum routing is chosen such that the VP does not depend on any of the external momenta $p_i$.

In addition to relating the hadronic VP to experimental data, the dispersive integral~\eqref{eq:hvp_dispersive} can also be used to calculate these non-factorisable contributions. To do so, we replace the VP in~\eqref{eq:fermionic_nonfactorisable} with the expression from~\eqref{eq:hvp_dispersive} and exchange the order of the two integrals. This \emph{dispersive approach}~\cite{PhysRev.124.1577} is thus based on the master formula
\begin{align}\label{eq:disperive_method_master}
	\mathcal{M}_\text{NF}^\text{had}
	= \frac{\alpha}{3\pi} \int_{4m_\pi^2}^\infty
	  \frac{\D z}{z} R_\gamma^\text{had}(z)  \langle \tilde{\mathcal{M}}_\text{NF} \rangle_\ell(z)\, .
\end{align}
The kernel of the dispersive integral is given by the original amplitude with the VP replaced by a photon of mass $z$,
\begin{align}\label{eq:dispersive_kernel}
	\langle \tilde{\mathcal{M}}_\text{NF} \rangle_\ell(z)
	= \int [\D \ell] 
	\frac{\tilde{\mathcal{M}}_\text{NF}}{\ell^2-z+i \delta}\, ,
\end{align}
and can therefore easily be computed with existing one-loop tools such as \verb|Package-X|. This approach was used to calculate the hadronic corrections to the muon decay~\cite{vanRitbergen:1998hn,Davydychev:2000ee} and Bhabha scattering~\cite{Actis:2007fs,Kuhn:2008zs,CarloniCalame:2011zq}. The dispersive formula~\eqref{eq:disperive_method_master} can of course also be used for leptons with $m_\pi \to m_\ell$ and the $R$-ratio replaced by the corresponding analytic expression for leptons
\begin{align}
	R_\gamma^\ell(z)
	= \Bigg( 1+\frac{2m_\ell^2}{z} \Bigg) \sqrt{1-\frac{4m_\ell^2}{z}}\, .
\end{align}
While this dispersive method is rather easy to use in principle, the numerical integration over $z$ is non-trivial. This is especially true for the hadronic contribution where the integration crosses narrowly peaked resonances in the time-like region. 

To avoid this issue one can resort to the \emph{hyperspherical method}. As explained in detail in the following, this approach casts the non-factorisable amplitude into the form\footnote{We will see in Section~\ref{sec:cont_physical} that this is not always possible.}
\begin{align}\label{eq:hyperspherical_masterformula}
	\mathcal{M}_\text{NF} 
	= -\int_0^\infty \D Q^2 \Pi(-Q^2) \langle \tilde{\mathcal{M}}_\text{NF} \rangle_\Omega (Q^2)
\end{align}
where the kernel $\langle \tilde{\mathcal{M}}_\text{NF} \rangle_\Omega$ is obtained by analytically integrating over the hyperspherical solid angle
\begin{align}
	\langle \tilde{\mathcal{M}}_\text{NF} \rangle_\Omega(Q^2)
	= \int \frac{\D \Omega}{2\pi^2} \tilde{\mathcal{M}}_\text{NF}\, .
\end{align}
Originally, this method was used in~\cite{Levine:1974xh,Levine:1975jz} to calculate the photonic three-loop corrections to the anomalous magnetic moment of the muon, an important contribution to $\alpha_\mu^\text{QED}$ in~\eqref{eq:g2_split}. In this calculation the remaining radial integration was performed analytically. The idea to use the same approach for HVP corrections, where the radial integration is done numerically, was only recently realised in~\cite{Fael:2018dmz} in the context of the MUonE experiment. As explained in detail in Section~\ref{sec:muone} the MUonE experiment aims at measuring the HVP in the space-like region. Contrary to the dispersive approach, the hyperspherical method provides a means to directly use this data to predict the subleading HVP corrections and therefore enables a completely independent extraction. The following discussion is heavily based on \cite{Fael:2018dmz} with the extension of the method to other processes in mind.

Analogously to standard loop calculations, the hyperspherical method is divided into the following steps: tensor reduction, loop integration, and UV/IR subtraction. For the reduction to scalar integrals the VP in~\eqref{eq:fermionic_nonfactorisable} can be ignored for the most part as long as
\begin{itemize}
	\item
	no loop momentum shifts are performed and
	\item
	scaleless integrals are not set to zero.
\end{itemize}
Shifts in the loop momentum would introduce in~\eqref{eq:fermionic_nonfactorisable} a dependence of the VP on the external momenta. The second restriction is due to the VP acting as a scale in the integrals. Apart from these two points, standard one-loop techniques can be used. All other steps are less trivial and are covered one-by-one in the following Sections~\ref{sec:loop_integration}, \ref{sec:uv_renormalisation}, and~\ref{sec:ir_subtraction}. The numerical integration over the radial variable $Q^2$ is then discussed in Section~\ref{sec:numerical_integration}. Finally, we comment in Section~\ref{sec:dispersive_vs_hyperspherical} on the advantages and disadvantages of the hyperspherical method compared to the more traditional dispersive approach.

\section{Loop integration}\label{sec:loop_integration}

As in the case of standard analytic loop integration (Section~\ref{sec:loop}), the loop momentum has to be transformed to a Euclidean vector in a first step. However, the presence of the VP in~\eqref{eq:fermionic_nonfactorisable} significantly complicates this procedure. We begin with the corresponding analytic continuation in Section~\ref{sec:cont_euclidean}. The subsequent analytic integration over the hyperspherical angles is presented in Section~\ref{sec:int_hyperspherical}. This yields the kernel function $\langle \tilde{\mathcal{M}}_\text{NF} \rangle_\Omega$ of~\eqref{eq:hyperspherical_masterformula} in the Euclidean region. The delicate continuation to the physical region is then discussed in Section~\ref{sec:cont_physical}. These three steps are illustrated with the two integrals
\begin{subequations}\label{eq:hyperspherical_examples}
\begin{align}
	I^\text{NF}_\triangle
	&= \int [\D\ell]
	\frac{\Pi(\ell^2)}{[\ell^2][(\ell+p_1)^2-m^2][(\ell+p_3)^2-m^2]}\, , \\
	I^\text{NF}_\square
	&= \int [\D\ell]
	\frac{\Pi(\ell^2)}{[\ell^2][(\ell+p_1)^2-m^2][(\ell+p_1-p_3)^2][(\ell-p_2)^2-M^2]}
\end{align}
\end{subequations}
corresponding to the diagrams shown in Figure~\ref{fig:muone_fermionic_vertex} and~\ref{fig:muone_fermionic_box}, respectively.

\subsection{Analytic continuation to the Euclidean region}
\label{sec:cont_euclidean}

\begin{figure}
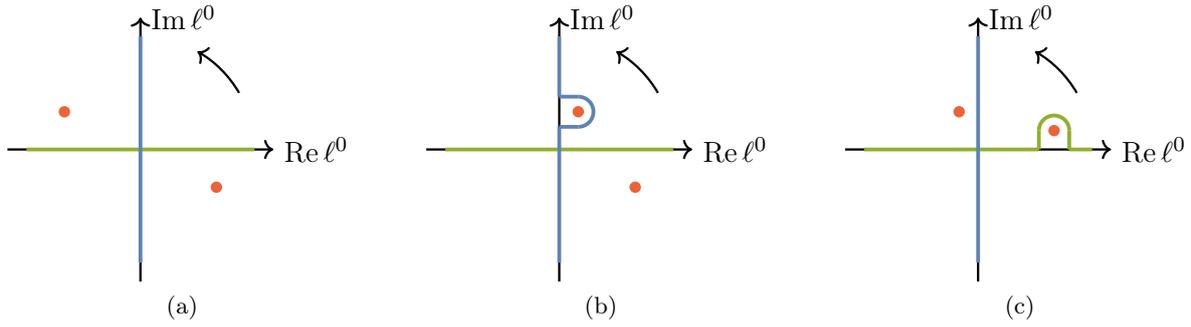

    \centering
    \subfloat[]{
    \begin{tikzpicture}[scale=1,baseline={(0,0)}]
        \input{tikz/wickrotation_normal}
    \end{tikzpicture}
    \label{fig:wickrotation_normal}
    }
    \hspace{0.5cm}
    \subfloat[]{
    \begin{tikzpicture}[scale=1,baseline={(0,0)}]
        \input{tikz/wickrotation_hyperspherical}
    \end{tikzpicture}
    \label{fig:wickrotation_hyperspherical}
    }
    \hspace{0.5cm}
    \subfloat[]{
    \begin{tikzpicture}[scale=1,baseline={(0,0)}]
        \input{tikz/wickrotation_hyperspherical_euclidean}
    \end{tikzpicture}
    \label{fig:wickrotation_hyperspherical_euclidean}
    }
\caption{Illustration of the pole structure during the Wick rotation. Figure~(a) shows the situation in the usual case. Figures (b) and (c) illustrate the location of the poles in the presence of VP before and after analytic continuation of the external momenta to the Euclidean region.}
\label{fig:cont_euclidean}
\end{figure}

As mentioned in Section~\ref{sec:loop} it is possible to transform Minkowskian loop momenta to Euclidean ones by means of a \textit{Wick rotation}. In a first step, the integration is mapped to the imaginary axis by virtue of the residue theorem as depicted in Figure~\ref{fig:wickrotation_normal}. In the standard case, it is possible to use loop momentum shifts to cast the integral to the simple form~\eqref{eq:loop_standard} where the corresponding poles
\begin{align}
	\pm \sqrt{|\vec{\ell}|^2+\Delta} \mp i \delta
\end{align}
do not interfere with the rotation. The poles are shown as red dots in Figure~\ref{fig:wickrotation_normal}. Next, the loop momentum is replaced by its Euclidean version via
\begin{align}
	\ell^0 \to i Q^0
\end{align}
implying
\begin{align}\label{eq:wick_rot}
	\ell^2 = \big(\ell^0\big)^2 -|\vec{\ell}|^2 \to - \big(Q^0\big)^2 - |\vec{Q}|^2=- Q^2 \, .
\end{align}

In the presence of the VP, however, the momentum routing has to be fixed to ensure that the VP itself does not depend on external momenta. We therefore have to take into account the poles of each propagator
\begin{align}
	\mathcal{P}_i = (\ell-p_i)^2-m_i^2+i\delta
\end{align}
 separately. They have the form 
\begin{align}
	p_i^0  \pm \sqrt{|\vec{\ell}-\vec{p}_i|^2+m_i^2} \mp i \delta
\end{align}
with $p_i$ and $m_i$ the external momentum and the mass of the propagator. For sufficiently large $p_i^0$ one pole enters the upper right quadrant, deforming the integration path after Wick rotation as depicted in Figure~\ref{fig:wickrotation_hyperspherical}. This can be avoided by analytically continuing the external lines to the Euclidean region via a rotation in the complex plane
\begin{align}\label{eq:wick_external}
	p_i^0 \to e^{i \phi} P_i^0
\end{align}
with $\phi \in [0,\pi/2]$. As a result we obtain a straight contour after the Wick rotation as shown in Figure~\ref{fig:wickrotation_hyperspherical_euclidean}. At the end point of the analytic continuation, $\phi = \pi/2$, the external momentum is completely Euclidean with
\begin{align}\label{eq:wick_external_euclidean}
	p_i^2 = \big(p_i^0\big)^2-|\vec{p_i}|^2 \to -\big(P_i^0\big)^2-|\vec{P_i}|^2=- P_i^2
\end{align}
in analogy to~\eqref{eq:wick_rot}. Being completely Euclidean at this point, we introduce hyperspherical coordinates with
\begin{align}
	\D^4 \ell \to i\D^4 Q = i\frac{Q^2}{2}\D Q^2 \D \Omega_Q\, .
\end{align}
This allows us to write the two integrals~\eqref{eq:hyperspherical_examples} as
\begin{align}\label{eq:radial_integration}
	I = -\int_0^\infty \D Q^2\, \Pi(-Q^2) \langle I \rangle_\Omega
\end{align}
with
\begin{subequations}
\begin{align}
	\langle I^\text{NF}_\triangle \rangle_\Omega
	&= \int \frac{\D\Omega_Q}{2\pi^2}
	\frac{(-1)^2}{[(Q+P_1)^2+m^2][(Q+P_3)^2+m^2]}
	\label{eq:euclidean_triangle}\, , \\
	\langle I^\text{NF}_\square \rangle_\Omega
	&= \int \frac{\D\Omega_Q}{2\pi^2}
	\frac{(-1)^3}{[(Q+P_1)^2+m^2][(Q+P_1-P_3)^2][(Q-P_2)^2+M^2]}
	\label{eq:euclidean_box}\, .
\end{align}
\end{subequations}
The external momenta are now all space-like and an analytic continuation back to the time-like region has to be performed eventually. This is discussed in detail in Section~\ref{sec:cont_physical} after the calculation of the kernels $\langle I \rangle_\Omega$ in the following Section~\ref{sec:int_hyperspherical}.

\subsection{Hyperspherical integration}
\label{sec:int_hyperspherical}

An elegant way to perform the hyperspherical integration is by means of Gegenbauer polynomials $C_n^{(\alpha)}$. They are defined via the generating function
\begin{align}
	\frac{1}{(1-2xt+t^2)^\alpha}
	\equiv \sum_{n=0}^\infty C_n^{(\alpha)}(x) t^n\, .
\end{align}
This allows us to express propagators in terms of Gegenbauer polynomials. In particular, we find
\begin{align}
	\frac{1}{(Q+P_i)^2+m_i^2}
	= \frac{Z_i}{|Q|\, |P_i|}
	\sum_{n=0}^{\infty} Z_i^n C_n^{(1)}(-\hat{Q} \cdot \hat{P}_i)
\end{align}
with
\begin{align}\label{eq:problematic_roots}
	Z_i = \frac{Q^2+P_i^2+m_i^2-\sqrt{\lambda(Q^2,P_i^2,-m_i^2)}}
		{2|Q|\, |P_i|}\, .
\end{align}
We have defined the normalised unit vectors $\hat{Q}=Q/|Q|$ and $\hat{P}_i=P_i/|P_i|$. The Källén function is denoted by $\lambda$. The Gegenbauer polynomials form an orthogonal basis of functions over the interval $[-1,1]$ w.r.t. the weight function $\sqrt{1-x^2}$. As a consequence, they satisfy the orthogonality relation
\begin{align}\label{eq:gegenbauer_orthogonality}
	\int \frac{\D\Omega_Q}{2\pi^2}
	C_n^{(1)}(\hat{Q}\cdot\hat{P_i}) C_m^{(1)}(\hat{Q}\cdot\hat{P_j})
	= \frac{\delta_{nm}}{n+1} C_n^{(1)}(\hat{P}_i\cdot\hat{P_j})\, .
\end{align}
This property makes the integration over the hyperspherical angles straightforward.

Applying this to $I^\text{NF}_\triangle$, we find
\begin{align}
	\langle I^\text{NF}_\triangle \rangle_\Omega
	= \frac{1}{Q^2|P_1|\, |P_3|}
	\sum_{n=0}^\infty \frac{(Z_1 Z_3)^{n+1}}{n+1} C_n^{(1)}(\hat{P_1}\cdot\hat{P_3})\, .
\end{align}
The remaining infinite sum of Gegenbauer polynomials can be evaluated by integrating
\begin{align}
	\frac{\D}{\D z} \sum_{n=0}^\infty \frac{z^{n+1}}{n+1} C_n^{(1)}(x)
	= \sum_{n=0}^\infty z^n C_n^{(1)}(x)
	= \frac{1}{1-2xz+z^2}
\end{align}
with the boundary condition
\begin{align}
	\sum_{n=0}^\infty \frac{z^{n+1}}{n+1} C_n^{(1)}(x) \, \Bigg|_{z=0} = 0\, .
\end{align}
We find
\begin{align}
	\langle I^\text{NF}_\triangle \rangle_\Omega
	= \frac{1}{Q^2|P_1|\,|P_3| \sqrt{1-\tau^2}}
	\arctan\Big(\frac{Z_1 Z_3 \sqrt{1-\tau^2}}{1-Z_1 Z_3 \tau}\Big)
\end{align}
with $\tau=\hat{P}_1\cdot\hat{P}_3$. Using the functional relation
\begin{align}
	\arctan(iz) 
	= \frac{i}{2} \log\Big(\frac{1+z}{1-z}\Big) \equiv i L(z)
\end{align}
this result can be transformed to
\begin{align}\label{eq:triangle_kernel_sol}
	\langle I^\text{NF}_\triangle \rangle_\Omega
	= \frac{1}{Q^2\sqrt{t(t-4m^2)}}
	\Big( L(z_1) - 2 L(z_2) \Big)
\end{align}
with
\begin{subequations}
\begin{align}
	z_1 &= \frac{\sqrt{t(t-4m^2)}}{2m^2-t}\, , \\
	z_2 &= \sqrt{1+\frac{4m^2}{Q^2}} \sqrt{1-\frac{4m^2}{t}}\, .
\end{align}
\end{subequations}
In this last step we have applied the on-shell conditions
\begin{align}
	P_i = -p_i^2 = -m^2, 
	\quad\quad (P_1-P_3)^2 = -(p_1-p_3)^2 = -t
\end{align}
which follows from~\eqref{eq:wick_external_euclidean}. This replacement is a delicate step in the calculation since it implicitly assumes that $P_i^2<0$. We are, however, still in the Euclidean region where $P_i^2>0$. In the above step we have therefore performed the corresponding analytic continuation without justification. This is discussed in detail in the next section.

The kernel function $\langle I^\text{NF}_\triangle \rangle_\Omega$ has to be real in the physical time-like region where $t<0$. However, the solution given in \eqref{eq:triangle_kernel_sol} has a non-vanishing imaginary part since $z_2>1$ and $0<z_1<1$. The above calculation therefore does not yield the correct imaginary part. Fortunately, this is unproblematic since the kernel multiplies in~\eqref{eq:radial_integration} the VP evaluated in the space-like domain and thus a manifestly real quantity. As a consequence, the imaginary part of the kernel does not impact the real part of the squared amplitude and does therefore not enter in the calculation of physical observables.

Let us turn to the box integral $I^\text{NF}_\square$. As can be seen from~\eqref{eq:euclidean_box}, we have three angular dependent propagators in this case instead of just two as for $I^\text{NF}_\triangle$. As a consequence, the orthogonality relation~\eqref{eq:gegenbauer_orthogonality} is not applicable here. Instead, the integration over the hyperspherical angles has to be done explicitely after Feynman parametrisation. This calculation was done in~\cite{Laporta:1994mb}. The corresponding result for a general box integral is given in Appendix~\ref{sec:hyperspherical_masterintegrals}. In the particular case considered here where 
\begin{align}
	P_1^2=-m^2,\quad\quad P_2^2=-M^2, \quad\quad (P_1+P_2)^2=-s
\end{align}
we find
\begin{align}\label{eq:box_kernel_sol}
	\langle I^\text{NF}_\square \rangle_\Omega
	= - \frac{1}{Q^2 |Q^2+t| \sqrt{\lambda(s,m^2,M^2)}}
	\Big( L(z_3) +  L(z_4) - L(z_5) \Big)
\end{align}
with
\begin{subequations}
\begin{align}
	z_3 &= \sqrt{1+\frac{4m^2}{Q^2}}
	      \sqrt{1-\frac{4sm^2}{(s-M^2+m^2)^2}} \text{sign}(Q^2+t)\, , \\
	z_4 &= \sqrt{1+\frac{4M^2}{Q^2}}
	      \sqrt{1-\frac{4sM^2}{(s-m^2+M^2)^2}} \text{sign}(Q^2+t)\, , \\
	z_5 &= \sqrt{1-\frac{4m^2M^2}{(s-M^2-m^2)^2}}\, .
\end{align}
\end{subequations}
Again, the analytic continuation to the physical region needs justification. As in the case of $I^\text{NF}_\triangle$ this is postponed to the next section.

In addition to the box kernel, Appendix~\ref{sec:hyperspherical_masterintegrals} also gives the corresponding results for a generic bubble and triangle integral. This thus provides all master kernels required for non-factorisable VP contributions of $2 \to 2$ processes at two loop.

\subsection{Analytic continuation to the physical region}
\label{sec:cont_physical}

\begin{figure}
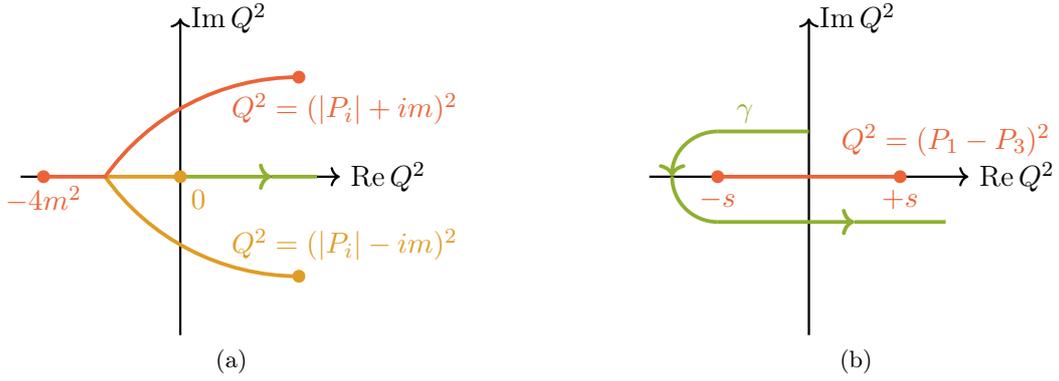

    \centering
    \subfloat[]{
    \begin{tikzpicture}[scale=1.2,baseline={(0,0)}]
        \input{tikz/branchpoint_ok}
    \end{tikzpicture}
    \label{fig:branchpoint_ok}
    }
    \hspace{2cm}
    \subfloat[]{
    \begin{tikzpicture}[scale=1.2,baseline={(0,0)}]
        \input{tikz/branchpoint_oha}
    \end{tikzpicture}
    \label{fig:branchpoint_oha}
    }
\caption{Analysis of the branchpoints when analytically continuing the kernel functions from the Euclidean to the physical region. The behaviour in (a) is unproblematic since the two branchpoints (orange/red) do not cross the integration path (green) of the radial integral along the positive real axis. Figure~(b), on the other hand, requires a deformation of the integration path to avoid the branch point.}
\label{fig:branchpoints}
\end{figure}

The final results for the kernel functions given in the previous section are a priori only valid in the Euclidean region. An analytic continuation to the physical region is required. This step is complicated by the fact that the continuation is performed for the full radial integral~\eqref{eq:radial_integration} and not just for the kernels. No branch points are thus allowed to cross the integration path. The square root $\lambda^{1/2}(Q^2,P_i^2,-m^2)$ in~\eqref{eq:problematic_roots} is problematic in this regard. Figure~\ref{fig:branchpoint_ok} shows the behaviour of the two branch points when varying $P_i^2$ from $P_i^2>0$ to its physical value $P_i^2=-m^2<0$. The positive real axis remains untouched. In the case of $I^\text{NF}_\triangle$ we can therefore conclude that the analytic continuation is unproblematic. The same analysis also applies to $I^\text{NF}_\square$ since we only encounter the additional possibly dangerous square root $\lambda^{1/2}(Q^2,(P_1-P_3)^2,0)$ in the corresponding calculation. In this case, the Euclidean region $(P_1-P_3)^2=-t>0$ already corresponds to the physical region where $t<0$.

It is still interesting to investigate what happens in the case where $t>0$. This is the situation for the box integral of the crossed s-channel process $e^+e^-\to\mu^+\mu^-$ where $t\to s>4m^2>0$. Figure~\ref{fig:branchpoint_oha} depicts the branch point behaviour in this scenario where it clearly interferes with the radial integration path. In~\cite{Levine:1974xh} this issue is solved (in a different context) by deforming the integration path as shown in Figure~\ref{fig:branchpoint_oha} in green. This has significant repercussions. Because of the deformation, the radial integral~\eqref{eq:radial_integration} is modified to
\begin{subequations}
\begin{align}
	I &= -\int_{\gamma} \D Q^2\, \Pi(-Q^2) \langle I \rangle_\Omega \\
	  &= -\int_{-s}^0 \D Q^2\, \Pi(-Q^2) \text{Disc}\big( \langle I \rangle_\Omega \big)
	     - \int_0^\infty \D Q^2\, \Pi(-Q^2) \langle I \rangle_\Omega
\end{align}
\end{subequations}
with $\text{Disc}\big( \langle I \rangle_\Omega \big)$ the branch cut discontinuity of the kernel function. Hence, we obtain an additional contribution with the VP evaluated in the time-like region $-Q^2>0$. The main advantage of the hyperspherical method of avoiding hadronic resonances is therefore lost. Furthermore, the VP acquires an imaginary part in this case, requiring the reconstruction of the imaginary part of the kernel functions. As discussed in the previous section, this is non-trivial. These considerations show that the hyperspherical method is only well-suited for $t$- and $u$-channel kinematics. For other processes, such as $e^+ e^- \to \mu^+ \mu^-$, the dispersive approach is clearly advantageous.

\section{UV renormalisation}\label{sec:uv_renormalisation}

In the hyperspherical method, UV divergences manifest themselves as non-integrable kernel functions in the region $Q^2\to\infty$. These singularities are not regulated in any way and have to be subtracted pointwise at the integrand level. In the following, we compute the corresponding wave function and mass counterterm in the hyperspherical approach.

The starting point is the self-energy diagram
\begin{align}
    -i \Sigma =
    \begin{tikzpicture}[scale=.8,baseline={(1,0)}]

	\draw[line width=.3mm]  (-1.5,0) node[left] {$p$} -- (1.5,0);
	\centerarc [line width=0.3mm,photon](0,0)(0:180:.9);
	\draw[line width=.3mm]  [fill=stuff] (0,1) circle (0.4);

	\centerarc [line width=0.3mm,->](0,0)(125:165:1.2);
	\node at (-1.2,.8) {$\ell$};

    \end{tikzpicture}
    = (-ie)^2\int \frac{\D^4\ell}{(2\pi)^4} \, \Pi(\ell^2)
	\frac{\gamma_\mu(\slashed{\ell}+\slashed{p}+m)\gamma^\mu}
	{[\ell^2][(\ell+p)^2-m^2]}\, .
\end{align}
After tensor decomposition this can be expressed as
\begin{align}
	\Sigma
	= \frac{\alpha}{4\pi}\Big\{
	4m S_1-\slashed{p}\Big(1+\frac{m^2}{p^2}\Big) S_1
	+\frac{\slashed{p}}{p^2} (S_2-S_3) \Big\}
\end{align}
with the scalar integrals
\begin{subequations}
\begin{align}
	S_1 &= \int[\D\ell]\, \frac{\Pi(\ell^2)}{[\ell^2][(\ell+p)^2-m^2]}, \\
	S_2 &= \int[\D\ell]\, \frac{\Pi(\ell^2)}{[(\ell+p)^2-m^2]}, \\
	S_3 &= \int[\D\ell]\, \frac{\Pi(\ell^2)}{[\ell^2]}\, .
\end{align}
\end{subequations}
The otherwise scaleless tadpole integral $S_3$ only contributes because of the presence of the VP. Based on the previous section and the master kernels given in Appendix~\ref{sec:hyperspherical_masterintegrals} we can calculate these integrals in the hyperspherical framework. We find
\begin{align}
	S_i = - \int_0^\infty \D Q^2\, \Pi(-Q^2) \langle S_i \rangle_\Omega
\end{align}
with the kernels
\begin{subequations}
\begin{align}
	\langle S_1 \rangle_\Omega 
	&= \frac{Q^2-p^2+m^2+\sqrt{\lambda(Q^2,-p^2,-m^2)}}{2p^2 Q^2}\, , \\
	\langle S_2 \rangle_\Omega &= -Q^2 \langle S_1 \rangle_\Omega\, , \\
	\langle S_3 \rangle_\Omega &= 1\, .
\end{align}
\end{subequations}
It is now straightforward to extract the on-shell counterterms
\begin{align}\label{eq:onshell_const}
	\delta m = - \Sigma(\slashed{p}=m)\, , \quad\quad 
	\delta Z = \frac{\D\Sigma}{\D\slashed{p}}\Bigg|_{\slashed{p}=m}
\end{align}
using the derivative relations
\begin{align}
	\frac{\D S_i(p^2)}{\D\slashed{p}}  = 2\slashed{p} \frac{\D S_i(p^2)}{\D p^2} , \quad\quad
	\frac{\D}{\D\slashed{p}} \frac{\slashed{p}}{p^2} = -\frac{1}{p^2} \, .
\end{align}
The corresponding kernel functions read
\begin{subequations}
\begin{align}\label{eq:counterterms_hyperspherical}
	\langle \delta m \rangle_\Omega
	&= \frac{Q^2(1-\kappa)+2m^2\kappa}{2m^3}\, , \\
	\langle \delta Z \rangle_\Omega
	&= \frac{Q^2(\kappa-1)(1-4\kappa+\kappa^2)
	+2m^2(1-\kappa+3\kappa^2-\kappa^3)}{4m^4\kappa}\, ,
\end{align}
\end{subequations}
with $\kappa=\sqrt{1+4m^2/Q^2}$.

\section{IR subtraction}\label{sec:ir_subtraction}

Even after UV renormalisation the hyperspherical kernel functions can have non-integrable singularities that are of IR origin. In the case of the box integral $I^\text{NF}_\square$ from Section~\ref{sec:loop_integration} this manifests as the $1/|Q^2+t|$ pole in~\eqref{eq:box_kernel_sol}. This singular behaviour of the integral comes from the soft IR divergence arising when the undressed photon from Figure~\ref{fig:muone_fermionic_box} becomes soft. Note on the contrary that the $1/Q^2$ pole does not lead to a singularity because of the on-shell renormalisation condition $\Pi(0)=0$. 

The unregulated divergence for $Q^2=-t>0$ can be dealt with by means of a suitable subtraction. In this specific case we write
\begin{align}\label{eq:ir_subtracted}
	I^\text{NF}_\square
	&= -\int_0^\infty \D Q^2
	\Big( \Pi(-Q^2)-\Pi(t) \Big) \langle I^\text{NF}_\square \rangle_\Omega
	+ \Pi(t) I_\square
\end{align}
where we have expressed the counterterm in terms of the undressed one-loop box integral
\begin{align}
	I_\square
	= \int [\D\ell]
	\frac{1}{[\ell^2][(\ell+p_1)^2-m^2][(\ell+p_1-p_3)^2][(\ell-p_2)^2-M^2]}
	= \int_0^\infty \D Q^2 \langle I^\text{NF}_\square \rangle_\Omega\, .
\end{align}
This makes the analytic calculation of the counterterm possible where the soft divergence is regulated in $d=4-2\epsilon$ dimensions. The first term in~\eqref{eq:ir_subtracted} is finite and can be integrated numerically.

\section{Radial integration}\label{sec:numerical_integration}

\begin{figure}
    \centering
    \subfloat[]{
	\includegraphics[scale=.7]{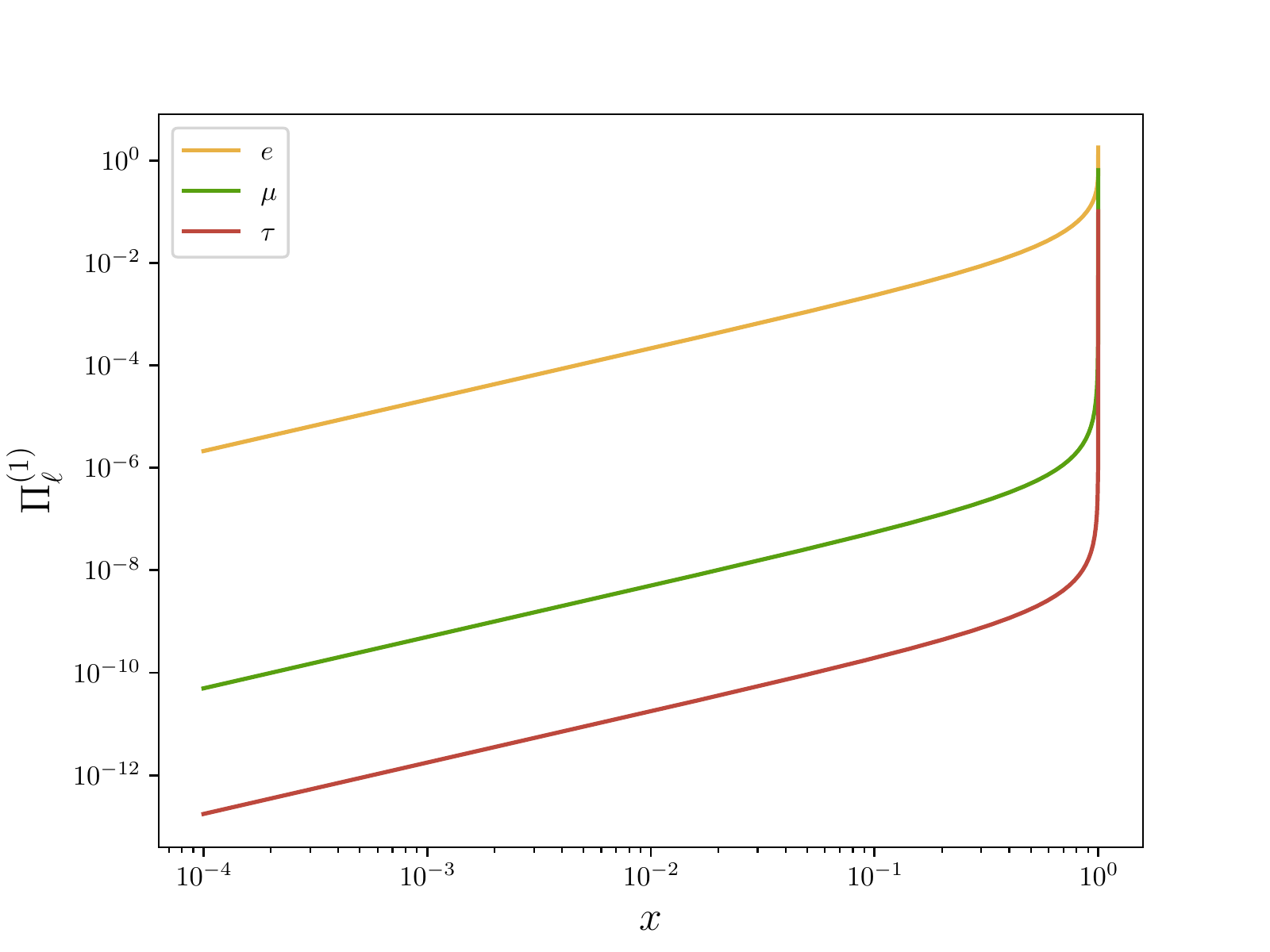}
    \label{fig:vp_normal}
    } \\
    \subfloat[]{
        \includegraphics[scale=.7]{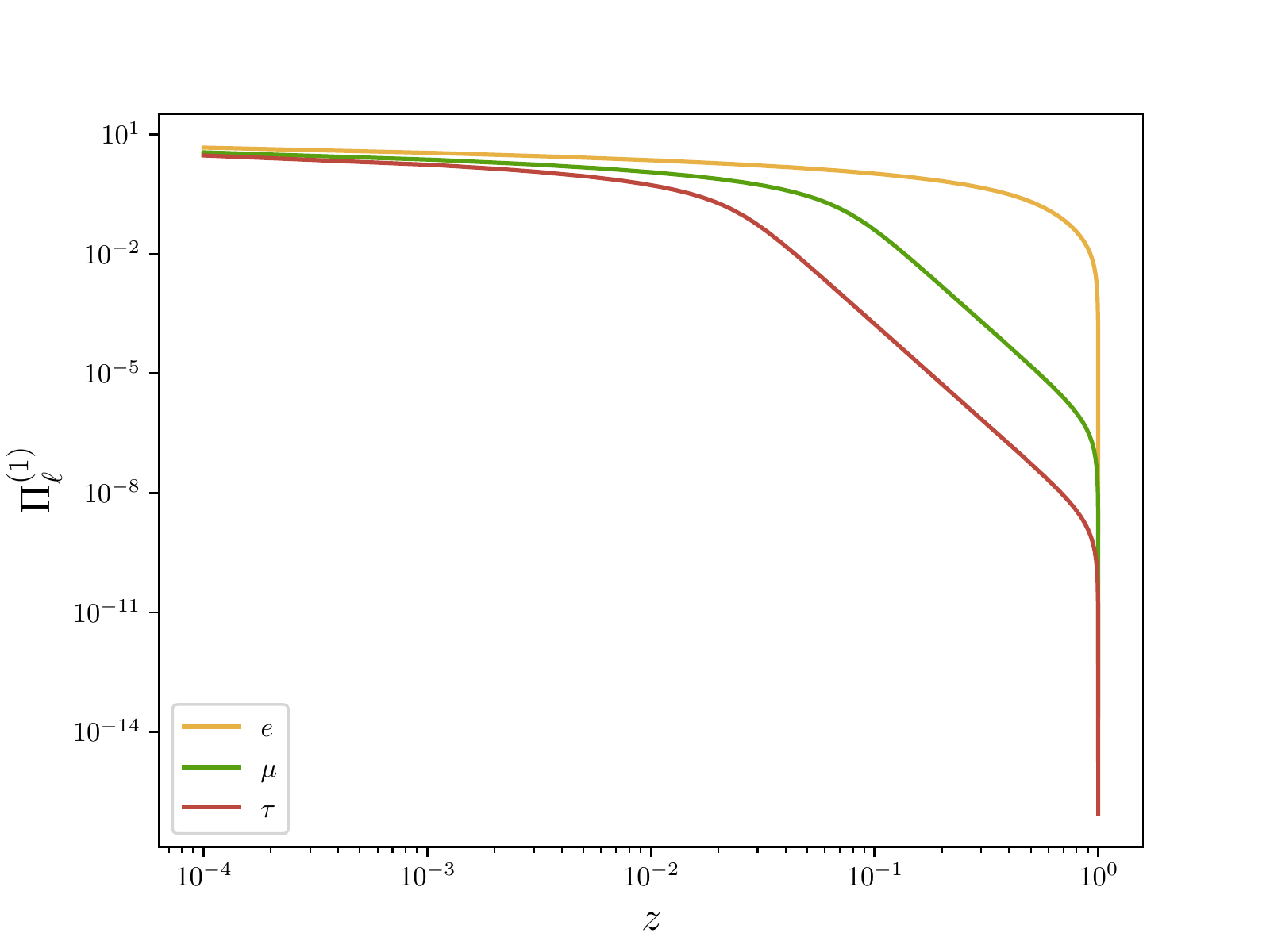}
    \label{fig:vp_sub}
    }
\caption{One-loop VP contribution from the three leptons $\ell=e,\mu,\tau$ evaluated in the space-like domain $\Pi^{(1)}_\ell(-Q^2)$ with $Q^2>0$. Figure~(a) shows the dependence on $x$ where $Q^2=m^2x/(1-x)$. The function is tightly squeezed at $x=1$, especially for the heavier leptons. The much more spread out dependence on $z=(1-x)^{1/5}$ is depicted in (b).}
\label{fig:vp_behaviour}
\end{figure}

With the methodology developed in the previous sections it is possible to write non-factorisable fermionic corrections (for $t$- and $u$-channel processes) in the form~\eqref{eq:hyperspherical_masterformula}. For leptonic contributions an analytic integration over the radial variable $Q^2$ could be envisaged. In the case of hadrons one is forced to solve the above integral numerically because the analytic form of the HVP is not known. It is therefore reasonable to also use the numerical approach for the leptonic part in order to treat all contributions simultaneously.

The radial integration can then conveniently be combined with the phase-space integration that is also done numerically. To do so, we map the integration domain $Q^2\in[0,\infty)$ to the unit interval $x\in[0,1)$ via the change of variables
\begin{align}
	Q^2 = \frac{m^2 x}{1-x}\, .
\end{align}
However, after this transformation the bulk of the VP for the heavier particles is tightly squeezed around $x=1$ as depicted in Figure~\ref{fig:vp_normal}. To facilitate the numerical integration it is therefore advantageous to perform an additional change of variables that zooms into this region. The transformation
\begin{align}
	x=1-z^5
\end{align}
turned out to be well-suited for this as shown in Figure~\ref{fig:vp_sub}.

An additional issue hampering the numerical integration are large cancellations among kernel functions. This significantly complicates the evaluation of the kernel amplitude with a sufficient precision. Particularly delicate in this regard are UV and IR regions where a high numerical precision is required for the respective subtractions to work. In addition to that, there are typically integrable logarithmic threshold singularities left in the kernel functions. One possibility to ensure the numerical stability is a careful expansion around the problematic points. Unfortunately, the corresponding convergence radii are often not sufficiently large for this approach to work. Instead, one can resort to quadruple precision arithmetic to ensure a stable numerical evaluation over the full domain of the radial integration.

We have successfully applied the hyperspherical method to calculate the NNLO VP corrections for the muon decay, M{\o}ller and $\mu$-e scattering, as well as $e^+e^-\to\gamma\gamma$. In the case of the muon decay and $\mu$-$e$ scattering, perfect agreement was found with the dispersive calculations of~\cite{Davydychev:2000ee} and~\cite{Fael:2019nsf}, respectively. The calculation for M{\o}ller scattering was verified by comparing to the exact electron loop result of~\cite{Bonciani:2004gi}. Only in the case of two-photon production no comparison with existing literature was possible. However, the calculation can be easily modified to the case where the VP is replaced by a massive photon
\begin{align}
	\Pi(q^2) \to \frac{q^2}{q^2-\lambda^2}\, .
\end{align}
The calculation reduces to a one-loop correction in this case and can thus be compared with automated tools such as \texttt{Package-X}. \footnote{This test can also be used to check intermediate expressions which can be very helpful.}

Unfortunately, we were not able to calculate the corresponding corrections for $\mu$-pair production ($e^+e^- \to \mu^+\mu^-$) and Bhabha scattering ($e^+e^- \to e^+e^-$) with the hyperspherical approach. The reason ist the non-trivial analytic continuation of the kernels from the Euclidean to the physical region. As discussed in Section~\ref{sec:cont_physical}, this is due to the $s$-channel contribution which is absent in the other processes. It is therefore clearly advantageous to follow the dispersive approach in this case.

\section{Dispersive vs. hyperspherical}\label{sec:dispersive_vs_hyperspherical}

The hyperspherical approach allows for a completely independent determination of the HVP by the MUonE experiment. It is therefore clearly the method of choice in this specific case. For all other applications, the situation is different. In this concluding section we therefore discuss the advantages and disadvantages of the two methods.

One of the main differences between the dispersive and the hyperspherical method is that the dispersive kernel~\eqref{eq:dispersive_kernel} corresponds to a standard one-loop amplitude. As a consequence, many publicly available tools can be used which significantly simplifies the implementation of the method. For example, similar numerical issues as in the hyperspherical method also arise in the dispersive integral~\eqref{eq:disperive_method_master}. In particular for large $z$, a sufficiently precise evaluation of the kernel becomes difficult due to large cancellations. To ameliorate this problem one can rely on the impressive numerical stability of the \texttt{Collier} library~\cite{Denner:2016kdg} to perform the tensor reduction as well as the evaluation of the scalar integrals. For extremely large values of $z$, the amplitude has to be expanded in order to obtain the cancellation analytically. To do so, one can use \texttt{Package-X} for the tensor reduction and the MoR for the expansion of the scalar integrals. Contrary to the hyperspherical method it is thus possible to avoid the use of quadruple precision arithmetic to ensure a sufficiently stable evaluation of the kernel. A significantly faster implementation is thus possible in this case.

Regarding the UV renormalisation the two methods are similar. Also in the case of the dispersive method the renormalisation is performed at the level of the kernel. The corresponding wave function and mass counterterm can be found in Appendix~\ref{sec:counterterms_dispersive}. Regarding the regularisation of IR singularities the two methods are completely different. As discussed in Section~\ref{sec:ir_subtraction}, suitable subtraction terms have to be constructed to render the hyperspherical kernel integrable. In the dispersive method, on the other hand, the soft divergences are automatically regularised in dimensional regularisation. This has clear advantages both from a technical point of view as well as for the numerical stability of the evaluation.

Based on these considerations, it is evident that the dispersive method is better suited for leptonic corrections. In the hadronic case the additional issue of hadronic resonances in the dispersive integral~\eqref{eq:disperive_method_master} has to be taken into consideration. This significantly complicates the numerical integration. Moreover, as discussed in the context of the physics case for the MUonE experiment in Section~\ref{sec:muone}, also the reliability of the experimental data has to be questioned. Once the corresponding high-quality space-like data is available, the hyperspherical method can be used to exploit the smooth behaviour of the hadronic VP in this region. To arrive at a robust prediction of the delicate hadronic contribution it is therefore advisable to implement both methods, yielding a reliable estimate of the uncertainty induced from experimental data.

\chapter{Collinear pseudo-singularities}\label{chap:cps}

After the subtraction of the soft singularities with FKS$^\ell$ as described in Section~\ref{sec:fks} the phase-space integration can be performed numerically. Since the corresponding integral is multi-dimensional, one typically relies on Monte Carlo methods to do so. These methods converge at a rate $\mathcal{O}(N^{-1/2})$ with the number of sampling points $N$ that is independent of the dimension of the integral and the smoothness of the integrand. In addition, Monte Carlo integrators are able to estimate the numerical error reliably even in the case of discontinuous integrands.

In a first step, we bring the $n$-particle phase-space measure
\begin{align}
	\D\Phi_n(P;p_1,...,p_n) 
	= \prod_{i=1}^n \frac{\D^3 p_i}{(2\pi)^3 2 E_i} 
	(2\pi)^4\delta^4\Big(P-\sum_{i=1}^n p_i\Big)
\end{align}
into a form that is suitable for Monte Carlo integration. The momenta and energies of the final-state particles are denoted by $p_i$ and $E_i$, respectively. The delta function enforces the conservation of the total intial-state momentum $P$. A parametrisation on the unit hypercube $x\in[0,1]^{3n-4}$ is required to rewrite the measure as
\begin{align}\label{eq:param}
	\D\Phi_n(P;p_1,...,p_n) = \D^{3n-4}x\, \mathcal{W}(P;x)
\end{align}
with the weight function $\mathcal{W}$ made up of the Jacobian of the parametrisation as well as cuts that ensure the physicality of the momentum configuration.
This gives rise to a mapping (\textit{phase-space generator})
\begin{align}
	x\in[0,1]^{3n-4} \to \big\{\{p_i\}, \mathcal{W} \big\}
\end{align}
that assigns each element on the hypercube a set of momenta $\{p_i\}$ and its weight $\mathcal{W}$. This parametrisation is not uniquely defined. In this chapter we describe how this freedom can be exploited to boost the performance of the Monte Carlo integration. 

Collinear divergences are regularised by finite fermion masses. As already discussed in Section~\ref{sec:pert_theory}, this can be seen from the pole structure of radiative amplitudes given in~\eqref{eq:matel_real}. In the collinear limit we find the behaviour
\begin{align}\label{eq:matel_cps}
	\mathcal{A}_{n+1} 
	\sim \frac{1}{E_i E_k(1-\beta_i  \cos \sphericalangle(p_i,k))}
	\overset{\sphericalangle(p_i,k)\to 0}{\sim} \frac{E_i^2}{m_i^2}\, .
\end{align}
The physical regularisation of collinear divergences leads to a significant simplification of the subtraction procedure. We have exploited this in the construction of the FKS$^\ell$ subtraction scheme discussed in Section~\ref{sec:fks}. However, since the mass $m_i$ is often small compared to the energy $E_i$, radiative amplitudes exhibit narrow peaks as remnants of the collinear singularities. These \emph{collinear pseudo-singularities} (CPS) significantly complicate a reliable numerical integration.

It is therefore important that the Monte Carlo method applied follows an adaptive algorithm. In our case we use \texttt{vegas}~\cite{Lepage:1980jk} that iteratively modifies the sampling grid during the integration. This approach is particularly effective if the parametrisation~\eqref{eq:param} is aligned with the CPS, i.e. if the photon-fermion angle, $\sphericalangle(p_i,k)$, is itself a \texttt{vegas} variable. This allows the algorithm to efficiently refine the integration grid in the collinear region ensuring a reliable estimate of the corresponding contribution to the integral.

It is not possible to find a parametrisation that is aligned with all CPS simultaneously. For this reason it is useful to split the phase space into multiple partitions that each contain one CPS only. This can be done by iteratively multiplying the integrand with 
\begin{align}
	1 = \Theta(c_{ij} \leq \{ c_{kl} \}) + \Theta(c_{ij}> \{c_{kl}\})
\end{align}
where $c_{ij}=p_i\cdot n_j$ and $k_j = \xi_j n_j$ as in~\eqref{eq:eikonal}. This \textit{multi-channeling} approach allows for a dedicated tuning of the phase-space parametrisation for each CPS separately.

Using $n_j$ instead of the full photon momentum $k_j$ ensures that the choice of the optimal phase-space partition is independent of the softness of the photons. Special care has to be taken, however, regarding the IR safety of the individual partitions. In the case where $c_{ij}=c_{kl}$ in the soft limit, the soft counterterm may be in a different partition as the limit. This  breaks the IR finiteness of the subtraction procedure.  Such a situation is encountered in the process $e^+(p_1)e^-(p_2) \to \gamma(k_1) \gamma(k_2) \gamma(k_3)$ where only $k_3$ can become soft. Tuning on the collinear emission of one of the hard photons then results in exactly this scenario since $p_1\cdot n_1 = p_2 \cdot n_2$ in the elastic case. One is therefore forced to introduce hard-coded cutoffs that specify when to switch the partition.

In the case of multiple photon emission amplitudes contain overlapping CPS. A simultaneous optimisation of the phase-space parametrisation could therefore be advantageous. The corresponding phase-space region is, however, much smaller compared to single CPS. As a consequence, it is sufficient to only take into account single CPS at low energies. We therefore restrict the following discussion to this case. Nevertheless, a tuning for overlapping CPS might become necessary when going to higher energies. Also relevant in this regard is an alternative approach to cope with the problem of CPS via a QCD-inspired subtraction scheme. The collinear factorisation of radiative one-loop amplitudes presented in Chapter~\ref{chap:coll} provides an important ingredient for such a procedure at NNLO.

In the following we describe how an optimal parametrisation can be found for single CPS. In the case of inital-state CPS this is straightforward and is discussed in Section~\ref{sec:tuning_isr}. The more delicate final-state tuning is presented in Section~\ref{sec:tuning_fsr}. Finally, we demonstrate the importance of a dedicated phase-space tuning in Section~\ref{sec:tuning_example} by means of a simple NLO example.

\section{Initial-state CPS}\label{sec:tuning_isr}

We specialise here to the case of initial-state CPS and show how an optimal phase-space parametrisation can be obtained. Consider a $(n+l)$-particle phase space with $l$ additionally radiated photons in the final state. The radiative part of the phase space can then be separated as
\begin{align}\label{eq:tuning_isr_start}
	\D\Phi_{n+l}(P;p_1,...,p_n,k_1,...,k_l)
	= \Bigg[ \prod_{i=1}^{l} \frac{\D^3 k_i}{(2\pi)^3 2 E_i} \Bigg]
	  \D\Phi_n(P-K;p_1,...,p_n)
\end{align}
with $K=\sum_i k_i$. We are completely free in the parametrisation of the photons since the momentum conserving $\delta$-function is absorbed in $\D\Phi_n$.\footnote{We restrict to the case $n>1$.} In particular, it is straightforward to find a parametrisation where the angle w.r.t. the beam axis is a variable. This ensures an optimal phase-space generation for initial-state CPS.

\begin{figure}
    \centering
    \begin{tikzpicture}[scale=1,baseline={(1,0)}]
    	\input{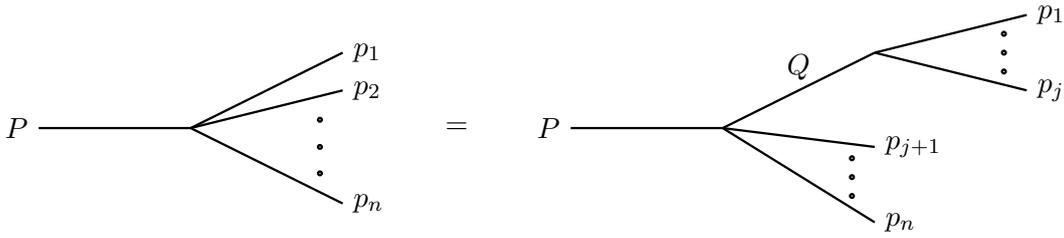}
    \end{tikzpicture}
    \caption{Schematic illustration of the phase-space decomposition formula~\eqref{eq:psfac}.}
    \label{fig:phasespace_decomposition}
\end{figure}

This leaves the remaining $n$-particle phase space without any particular constraint regarding the parametrisation. We therefore follow the most convenient approach based on the factorising property
\begin{align}\label{eq:psfac}
	\D\Phi_n(P;p_1,...,p_n)
	= \frac{1}{2\pi} \D Q^2 \D\Phi_j(Q;p_1,...,p_j) \D\Phi_{n-j+1}(P;Q,p_{j+1},...,p_n)
\end{align}
with $Q=\sum_{i=1}^j p_i$. This can easily be derived by multiplying the phase space with
\begin{subequations}
\begin{align}
	1
	&= \int \frac{\D^4 Q}{(2\pi)^4} (2\pi)^4 \delta^4\Big(Q-\sum_{i=1}^j p_i \Big)
	   \int \frac{\D Q'^2}{2\pi} (2\pi) \delta\Big(Q'^2-Q^2 \Big) \\
	&= \int \frac{\D Q^2}{2\pi} \frac{\D^3 Q}{(2\pi)^3 2 E_Q}
	   (2\pi)^4 \delta^4\Big(Q-\sum_{i=1}^j p_i \Big)\, .
\end{align}
\end{subequations}
A schematic illustration of the phase-space decomposition formula~\eqref{eq:psfac} is shown in Figure~\ref{fig:phasespace_decomposition}. The iterative application of the decomposition then yields the phase space in terms of sequential two-body decays~\cite{Weinzierl:2000wd}
\begin{align}\label{eq:seq_twobody_decay}
	\D\Phi_n 
	= \frac{1}{(2\pi)^{n-2}} \D M_{n-1}^2 ...\D M_{2}^2 \D\Phi_2(n) ... \D\Phi_2(2)
\end{align}
with $M_i=Q_i^2$, $Q_i = \sum_{j=1}^i p_j$, and $\D\Phi_2(i)=\D\Phi_2(Q_i;Q_{i-1},p_i)$. The momenta can now be generated in the respective rest frame of the decay where
\begin{align}
	\D\Phi_2(i) 
	= \frac{1}{(2\pi)^2} \frac{\sqrt{\lambda(Q_i^2,Q_{i-1}^2,m_i^2)}}{8Q_i^2}
	  \D\Omega_i \, \Theta\Big(\sqrt{Q_i^2}>\sqrt{Q_{i-1}^2}\Big) \, .
\end{align}
The $\Theta$-function ensures that sufficient energy is available for the decay. A suitable Lorentz transformation $q_i \to \Lambda q_i$ can then be used to boost the resulting momenta back to a common frame. This does not generate a Jacobian due to the defining property $|\text{det}\Lambda|=1$ of Lorentz transformations.

In the case of initial-state CPS it is therefore possible to decouple the part that requires a specific optimisation from the rest of the phase space. As we will see in the following, this is not possible for final-state CPS, leading to additional complications.

\section{Final-state CPS}\label{sec:tuning_fsr}

Let us consider (without loss of generality) the tuning for the final-state CPS $p_1\cdot k_i$. The approach presented in the previous section is not helpful in this case since the Lorentz boost results in a complicated parametrisation of the final-state fermion momenta. Instead, we write the phase space as
\begin{align}\label{eq:fsr_start}
	\D\Phi_{n+l}(P;p_1,...,p_n,k_1,...,k_l)
	= \prod_{i=1}^{l} \frac{\D^3 k_i}{(2\pi)^3 2 E_i}
	  \frac{\D^3 p_1}{(2\pi)^3 2 E_1}
	  \D\Phi_{n-1}(P-K-p_1;p_2,...,p_n)\, ,
\end{align}
which now allows for a suitable parametrisation of $k_i$ and $p_1$. This is done by first aligning $p_1$ with the $z$-axis and generating all photon momenta relative to this, i.e.
\begin{subequations}\label{eq:fsr_param}
\begin{align}
	p_1 &= E_1 (1,0,0,\beta_1),\,  \\
	k_i &= \frac{\sqrt{s}}{2} \xi_i 
	       (1,\sin{\theta_i} \sin{\phi_i}, \sin{\theta_i}\cos{\phi_i},\cos{\theta_i})\, ,
\end{align}
\end{subequations}
with $\phi_1=0$. Afterwards an Euler rotation $R(\phi,\theta,\psi)$ can be used to rotate to a completely general configuration. Since this leaves the scalar products invariant, the simple relation
\begin{align}
	p_1 \cdot k_i = \frac{\sqrt{s}}{2} \xi_i E_1 (1-\beta_1 \cos{\theta_i})
\end{align}
is preserved.

We are therefore left to evaluate the remaining $(n-1)$-particle phase space in~\eqref{eq:fsr_start}. For $n>2$ this is trivial since the sequential two-body decomposition~\eqref{eq:seq_twobody_decay} can be applied. An example for this is the muon decay. For $2 \to 2$ scattering processes this is, however, not possible. In this case there are not sufficient integration variables left in $\D\Phi_{n-1}$ to solve the momentum conserving $\delta$-function. As a consequence, it is not possible to decouple $\D\Phi_{n-1}$ from the rest of the phase space in \eqref{eq:fsr_start}. In the remainder of this section we therefore specialise to the case $n=2$. 

To eliminate the four-dimensional delta function we use the identity
\begin{align}
	 \frac{\D^3 p_2}{(2\pi)^3 2 E_2}
	 = \frac{\D^4 p_2}{(2\pi)^3} \delta(p_2^2-m_2^2)\Theta(p_2^0)\, ,
\end{align}
which yields
\begin{align}\label{eq:fsr_2to2}
	\D\Phi_{2+l}
	= \prod_{i=1}^{l} \frac{\D^3 k_i}{(2\pi)^3 2 E_i}
	  \frac{\D^3 p_1}{(2\pi)^3 2 E_1}
	  (2\pi)\delta\Big( (P-K-p_1)^2-m_2^2\Big) \Theta(P^0-K^0-p_1^0)\, .
\end{align}
Significantly more work is required to solve the remaining delta function. In the CMS frame we find
\begin{align}
	\frac{(P-K-p_1)^2-m_2^2}{\sqrt{s}}
	= -\big(\sum_i \xi_i \cos{\theta_i} \big)  \sqrt{E_1^2-m_1^2}
	   + \Big(\sum_i \xi_i - 2 \Big) E_1
	   + C
\end{align}
where
\begin{align}
	C = \frac{s\big(1-\sum_i \xi_i\big)+K^2+m_1^2-m_2^2}{\sqrt{s}}
\end{align}
is independent of the fermion energy $E_1$. As a consequence, $E_1$ satisfies
\begin{align}
	\sqrt{E_1^2-m_1^2} 
	= \frac{\big(\sum_i \xi_i-2 \big) E_1 + C}{\sum_i \xi_i \cos{\theta_i}}
	\equiv g(E_1)\, .
\end{align}
For $m_1 \neq 0$ this is a radical equation that can be solved by squaring both sides. We obtain the two solutions
\begin{align}\label{eq:fsr_2to2_sol}
	E_1^{\pm}
	= \frac{C\big(2-\sum_i\xi_i\big)\pm \big|\sum_i\xi_i\cos{\theta_i}\big| \sqrt{C^2-m_1^2 Y}}{Y}
\end{align}
with $Y=\big(2-\sum_i\xi_i\big)^2-\big(\sum_i\xi_i\cos{\theta_i\big)^2}$. The squaring of the above equation introduces extraneous solutions if $g(E_1^\pm)<0$. Together with the $\Theta$-function in~\eqref{eq:fsr_2to2} and the positivity of the square root argument in~\eqref{eq:fsr_2to2_sol} this ensures the physicality of the generated momenta. Due to the non-vanishing fermion mass, each set of random variables gives rise to up to two sets of physical momenta. This leads to an unexpected doubling of the computing time.

The overhead can be partially compensated by exploiting the typical CPS hierarchy $m_1\ll E_1$. To do so, we compare~\eqref{eq:fsr_2to2_sol} to the massless solution
\begin{align}
	E_1^{m=0}
	= \frac{C}{\big(2-\sum_i \xi_i\big)+\sum_i \xi_i \cos{\theta_i}}\, .
\end{align}
This motivates the reorganisation of the massive solutions as
\begin{align}\label{eq:fsr_2to2_sol_new}
	\tilde{E}_1^{\pm}
	= \frac{C\big(2-\sum_i\xi_i\big)\pm \sum_i\xi_i\cos{\theta_i} \sqrt{C^2-m_1^2 Y}}{Y}
\end{align}
which satisfies $\tilde{E_1}^{-} \to E_1^{m=0}$ in the massless limit. The contribution from $\tilde{E}_1^+$ is suppressed for small masses. Most of the computing time can thus be spent on the \textit{bulk region} $\tilde{E}_1^-$ at the expense of the less relevant \textit{corner region} $\tilde{E}_1^+$. As we will see in the example presented in the following section, the corner region turns out to be negligible in most practical applications.

We can now rewrite the $\delta$-function in~\eqref{eq:fsr_2to2} as
\begin{align}
	 \delta\Big((P-K-p_1)^2-m_2^2\Big)
	 = \frac{\delta(E_1-\tilde{E}_1^+)}{|J(\tilde{E}_1^+)|}
	   +  \frac{\delta(E_1-\tilde{E}_1^-)}{|J(\tilde{E}_1^-)|}
\end{align}
with the Jacobian
\begin{align}
	J(E_1) 
	= \sqrt{s}\Bigg(
	  \sum_i \xi_i-2
	  -\frac{E_1 \sum_i \xi_i\cos{\theta_i}}{\sqrt{E_1^2-m_1^2}}
	  \Bigg)\, .
\end{align}
This yields for the phase space
\begin{align}\label{eq:fsr_2to2_split}
	\D\Phi_{2+l}(P;p_1,p_2,k_1,...,k_l)
	= \D\Psi_{2+l}(\tilde{E}_1^+) 
	  + \D\Psi_{2+l}(\tilde{E}_1^-)
\end{align}
with
\begin{align}
\begin{split}
	\lefteqn{\D\Psi_{2+l}(E_1) = } \\ & \quad 
	\prod_{i=1}^{l} \frac{\D^3 k_i}{(2\pi)^3 2 E_i}
	  \frac{\D^2 \Omega_1}{(2\pi)^2 2 |J(E_1)|} 
	  \sqrt{E_1^2-m_1^2}
	  \Theta\big(P^0-K^0-E_1\big)\Theta\big(C^2-m_1^2 Y\big)\Theta\big(g(E_1)\big)\, .
\end{split}
\end{align}
As a last step, we map the remaining integration variables onto the hypercube. We find
\begin{align}
	\D\Psi_{2+l}(E_1,x_{i,j},\xi_i,y_j)
	= \Bigg(\prod_{i=1}^{l}\D x_{i,1} \D x_{i,2} \D\xi_i\Bigg)\D y_1 \D y_2
	  \mathcal{W}(E_1,x_{i,j},\xi_i,y_j)
\end{align}
with the weight function
\begin{align}\label{eq:fsr_tuning_weight}
	\mathcal{W}
	= \Bigg(\prod_{i=1}^{l} \frac{s\, \xi_i}{4(2\pi)^2}\Bigg)
	   \frac{\sqrt{E_1^2-m_1^2}}{2\pi |J(E_1)|} 
	    \Theta\big(P^0-K^0-E_1\big)\Theta\big(C^2-m_1^2 Y\big)\Theta\big(g(E_1)\big)
\end{align}
and $E_1\equiv E_1(x_{i,j},\xi_i,y_j)$. The parameters $x_{i,1}$ and $x_{i,2}$ for $i\neq 1$ are related to the parametrisation in~\eqref{eq:fsr_param} via $\cos{\theta_i}=2x_{i,1}-1$ and $\phi_i=2\pi x_{i,2}$. Furthermore, we have $\cos{\theta_1}=2x_{1,1}-1$. The remaining three parameters (apart from the photon energies $\xi_i$) then correspond to the Euler angles with $\psi=2\pi x_{1,2}$, $\cos{\theta}=2y_1-1$, and $\phi=2\pi y_2$.

\section{Demonstration at NLO}\label{sec:tuning_example}

\begin{figure}
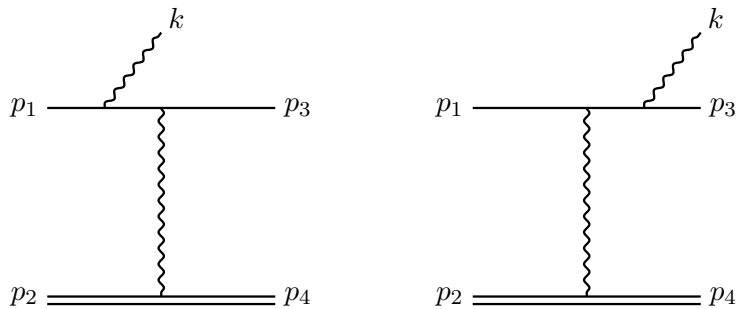

    \centering
    \subfloat{
	\begin{tikzpicture}[scale=1,baseline={(0,0)}]
    		\input{tikz/muone_isr}
    	\end{tikzpicture}
    \label{fig:muone_irs}
    }
    \hspace{1cm}
    \subfloat{
        \begin{tikzpicture}[scale=1,baseline={(0,0)}]
    		\input{tikz/muone_fsr}
    	\end{tikzpicture}
    \label{fig:muone_fsr}
    }
\caption{Electronic real-emission diagrams that contribute to $\mu$-$e$ scattering at NLO.}
\label{fig:muone_nlo_real}
\end{figure}

We demonstrate the importance of the phase-space tuning with the NLO corrections to $\mu$-$e$ scattering
\begin{align}
	e^-(p_1)\mu^-(p_2)\to e^-(p_3)\mu^-(p_4)\{\gamma(p_5)\} \, .
\end{align}
We restrict to electron line corrections where the tuning is most important. In addition to the virtual contribution of Figure~\ref{fig:muone_triangle} we have to take into account the two real-emission diagrams shown in Figure~\ref{fig:muone_nlo_real}. The corresponding squared amplitude has the structure
\begin{align}
	\mathcal{M}_{n+1}^{(0)}
	= \frac{A_1}{(k\cdot p_1)^2} 
	  + \frac{A_2}{(k\cdot p_1)(k\cdot p_3)}
	  + \frac{A_3}{(k\cdot p_3)^2} 
\end{align}
with the initial-state and final-state CPS $k\cdot p_1=\xi n \cdot p_1$ and $k\cdot p_3=\xi n\cdot p_3$. To allow for a separate phase-space tuning we use the multi-channeling split
\begin{align}
	\mathcal{M}_{n+1}^{(0)}
	= \Big(
	   \Theta(n\cdot p_1 < n\cdot p_3) 
	   + \Theta(n\cdot p_1 > n\cdot p_3)
	  \Big) \mathcal{M}_{n+1}^{(0)}
	\equiv \mathcal{M}_{n+1}^{n\cdot p_1} + \mathcal{M}_{n+1}^{n\cdot p_3}\, ,
\end{align}
which only leaves one CPS per partition.

For $\mathcal{M}_{n+1}^{n\cdot p_1}$ we use the phase space for initial-state CPS discussed in Section~\ref{sec:tuning_isr}. In this simple $2\to 3$ example, the $2$-body decomposition is trivial with~\eqref{eq:tuning_isr_start} given by
\begin{align}\label{eq:tuning_example_2body}
	\D\Phi_3(P;p_1,p_2,k) 
	= \frac{\D^3 k}{(2\pi)^3 2 E_k} \D\Phi_2(P-k;p_1,p_2)\, .
\end{align}

In order to tune for the final-state CPS, we apply the phase space generation of Section~\ref{sec:tuning_fsr} for $\mathcal{M}_{n+1}^{n\cdot p_3}$. In particular, we use the split of~\eqref{eq:fsr_2to2_split} into the bulk and the corner region. The corresponding weight function is given by~\eqref{eq:fsr_tuning_weight} with $l=1$ and $C=\big(s(1-\xi)+m_1^2-m_2^2\big)/\sqrt{s}$.  

We then consider the following MuonE observable. We assume a muon beam of energy $E_\text{beam}=150\,\mathrm{GeV}$ incident on an electron target at rest. The measurement function selects only scattered electrons with an energy greater than $1\,\mathrm{GeV}$ and muons with a minimal scattering angle of $0.3\,\mathrm{mrad}$. Furthermore, to exacerbate the problem of CPS we scale the electron mass down by an order of magnitude, i.e. $m_e\to m_e/10$. To observe the impact of the developed methodology, we calculate the total cross section with and without a dedicated tuning of the phase-space parametrisation. In the latter case, the 2-body decomposition~\eqref{eq:tuning_example_2body} is used for the entire phase space.  The \texttt{vegas} integration is done in both cases using $20$ iterations with $10^7$ evaluation points each.  In order to test the robustness of the untuned calculation we perform an additional high-statistics run where the number of evaluation points is increased to $10^8$.  

Furthermore, in order to see whether the corner region gives a contribution, we run the corresponding integration for one iteration with $10^9$ points. We observe that \texttt{vegas} is not able to find a non-vanishing phase-space region. At the level of the achieved precision we can therefore conclude that the corner region does not contribute. This has been the case for all practical applications so far. Nevertheless, it is still recommended to explicitly check this for each observable. In particular for processes with larger masses the corner region is expected to become relevant. However, a tuning for the corresponding CPS might not be required in this case.

\begin{figure}
    \centering
    \subfloat[untuned, $\chi^2=11.3$]{
        \includegraphics[width=.55\textwidth]{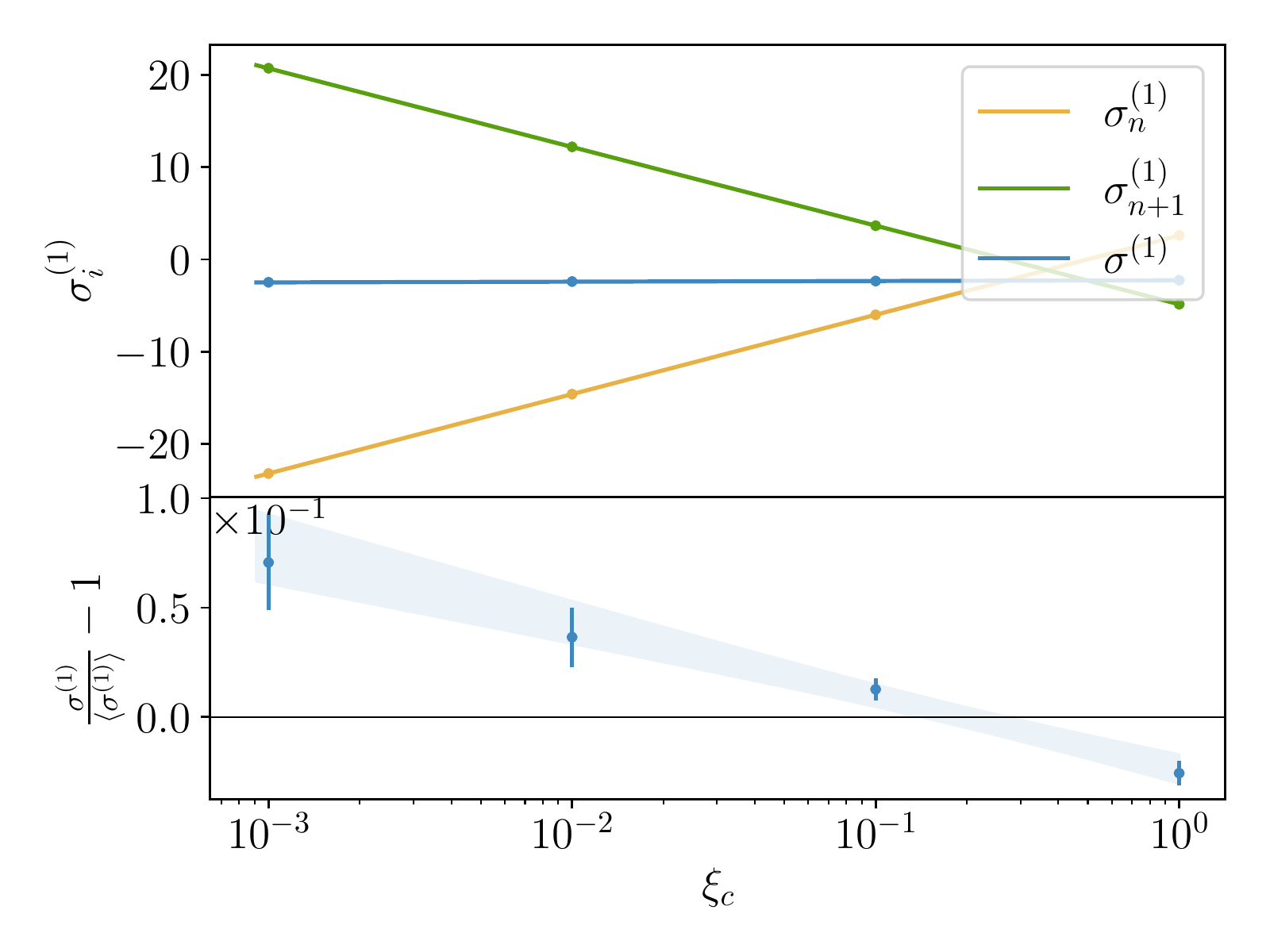}
    \label{fig:xicut_untuned}
    }\\ \vspace{-.4cm}
    \subfloat[untuned with high statistics, $\chi^2=2.1$]{
        \includegraphics[width=.55\textwidth]{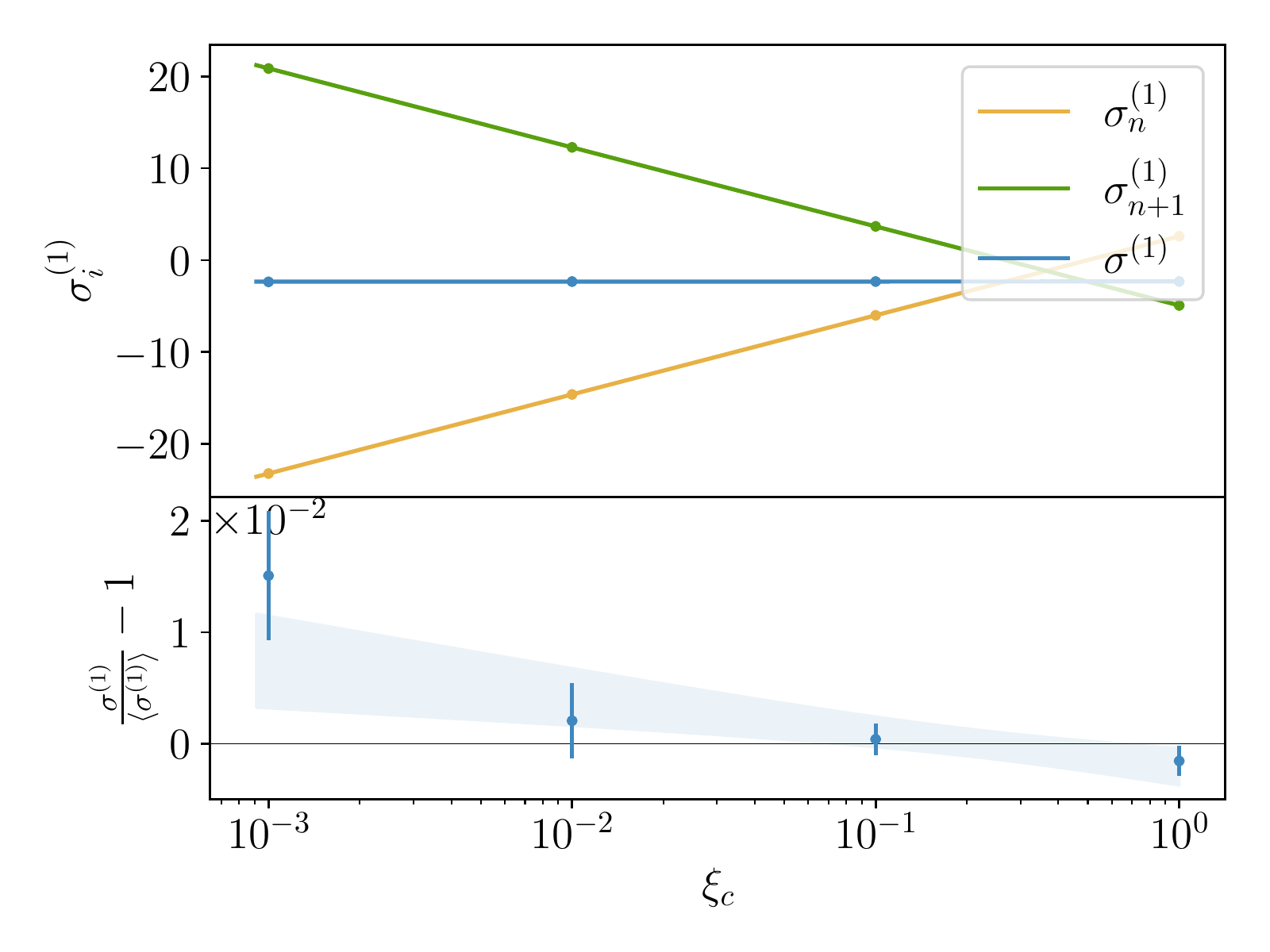}
    \label{fig:xicut_untuned_high}
    }\\ \vspace{-.4cm}
    \subfloat[tuned, $\chi^2=0.55$]{
        \includegraphics[width=.55\textwidth]{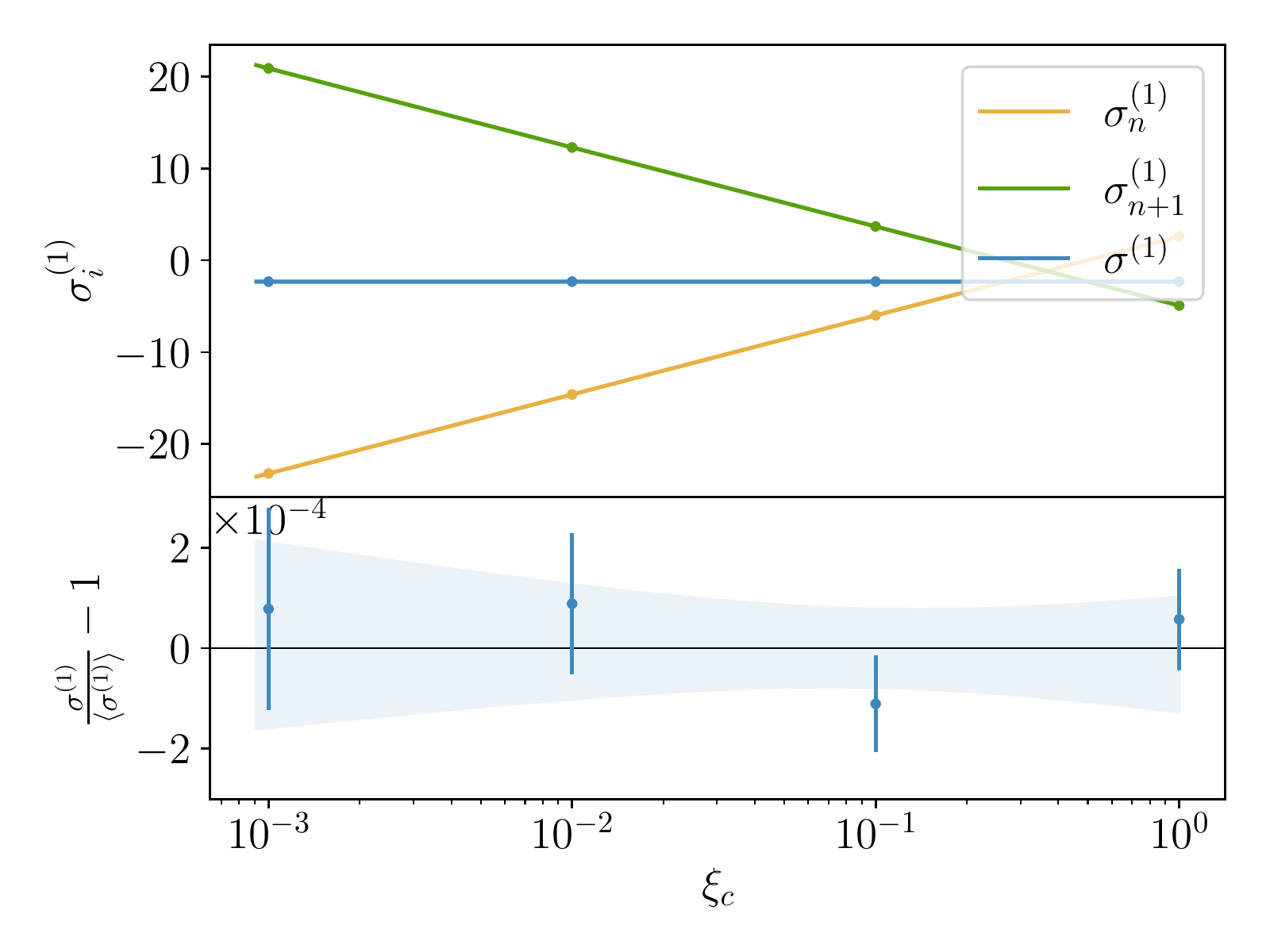}
    \label{fig:xicut_tuned}
    }
\caption{The electronic NLO correction to the $\mu$-$e$ scattering cross section and its dependence on the unphysical $\xi_c$ parameter. The error band shows the $68\%$ confidence level of the fit. We further give the chi-square as an unbiased measure of the $\xi_c$ (in)dependence. The electron mass is taken to be $m_e\to m_e/10$ in order to enhance the CPS.}
\label{fig:xicut_cps}
\end{figure}

Figure~\ref{fig:xicut_cps} shows the comparison of the three results. In the upper panels, both the $n$- as well as the $(n+1)$-particle contribution defined in~\eqref{eq:fks_nlo} are shown individually for different values of the $\xi_c$ parameter. In Section~\ref{sec:fks} we have emphasised in this context that a crucial check for the correct implementation and the numerical stability of the integration is the exact cancellation of this dependence. For this reason the $\xi_c$ (in)dependence of the total cross section is separately shown in the lower panels. Furthermore, it is helpful to use the \textit{chi-square} of the combination of the results with different $\xi_c$ values as an unbiased measure of their compatibility. This statistical quantity is defined as\footnote{We use the normalisation of a chi-square distribution with $N-1$ degrees of freedom since the $N$ values are used to estimate the mean.}
\begin{align}
	\chi^2 =
	 \frac{1}{N-1}\sum_{n=1}^N
	  \frac{\big(\sigma(\xi_c^n)-\langle \sigma(\xi_c^n) \rangle\big)^2}{d(\xi_c^n)^2}
\end{align}
with $N$ different choices of $\xi_c=\xi_c^n$, $\sigma(\xi_c^n)$ the corresponding cross sections with the Monte Carlo errors $d(\xi_c^n)$, and $\langle \sigma(\xi_c^n) \rangle$ the mean value. Compatibility of the different cross section values is indicated by a chi-square of $\chi^2 \sim 1$.

The untuned result of Figure~\ref{fig:xicut_untuned} exhibits a clear $\xi_c$ dependence, resulting in a chi-square of $\chi^2=11.3$. Furthermore, the error of the Monte Carlo integration is at the level of $10\%$ and therefore sizable. This can be significantly improved with a tenfold increase in the number of evaluation points as shown in Figure~\ref{fig:xicut_untuned_high}. The tuned phase space, on the other hand, ensures perfect $\xi_c$ independence ($\chi^2=0.5$) and a much smaller error even with lower statistics. The corresponding result is given in Figure~\ref{fig:xicut_tuned}. We can thus conclude that the \texttt{vegas} integration is not able to correctly estimate the contribution from the CPS in the untuned scenario with reasonable statistics. A dedicated tuning of the phase-space parametrisation for CPS is therefore compulsory not only to get a satisfactory convergence of the Monte Carlo error but also to get correct results in the first place.

\chapter{Next-to-soft stabilisation}\label{chap:nts}

\begin{figure}
    \centering
    \subfloat[Arbitrary phase-space point]{
        \includegraphics[width=.7\textwidth]{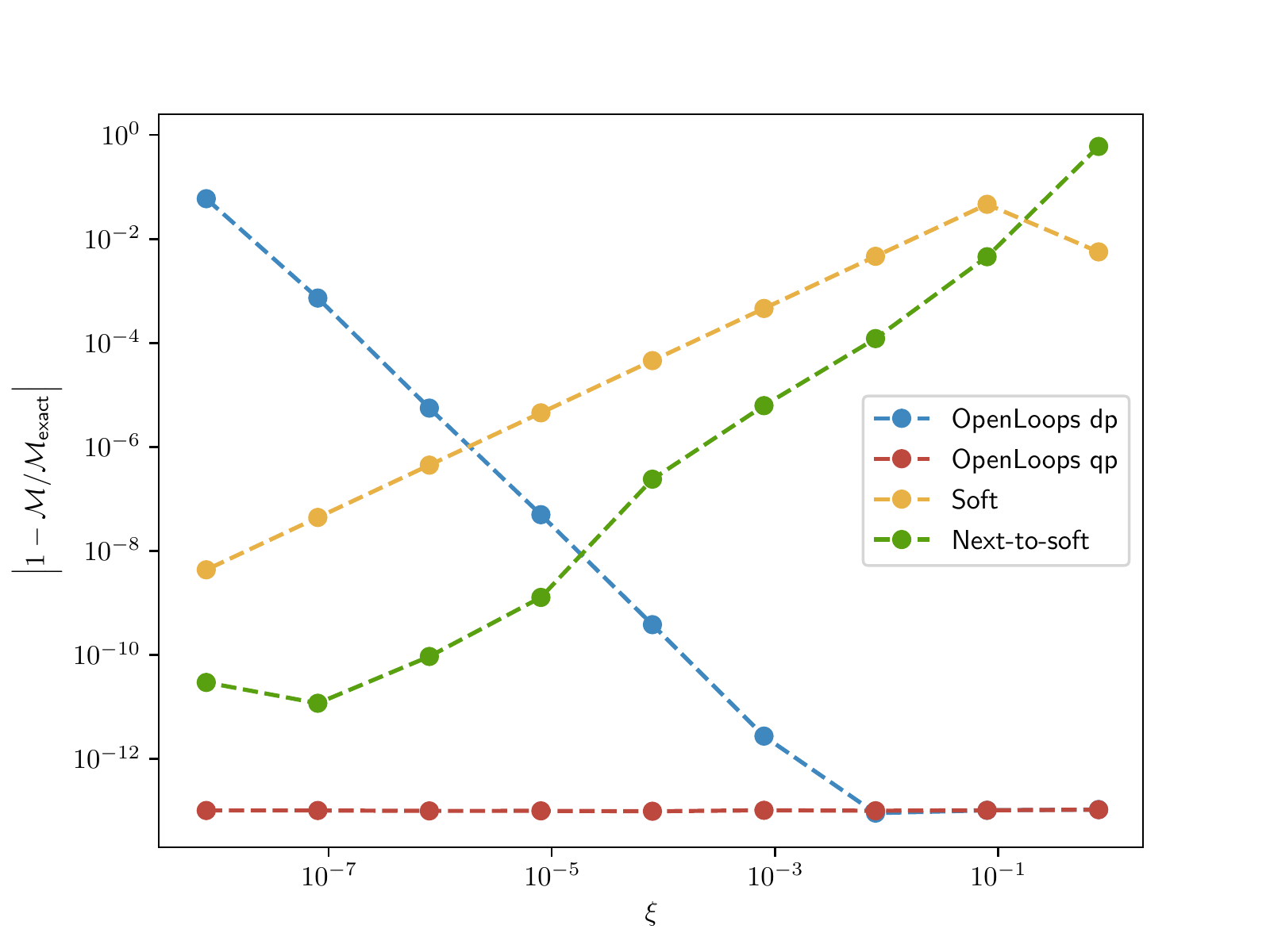}
    \label{fig:softlimit_rand}
    } \\
    \subfloat[Initial-state collinear phase-space point]{
        \includegraphics[width=.7\textwidth]{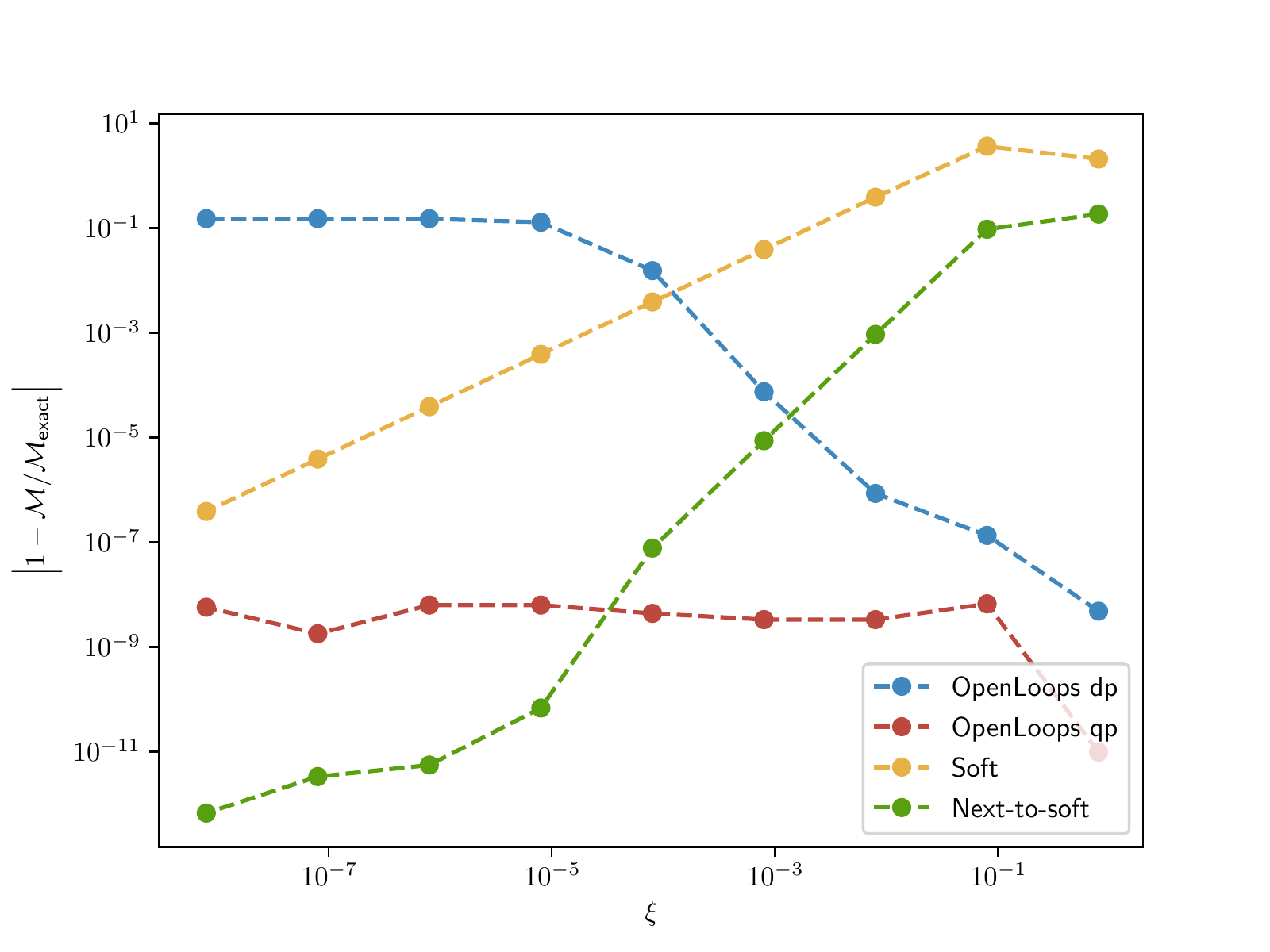}
    \label{fig:softlimit_coll}
    }
\caption{Behaviour of the soft approximations and of OpenLoops compared to the `exact' real-virtual squared amplitude for Bhabha scattering in the soft limit $E_k=\xi\times\sqrt{s}/2 \to 0$. In the case of OpenLoops running in quadruple precision mode we observe perfect agreement.}
\label{fig:softlimit_bhabha}
\end{figure}

One of the main challenges in fully differential NNLO calculations is the numerical stability of the real-virtual amplitude. There has therefore been a massive effort to build automated one-loop tools that pay special attention to a numerically stable evaluation. Particularly successful in this regard is OpenLoops~\cite{Buccioni:2017yxi,Buccioni:2019sur}. However, most of these automated codes are tailored to QCD with massless fermions. When used for QED, additional numerical problems often arise due to its multi-scale nature and large scale hierarchies ($m^2 \ll S$). Typically, these issues occur when the emitted photon becomes soft and are further exacerbated in the collinear region. This behaviour is illustrated in Figure~\ref{fig:softlimit_bhabha} in the case of radiative Bhabha scattering
\begin{align}
e^-(p_1) e^+(p_2)\to e^-(p_3) e^+(p_4)\gamma(k)\, .
\end{align}
We have used a CMS energy of $\sqrt{s}=1020\,\text{MeV}$ tailored to $\phi$ factories. The `exact' reference value is calculated with \texttt{Mathematica} in arbitrary precision and agrees perfectly with OpenLoops running in quadruple precision mode~\cite{max} shown in red. The corresponding deviation for OpenLoops in its standard mode is shown in blue for an arbitrary as well as an initial-state collinear phase-space point. For the former, at $\xi=2 E_k/\sqrt{s}=10^{-5}$ the relative error is $10^{-8}$. In the collinear case, the numerical instabilities are strongly enhanced with a relative difference of $10^{-1}$ for $\xi=10^{-5}$.

An obvious idea to cope with this problem is to expand the real-virtual amplitude for small photon energies and to switch to this approximation for sufficiently small $\xi$. In analogy to Section~\ref{sec:yfs}, we introduce the soft book-keeping parameter $\lambda_s$ with the scaling behaviour
\begin{align}\label{eq:soft_powercounting}
\begin{split}
	p_i &\to p_i\, , \\
	k &\to \lambda_s\, k\, , \\
	m &\to m\, .
\end{split}
\end{align}
The soft expansion of the squared amplitude then takes the form
\begin{align}
	\mathcal{M}_{n+1} 
	= \frac{1}{\lambda_s^2} \mathcal{M}_{n+1}^\text{sLP}
	  +\frac{1}{\lambda_s} \mathcal{M}_{n+1}^\text{sNLP}
	  + \mathcal{O}(\lambda_s^0)\, .
\end{align}
The leading $\mathcal{O}(\lambda_s^{-2})$ term in this expansion, $\mathcal{M}_{n+1}^\text{sLP}$, is given by the eikonal approximation~\eqref{eq:soft_limit}  as explicitly proven in Section~\ref{sec:yfs}. At one loop, it can thus easily be calculated based on the non-radiative correction $\mathcal{M}_n^{(1)}$. As can be seen from the yellow line in Figure~\ref{fig:softlimit_bhabha} this approach is insufficient in the collinear region. If an accuracy below $10^{-3}$ is to be aimed at, in this case one has to switch to the expansion at $\xi\sim10^{-3}$. However, the exact result is not sufficiently well approximated by the leading soft contribution in this region. Hence, to ensure a decent approximation we have to include the $\mathcal{O}(\lambda_s^{-1})$ term $\mathcal{M}_{n+1}^\text{sNLP}$.

It has been shown a long time ago by Low, Burnett, and Kroll~\cite{Low:1958sn,Burnett:1967km} that also this subleading term is related to the non-radiative process at tree level via a differential operator.\footnote{In the case of gravity this even holds true up to sub-subleading power~\cite{Cachazo:2014fwa,Bern:2014vva,Beneke:2021umj}.} This so-called \textit{LBK theorem} was later extended to massless particles~\cite{DelDuca:1990gz} where a universal radiative jet function was introduced to take into account collinear effects. More recently, the massless version of the theorem has attracted some attention in the context of resummation of next-to-leading power threshold logarithms. To this end the theorem has been extended to also include loop corrections in the framework of diagrammatic factorisation~\cite{Bonocore:2015esa,Bonocore:2016awd,Laenen:2020nrt} as well as in SCET~\cite{Larkoski:2014bxa,Beneke:2019oqx,Liu:2021mac}. In our case, however, we are interested in QED where all fermion masses and all other scales are considered to be much larger than the energy of the emitted photon. These recent loop-level extensions are thus not applicable since the underlying EFT is heavy-quark effective theory (\ac{HQET}) instead of SCET. In particular, there is no radiative jet function in this case due to the absence of any collinear scale. This leaves hard and soft modes as the only relevant degrees of freedom. This is completely analogous to the YFS proof of Section~\ref{sec:yfs}. 

One option is therefore to calculate the subleading term $\mathcal{M}_{n+1}^\text{sNLP}$ with a brute force calculation. Rational coefficients and simple Passarino-Veltman functions can thereby be expanded in Mathematica. More complicated triangle- and box-functions, on the other hand, can be expanded at the loop-integrand level using the MoR (Section~\ref{sec:mor}) and calculated using Mellin-Barnes techniques (Section~\ref{sec:loop}). The impact of the inclusion of the corresponding result in the soft expansion is shown in green in Figure~\ref{fig:softlimit_bhabha} where a significant improvement of the approximation is observed. This allows us to switch to a reliable expansion as early as $\xi\sim 10^{-3}$. This \emph{next-to-soft stabilisation} ensures the numerical stability of the real-virtual amplitude for small photon energies which is a prerequisite for the IR subtraction to work. Additionally, since \texttt{vegas} tends to sample predominantly in the soft and collinear region, it also results in a significant speed-up in the integration. Based on this, we were able to compute for the first time the fully differential NNLO corrections to Bhabha~\cite{Banerjee:2021mty} and M{\o}ller~\cite{Banerjee:2021qvi} scattering.

While the above brute force approach was successful, the corresponding calculation is cumbersome. A loop-level extension of the LBK theorem for massive fermions is therefore desirable. This would allow us to apply next-to-soft stabilisation to other processes without the need of explicit computations. It turns out that this is indeed possible at one loop using the same approach as for the YFS proof of Section~\ref{sec:yfs}. In the remainder of this chapter we therefore present this extension. We first start in Section~\ref{sec:lbk} with a short review of the tree-level derivation of the LBK theorem. The one-loop extension is then discussed in detail in the following Section~\ref{sec:lbk_oneloop} with the main result given in~\eqref{eq:lbk_oneloop}. Finally, we present in Section~\ref{sec:lbk_validation} a highly non-trivial validation of this result by means of the $2\to 4$ process $e^-e^+\to e^- e^+ \gamma \gamma$.

\section{The LBK theorem}
\label{sec:lbk}

Following the YFS approach of Section~\ref{sec:yfs}, we split the radiative tree-level amplitude $\mathcal{A}^{(0)}_{n+1}$ into contributions due to external and internal emission. Since we restrict here to tree level, we set $\Lambda=1$ and $S=i/(\slashed{p_i}-\slashed{k}-m)$ in~\eqref{eq:hard_split}. The diagrammatic representation of the split then simplifies to
\begin{align}\label{eq:lbk_split_tree}
    \mathcal{A}_{n+1}^{(0)} =
    \sum_i \Bigg(
    \begin{tikzpicture}[scale=.8,baseline={(1,0)}]
    	\input{tikz/yfs_external_tree}
    \end{tikzpicture}
    \hspace{0.185cm} \Bigg)
    +
    \begin{tikzpicture}[scale=.8,baseline={(1,0)}]
    	\input{tikz/yfs_internal}
    \end{tikzpicture}
    =
    \mathcal{A}_{n+1}^{(0),\text{ext} }+ \mathcal{A}_{n+1}^{(0),\text{int}}.
\end{align}
Contrary to the leading soft limit, we are not allowed to neglect internal emission at NLP. Furthermore, particular care has to be taken regarding radiative and non-radiative kinematics. For this reason we define in addition to the on-shell momenta $\{p\}=\{p_1,\,...\,,p_i,\,...\,,p_n\}$ the sets of momenta $\{p\}_i=\{p_1,\,...\,,p_i-k,\,...\,,p_n\}$ that are adapted to emission from line $i$. Taking all particles apart from the emitted photon to be incoming (but ignoring the complex conjugation of the polarisation vector $\epsilon$) allows us to write
the soft expansion of $\mathcal{A}_{n+1}^{(0),\text{ext} }$ as
\begin{subequations}\label{eq:ext_expansion}
\begin{align}
    \mathcal{A}_{n+1}^{(0),\text{ext} }
    \,=\,& \sum_i Q_i 
    \frac{\Gamma^\text{ext}(\{p\}_i)(\slashed{p}_i-\slashed{k}+m)\gamma^\mu u(p_i) \epsilon_\mu(k)}{2 k\cdot p_i} \\
    \,=\,& \sum_i Q_i \Big(
    \frac{\epsilon\cdot p_i}{k\cdot p_i}\Gamma^\text{ext}(\{p\}_i)
    - \frac{\Gamma^\text{ext}(\{p\}_i)\slashed{k}\slashed{\epsilon}}{2 k\cdot p_i}
    \Big) u(p_i) \\
    \wideeq{k\sim\lambda_s}& \sum_i Q_i \Big(
    \frac{1}{\lambda_s}\frac{\epsilon\cdot p_i}{k\cdot p_i} \Gamma^\text{ext}(\{p\})
    -\frac{\epsilon\cdot p_i}{k\cdot p_i} k\cdot\frac{\partial}{\partial p_i} \Gamma^\text{ext}(\{p\}) 
    -\frac{\Gamma^\text{ext}(\{p\})\slashed{k}\slashed{\epsilon}}{2k\cdot p_i}
    \Big) u(p_i)
    + \mathcal{O}(\lambda_s)\, .
\end{align}
\end{subequations}
Since $\{p\}$ satisfies the radiative momentum conservation $\sum_i p_i = k$ this is not a strict expansion in $\lambda_s$. Following~\cite{Adler:1966gc} we can make the above split gauge invariant (up to subleading power) via the modification
\begin{subequations}
\begin{align} \label{eq:ext-ginv}
    &\mathcal{A}^{(0),\text{ext} }_{n+1} 
    \to \mathcal{A}^\text{I}_{n+1}
    \equiv \epsilon\cdot A^\text{I}_{n+1}
    = \mathcal{A}^{(0),\text{ext} }_{n+1} 
    + \sum_i Q_i \epsilon\cdot\pder{p_i} \Gamma^\text{ext}(\{p\}) u(p_i)\, , \\
    &\mathcal{A}^{(0),\text{int} }_{n+1} 
    \to \mathcal{A}^\text{II}_{n+1}
    \equiv \epsilon\cdot A^\text{II}_{n+1}
    = \mathcal{A}^{(0),\text{int} }_{n+1} 
    - \sum_i Q_i \epsilon\cdot\pder{p_i} \Gamma^\text{ext}(\{p\}) u(p_i)\, .
\end{align}
\end{subequations}
Indeed, $k\cdot  A^\text{I}_{n+1} \sim \mathcal{O}(\lambda_s^2)$. The leading contributions in $\lambda_s$ vanish due to $\sum_i Q_i=0$ and the subleading contributions cancel between the two terms of the last expression in \eqref{eq:ext-ginv}. Because the full amplitude is gauge invariant we also have  $k\cdot  A^\text{II}_{n+1} \sim \mathcal{O}(\lambda_s^2)$. This does not directly imply that $\mathcal{A}^\text{II}_{n+1} \sim \mathcal{O}(\lambda_s)$ due to possible $\sim k^\mu$ terms in $\mathcal{A}^\text{II}_{n+1}$. However, the leading $\mathcal{O}(\lambda_s^0)$ term in $A^\text{II}_{n+1}$ must be independent of $k$ due to the lack of $1/k$ poles in $\mathcal{A}_{n+1}^{(0),\text{int} }$. As a consequence, we indeed find that $\mathcal{A}^\text{II}_{n+1}$ does not contribute at NLP. As a consequence, the soft expansion of the total amplitude can be written as
\begin{equation}
    \mathcal{A}_{n+1}^{(0)} = \sum_i Q_i \Big(
    \frac{1}{\lambda_s}\frac{\epsilon\cdot p_i}{k\cdot p_i}\Gamma^\text{ext}(\{p\})
    -\frac{\Gamma^\text{ext}(\{p\})\slashed{k}\slashed{\epsilon}}{2k\cdot p_i}
    -\big[\epsilon\cdot D_i \Gamma^\text{ext}(\{p\})\big]
    \Big) u(p_i)+\mathcal{O}(\lambda_s)
\end{equation}
with the \textit{LBK operator}
\begin{equation}
    D_i^\mu = \frac{p_i^\mu}{k\cdot p_i}k
    \cdot\frac{\partial}{\partial p_i}
    - \frac{\partial}{\partial p_{i,\mu}}\, .
\end{equation}
Squaring the amplitude, summing over spins and polarisations, and using the identity
\begin{align}
    \frac{(\slashed{p_i}+m)\slashed{\epsilon}\slashed{k}
    +\slashed{k}\slashed{\epsilon}(\slashed{p_i}+m)}{2k\cdot p_i}
    =\frac{\epsilon\cdot p_i}{k\cdot p_i}\slashed{k}-\slashed{\epsilon}
    =\epsilon\cdot D_i (\slashed{p_i}+m)
\end{align}
then yields
\begin{equation}\label{eq:lbk}
    \mathcal{M}_{n+1}^{(0)}(\{p\},k)
    =\sum_{ij} Q_i Q_j \Big(
    -\frac{1}{\lambda_s^2} \frac{p_i\cdot p_j}{(k\cdot p_i)(k\cdot p_j)}
    + \frac{1}{\lambda_s} \frac{p_j\cdot D_i}{k\cdot p_j}
    \Big) \mathcal{M}_n^{(0)}(\{p\})
    + \mathcal{O}(\lambda_s^0)\, .
\end{equation}
This shows that not only is the leading term in the soft expansion related to the non-radiative process but that this is also true at subleading power at tree level. However, the non-radiative squared amplitude in \eqref{eq:lbk} is evaluated with a set of momenta $\{p\}$ that does not satisfy momentum conservation. This is unproblematic at tree level. If, on the other hand, loop corrections are taken into account (Section~\ref{sec:lbk_oneloop}) this significantly complicates the evaluation of the corresponding integrals. In this case a different formulation of the LBK theorem is helpful. To this end, we reabsorb the first term of the LBK operator to undo the expansion and write
\begin{equation}\label{eq:lbk_interm}
    \mathcal{M}_{n+1}^{(0)}(\{p\},k)
    = -\sum_{ij} Q_i Q_j \Big(
    \frac{1}{\lambda_s^2} \frac{p_i\cdot p_j}{(k\cdot p_i)(k\cdot p_j)}
    +\frac{1}{\lambda_s}\frac{1}{k\cdot p_j}p_j\cdot \pder{p_i} \Big)
    \mathcal{M}_n^{(0)}(\{p\}_i) + \mathcal{O}(\lambda_s^0)\, .
\end{equation}
Since $\{p\}_i$ satisfies momentum conservation we can now express the non-radiative squared amplitude in terms of invariants
\begin{align}
    \mathcal{M}_{n}^{(0)}(\{p\}_i)
    = \mathcal{M}_{n}^{(0)} (\{s\}_i,\{m^2\}_i)
\end{align}
with $\{s\}_i=\big\{s(\{p\}_i,\{m^2\}_i)\big\}$ and $\{m^2\}_i=\{p_1^2=m_1^2,\,...\,,(p_i-k)^2,\,...\,,p_n^2=m_n^2\,\}$. The corresponding expansion in $k$ can then be written  as
\begin{align}
\begin{split}
    \lefteqn{\mathcal{M}_{n}^{(0)} (\{s\}_i,\{m^2\}_i)
    =} \\ & \Bigg(
    1-\lambda_s
    \sum_L \Big( k\cdot\pder[s_L]{p_i}+2k\cdot p_i\pder[s_L]{m_i^2}\Big)\pder{s_L}
    -\lambda_s\, 2k\cdot p_i\pder{m_i^2}
    \Bigg)
    \mathcal{M}_{n}^{(0)} (\{s\},\{m^2\})
    + \mathcal{O}(\lambda_s^2)\, ,
    \label{eq:deriv1}
\end{split}
\end{align}
where the sum $L$ is over the set of independent invariants $\{s\}=\big\{s(\{p\},\{m^2\})\big\}$ expressed in terms of the momenta $\{p\}$ and the on-shell masses $\{m^2\}$.  Similarly, we can write
\begin{align}
\begin{split}
    \lefteqn{p_j\cdot \pder{p_i} \mathcal{M}_{n}^{(0)} (\{s\}_i,\{m^2\}_i)
    = } \\ & \Bigg(
    \sum_L\Big(p_j\cdot\pder[s_L]{p_i}+2p_i\cdot p_j\pder[s_L]{m_i^2} \Big) \pder{s_L}
    + 2p_i\cdot p_j\pder{m_i^2}
    \Bigg)
    \mathcal{M}_{n}^{(0)} (\{s\},\{m^2\})
    + \mathcal{O}(\lambda_s)\, .
    \label{eq:deriv2}
\end{split}
\end{align}
Inserting~\eqref{eq:deriv1} and~\eqref{eq:deriv2} into~\eqref{eq:lbk_interm}, all derivatives with respect to the masses cancel and we obtain the simple formulation of the LBK theorem in terms of invariants
\begin{align}\label{eq:lbk-inv}
    \mathcal{M}_{n+1}^{(0)}(\{p\},k)
    =  \sum_{ij} Q_i Q_j \Big(
    -\frac{1}{\lambda_s^2} \frac{p_i\cdot p_j}{(k\cdot p_i)(k\cdot p_j)}
    + \frac{1}{\lambda_s} \frac{p_j\cdot\tilde{D}_i}{k\cdot p_j}
    \Big) \mathcal{M}_{n}^{(0)} (\{s\},\{m^2\})
    + \mathcal{O}(\lambda_s^0)
\end{align}
with the modified LBK operator
\begin{align}\label{eq:lbkop}
    \tilde{D}_i^\mu 
    = \sum_L \Big(
    \frac{p_i^\mu}{k\cdot p_i} k\cdot \pder[s_L]{p_i}-\pder[s_L]{p_{i,\mu}}
    \Big) \pder{s_L}\, .
\end{align}
The advantage of \eqref{eq:lbk-inv} over \eqref{eq:lbk} is that conventional one-loop techniques can be applied in this case. We emphasise that the choice of $\{s\}=\big\{s(\{p\},\{m^2\})\big\}$ is ambiguous since the momenta $\{p\}$ do not satisfy momentum conservation. This is however not an issue as long as the same definition is used in the calculation of the derivatives $\partial s_L/\partial p_i^\mu$. The above formula can therefore be conveniently used to analytically compute the soft limit of tree-level amplitudes up to subleading power. An alternative approach that is particularly suitable for the numerical evaluation of the LBK theorem was recently presented in~\cite{Bonocore:2021cbv}.

The above formula assumes all particles apart from the photon to be incoming. In the case of outgoing particles the corresponding momentum $p$ has to be replaced with $-p$. In particular, this also implies $\partial/\partial p^\mu\to-\partial/\partial p^\mu$.

\section{One-loop generalisation of the LBK theorem}
\label{sec:lbk_oneloop}

The derivation of the previous section cannot be naively applied in the presence of loop corrections due to contributions from regions where additional scales become small and thus modify the power counting~\eqref{eq:soft_powercounting}. The MoR, introduced in Section~\ref{sec:mor}, can be used to disentangle these regions. As already observed in Section~\ref{sec:yfs}, only hard ($\ell \sim \lambda_s^0$) and soft ($\ell \sim \lambda_s$) scalings contribute in this simple limit. The collinear regions, on the other hand, all vanish due to the absence of collinear scales. In the case of the hard region the LBK formula still holds and we can use~\eqref{eq:lbk-inv} to compute this contribution. A proof of this fact is presented in the following Section~\ref{sec:lbk_hard}. We are then left to evaluate the soft region. In Section~\ref{sec:yfs} we have shown that this contribution vanishes at LP to all orders in perturbation theory. We make use of this cancellation to evaluate the soft region at NLP in a generic way. The combination of hard and soft contributions, that will be given in Section~\ref{sec:lbk_full}, therefore generalises the LBK theorem to one loop.

\subsection{The hard region}\label{sec:lbk_hard}

The derivation of the LBK theorem presented in Section~\ref{sec:lbk} relies on gauge invariance as well as on the fact that diagrams with internal emission do not contain any $1/k$ poles. While gauge invariance is satisfied for the soft and hard region separately, it is the second property that is spoiled beyond tree level. It is not possible to avoid this issue for the soft region and its contribution has to be evaluated in a different way (see Sections~\ref{sec:soft_general}, \ref{sec:softLP}, and \ref{sec:softNLP}). In the case of the hard region, on the other hand, we now show that the diagrams that violate this condition cancel at the level of the spin-summed squared amplitude. As a consequence, the LBK theorem remains valid and can therefore be used to compute the complete hard contribution.

At one loop it is obviously not permissible to set $\Lambda=1$ and $S=i/(\slashed{p_i}-\slashed{k}-m)$ in~\eqref{eq:hard_split} as was done in~\eqref{eq:lbk_split_tree}. We thus have the additional external leg corrections
\begin{align}\label{eq:lbk_split_oneloop}
    \mathcal{A}_{\text{ext},i}^{(1)} =
    \begin{tikzpicture}[scale=.8,baseline={(1,0)}]
    	\input{tikz/yfs_external_vertex}
    \end{tikzpicture}
    +
     \begin{tikzpicture}[scale=.8,baseline={(1,0)}]
    	\input{tikz/yfs_external_bubble}
    \end{tikzpicture}
     +
     \begin{tikzpicture}[scale=.8,baseline={(1,0)}]
    	\input{tikz/yfs_external_massct}
    \end{tikzpicture}
\end{align}
that spoil the basic assumptions of the LBK proof. The vertex correction is an internal emission diagram with a $1/k$ pole. The self energy contributions, on the other hand, classify as external and can thus be expanded following~\eqref{eq:ext_expansion}. Naively, one would therefore expect that this yields additional contributions that do not reduce to the non-radiative amplitude.  In the following, we show that all of this is unproblematic due to a cancellation of these corrections up to NLP. The cancellation at LP occurs at the amplitude level and is a consequence of the Ward identity. This is shown in Section~\ref{sec:yfs} at all orders in perturbation theory. At NLP the mass counterterm diagram only gives terms $\slashed{k}\slashed{k}=k^2=0$ and thus vanishes. As a consequence, only the first two diagrams in~\eqref{eq:lbk_split_oneloop} have to be taken into account. After expanding the integrals in the hard momentum region, the amplitude can be straightforwardly computed with \texttt{Package-X}. We find
\begin{align}
	\mathcal{A}_{\text{ext},i}^{(1)} = Q_i^3\, \Gamma^\text{ext} \epsilon\cdot H u(p_i) + \mathcal{O}(\lambda_s)
\end{align}
with
\begin{align}\label{eq:hardtensor}
	H^\mu
	= \frac{1}{m} \gamma^\mu
	- \frac{1}{m (k \cdot p_i)} \slashed{k} p_i^\mu
	- \frac{1}{k \cdot p_i} \gamma^\mu \slashed{k}\, .
\end{align}
Hence, the NLP contribution of $\mathcal{A}_{\text{ext},i}^{(1)}$ does not vanish at the amplitude level.

At the level of the squared amplitude this contribution enters via interference with the eikonal approximation
\begin{align}
	\mathcal{A}_{n+1}^{(1)}
	= \sum_j Q_j \frac{\epsilon \cdot p_j}{k \cdot p_j}\, \Gamma^\text{ext}u(p_i)
	+ \mathcal{O}(\lambda_s^0)
\end{align}
and gives after summing over the spin of the external leg
\begin{align}
	\mathcal{M}_{\text{ext},i}^{(1)} 
	= \sum_\text{spin} \mathcal{A}_{\text{ext},i}^{(1)}{ \mathcal{A}_{n+1}^{(1)}}^\dagger + h.c.
	= \sum_j Q_j \frac{\epsilon^* \cdot p_j}{k \cdot p_j} \,
	 \Gamma^\text{ext} \epsilon \cdot H (\slashed{p_i}+m){\Gamma^\text{ext}}^\dagger  + h.c.
	  + \mathcal{O}(\lambda_s^0) \, .
\end{align}
We can now use basic Dirac algebra to rewrite the three terms originating from~\eqref{eq:hardtensor} as follows. The first term satisfies the identity
\begin{align}
	\frac{1}{m}\Gamma^\text{ext} \slashed{\epsilon} (\slashed{p_i}+m) {\Gamma^\text{ext}}^\dagger + h.c.
	= \frac{\epsilon \cdot p_i}{m} \Gamma^\text{ext}{\Gamma^\text{ext}}^\dagger
	+\Gamma^\text{ext}\slashed{\epsilon}{\Gamma^\text{ext}}^\dagger+ h.c.
\end{align}
due to the hermitian conjugate. Similarly, we can show that
\begin{align}
         -\frac{\epsilon \cdot p_i}{m (k\cdot p_i)}
         \Gamma^\text{ext} \slashed{\slashed{k}}( \slashed{p_i} +m){\Gamma^\text{ext}}^\dagger + h.c.
	= -\frac{\epsilon \cdot p_i}{m} \Gamma^\text{ext}{\Gamma^\text{ext}}^\dagger 
	-\frac{\epsilon \cdot p_i}{k \cdot p_i} \Gamma^\text{ext}\slashed{k}{\Gamma^\text{ext}}^\dagger + h.c.
\end{align}
and
\begin{align}
       -\frac{1}{k\cdot p_i}\Gamma^\text{ext} \slashed{\epsilon} \slashed{k} (\slashed{p_i}+m) {\Gamma^\text{ext}}^\dagger + h.c.
	= - \Gamma^\text{ext} \slashed{\epsilon}{\Gamma^\text{ext}}^\dagger
	+ \frac{ \epsilon \cdot p_i}{k \cdot p_i} \Gamma^\text{ext} \slashed{k} {\Gamma^\text{ext}}^\dagger
	+ h.c.
\end{align}
where we have used $\epsilon(k) \cdot k = 0$ for the latter identity. The cancellation is now manifest and we find
\begin{align}
	\mathcal{M}_{\text{ext},i}^{(1)} = \mathcal{O}(\lambda_s^0) \, .
\end{align}

We conclude that the problematic diagrams of~\eqref{eq:lbk_split_oneloop} vanish at NLP at the level of the spin-summed squared amplitude. The remaining hard contributions satisfy the properties that the LBK proof relies on. The complete hard region is thus given by the LBK theorem~\eqref{eq:lbk-inv} with $\mathcal{M}_{n}^{(0)}\to \mathcal{M}_{n}^{(1)}$. In order to arrive at a complete generalisation of the LBK theorem at one loop we have to evaluate also the soft region in a generic way. This is the subject of the following sections.

\subsection{General considerations regarding the soft contribution}\label{sec:soft_general}

\begin{figure}
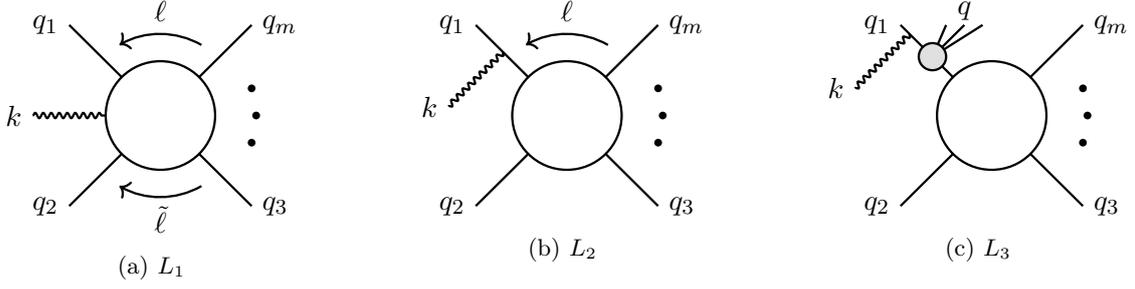

    \centering
    \subfloat[$L_1$]{
        \begin{tikzpicture}[scale=1.2,baseline={(0,0)}]
        \input{tikz/loop1}
    \end{tikzpicture}
    \label{fig:loop1}
    }
    \hspace{1cm}
    \subfloat[$L_2$]{
        \begin{tikzpicture}[scale=1.2,baseline={(0,0)}]
        \input{tikz/loop2}
    \end{tikzpicture}
    \label{fig:loop2}
    }
    \hspace{1cm}
    \subfloat[$L_3$]{
        \begin{tikzpicture}[scale=1.2,baseline={(0,0)}]
        \input{tikz/loop3}
    \end{tikzpicture}
    \label{fig:loop3}
    }
\caption{Classification of one-loop integrals encountered in the calculation of radiative QED amplitudes. The legs that connect to the loop can either be on shell or off shell.}
\label{fig:loops}
\end{figure}

In order to systematically analyse possible origins of the soft contribution we follow a similar strategy as in Section~\ref{sec:yfs} where the diagrammatic representation~\eqref{eq:soft_rep_yfs} was introduced at LP. To extend this to NLP we make these considerations more precise at one loop. To do so, we classify the one-loop integrals as illustrated in Figure~\ref{fig:loops} where the momenta $q_i$ can be off shell (internal) or on shell (external). The circle symbolises the one-loop integral associated with the 1PI part of a particular Feynman diagram. The first class, $L_1$, includes $(m+1)$-point integrals from diagrams where the photon is directly attached to this 1PI part. Class $L_2$ includes $m$-point integrals from diagrams where the photon is attached to a leg that directly connects to the 1PI part with momentum $q_1-k$. As we will see, the treatment of these integrals depends on whether the momentum of the adjacent leg $q_m$ is on shell or off shell. Finally, for integrals of the type $L_3$ the photon is attached indirectly to the $m$-point 1PI part such that the momentum flowing into the loop integral is $q_1-k+\tilde{q}$ with a non-zero $\tilde{q}$. 

For integrals to have a non-vanishing soft contribution the momentum routing has to be chosen such that the loop momentum $\ell$ is aligned with a photon propagator. All other choices lead only to linear propagators in the soft momentum expansion and therefore vanish as a consequence of the residue theorem. There can thus be at most as many soft regions as the number of photons in the loop. However, most of them yield scaleless integrals and vanish in dimensional regularisation. This is in particular the case for all possible routings of $L_3$. The presence of the momentum $\tilde{q}$ allows to set $\ell=0$ for the soft contribution in all propagators except for the photon propagator with momentum $\ell$. Hence, loop integrals of the form $L_3$ do not contribute to the soft region. For the second class, on the other hand, there is one non-vanishing soft contribution indicated by the momentum routing in Figure~\ref{fig:loop2} if the corresponding internal propagator is given by a photon and if in addition $q_1$ is on shell. In the case where $q_m$ is off shell the soft expansion starts at $\mathcal{O}(\lambda_s)$ and is given by
\begin{equation}
   I_{L_2}(q_m^2\neq m_m^2) = \int \frac{\text{d}^d \ell}{(2\pi)^d} \frac{1}{
      [\ell^2]\ [2\ell\cdot q_1-2k\cdot q_1]\ [(q_1+q_2)^2-m_2^2]\ ...\ [q_m^2-m_m^2]}
      .
\end{equation}
For on-shell $q_m$ the leading integral reads instead
\begin{equation}
   I_{L_2}(q_m^2 = m_m^2) = \int \frac{\text{d}^d \ell}{(2\pi)^d} \frac{1}{
      [\ell^2]\ [2\ell\cdot q_1-2k\cdot q_1]\ [(q_1+q_2)^2-m_2^2]\ ...\ [-2\ell\cdot q_m]}
\end{equation}
which already contributes at $\mathcal{O}(\lambda_s^0)$. Finally, the first class of loop integrals $L_1$ gives rise to up to two non-vanishing soft contributions given by the two momentum routings $\ell$ and $\tilde{\ell}$ in Figure~\ref{fig:loop1} if the corresponding propagators are photons. The integral for routing $\ell$
\begin{align}
   \lefteqn{I_{L_1} = } \notag\\&
   \int \frac{\text{d}^d \ell}{(2\pi)^d} \frac{1}{
      [\ell^2]\ [2\ell\cdot q_1+q_1^2-m_1^2] \ [2\ell\cdot q_1-2k\cdot q_1+q_1^2-m_1^2]\ ...\ [-2\ell\cdot q_m+q_m^2-m_m^2]}
\end{align}  
is only non-zero if $q_1$ is on shell and it starts to contribute at $\mathcal{O}(\lambda_s^{-1})$ if $q_m$ is on shell and at $\mathcal{O}(\lambda_s^0)$ otherwise. The analogous statements hold for the $\tilde{\ell}$ momentum routing.

\begin{figure}
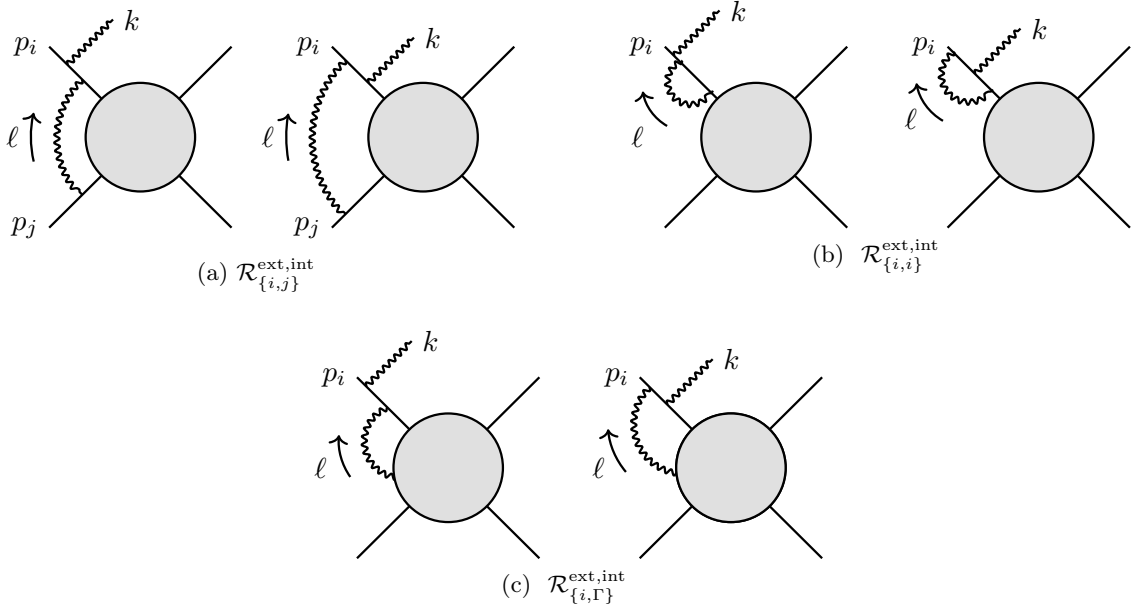

    \centering
    \subfloat[$\mathcal{R}_{\{i,j\}}^{\text{ext},\text{int}}$]{
        \begin{tikzpicture}[scale=1.2,baseline={(0,0)}]
        \input{tikz/softdiag1}
    \end{tikzpicture}
    \label{fig:softdiag1}
    }
    \hspace{1cm}
    \subfloat[
    $\mathcal{R}_{\{i,i\}}^{\text{ext},\text{int}}$]{
        \begin{tikzpicture}[scale=1.2,baseline={(0,0)}]
        \input{tikz/softdiag2}
    \end{tikzpicture}
    \label{fig:softdiag2}
    }
    \hspace{1cm}
    \subfloat[
    $\mathcal{R}_{\{i,\Gamma\}}^{\text{ext},\text{int}}$]{
        \begin{tikzpicture}[scale=1.2,baseline={(0,0)}]
        \input{tikz/softdiag3}
    \end{tikzpicture}
    \label{fig:softdiag3}
    }
\caption{Diagrammatic classification of the soft contributions each
corresponding to a particular momentum routing of a Feynman diagram.}
\label{fig:softdiags}
\end{figure}

The above reasoning allows to represent every possible soft contribution according to the three pairs of diagrams shown in Figure~\ref{fig:softdiags} where the external legs are now all on shell. These diagrams are one-loop specialisations of the representation~\eqref{eq:soft_rep_yfs} that was used to derive the LP soft limit. Every $\mathcal{R}_{\{e,a\}}^\text{int, ext}$ corresponds to an amplitude with a specific choice of the momentum routing where the soft contribution does not vanish. The labels for emission and absorption $\{e,a\}$ take on the values $1,\ldots,n$ or $\Gamma$. The superscript $\text{int, ext}$ indicates whether the photon $k$ is attached internally or externally. In the notation~\eqref{eq:soft_rep_yfs} used for the case of an arbitrary number of soft virtual photons we have the identifications $\mathcal{R}^\text{int}_{\{e,a\}}=\mathcal{R}^{1,0}_{\{e,a\}}$ and $\mathcal{R}^\text{ext}_{\{e,a\}}=\mathcal{R}^{0,1}_{\{e,a\}}$. In the former (latter) case we are dealing with integrals of the type $L_1$ ($L_2$). As mentioned in connection with $L_1$, it is possible that one amplitude contributes to two soft representations. Taking $e=i$ and assuming $p_i$ to be incoming, we can write the corresponding expressions generically as 
\begin{subequations}\label{eq:softrep}
\begin{align}
    &i \mathcal{R}_{\{e,a\}}^\text{ext}
    = \frac{
    Q_i^2 \Gamma_{\{e,a\}}^\mu(\slashed{\ell}+\slashed{p_i}-\slashed{k}+m)
    \gamma_\mu(\slashed{p_i}-\slashed{k}+m)\slashed{\epsilon}u(p_i)
    }{
    -2 k\cdot p_i [\ell^2][\ell^2+2\ell\cdot(p_i-k)-2k\cdot p_i]
    }\, , \\
    &i \mathcal{R}_{\{e,a\}}^\text{int}
    = \frac{
    Q_i^2 \Gamma_{\{e,a\}}^\mu(\slashed{\ell}+\slashed{p_i}-\slashed{k}+m)
    \slashed{\epsilon}(\slashed{\ell}+\slashed{p_i}+m)\gamma_\mu u(p_i)
    }{
    [\ell^2][\ell^2+2\ell\cdot(p_i-k)-2k\cdot p_i][\ell^2+2\ell\cdot p_i]
    }\, .
\end{align}
\end{subequations}
All terms related to the emission from leg  $e=i$ and the soft photon propagator are given explicitly in \eqref{eq:softrep}. The vertex and fermion propagator related to the absorption is common to $\mathcal{R}_{\{e,a\}}^\text{ext}$ and $\mathcal{R}_{\{e,a\}}^\text{int}$ and is included in $\Gamma_{\{e,a\}}^\mu$. This implies the scalings $\Gamma_{\{i,j\}}^\mu
\sim \Gamma_{\{i,i\}}^\mu \sim \lambda_s^{-1}$ and $\Gamma_{\{i,\Gamma\}}^\mu \sim \lambda_s^0$. We then write the expansion in the soft region of the sum of the diagram pairs as
\begin{subequations}
\begin{align}
    i \mathcal{R}_{\{e,a\}}^\text{soft}
    \wideeq{k\sim\lambda_s} &\frac{1}{\lambda_s}S_{\{e,a\}}^\text{LP}+S_{\{e,a\}}^\text{NLP}+\mathcal{O}(\lambda_s) \\
    \,=\,& \frac{1}{\lambda_s} \Big(
    S_{\{e,a\}}^{\text{LP},\text{ext}} + S_{\{e,a\}}^{\text{LP},\text{int}} \Big) + S_{\{e,a\}}^{\text{NLP},\text{ext}} + S_{\{e,a\}}^{\text{NLP},\text{int}}+\mathcal{O}(\lambda_s)\, ,
\end{align}
\end{subequations}
with the leading and subleading power terms denoted by $S_{\{e,a\}}^\text{LP}$ and $S_{\{e,a\}}^\text{NLP}$, respectively. Based on the previously discussed power counting of the integrals $L_1$ and $L_2$ we can deduce that $S_{\{i,\Gamma\}}^\text{LP}=S_{\{i,\Gamma\}}^{\text{LP},\text{ext}}=S_{\{i,\Gamma\}}^{\text{LP},\text{int}}=0$.

\subsection{Vanishing of the soft contribution at leading power}\label{sec:softLP}

In Section~\ref{sec:yfs} it is proven that the LP soft contribution vanishes to all orders in perturbation theory. To prepare for the NLP discussion in the following section, we repeat the argument in the one-loop case. The LP soft contribution to~\eqref{eq:softrep} is given by
\begin{subequations}
\begin{align}
    &S_{\{e,a\}}^{\text{LP},\text{ext}}
    = \frac{
    Q_i^2 \Gamma_{\{e,a\}}^\mu(\slashed{p_i}+m)
    \gamma_\mu(\slashed{p_i}+m)\slashed{\epsilon}u(p_i)
    }{
    -4 k\cdot p_i [\ell^2][\ell\cdot p_i-k\cdot p_i]
    }
    =
    -\frac{
    Q_i^2 \Gamma_{\{e,a\}}^\mu u(p_i) p_{i,\mu} p_i\cdot\epsilon
    }{
    [\ell^2][\ell\cdot p_i-k\cdot p_i][k\cdot p_i]
    }
    \, , \\
    &S_{\{e,a\}}^{\text{LP},\text{int}}
    = \frac{
    Q_i^2 \Gamma_{\{e,a\}}^\mu (\slashed{p_i}+m)
    \slashed{\epsilon}(\slashed{p_i}+m)\gamma_\mu u(p_i)
    }{
    4[\ell^2][\ell\cdot p_i-k\cdot p_i][\ell\cdot p_i]
    }
    =
    +\frac{
    Q_i^2 \Gamma_{\{e,a\}}^\mu u(p_i) p_{i,\mu} p_i\cdot\epsilon
    }{
    [\ell^2][\ell\cdot p_i-k\cdot p_i][\ell\cdot p_i]
    }\, .
\end{align}
\end{subequations}
We bring the propagators to the same form using the partial fraction identity~\eqref{eq:partialfraction_yfs} for $m=1$ given by
\begin{equation}
    \frac{1}{[\ell\cdot p_i-k\cdot p_i][\ell\cdot p_i]}
    =\frac{1}{k\cdot p_i} \Big(
    \frac{1}{[\ell\cdot p_i-k\cdot p_i]}-\frac{1}{[\ell\cdot p_i]} \Big)\, .
\end{equation}
The second term in the curly brackets can be neglected up to
scaleless integrals. We then see immediately that
\begin{equation}
    S_{\{e,a\}}^\text{LP}=S_{\{e,a\}}^{\text{LP},\text{ext}}+S_{\{e,a\}}^{\text{LP},\text{int}}=0\, .
\end{equation}
We have thus again reproduced at one loop the all-order result of Section~\ref{sec:yfs} that the the eikonal approximation in QED does not receive genuine loop corrections. Furthermore, we have also shown that $S_{\{i,\Gamma\}}^\text{NLP}=0$ since it effectively corresponds to a leading-power contribution. The third class of soft contributions, $\mathcal{R}_{\{i,\Gamma\}}^\text{soft}$, can therefore be omitted in the following discussion.

\subsection{Soft contribution at subleading power}\label{sec:softNLP}

At subleading power there are contributions in~\eqref{eq:softrep} from either the higher-order expansion of propagators (denominator) or from numerator terms proportional to $\ell$ or $k$. We therefore write
\begin{subequations}
\begin{align}
    S_{\{e,a\}}^\text{NLP} 
    &= S_{\{e,a\}}^{(\text{NLP},\text{D})}+S_{\{e,a\}}^{(\text{NLP},\text{N})} \\
    &= \Big(
       S_{\{e,a\}}^{(\text{NLP},\text{D}),\text{ext}}+S_{\{e,a\}}^{(\text{NLP},\text{D}),\text{int}}
    \Big) + \Big(
       S_{\{e,a\}}^{(\text{NLP},\text{N}),\text{ext}}+S_{\{e,a\}}^{(\text{NLP},\text{N}),\text{int}}
    \Big)\, .
\end{align}
\end{subequations}

For the denominator type the leading-power cancellation of Section~\ref{sec:softLP} occurs if propagators other than $[\ell^2+2\ell\cdot(p_i-k)-2 k\cdot p_i]$ or $[\ell^2+2\ell\cdot p_i]$ are expanded. Furthermore, expansion in $\ell^2$ of these two propagators results only in linear propagators. Consequently, we only have to consider the expansion in $\ell\cdot k$ of the propagator $[\ell^2+2\ell\cdot(p_i-k)-2 k\cdot p_i]$ in~\eqref{eq:softrep}. Using partial fraction then yields the simple contribution
\begin{equation}
    S_{\{e,a\}}^{(\text{NLP},\text{D})} 
    = S_{\{e,a\}}^{(\text{NLP},\text{D}),\text{ext}}+S_{\{e,a\}}^{(\text{NLP},\text{D}),\text{int}} 
    = -\frac{
    Q_i^2 \Gamma_{\{e,a\}}^\mu u(p_i) p_{i,\mu} p_i\cdot\epsilon\, \ell\cdot k
    }{
    (k\cdot p_i)^2 [\ell^2][\ell\cdot p_i-k\cdot p_i]
    }\, .
\end{equation}
The numerator type can be written as
\begin{equation}
    S_{\{e,a\}}^{(\text{NLP},\text{N})} 
    = \frac{Q_i^2 \Gamma_{\{e,a\}}^\mu}{4 k\cdot p_i[\ell^2][\ell\cdot p_i-k\cdot p_i]}
    \big(T^\text{ext}_\mu+K^\text{ext}_\mu+T^\text{int}_\mu+K^\text{int}_\mu \big)u(p_i)
\end{equation}
with
\begin{subequations}
\begin{align}
    T^\text{ext}_\mu
    &= -(\slashed{\ell}-\slashed{k})\gamma_\mu(\slashed{p}_i+m)\slashed{\epsilon}\, , \\
    K^\text{ext}_\mu
    &= (\slashed{p}_i+m)\gamma_\mu\slashed{k}\slashed{\epsilon}\, , \\
    T^\text{int}_\mu
    &= (\slashed{\ell}-\slashed{k})\slashed{\epsilon}(\slashed{p}_i+m)\gamma_\mu\, , \\
    K^\text{int}_\mu
    &= (\slashed{p}_i+m)\slashed{\epsilon}\slashed{\ell}\gamma_\mu\, .
\end{align}
\end{subequations}
Due to various cancellations among $T_\mu^{\text{ext},\text{int}}$ and $K_\mu^{\text{ext},\text{int}}$ we obtain the simple result
\begin{equation}
    S_{\{e,a\}}^{(\text{NLP},\text{N})}
    = \frac{Q_i^2 \Gamma_{\{e,a\}}^\mu u(p_i)}{k\cdot p_i [\ell^2][\ell\cdot p_i-k\cdot p_i]}
    \big( k_\mu p_i\cdot\epsilon-\epsilon_\mu k\cdot p_i+p_{i,\mu} \ell\cdot\epsilon \big)\, ,
\end{equation}
where we have used the replacement $\ell\cdot p_i\to k\cdot p_i$ in the numerator which holds up to scaleless integrals.

To make further progress at this point we need to treat $S_{\{i,j\}}^\text{NLP}$ and $S_{\{i,i\}}^\text{NLP}$ separately in order to specify the form of $\Gamma_{\{e,a\}}^\mu$. In the case where the photon is reabsorbed by the emitting leg, i.e. $S_{\{i,i\}}^\text{NLP}$, we have
\begin{equation}
    \Gamma_{\{i,i\}}^\mu u(p_i)
    = -\frac{Q_i \Gamma^{(0)} (\slashed{p_i}-\slashed{k}+m)\gamma^\mu u(p_i)}{-2 k\cdot p_i}
    \wideeq{k\sim\lambda_s}  \frac{1}{\lambda_s} Q_i \mathcal{A}_n^{(0)} \frac{p_i^\mu}{k\cdot p_i} + \mathcal{O}(\lambda_s^0)
\end{equation}
where $\mathcal{A}_n^{(0)}=\Gamma^{(0)} u(p_i)$ corresponds to the non-radiative tree-level amplitude. In this case, we further have the simple Passarino-Veltman decomposition
\begin{equation}
    \ell^\rho \to \frac{\ell\cdot p_i}{m_i^2} p_i^\rho
\end{equation}
where we can again replace $\ell\cdot p_i$ with $k\cdot p_i$. It is then straightforward to see that
\begin{equation}
    S_{\{i,i\}}^\text{NLP}=S_{\{i,i\}}^{(\text{NLP},\text{D})}+S_{\{i,i\}}^{(\text{NLP},\text{N})} = 0\, .
\end{equation}
Hence, diagrams where the loop corrects only the emitting leg do not contribute at subleading power.

In the case of $S_{\{i,j\}}^\text{NLP}$, on the other hand, we find for an incoming particle $p_j$ that
\begin{equation}
    \Gamma_{\{i,j\}}^\mu u(p_i) 
    \wideeq{k\sim\lambda_s} - \frac{1}{\lambda_s} Q_j \mathcal{A}_n^{(0)} \frac{p_j^\mu}{[-\ell\cdot p_j]} + \mathcal{O}(\lambda_s^0)\, .
\end{equation}
This in turn results after the tensor decomposition
\begin{align}
    \ell^\rho \to
    \frac{\ell \cdot p_j\, p_i\cdot p_j-\ell \cdot p_i\, m_j^2}{(p_i \cdot p_j)^2-m_i^2 m_j^2}\,p_i^\rho
    + \frac{\ell \cdot p_i\, p_i\cdot p_j-\ell \cdot p_j\, m_i^2}{(p_i \cdot p_j)^2-m_i^2 m_j^2}\,p_j^\rho
\end{align}
in the only non-vanishing subleading power contribution of the form
\begin{equation}\label{eq:amp_soft}
    S_{\{i,j\}}^\text{NLP}
    \equiv S_{\{i,j\}}^\text{NLP}(p_i,p_j,Q_i,Q_j)
    = - Q_i^2 Q_j (i\mathcal{A}_n^{(0)})
     \Big( \frac{p_i\cdot\epsilon}{k\cdot p_i}-\frac{p_j\cdot\epsilon}{k\cdot p_j} \Big)
     S(p_i,p_j,k)\, ,
\end{equation}
where we have defined the function
\begin{equation}\label{eq:softfunc}
    S(p_i,p_j,k)
    = \frac{m_i^2 k\cdot p_j}{\big((p_i\cdot p_j)^2-m_i^2 m_j^2\big)k\cdot p_i}
      \Big( p_i\cdot p_j I_1(p_i,k)+m_j^2 k\cdot p_i I_2(p_i,p_j,k) \Big)\, .
\end{equation}
The analytic results for the integrals
\begin{subequations}
\begin{align}
    I_1(p_i,k) &= i \mu^{2\epsilon}\int \frac{\text{d}^d \ell}{(2\pi)^d} \frac{1}{[\ell^2+i\delta][\ell\cdot p_i-k\cdot p_i+i\delta]}\, , \label{eq:softinta}\\
    I_2(p_i,p_j,k) &= i \mu^{2\epsilon} \int \frac{\text{d}^d \ell}{(2\pi)^d} \frac{1}{[\ell^2+i\delta][-\ell\cdot p_j+i\delta][\ell\cdot p_i-k\cdot p_i+i\delta]}\label{eq:softintb}
\end{align}
\end{subequations}
can be found in Appendix~\ref{sec:softints}. The causal $+i\delta$ prescription is given explicitly in the above integrals.

The result~\eqref{eq:amp_soft} is also valid for incoming antiparticles with the overall sign difference parametrised by the fermion charges $Q_i$ and $Q_j$. The total soft contribution can thus be obtained by summing the above expression over all external charged fermions, i.e.
\begin{equation}\label{eq:amp_soft_full}
    \mathcal{A}_{n+1}^{(1),\text{soft}}
    = \sum_{i\neq j} S_{\{i,j\}}^\text{NLP}(p_i,p_j,Q_i,Q_j) + \mathcal{O}(\lambda_s)\, .
\end{equation}
The corresponding expression for the squared amplitude can be obtained by interfering with the eikonal approximation of the tree-level amplitude. The resulting formula is given in the following section.

\subsection{One-loop extension of the LBK theorem}\label{sec:lbk_full}

\begin{figure}
    \centering
    \begin{tikzpicture}[scale=1,baseline={(1,0)}]
    	\input{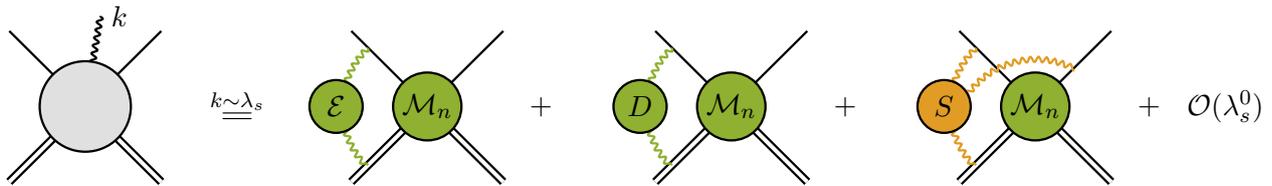}
    \end{tikzpicture}
    \caption{Schematic illustration of the soft factorisation
    of~\eqref{eq:lbk_oneloop} at subleading power and at one loop.}
    \label{fig:soft_fac}
\end{figure}

Based on the previous discussion we find that the one-loop correction to a generic radiative process in QED in the limit where the emitted photon becomes soft satisfies the expansion
\begin{subequations}\label{eq:lbk_oneloop}
\begin{equation}
    \mathcal{M}_{n+1}^{(1)}(\{p\},k)
    \wideeq{k\sim\lambda_s} 
    \mathcal{M}_{n+1}^{(1),\text{hard}} + \mathcal{M}_{n+1}^{(1),\text{soft}}
\end{equation}
with the hard contribution
\begin{align} \label{eq:lbk_oneloop_hard} 
    \mathcal{M}_{n+1}^{(1),\text{hard}} =
     \sum_l \sum_i Q_i Q_l \Big(
    - \frac{1}{\lambda_s^2}
    \frac{p_i\cdot p_l}{(k\cdot p_i)(k\cdot p_l)}
    + \frac{1}{\lambda_s}
    \frac{p_l\cdot \tilde{D}_i}{k\cdot p_l}
    \Big)
    \mathcal{M}_n^{(1)} (\{s\},\{m^2\})
    + \mathcal{O}(\lambda_s^0)
\end{align}
and the soft region generically given by
\begin{align} \label{eq:lbk_oneloop_soft}
 \begin{split}
   \mathcal{M}_{n+1}^{(1),\text{soft}} 
   &= \frac{1}{\lambda_s} \sum_l \sum_{i \neq j} Q_i^2 Q_j Q_l
    \Big( \frac{p_i\cdot p_l}{(k\cdot p_i)(k\cdot p_l)} 
    - \frac{p_j\cdot p_l}{(k\cdot p_j)(k\cdot p_l)} \Big) 2
    S(p_i,p_j,k) 
    \\ & \hspace{8.1cm} \times
     \mathcal{M}_{n}^{(0)}(\{s\},\{m^2\}) + \mathcal{O}(\lambda_s^0)\, .
\end{split}
\end{align}
\end{subequations}
This is the generalisation of the LBK theorem at one loop. We emphasise that the above result assumes all particles to be incoming. For outgoing particles one can simply replace the corresponding momentum $p$ with $-p$. Furthermore, the LBK operator $\tilde{D}_i$ and the function $S(p_i,p_j,k)$ are defined in~\eqref{eq:lbkop} and~\eqref{eq:softfunc}, respectively. 

A conceptual illustration of the factorisation formula~\eqref{eq:lbk_oneloop} is shown in Figure~\ref{fig:soft_fac}. Contributions with hard and soft origin are depicted in green and orange, respectively. The first two diagrams on the r.h.s correspond to the hard sector given by~\eqref{eq:lbk_oneloop_hard}. The factorisation of~\eqref{eq:lbk_oneloop_soft} into a universal soft function - connecting three external legs simultaneously - and the non-radiative squared amplitude is illustrated in the third diagram.
Based on this, a naive extrapolation to higher orders in perturbation theory is possible by interpreting Figure~\ref{fig:soft_fac} as an all-order statement. First of all, this would imply that the LBK operator yields the hard contribution also beyond one loop. More interestingly, however, it would significantly constrain the mixed hard-soft structure. At two loops, for example, the hard-soft region would be fixed through objects that already enter in~\eqref{eq:lbk_oneloop}. In particular, it would correspond to~\eqref{eq:lbk_oneloop_soft} with $\mathcal{M}_n^{(0)}\to\mathcal{M}_n^{(1)}$. The only new contribution in the factorisation formula would therefore be the two-loop soft function corresponding to the purely soft region.

\section{Validation}\label{sec:lbk_validation}

To demonstrate the correctness and applicability of equation~\eqref{eq:lbk_oneloop} we consider the soft limit in the process
\begin{align}\label{eq:dummyprocess}
e^-(p_1)e^+(p_2) \to e^-(p_3) e^+(p_4) \gamma(p_5)\gamma(k)
\end{align}
at one loop where $k$ becomes soft. This $2\to4$ process is a highly non-trivial test of our formalism since the full one-loop amplitude is rather involved and contains hexagon functions. We compare our approximations to OpenLoops running in quadruple precision mode. The process~\eqref{eq:dummyprocess} could also be considered to be the real-real-virtual contribution to the N$^3$LO corrections to Bhabha scattering. Hence, implementing this amplitude in a way that remains sufficiently stable for soft emission would be essential for any future N$^3$LO calculation. We use a CMS energy of $\sqrt{s}=10.583\, \text{GeV}$, tailored to the beam energy of the Belle II experiment~\cite{Belle-II:2018jsg} running at the $\Upsilon(4S)$ resonance.

\begin{figure}
    \centering
    \includegraphics[width=0.8\textwidth]{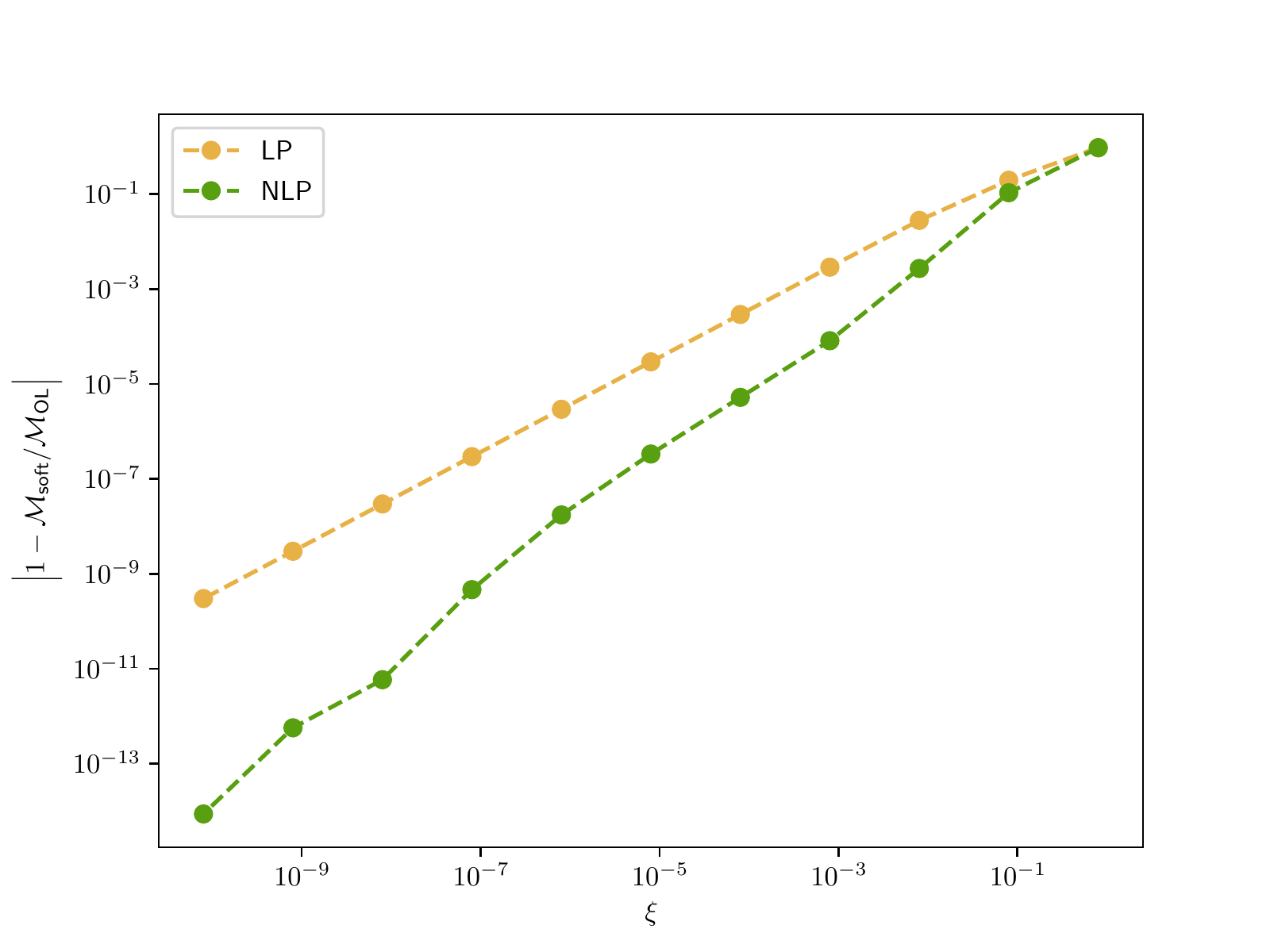}
    \caption{Convergence of the soft limit $\mathcal{M}_\text{soft}$ at LP and NLP of the one-loop correction to $e^-e^+\to e^- e^+ \gamma\gamma$. The reference value $\mathcal{M}_\text{OL}$ is calculated with OpenLoops using
    quadruple precision.}
\label{fig:ee2eegg_nts}
\end{figure}

We consider the limit where one of the two photons becomes soft while the other photon remains hard, i.e. $k\to 0$.  Looking at~\eqref{eq:lbk_oneloop}, we have
\begin{subequations}
\begin{alignat}{10}
&p_1&\to&+p_1\, ,&\qquad   &p_2&\to&+p_2\, ,&\qquad
&p_3&\to&-p_3\, ,&\qquad   &p_4&\to&-p_4\, ,&\qquad
&p_5&\to&-p_5\, ,
\\
&Q_1&=  &- e\, ,&   &Q_2&=  &+ e\, ,&
&Q_3&=  &+ e\, ,&   &Q_4&=  &- e\, ,&
&Q_5&=   &\ 0\, .
\end{alignat}
\end{subequations}
Of course the above sign convention for the outgoing particles also has to be taken into account in the case of the derivatives $\partial/\partial p_{i,\mu}$. Furthermore, we define the set of invariants $\{s\}$ as
\begin{align}
    \{s\}=\{s=(p_1+p_2)^2,
            t=(p_2-p_4)^2,
            s_{15}=2p_1 \cdot p_5, 
            s_{25}=2p_2 \cdot p_5,
            s_{35}=2p_3 \cdot p_5\}\, .
\end{align}
We emphasise again that this choice is not unique. It is therefore crucial to use the same definition both in the evaluation of the non-radiative amplitude as well as for the derivatives $\partial/\partial p_{i,\mu}$ in the LBK operator~\eqref{eq:lbkop}. Since already the one-loop correction for $ee\to ee\gamma$ is rather complicated, we have implemented the corresponding derivatives numerically to a very high precision in \texttt{Mathematica}. Combining this with the soft contribution from~\eqref{eq:lbk_oneloop_soft} then yields the complete NLP approximation. The corresponding $\lambda_s^{-2}$ and $\lambda_s^{-1}$ terms can then be compared to OpenLoops as a function of $\xi=2E_k/\sqrt{s}$. The result is shown in Figure~\ref{fig:ee2eegg_nts} down to values of $10^{-10}$. It is clearly visible that including the $\lambda_s^{-1}$ (NLP) terms significantly improves the precision of the approximation. This behaviour clearly validates our one-loop generalisation of the LBK theorem presented in~\eqref{eq:lbk_oneloop}.

\chapter{Collinear factorisation}\label{chap:coll}

 As explained in Section~\ref{chap:cps}, radiative QED amplitudes suffer from CPS that hamper the numerical integration over the phase space. This problem can be alleviated by using a multi-channeling approach combined with a dedicated tuning of the phase-space parametrisation. For low and intermediate energies this has proven successful. For high energies, such as for the FCC-ee, the problem of CPS is significantly exacerbated. As a consequence, additional methods have to be developed in this case that can be applied in combination with the tuning.

One option in this regard is a QCD-inspired subtraction scheme that relocates the problem of CPS to simpler counterterms. Such an approach was presented in~\cite{Dittmaier:1999mb} where the dipole formalism~\cite{Catani:1996jh,Catani:1996vz} was extended to QED with massive fermions. It was later shown in~\cite{Dittmaier:2008md} how non-collinear safe observables can be incorporated in this framework. As explained in the context of the FKS$^\ell$ scheme presented in Section~\ref{sec:fks}, a prerequisite for an efficient subtraction procedure is the construction of process independent counterterms that mimic the singular behaviour of the amplitudes. In the case of CPS, this requires the study of the collinear behaviour of radiative amplitudes for small but non-vanishing fermion masses.

To facilitate the power counting in this small-mass collinear limit we introduce the book-keeping parameter $\lambda_c$ and assume the mass to scale as $p^2 = m^2 \sim \lambda_c^2$. This in turn gives the collinear behaviour 
\begin{align}
	k \cdot p 
	= E_k E_p \big(1-\beta \cos\sphericalangle(k,p)\big)
	\widesim{\sphericalangle(k,p)=0} \lambda_c^2\, .
\end{align}
It is a well-known feature of gauge theories that the amplitudes scale as $\mathcal{A}_{n+1} \sim \lambda_c^{-1}$ in the collinear limit and not as $\sim \lambda_c^{-2}$ as one would naively expect since the numerator compensates one power of $\lambda_c$. At the level of the squared amplitude we therefore have the expansion
\begin{align}
    \mathcal{M}_{n+1}
    \wideeq{k\cdot p \sim \lambda_c^2}
    \frac{1}{\lambda_c^2}\mathcal{M}_{n+1}^\text{cLP} 
    + \mathcal{O}(\lambda_c^{-1})\, .
\end{align}

The collinear limit has been extensively investigated in the context of QCD with massless quarks where it gives rise to IR singularities. The factorisation into a process-independent splitting function multiplying the non-radiative amplitude has therefore been known for some time now~\cite{Bern:1994zx,Kosower:1999rx}. While the splitting functions correspond to the Altarelli-Parisi kernels at tree level, this is no longer true if loop corrections are taken into account. The two-loop corrections to the QCD splitting functions have been calculated in~\cite{Bern:2004cz,Badger:2004uk}. Much less is known, however, in the case of QED where collinear divergences are regularised by finite fermion masses. The corresponding splitting function is currently only known at tree level where it coincides with the QCD version up to a polynomial mass term~\cite{Baier:1973ms,Berends:1981uq,Kleiss:1986ct}.

Nevertheless, a similar factorising structure can also be expected beyond tree level. This is due to the applicability of SCET to the case of small but non-vanishing fermion masses, as we have seen in the context of massification discussed in Section~\ref{sec:massification}. We can therefore adjust the SCET derivation of Section~\ref{sec:factorisation} to the case considered here. The correlator~\eqref{eq:massification_correlator} now reads
\begin{align}
	\mathcal{A}_{n+1} \sim 
	\langle 0 | \psi_{e_1}…\psi_{e_i}A_{e_i}^\mu…\psi_{e_n} e^{i S_{n+1}} |0\rangle
\end{align}
with the additional collinear photon $A_{e_i}^\mu$ in the external state. In analogy to~\eqref{eq:massification_amp}, we then expect the amplitude to factorise as
\begin{align}\label{eq:coll_amp}
	\mathcal{A}_{n+1}
	\sim
	\langle 0 | \tilde{\psi}_{e_i} \tilde{A}_{e_i}^\mu \bar{\mathcal{O}}^i_{n+1} e^{i S^i_\text{coll}} |0\rangle
	\, \Big( \prod_{j \neq i} \langle 0 | \tilde{\psi}_{e_j} \bar{\mathcal{O}}^j_{n+1} e^{i S^j_\text{coll}} |0\rangle \Big)
	\, \langle 0 | \bar{\mathcal{O}}_{n+1}^s e^{i S_\text{soft}} | 0 \rangle 
	\, C_{n+1} 
	+ \mathcal{O}(\lambda_c^{0})\, .
\end{align}
In addition to the collinear massification constants
\begin{align}
 \langle 0 | \tilde{\psi}_{e_j} \bar{\mathcal{O}}^j_{n+1} e^{i S^j_\text{coll}} |0\rangle
 \end{align}
 we therefore encounter the new \textit{universal} object
 \begin{align}
\langle 0 | \tilde{\psi}_{e_i} \tilde{A}_{e_i}^\mu \bar{\mathcal{O}}^i_{n+1} e^{i S^i_\text{coll}} |0\rangle\, .
 \end{align}
This term corresponds to the \textit{massive splitting function} and is the main object of study in this chapter. As in the case of massification, we do not directly work within SCET but instead apply the MoR to disentangle universal collinear contributions from the process dependent hard part.

In the following, we calculate the different terms of the factorisation formula~\eqref{eq:coll_amp} at one loop. This allows to calculate $\mathcal{M}_{n+1}^\text{cLP}$ based on the non-radiative amplitude without the need of additional process dependent computations. This, in turn, lays the foundation for the NNLO extension of the aforementioned subtraction scheme for CPS. We first start in Section~\ref{sec:coll_tree} by reproducing the tree-level derivation from~\cite{Baier:1973ms} and then discuss the one-loop extension in Section~\ref{sec:coll_loop}. The final factorisation formula is then given in~\eqref{eq:collfac_isr} for initial-state radiation (\ac{ISR}) and in~\eqref{eq:collfac_fsr} for final-state radiation (\ac{FSR}). Section~\ref{sec:softcoll} studies the behaviour of these results in the double soft-collinear limit, drawing a connection to the LBK theorem from the previous chapter. Finally, the collinear factorisation formulas are validated in Section~\ref{sec:coll_validation} by means of the $2\to 4$ process $e^- e^+ \to e^- e^+ \gamma\gamma$.

\section{Collinear factorisation at tree level}
\label{sec:coll_tree}

Contrary to the soft limit, care has to be taken in the collinear case when treating the gauge dependence of the emitted photon. Only axial gauge, where the sum over photon polarisations is given by
\begin{equation}\label{eq:axialgauge}
    \sum_{\text{pol}} 
    \epsilon_\mu(k) \epsilon_\nu(k)
    = -g_{\mu\nu}+\frac{k_\mu r_\nu+k_\nu r_\mu}{k\cdot r}\, ,
\end{equation}
does not mix up the power counting of individual diagrams. At tree level a convenient choice for the gauge vector $r$ is $\bar{k}\equiv(E_k,-\vec{k})$. At leading power in the collinear limit we therefore only need to consider diagrams where the photon is emitted from the collinear fermion leg. Restricting the discussion for the moment to ISR, we have
\begin{equation}
    \mathcal{A}^{(0)}_{n+1} 
    \wideeq{k\cdot p\sim \lambda_c^2} \begin{tikzpicture}[scale=.8,baseline={(0,-.1)}]
        
   \draw[line width=.3mm]  (-1,0) node[left]{\footnotesize${p}$} -- (1,0);
    \draw[line width=0.3mm,photon]  (-.3,0) -- (0,.65) node[right]{\footnotesize${k}$};
    
    \draw[line width=.3mm]  [fill=stuff] (1,0) circle (0.45) node[black]{$\Gamma_0$};
       \end{tikzpicture} + \mathcal{O}(\lambda_c^{0})
    = - Q \Gamma_0(p-k)
    \frac{\slashed{p}-\slashed{k}+m}{-2 k\cdot p}
    \gamma^\mu u(p) \epsilon_\mu(k) + \mathcal{O}(\lambda_c^{0})\, .
\end{equation}
We then write the fermion propagator in terms of quasi-real spinors~\cite{Baier:1973ms} with energy
\begin{equation}
    E_{p-k} 
    \equiv \sqrt{(\vec{p}-\vec{k})^2+m^2}
    =E_{p}-E_k + \lambda_c^2\frac{k\cdot p}{E_{p}-E_k}+\mathcal{O}(\lambda_c^4)
\end{equation}
as
\begin{subequations}
\begin{align}
    \frac{\slashed{p}-\slashed{k}+m}{-2 k\cdot p}
    & = \frac{1}{2 E_{p-k}} \sum_s \Big(
        \frac{u^s(p_{ik}) \bar{u}^s(p_{ik})}{E_{p}-E_k-E_{p-k}}
      + \frac{v^s(\bar{p}_{ik}) \bar{v}^s(\bar{p}_{ik})}{E_{p}-E_k+E_{p-k}}
    \Big) \\
    & \wideeq{k\cdot p\sim\lambda_c^2}
    -\frac{1}{\lambda_c^2}\frac{1}{2 k\cdot p}\sum_s 
        u^s(p_{ik}) \bar{u}^s(p_{ik})
      + \mathcal{O}(\lambda_c^{-1})
\end{align}
\end{subequations}
with $p_{ik} = (E_{p-k},\vec{p}-\vec{k})$ and $\bar{p}_{ik} =(E_{p-k},-\vec{p}+\vec{k})$. It is then straightforward to derive the factorised result for the squared amplitude
\begin{equation}\label{eq:collfac_treelevel}
    \mathcal{M}^{(0)}_{n+1} 
    \wideeq{k\cdot p \sim \lambda_c^2} 
    \frac{1}{\lambda_c^2} J^{(0)}_\text{ISR}(x,m)\ \mathcal{M}^{(0)}_n(p-k,m=0) + \mathcal{O}(\lambda_c^{-1})\, ,
\end{equation}
where the tree-level splitting function for ISR in its standard form~\cite{Dittmaier:1999mb} reads
\begin{equation}
    J^{(0)}_\text{ISR}(x,m)
    = \frac{Q^2}{x (k\cdot p)} \Big(
    \frac{1+x^2}{1-x}-\frac{x\,m^2}{k\cdot p} \Big)\, , \quad
    x = \frac{E_p-E_k}{E_p}\, .
\end{equation}
The analogous derivation for FSR yields
\begin{equation}
    \mathcal{M}^{(0)}_{n+1} 
    \wideeq{k\cdot p \sim \lambda_c^2} 
    \frac{1}{\lambda_c^2} J^{(0)}_\text{FSR}(z,m)\ \mathcal{M}^{(0)}_n(p+k,m=0) + \mathcal{O}(\lambda_c^{-1})
\end{equation}
with
\begin{equation}
    J^{(0)}_\text{FSR}(z,m)
    = \frac{Q^2}{k\cdot p} \Big(
    \frac{1+z^2}{1-z}-\frac{m^2}{k\cdot p} \Big)\, , \quad
    z = \frac{E_p}{E_p+E_k}\, .
\end{equation}
Alternatively, $J_\text{FSR}^{(0)}(z,m)$ can be derived from $J_\text{ISR}^{(0)}(x,m)$ via the crossing relation $p\to -p$, i.e. by replacing $k\cdot p \to -k\cdot p$ and $x\to z^{-1}$.

\section{Collinear factorisation at one loop}
\label{sec:coll_loop}

\begin{figure}
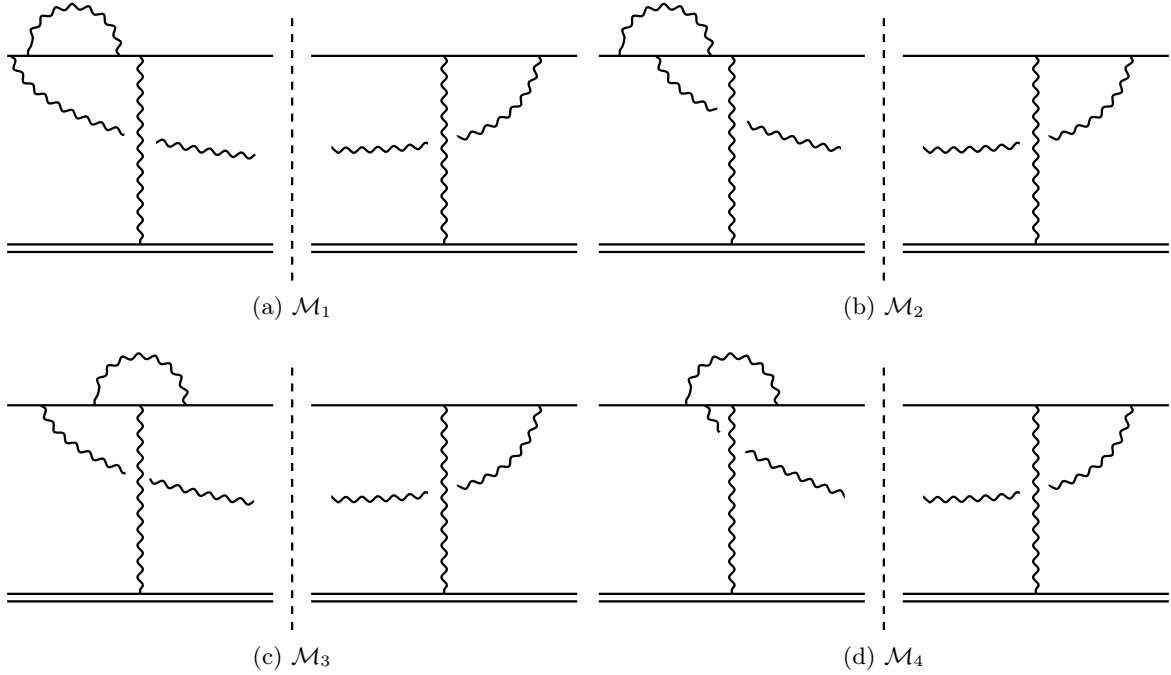

    \centering
    \subfloat[$\mathcal{M}_1$]{
        \begin{tikzpicture}[scale=1,baseline={(0,0)}]
        \input{tikz/mue_diag1}
    \end{tikzpicture}
    \label{mue_diag1}
    }
    \subfloat[$\mathcal{M}_2$]{
        \begin{tikzpicture}[scale=1,baseline={(0,0)}]
        \input{tikz/mue_diag2}
    \end{tikzpicture}
    \label{mue_diag2}
    }\\
    \subfloat[$\mathcal{M}_3$]{
        \begin{tikzpicture}[scale=1,baseline={(0,0)}]
        \input{tikz/mue_diag3}
    \end{tikzpicture}
    \label{mue_diag3}
    }
    \subfloat[$\mathcal{M}_4$]{
        \begin{tikzpicture}[scale=1,baseline={(0,0)}]
        \input{tikz/mue_diag4}
    \end{tikzpicture}
    \label{mue_diag4}
    }
\caption{Interference terms that contribute at LP in the limit where the emitted photon becomes
collinear to the initial-state electron.}
\label{fig:mue_diags}
\end{figure}

The \textit{quasi-real electron method} from the previous section does not work anymore if loop corrections are taken into account due to non-factorisable diagrams where the photon is emitted from a loop. However, the MoR can be applied in this case to disentangle universal collinear contributions from the process-dependent hard part. This is completely analogous to the method of massification discussed in Section~\ref{sec:massification}. Also in this case, it is therefore possible to determine the splitting function based on a specific process. To this end we again use $\mu$-$e$ scattering. This time we consider the radiative process  
\begin{equation}
    e^-(p_1)\mu^-(q_1)\to e^-(p_2)\mu^-(q_2)\gamma(k)
\end{equation}
and calculate the small-mass collinear limit of the one-loop corrections to the electron line. In the case of ISR, the scale hierarchy reads
\begin{equation}
    k\cdot p_1,\ p_1^2=p_2^2=m^2 \sim \lambda_c^2 \ll q_1^2=q_2^2=M^2,\ S \sim \lambda_c^0
\end{equation}
where $S$ once again denotes all hard invariants in the process. Working at leading power and in axial gauge, we only need to take the four interference terms shown in Figure~\ref{fig:mue_diags} into account. In particular, we have used the convenient choice $r=p_2$ for the gauge vector. This choice is allowed since $p_2^2=m^2$ is small. Note that $r=p_1$ is not permissible in this case since the small numerator $k\cdot p_1$ in~\eqref{eq:axialgauge} would disturb the power counting.

The calculation then follows the same steps as for the one-loop massification constant of Section~\ref{sec:massification_oneloop}. Also here the external momenta can be decomposed into large and small components via the light-cone decompostion~\eqref{eq:lightcone_decomposition}. The resulting scaling of the individual components in the light-cone bases then
reads
\begin{subequations}
\begin{align}
    &p_1 \sim k \sim (\lambda_c^2,1,\lambda_c)_1\sim (1,1,1)_2\, , \\
    &p_2 \sim (1,1,1)_1\sim (\lambda_c^2,1,\lambda_c)_2\, , \\
    &q_1 \sim q_2 \sim (1,1,1)_1 \sim (1,1,1)_2\, .
\end{align}
\end{subequations}

Based on the achieved disentanglement of scales it is possible to identify the contributing momentum regions. In order to do so it is helpful to use the formulation of the MoR in the parametric representation introduced in Section~\ref{sec:mor_param} and automatised in the public code \texttt{asy.m}. The following four regions are then found to contribute to the individual interference terms $\mathcal{M}_i$:
\vspace{-.4cm}
\begin{subequations}
\begin{alignat}{3}
    &\mbox{hard:}                  \quad &\ell& \sim (1,1,1)_1               \ \ &\sim&\ \  (1,1,1)_2 \\
    &\mbox{$p_1$-collinear:}       \quad &\ell& \sim (\lambda_c^2,1,\lambda_c)_1 \ \ &\sim&\ \  (1,1,1)_2 \\
    &\mbox{$p_2$-collinear:}       \quad &\ell& \sim (1,1,1)_1               \ \ &\sim&\ \  (\lambda_c^2,1,\lambda_c)_2\\
    &\mbox{$p_2$-ultra-collinear:} \quad &\ell& \sim (1,1,1)_1               \ \ &\sim&\ \  (\lambda_c^4,\lambda_c^2,\lambda_c^3)_2
\end{alignat}
\end{subequations}
The terms that correct the incoming electron line, i.e. $\mathcal{M}_1$ and $\mathcal{M}_2$, get only contributions from the $p_1$-collinear region. Furthermore, at leading power the hard region only contributes to the factorisable diagram $\mathcal{M}_3$.  Since we can apply the quasi-real electron method in this case it follows immediately that
\begin{equation}\label{eq:fac_hard}
    \mathcal{M}_{n+1}^{(1),\text{hard}} 
    = \frac{1}{\lambda_c^2}J^{(0)}_\text{ISR}(x,m) \mathcal{M}^{(1)}_n(p_1-k,m=0) + \mathcal{O}(\lambda_c^{-1})\, .
\end{equation}
In addition to the hard region, all other three scalings contribute to $\mathcal{M}_3$. In the case of $\mathcal{M}_4$, on the other hand, only the $p_1$-collinear and $p_2$-ultra-collinear regions are present at leading power. As can be expected, the unphysical ultra-collinear region cancels between $\mathcal{M}_3$ and $\mathcal{M}_4$. We are then left with the two collinear contributions. They can be computed with the same methods used for the one-loop calculation of the massification constant presented in Section~\ref{sec:massification_oneloop}. Also in this case it is of paramount importance to perform the expansion in the respective region before tensor reduction. Otherwise, it is not possible to correctly separate the individual momentum regions. Using the simplifying relation
\begin{align}\label{eq:collmomrel}
	k = \frac{E_k}{E_{p_i}} p_i + \mathcal{O}(\lambda_c) 
	= (1-x) p_i + \mathcal{O}(\lambda_c)
\end{align}
we then find that the collinear contributions factorise according to
\begin{subequations}
\begin{align}
    \mathcal{M}^{(1),p_1\text{-coll}}_{n+1} 
    &= \frac{1}{\lambda_c^2} J^{(1)}_\text{ISR}(x,m)\ \mathcal{M}_n^{(0)}(p_1-k,m=0) + \mathcal{O}(\lambda_c^{-1})\, , \\
    \mathcal{M}^{(1),p_2\text{-coll}}_{n+1} 
    &= \frac{1}{\lambda_c^2} Z^{(1)}(m) \ J^{(0)}_\text{ISR}(x,m)\ \mathcal{M}_n^{(0)}(p_1-k,m=0) + \mathcal{O}(\lambda_c^{-1})\, .
\end{align}
\end{subequations}
Apart from the interference terms $\mathcal{M}_i$ we also need to take into account mass and wave function renormalisation. All counterterms connected to the heavy particles (muon) enter in~\eqref{eq:fac_hard} in the renormalisation of the non-radiative massless one-loop correction $\mathcal{M}_n^{(1)}$. The counterterms for the emitting electron, on the other hand, renormalise the one-loop splitting function $J^{(1)}_\text{ISR}$, while the ones for the other light particle (outgoing electron) contributes to $Z^{(1)}$. The renormalised results for $J^{(1)}_\text{ISR}$ and $Z^{(1)}$ are given in Appendix~\ref{sec:splitfunc}. The factor $Z^{(1)}$ corresponds to the one-loop massification constant~\eqref{eq:massification_constant}. The $p_2$-collinear contribution therefore takes the leading-order mass effects of the outgoing electron into account. The one-loop splitting function $J^{(1)}_\text{ISR}$ contains both the corresponding mass terms as well as leading-power corrections due to the collinear emission. The small-mass collinear limit considered here can thus be viewed as a generalisation of massification.  

\begin{figure}
    \centering
    \begin{tikzpicture}[scale=1,baseline={(1,0)}]
    	\input{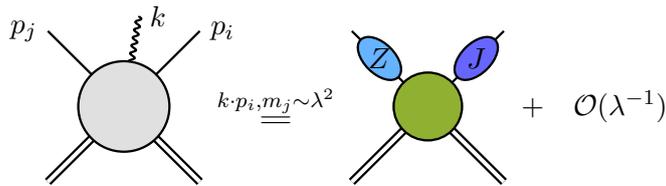}
    \end{tikzpicture}
    \caption{Schematic illustration of the collinear factorisation
    formulas of~\eqref{eq:collfac_isr} and~\eqref{eq:collfac_fsr}.}
    \label{fig:coll_fac}
\end{figure}

As already mentioned previously the collinear contributions are expected to be process independent. Thus, one-loop diagrams for $\mu$-$e$ scattering other than those shown in Figure~\ref{fig:mue_diags} are note expected to lead to such contributions. We have explicitly checked that this is the case due to a cancellation between diagram pairs that are related (up to a sign) through the crossing $q_1\leftrightarrow-q_2$. This is completely analogous to the case of massification (see Section~\ref{sec:massification_oneloop}). The only additional contribution is therefore the hard one originating from factorisable diagrams that trivially exhibit the factorising structure of~\eqref{eq:fac_hard}. If, on the other hand, we take the muon to be light as well, i.e. $M^2\sim m^2\sim \lambda_c^2$, there are two additional collinear contributions with exactly the same structure as for the outgoing electron
\begin{equation}
    \mathcal{M}_{n+1}^{(1),q_1\text{-coll}}=
    \mathcal{M}_{n+1}^{(1),q_2\text{-coll}}
    =\frac{1}{\lambda_c^2} Z^{(1)}(M) J_\text{ISR}^{(0)}(x,m) \mathcal{M}_n^{(0)}(p_1-k,m=0,M=0)
    +\mathcal{O}(\lambda_c^{-1})
\end{equation}
consistent with the expectation based on massification. We are thus lead to the main result of this section that at one loop can be written through the factorisation formula
\begin{equation}\label{eq:collfac_isr}
    \mathcal{M}_{n+1} \wideeq{k\cdot p_i, m_j^2 \sim \lambda_c^2}
    \frac{1}{\lambda_c^2}J_\text{ISR}(x,m_i) \Bigg( \prod_{j\neq i} Z(m_j) \Bigg) \mathcal{M}_n(p_i-k,m_i=0,m_j=0)
    + \mathcal{O}(\lambda_c^{-1}),
\end{equation}
where we have defined the all-order quantities
\begin{subequations}
\begin{align}
    &J_\text{ISR} = J_\text{ISR}^{(0)} + J_\text{ISR}^{(1)} + \mathcal{O}(\alpha^3)\, , \\
    &Z = 1 + Z^{(1)} + \mathcal{O}(\alpha^2)\, .
\end{align}
\end{subequations}
In \eqref{eq:collfac_isr} the product is over all external fermion lines $j\neq i$ with a small mass $m_j^2\sim\lambda_c^2$.  
We therefore indeed confirm the factorising structure given in \eqref{eq:coll_amp} that was predicted from first-principle considerations in SCET. In particular, we find that the Wilson coefficient $C_{n+1}$ in~\eqref{eq:coll_amp} is related to the non-radiative process with shifted kinematics and that there is no soft contribution at one loop. Furthermore, the same calculation with only minor modifications can also be applied to the case of FSR yielding the analogous formula
\begin{equation}\label{eq:collfac_fsr}
    \mathcal{M}_{n+1} \wideeq{k\cdot p_i, m_j^2 \sim \lambda_c^2}
    \frac{1}{\lambda_c^2}J_\text{FSR}(z,m_i) \Bigg(\prod_{j\neq i} Z(m_j)\Bigg) \mathcal{M}_n(p_i+k,m_i=0,m_j=0)
    + \mathcal{O}(\lambda_c^{-1})\, .
\end{equation}
A schematic illustration of these factorisation formulas is given in Figure~\ref{fig:coll_fac}. Furthermore, the corresponding expressions for $J_\text{ISR}$, $J_\text{FSR}$, and $Z$ can be found in Appendix~\ref{sec:splitfunc}. As expected, we find that the ISR and FSR splitting functions are related via crossing symmetry.

It is useful to compare our findings to the corresponding factorisation formula for massless fermions that can be extracted from the QCD results of~\cite{Bern:1994zx}. Suppressing the separation into ISR and FSR, the massless collinear limit can be written as
\begin{align}
    \mathcal{M}_{n+1} \wideeq{k\cdot p_i \to 0} \bar{J}(y) \mathcal{M}_n\, ,
\end{align}
where the only process-independent contribution $\bar{J}$ comes from the collinear fermion ($p_i$) and $y\in\{x,z\}$. The corresponding expressions at tree level and at one loop are given in Appendix~\ref{sec:splitfunc}. For massive particles, on the other hand, every light fermion contributes an additional factor in the factorisation formula taking into account the corresponding small-mass effects. This results in the more complex collinear structure of~\eqref{eq:collfac_isr} and~\eqref{eq:collfac_fsr} than one would have naively expected based on the known QCD formula. Nevertheless, it turns out that there is a relation between the massive and massless splitting functions. In particular, we find
\begin{subequations}\label{eq:massless_splittings}
\begin{align}
    &J^{(0)}(y,m) \wideeq{m\to 0} \bar{J}^{(0)}(y) + \mathcal{O}(m)\, , \\
    &J^{(1)}(y,m) \wideeq{m\to 0}
    \bar{J}^{(1)}(y) + Z^{(1)}(m)\bar{J}^{(0)}(y)+ \mathcal{O}(m)\, ,
\end{align}
\end{subequations}
where the massive splitting function reduces in the massless limit to the massless one plus singular corrections from massification. It is conceivable that the same relation will also hold beyond one loop. In this case, however, there will be non-vanishing soft contributions from closed fermion loops as discussed in Section~\ref{sec:massification_twoloop}. In addition to being an interesting result in its own right, this represents a strong check for the validity of the results presented in this section and in Appendix~\ref{sec:splitfunc}.

\section{Next-to-soft collinear limit}\label{sec:softcoll}

A highly non-trivial consistency check of both the collinear factorisation formulas~\eqref{eq:collfac_isr} and~\eqref{eq:collfac_fsr} as well as the LBK theorem~\eqref{eq:lbk_oneloop} is the comparison of these results in the next-to-soft collinear limit. In this double limit we expect the two results to coincide.

We start with the expansion of the initial-state formula~\eqref{eq:collfac_isr} in the soft limit. Based on the results given in Appendix~\ref{sec:splitfunc}, we find for the initial-state splitting function
\begin{align}
	J_\text{ISR} 
	\wideeq{k\sim\lambda_s}
	\frac{1}{\lambda_s^2} J_\text{ISR}^\text{sLP} 
	+ \frac{1}{\lambda_s} J_\text{ISR}^\text{sNLP} 
	+ \mathcal{O}(\lambda_s^0)
\end{align}
with
\begin{subequations}
\begin{align}
	J_\text{ISR}^\text{sLP} &= \eik_\text{coll} \, Z(m_i)\, , \\
	J_\text{ISR}^\text{sNLP} &= \eik_\text{coll} \, \frac{Q_i^2 (1-x)}{ 4\pi^2}
	\Bigg(\frac{1}{\epsilon}+2+2\log\Big(\frac{m^2}{2k\cdot p_i}\Big)  \Bigg)
	\equiv \eik_\text{coll} \, \mathcal{S}_\text{coll}
\end{align}
\end{subequations}
and
\begin{align}
	\eik_\text{coll} 
	= \frac{Q_i^2}{k\cdot p_i} 
	\Bigg(\frac{2}{1-x}-\frac{m^2}{k \cdot p_i}\Bigg)\, .
\end{align}
The notation is chosen such as to indicate that $\eik_\text{coll}$ corresponds to the eikonal~\eqref{eq:eikonal} expanded in the collinear limit. Inserting this into the initial-state factorisation formula~\eqref{eq:collfac_isr} then yields for the next-to-soft collinear limit\footnote{As for the collinear factorisation formulas~\eqref{eq:collfac_isr} and~\eqref{eq:collfac_fsr} this has only been shown to hold true up to the one-loop level.}
\begin{align}\label{eq:ntscoll}
	\mathcal{M}_{n+1}
	\overset{k\cdot p_i,m_j^2\sim\lambda_c^2}{ \underset{k\sim\lambda_s}{=}}
	\frac{1}{\lambda_c^2 \lambda_s^2} \eik_\text{coll}
	\Big(1-\lambda_s\, k\cdot\frac{\partial}{\partial p_i}
	+\lambda_s\, S_\text{coll} \Big)
	\mathcal{M}_n^\text{massified}(p_i,m_j=0)
	+\mathcal{O}(\lambda_c^{-1}\lambda_s^0)
\end{align}
with the massified squared amplitude, $\mathcal{M}_n^\text{massified}$, given by~\eqref{eq:massification}.

We can then compare this to the LBK theorem~\eqref{eq:lbk_oneloop} expanded in the collinear limit. To do so we use of the relation~\eqref{eq:collmomrel} 
combined with charge conservation
\begin{align}
	\sum_{l\neq i} Q_l = -Q_i
\end{align}
to solve the sum over the external fermions. The hard contribution~\eqref{eq:lbk_oneloop_hard} then reduces to the first two terms in~\eqref{eq:ntscoll}, while the third term is reproduced by the soft part~\eqref{eq:lbk_oneloop_soft}.

We can therefore conclude that both the collinear as well as the next-to-soft limit yield consistent expressions in the soft-collinear region. This is a strong check for both results. At one loop, this is not particularly relevant since the limits can be checked against exact calculations (Section~\ref{sec:lbk_validation} and Section~\ref{sec:coll_validation}). However, when trying to generalise these formulas beyond one loop this provides an important consistency check.

\section{Validation}\label{sec:coll_validation}

Analogously to the validation of the next-to-soft limit in Section~\ref{sec:lbk_validation} we demonstrate the correctness and applicability of the collinear factorisation formulas~\eqref{eq:collfac_isr} and~\eqref{eq:collfac_fsr} with 
the process~\eqref{eq:dummyprocess} (double radiative Bhabha scattering). Again, we use a CMS energy of $\sqrt{s}=10.583\,\mathrm{GeV}$ and compare our approximations to OpenLoops running in quadruple precision mode. 

Once the massless one-loop correction for the process
\begin{align}
e^-(p_1)e^+(p_2) \to e^-(p_3) e^+(p_4) \gamma(p_5)
\end{align}
is known, the application of the factorisation formulas~\eqref{eq:collfac_isr} and~\eqref{eq:collfac_fsr} is rather straightforward. As an example we consider the case of the photon $k$ becoming collinear to $p_1$ (ISR) or the case of it becoming collinear to $p_3$ (FSR). The cases of $p_2$ and $p_4$ are completely analogous.

The massified approximation~\eqref{eq:massification} for the squared amplitude reads
\begin{align}
\mathcal{M}_{n+1} \wideeq{m_i\sim\lambda}
    Z(m_1) Z(m_2) Z(m_3) Z(m_4)
    \mathcal{M}_{n+1}(p_1, p_2, p_3, p_4, p_5, k; m_i=0)
    +\mathcal{O}(\lambda) 
\end{align}
which is valid for the bulk of the phase space, i.e. assuming $k$ is neither soft nor collinear. Note that the masses are only given indices so that the different $Z$ can be better disentangled once $k$ becomes collinear. Of course all $m_i$ are equal.

\begin{figure}
    \centering
    \includegraphics[width=0.8\textwidth]{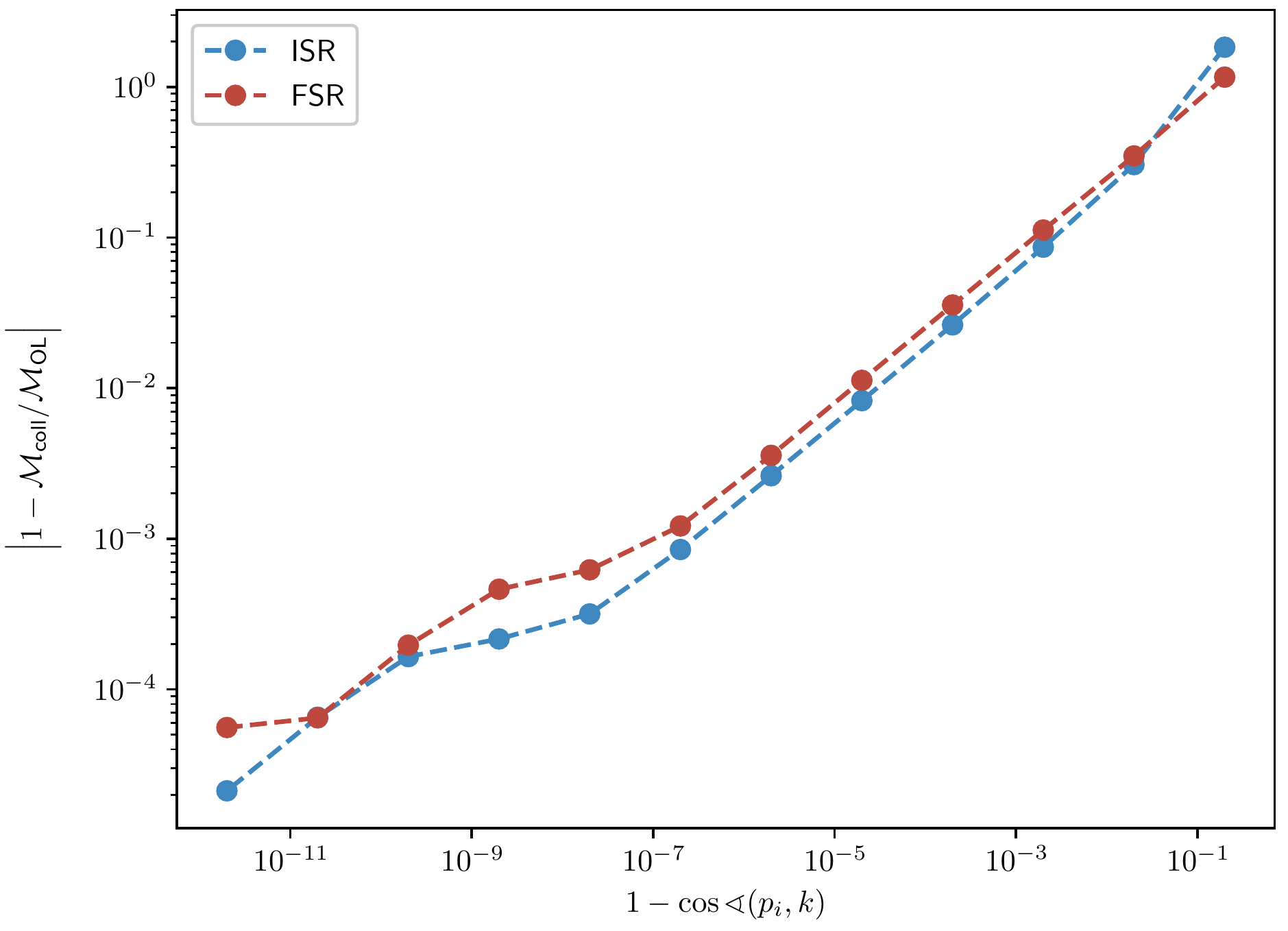}
    \caption{Convergence of the collinear limit
    $\mathcal{M}_{\text{coll}}$ at LP for ISR
    \eqref{eq:rrv:isr} and FSR \eqref{eq:rrv:fsr} as a function of the
    collinearity for the one-loop correction to $e^-e^+\to e^- e^+ \gamma\gamma$. The reference value $\mathcal{M}_{\text{OL}}$ is
    calculated with OpenLoops in quadruple precision.}
\label{fig:ee2eegg_pcl}
\end{figure}

In the ISR limit we replace $Z(m_1)$ with $J_\text{ISR}$, reducing the number of particles in the massless matrix element
\begin{align}
\begin{split}
  \lefteqn{\mathcal{M}_{n+1} \wideeq{k\cdot p_1,m_i^2\sim\lambda_c^2} } \\ & \quad
   \frac{1}{\lambda_c^2}J_\text{ISR}(x, m_1)  Z(m_2) Z(m_3) Z(m_4)
    \mathcal{M}_{n}(p_1-k, p_2, p_3, p_4, p_5; m_i=0)
    +\mathcal{O}(\lambda_c^{-1})\, .\label{eq:rrv:isr}
\end{split}
\end{align}
In complete analogy the FSR limit is given by
\begin{align}
\begin{split}
  \lefteqn{\mathcal{M}_{n+1} \wideeq{k\cdot p_3,m_i^2\sim\lambda_c^2}} \\ & \quad
    \frac{1}{\lambda_c^2}Z(m_1) Z(m_2) J_\text{FSR}(z, m_3)  Z(m_4)
    \mathcal{M}_{n}(p_1, p_2, p_3+k, p_4, p_5; m_i=0)
    +\mathcal{O}(\lambda_c^{-1})\, .\label{eq:rrv:fsr}
\end{split}
\end{align}
All that is left to do before we can compare to OpenLoops is to multiply out the terms in \eqref{eq:rrv:isr} and \eqref{eq:rrv:fsr}. The result of this comparison is shown in Figure~\ref{fig:ee2eegg_pcl} for ISR and FSR as a function of the `collinearity' $1-\cos\sphericalangle(p_i,k)$. To understand the observed convergence behaviour it is important to realise that the expansion is not performed in the collinearity but in $k\cdot p_i, m_i^2 \to 0$. The approximation thus only improves while $k\cdot p_i$ gets smaller. At the point, however, where 
\begin{align}
 k\cdot p_i = E_i E_k \frac{m_i^2}{E_i^2} + \mathcal{O}(\lambda_c^4)
\end{align}
is approximately satisifed, the limit saturates since $m_i$ is kept constant. This explains the kink at $10^{-7}$. We can therefore conclude that Figure~\ref{fig:ee2eegg_pcl} represents a strong validation of our factorisation formulas given in~\eqref{eq:collfac_isr} and~\eqref{eq:collfac_fsr}.

\chapter{Results}\label{chap:results}

This section presents for the first time the complete set of fully differential NNLO corrections to $\mu$-$e$ scattering
\begin{align}\label{eq:muone_labelling}
	e^-(p_1) \mu^-(p_2) \to e^-(p_3) \mu^-(p_4)\{\gamma(p_5)\gamma(p_6)\}
\end{align}
that will be published in~\cite{muonennlo}. We give separate results for photonic and fermionic contributions as defined at the beginning of Chapter~\ref{chap:fermionic}. In both cases, the UV and IR divergences are regularised in $d=4-2\epsilon$ dimensions and the renormalisation is performed in the on-shell scheme. All photonic amplitudes are computed analytically. The fermionic corrections are entirely due to VP and are calculated with the semi-numerical hyperspherical method (Chapter~\ref{chap:fermionic}). In particular, this also includes non-perturbative hadronic contributions.

All photonic tree-level and one-loop diagrams (apart from the real-virtual contribution) are generated using QGraf~\cite{NOGUEIRA1993279} and calculated with \texttt{Package-X}~\cite{Patel:2015tea}. Compared to automated one-loop tools, this allows us to implement these contributions more efficiently. While for most of the obtained expressions a sufficiently stable and efficient implementation is possible, we use a different approach for the numerically delicate real-virtual amplitude. We combine next-to-soft stabilisation (Chapter~\ref{chap:nts}) with the remarkable numerical stability of OpenLoops~\cite{Buccioni:2017yxi,Buccioni:2019sur}. The one-loop generalisation of the LBK theorem given in~\eqref{eq:lbk_oneloop} can be used to calculate the corresponding soft expansion up to NLP. This approach yields a stable and fast implementation of this problematic contribution.

The photonic corrections are further split into gauge invariant subsets according to~\eqref{eq:twoloop_split}. The pure electronic and muonic two-loop amplitudes can be calculated with full mass dependence based on the heavy quark form factor~\cite{Mastrolia:2003yz,Bonciani:2003ai,Bernreuther:2004ih,Gluza:2009yy}. The corresponding result is given in terms of HPLs that are evaluated with the \texttt{FORTRAN} subroutine \texttt{hplog}~\cite{Gehrmann:2001pz}. In the case of the remaining (mixed) corrections such an exact computation is highly challenging. Instead, we apply the method of massification (Section~\ref{sec:massification}) to the recently completed massless two-loop amplitude~\cite{Bonciani:2021okt}. 
Contrary to the form factor corrections, the result cannot be expressed in HPLs alone but instead is written in terms of generalised polylogarithms (\ac{GPL})~\cite{Goncharov:1998kja}. These functions can be efficiently evaluated using \texttt{handyG}~\cite{Naterop:2019xaf}, a \texttt{FORTRAN} implementation of the algorithm presented in~\cite{Vollinga:2004sn}. Very recently, the beta version v0.2.0b was completed which includes an improved version of the cache system. The resulting speed up is crucial due to the large number of GPLs (5044) that have to be evaluated. With the improved code we arrive at a reasonable evaluation speed of $\mathcal{O}(1\, \text{s})$ per phase-space point.

All corrections are implemented in the \mcmule{} framework, which allows to calculate any (IR-safe) observable fully differentially. The soft singularities arising in the phase-space integration of radiative contributions are dealt with using the FKS$^\ell$ subtraction scheme (Section~\ref{sec:fks}). Collinear divergences, on the other hand, are naturally regularised by the finite masses of the fermions. Since the electron mass is small this results in CPS that hamper the reliability of the phase-space integration. To address this issue we use a dedicated tuning of the phase-space parametrisation (Chapter~\ref{chap:cps}) to help the \texttt{vegas} integration~\cite{Lepage:1980jk} find and deal with these problematic regions.

The following comparisons have been conducted to verify the correctness of the calculation to the extent possible. In case of the fermionic corrections, we have compared with the dispersive calculation of~\cite{Fael:2019nsf}. Perfect agreement was obtained for the leptonic contributions where the only uncertainty is due to the precision of the numerical integration. In case of the hadronic corrections a deviation of $\sim 1\%$ in the NNLO coefficient was observed consistent with the expected uncertainty of the HVP. 
Furthermore, we have verified the calculation of the complete set of electronic corrections that we have published in~\cite{Banerjee:2020rww} with the \textsc{MESMER} result of~\cite{CarloniCalame:2020yoz}. In the case of the mixed corrections, such a full check is currently not possible since no other calculation exists. Instead, we have performed a dedicated comparison with the \textsc{MESMER} collaboration~\cite{carlo} for the radiative process at NLO. A small photon energy cut of $\xi > 10^{-6}$ was used to test the real-virtual contribution in the soft region. Perfect agreement was found both at the integrated as well as at the differential level.

In the following, we present fully differential NNLO predictions for observables tailored to the MUonE experiment. With the momenta of the particles labelled as in~\eqref{eq:muone_labelling} we define the invariants $t_e=(p_1-p_3)^2$ and $t_\mu=(p_2-p_4)^2$. In the elastic case we have $t_e=t_\mu$. The energy of the outgoing electron and muon are denoted by $E_e$ and $E_\mu$, respectively. Additionally, we use $\theta_e$ and $\theta_\mu$ as the corresponding scattering angles relative to the beam axis. We further assume a $150\,\text{GeV}$ muon beam, consistent with the M2 beam line at CERN North Area~\cite{Abbiendi:2677471}, incident on an electron at rest.

As mentioned in the introduction, the total cross section for $\mu$-$e$ scattering is ill-defined due to the behaviour $\D\sigma/\D t\sim t^2$ in~\eqref{eq:mue_born} with $t_\text{min}\leq t \leq 0$. We therefore have to apply a cut on the maximal value of $t$ or equivalently on the minimal energy of the outgoing electron. In all of the results below we choose $E_e>1\,\text{GeV}$. To model the geometry of the detector we require in addition that $\theta_\mu > 0.3\,\text{mrad}$.

The MUonE experiment aims at measuring the elastic scattering of muons and electrons. The elasticity requirement is needed in order to be able to reconstruct the momentum flowing through the HVP for a given event. To implement this in the calculation we use the condition~\eqref{eq:elasticity_band} and apply the elasticity cut
\begin{align}
	0.9 < \frac{\theta_\mu}{\theta_\mu^\text{el}} < 1.1 \, .
\end{align}
In the following, we present results with and without this additional cut, in order to analyse its impact on the radiative corrections. We therefore consider the two scenarios
\begin{itemize}
	\item
	\texttt{S1}: $E_e>1\,\text{GeV}$, $\theta_\mu > 0.3\,\text{mrad}$,
	\item
	\texttt{S2}: $E_e>1\,\text{GeV}$, $\theta_\mu > 0.3\,\text{mrad}$, $0.9<\theta_\mu/\theta_\mu^\text{el}<1.1$.
\end{itemize}
All of the presented results use the input parameters~\cite{PhysRevD.98.030001}
\begin{align}
\begin{split}
	&\alpha = 1/137.035999084, \quad\quad\,\, \, m_e = 0.510998950\, \text{MeV}, \\
	& m_\mu = 105.658375\, \text{MeV}, \quad\quad m_\tau = 1776.86\, \text{MeV}\, .
\end{split}
\end{align}
Furthermore, we rely on \texttt{alphaQED}~\cite{Jegerlehner:2001ca,Jegerlehner:2006ju,Jegerlehner:2011mw} for the HVP. In particular, we use the most recent version \texttt{alphaQEDc19}.

\begin{figure}
    \centering
    \subfloat[electronic, $\chi^2=1.3$]{
        \includegraphics[width=.55\textwidth]{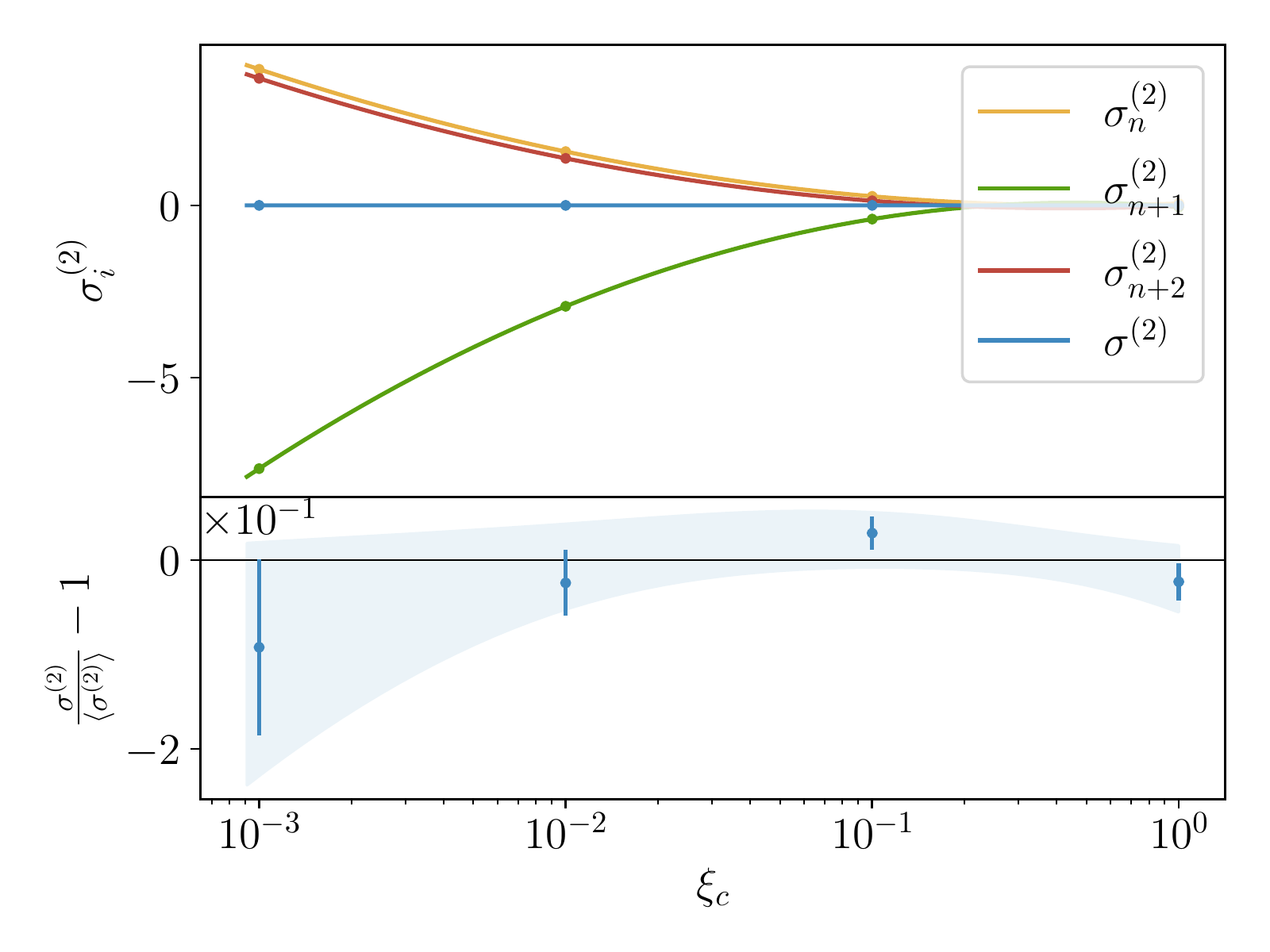}
    \label{fig:xicut_eeee_nobandcu}
    }\\ \vspace{-.4cm}
    \subfloat[mixed, $\chi^2=0.8$]{
        \includegraphics[width=.55\textwidth]{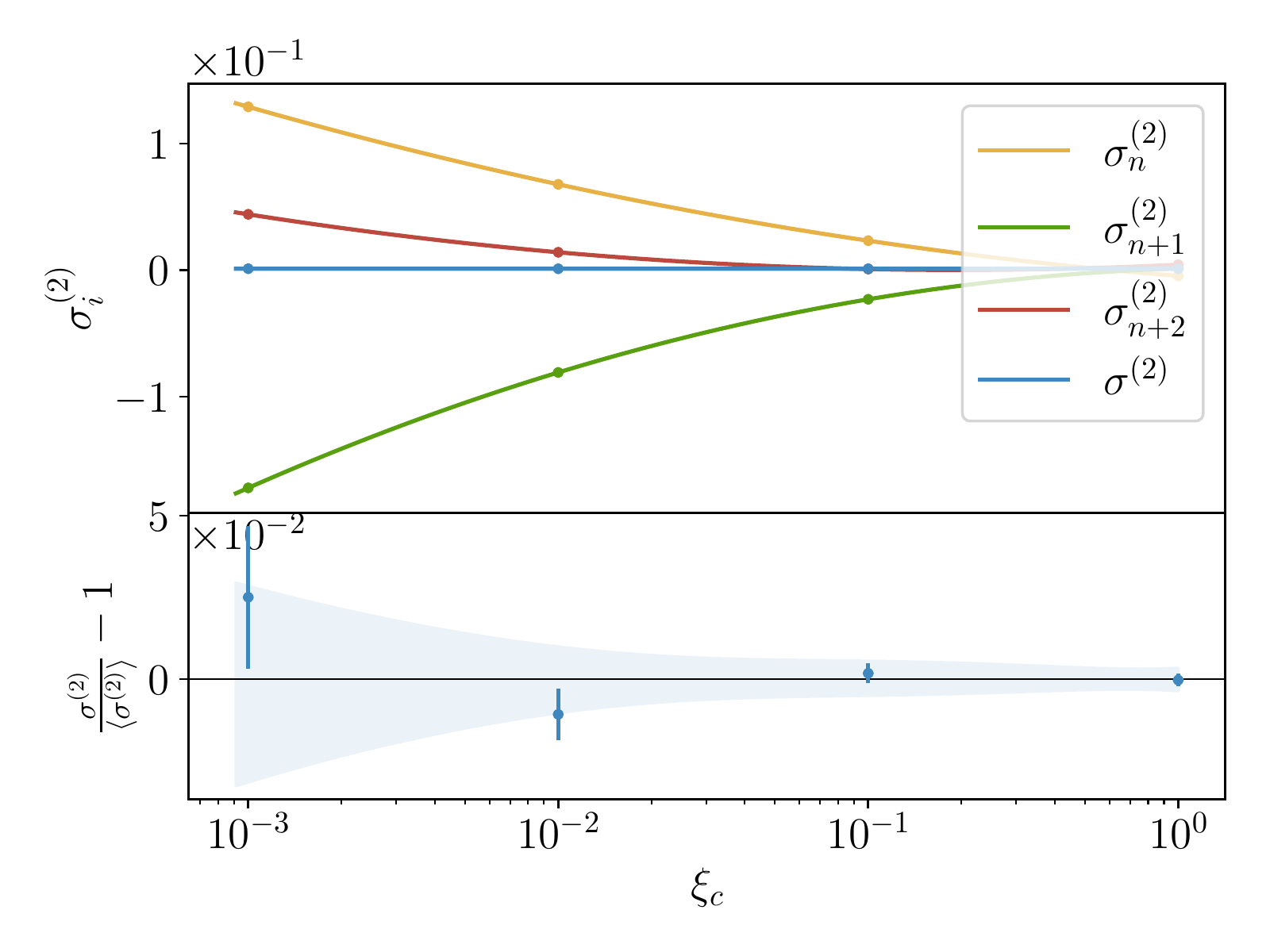}
    \label{fig:xicut_mixd_nobandcut}
    }\\ \vspace{-.4cm}
    \subfloat[muonic, $\chi^2=0.5$]{
        \includegraphics[width=.55\textwidth]{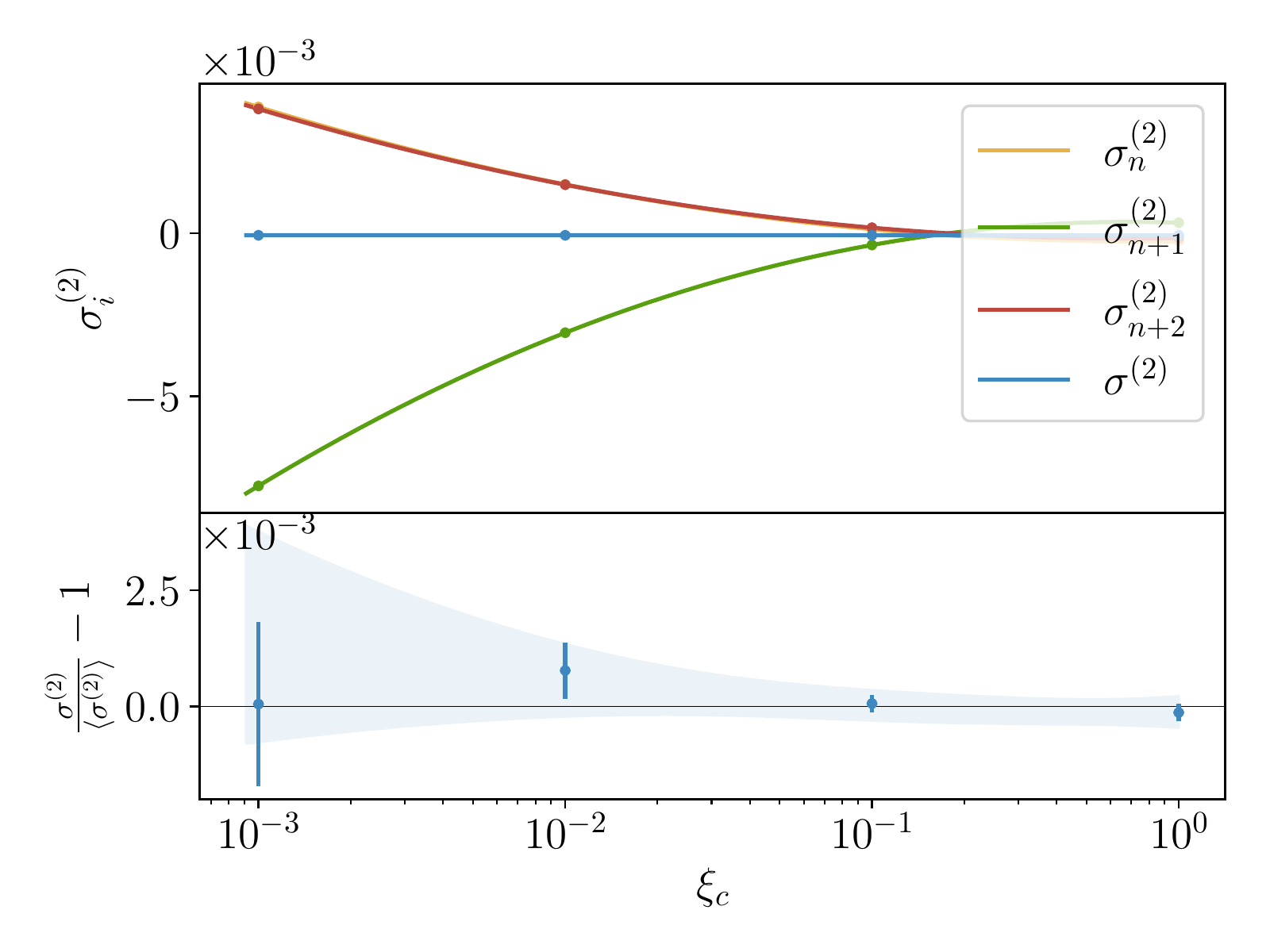}
    \label{fig:xicut_mmmm_nobandcut}
    }\caption{The $\xi_c$ (in)dependence of the photonic NNLO corrections to the integrated cross section for \texttt{S1}. Separate results are shown for purely electronic and muonic as well as mixed corrections. The error band shows the $68\%$ confidence level of the fit. We further give the chi-square as an unbiased measure of the (in)dependence.}
\label{fig:xicut_muone_nnlo}
\end{figure}

As explained in Section~\ref{sec:fks}, a crucial feature of the FKS$^\ell$ subtraction scheme is the exact cancellation of the $\xi_c$ dependence among the individual pieces in~\eqref{eq:fks_nnlo}. This serves as a strong check for the correctness of the implementation as well as the numerical stability of the integration. We have found perfect $\xi_c$ independence for all contributions at NLO and at NNLO. This is shown in Figure~\ref{fig:xicut_muone_nnlo} for the most delicate photonic NNLO corrections in scenario \texttt{S1}. The larger Monte Carlo error for smaller $\xi_c$ is due to the cancellation between the three different contributions behaving as $\sim \log^2{\xi_c}$. The smaller values are thus only used to check the $\xi_c$ independence. For actual predictions one typically chooses $\xi_c \gtrsim 0.1$. The observed $\xi_c$ independence thus implies that a reliable sampling of the CPS is ensured by the dedicated tuning of the phase-space parametrisation. This is analogous to the discussion of Section~\ref{sec:tuning_example}. Furthermore, it also strongly suggests that the numerical instabilities encountered in the real-virtual corrections are under good control due to the next-to-soft stabilisation.

\begin{table}
\centering
 \begin{tabular}{c|r r|| r r} 
 & \multicolumn{2}{c||}{$\sigma/\rm \upmu b$} & \multicolumn{2}{c}{$\delta K^{(i)}/\%$} \\
  & \multicolumn{1}{c}{\tt S1} & \multicolumn{1}{c||}{\tt S2} & \multicolumn{1}{c}{\tt S1} & \multicolumn{1}{c}{\tt S2} \\[0.3ex] 
 \hline
 \rule{0pt}{3ex}
 $\sigma^{(0)}$ & \tt 121.42288  &\tt 121.42288 & & \\[0.3ex]
 \hline
 \rule{0pt}{3ex}
 $\sigma^{(1)}_e$ & \tt   -0.73604(5)  &\tt  -5.28211(5)& \tt -0.60618(5)&\tt -4.35018(4)\\
 $\sigma^{(1)}_{e\mu}$ & \tt   -0.23986  &\tt  -0.18200& \tt -0.19754&\tt -0.14989\\
 $\sigma^{(1)}_\mu$ & \tt   -0.03514  &\tt  -0.16824& \tt -0.02894&\tt -0.13856\\
 $\sigma^{(1)}_\mathrm{VP}$ & \tt   1.57105  &\tt  1.57105& \tt 1.29386&\tt 1.29386\\[0.5ex]
 $\sigma^{(1)}_\mathrm{had}$ & \tt   0.01597  &\tt  0.01597& \tt 0.01315&\tt 0.01315\\[0.5ex]
 \hline
 \rule{0pt}{3ex}
 $\sigma^{(2)}_e$ & \tt  0.00159(2)  &\tt  0.07426(1)& \tt 0.00131(2) &\tt  0.06327(1) \\
  $\sigma^{(2)}_{e\mu}$ & \tt  0.0011  &\tt  0.02066& \tt 0.00091 &\tt  0.01761 \\
   $\sigma^{(2)}_\mu$ & \tt  -0.00006  &\tt  0.000005& \tt -0.00004 &\tt  0.000004 \\
 $\sigma^{(2)}_\mathrm{VP}$ & \tt  -0.01358  &\tt  -0.07341& \tt -0.01113 &\tt  -0.06255 \\[0.5ex]
 \hline\hline
 \rule{0pt}{2.5ex}
 $\sigma_{2}$   & \tt   121.97194(6) &\tt 117.38309(5)& & \\
\end{tabular}
\caption{\label{tab:xsection}
The integrated cross section for \texttt{S1} and \texttt{S2} at LO,
NLO, and NNLO. The results are split into purely electronic and muonic, mixed, and VP corrections. All three leptons as well as the hadronic contribution are included in the VP. Furthermore, the hadronic correction at NLO, $\sigma_\text{had}^{(1)}$, is also given separately. Where no error is given, all digits are significant compared to the precision of the numerical integration.}
\end{table}

The order-by-order contributions, $\sigma^{(i)}$, to the integrated cross section, $\sigma_2 = \sigma^{(0)} +  \sigma^{(1)}+ \sigma^{(2)}$, for both scenarios are presented in Table~\ref{tab:xsection}. Electronic, muonic, mixed, and VP corrections are given separately and are denoted by $\sigma_e^{(i)}$, $\sigma_{e\mu}^{(i)}$,  $\sigma_\mu^{(i)}$, and $\sigma_\text{VP}^{(i)}$, respectively. All three leptons as well as the hadronic contribution are included in the VP. Furthermore, the hadronic correction at NLO, $\sigma_\text{had}^{(1)}$, is also given separately. Additionally, we show the corresponding $K$ factors defined as
\begin{align}\label{eq:kfac}
	K^{(i)} = 1 + \delta K^{(i)} = \frac{\sigma_i}{\sigma_{i-1}}\, .
\end{align}

Before discussing these results, we comment on the behaviour of these corrections when going from negatively charged muons in~\eqref{eq:muone_labelling} to positive ones. The two processes are related via the crossing relation $p_2 \leftrightarrow- p_4$. Alternatively, one can also replace $Q_\mu \to -Q_\mu$ in the gauge invariant split~\eqref{eq:twoloop_split}. This, in turn, implies that the purely electronic and muonic corrections are the same for positively and negatively charged muons, while the NLO mixed contribution, $\sigma_{e\mu}^{(1)}$, changes sign. Only the mixed NNLO correction, $\sigma_{e\mu}^{(2)}$, has no definite behaviour under this transformation since it includes multiple gauge invariant subsets.

We observe moderate NLO and NNLO corrections of around $1 \%$ and $0.01 \%$ for \texttt{S1} (without elasticity cut) and  $5 \%$ and $0.05 \%$ for \texttt{S2} (with elasticity cut). Based on a naive extrapolation, the error due to missing higher-order corrections is estimated to be well below MUonE's $10\, \text{ppm}$ target precision. This statement is, however, only true for the integrated cross section. We will see below that the corrections can be much larger at the differential level. As discussed in Section~\ref{sec:10ppm}, the elasticity cut forces additional radiation to be soft and therefore introduces additional large logarithms. This results in larger corrections for \texttt{S2} compared to \texttt{S1}. The only exception is the purely muonic NNLO contribution, $\sigma_\mu^{(2)}$, where the correction changes sign. Furthermore, the fermionic contributions completely dominate the corrections in the case of \texttt{S1}. The situation is more balanced for \texttt{S2} where the soft enhancement is more pronounced for the photonic corrections where up to two photons are emitted. Interestingly, a similar effect can be observed when comparing the purely electronic corrections - which are expected to be dominant - with the mixed contributions. While they are of the same size for \texttt{S1}, the expected hierarchy is better satisfied for \texttt{S2}.

\begin{figure}
    \centering
    \subfloat[\texttt{S1}]{
        \includegraphics[width=.8\textwidth]{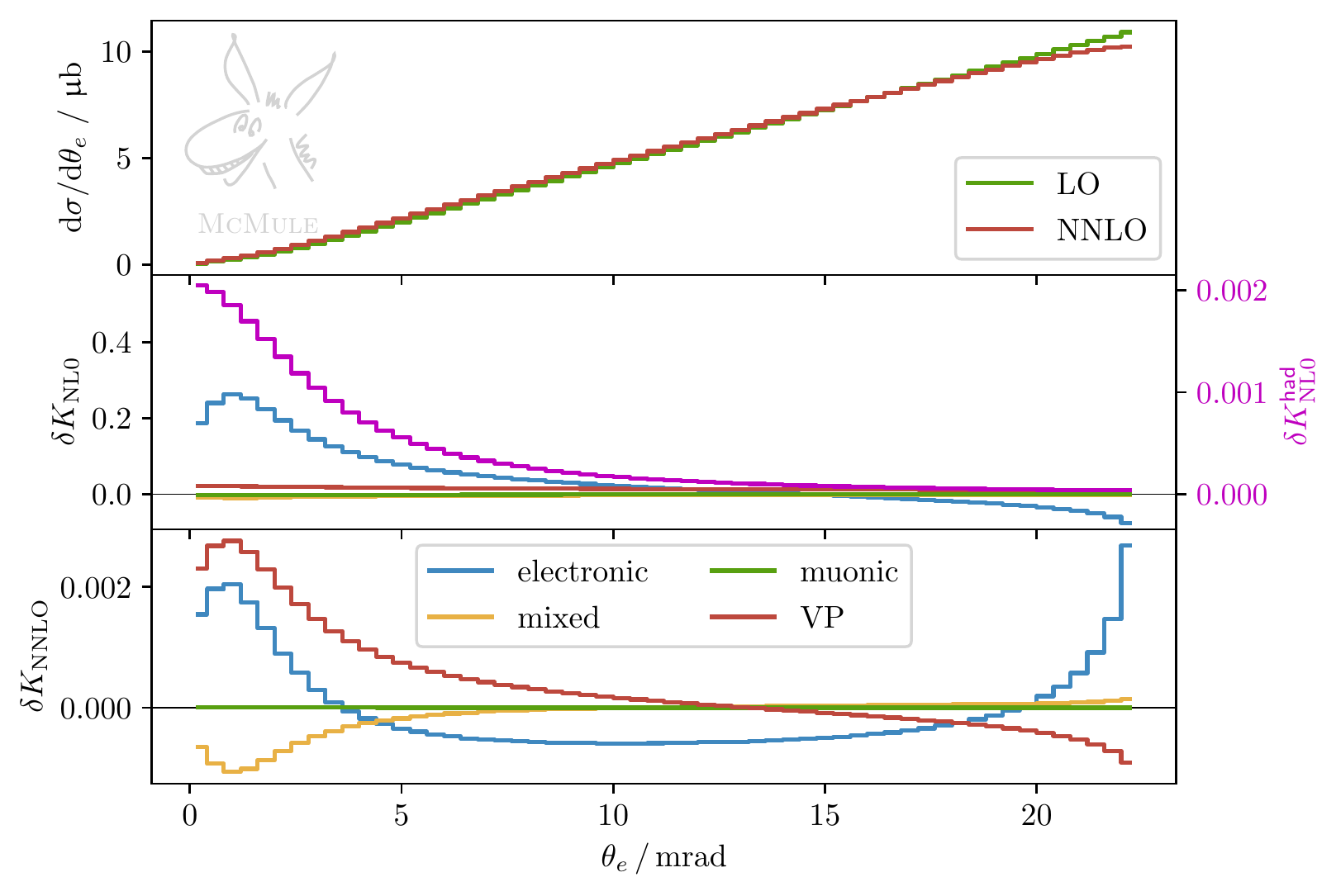}
    \label{fig:thetae_nobandcut}
    } \\ 
    \subfloat[\texttt{S2}]{
        \includegraphics[width=.8\textwidth]{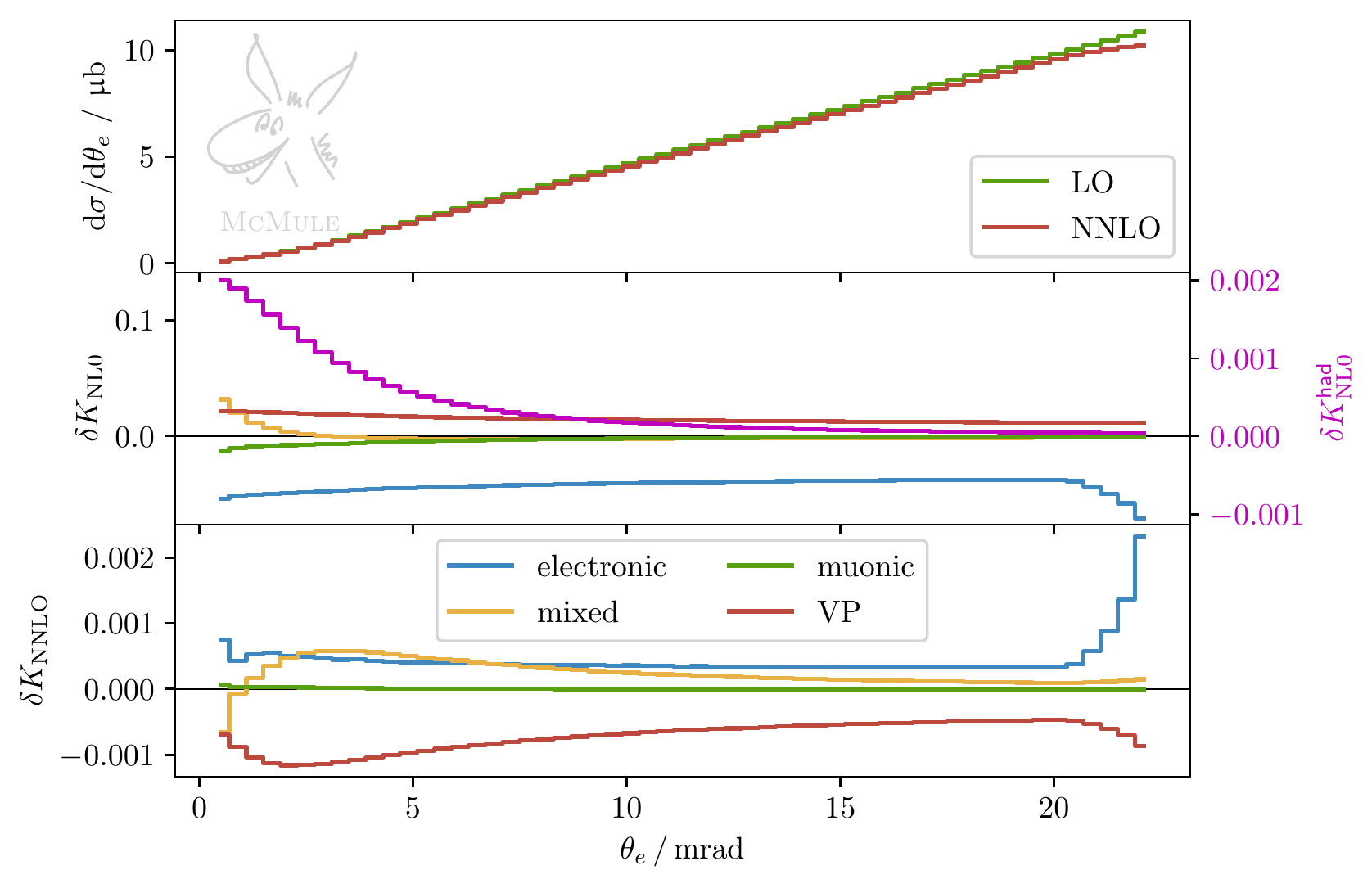}
    \label{fig:thetae_bandcut}
    }\caption{The upper panels show the differential cross section w.r.t. $\theta_e$ for \texttt{S1} and \texttt{S2} at LO (green) and NNLO (red). The middle and the lower panels provide the NLO and NNLO $K$ factors, respectively. The correction is split into purely electronic and muonic, mixed, and VP contributions. All three leptons as well as the hadronic contribution are included in the VP. The hadronic correction at NLO corresponds to the signal of the experiment and is shown separately in pink.}
\label{fig:thetae}
\end{figure}

\begin{figure}
    \centering
    \subfloat[\texttt{S1}]{
        \includegraphics[width=.8\textwidth]{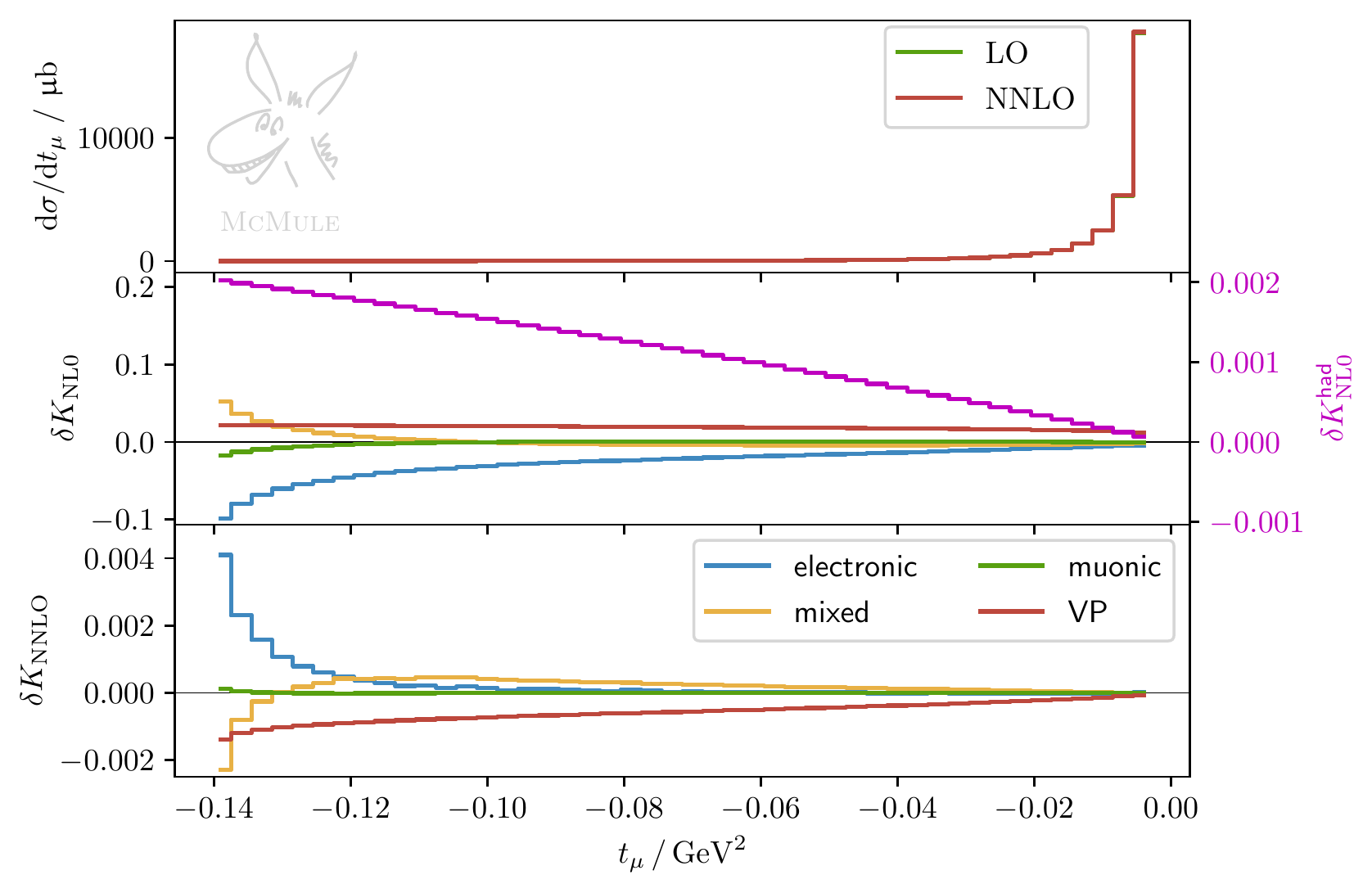}
    \label{fig:tm_nobandcut}
    } \\ 
    \subfloat[\texttt{S2}]{
        \includegraphics[width=.8\textwidth]{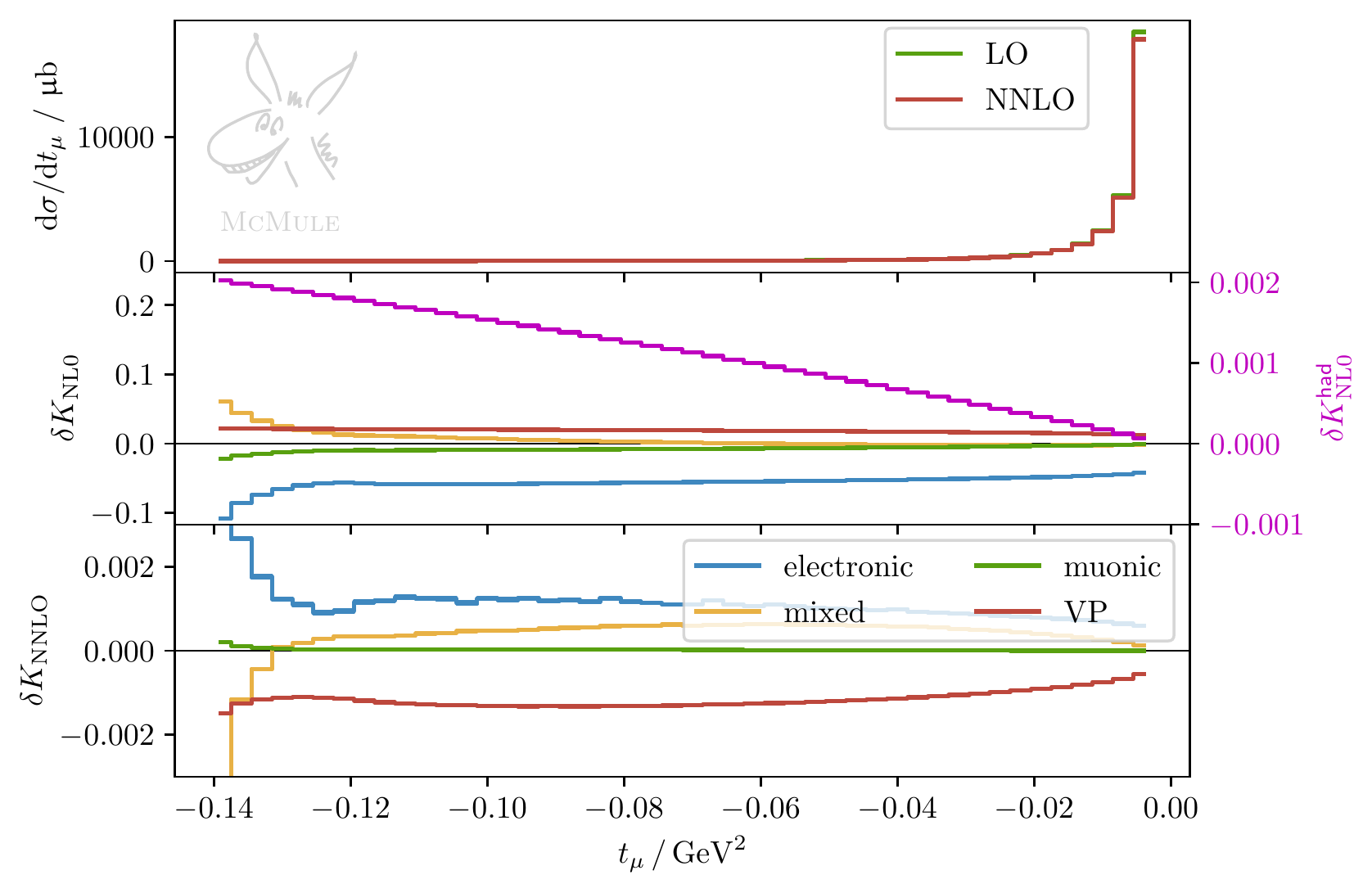}
    \label{fig:tm_bandcut}
    }\caption{The upper panels show the differential cross section w.r.t. $t_\mu$ for \texttt{S1} and \texttt{S2} at LO (green) and NNLO (red). The middle and the lower panels provide the NLO and NNLO $K$ factors, respectively. The correction is split into purely electronic and muonic, mixed, and VP contributions. All three leptons as well as the hadronic contribution are included in the VP. The hadronic correction at NLO corresponds to the signal of the experiment and is shown separately in pink.}
\label{fig:tmm}
\end{figure}

As a Monte Carlo integrator, \mcmule{} allows for the calculation of any number of differential observables in the same run. Here, we only show differential results that are of particular interest to the MUonE experiment. In particular, Figure~\ref{fig:thetae} and Figure~\ref{fig:tmm} present distributions w.r.t. $\theta_e$ and $t_\mu$. The differential cross section at LO as well as at NNLO are displayed in the upper panels. In addition, the middle and the lower panels show the differential $K$ factors defined in~\eqref{eq:kfac} at NLO and NNLO, respectively. We again provide separate results for purely electronic and muonic, mixed, and VP corrections. The signal of the experiment, $\delta K^\text{had}_\text{NLO}$, is shown in pink.

We first remark on the numerical error of the electronic NNLO corrections compared to older results that we have pusblished in~\cite{Banerjee:2020rww}. In the original calculation we have not used any special tools for the real-virtual contribution but instead calculated the amplitude with \texttt{Package-X}. On the one hand, this allowed us to obtain an extremely performant implementation speed-wise. At the same time, however, the evaluation was plagued by numerical instabilities that hampered the reliable integration over the phase space. As a consequence, some of the differential results in~\cite{Banerjee:2020rww} suffer from large numerical errors as well as discontinuous jumps. The new calculation, on the other hand, relies on next-to-soft stabilisation in combination with OpenLoops for this numerically delicate contribution. This ensures a much more reliable calculation of distributions compared to the previous implementation, which is clearly supported by the presented results.

Based on Figure~\ref{fig:thetae} we can confirm the claim made in Section~\ref{sec:muone} that the MUonE signal changes the differential cross section by up to $\mathcal{O}(10^{-3})$ in the region of small electron scattering angles ($\theta_e \lesssim 5\, \text{mrad}$). For the other contributions sizable NLO and NNLO corrections of up to $30\%$ and $0.3\%$ can be observed. Naively, one could therefore conclude that the target precision of $10\,\text{ppm}$ of MUonE is far out of reach. First of all, however, it has to be noted that the enhancement of the corrections at the end points of the distributions is due to soft photon emission. For a reliable description in this region, the corresponding logarithms need to be resummed. As mentioned in Section~\ref{sec:pert_theory}, a parton shower approach can be used to do this at LL accuracy. Secondly, the elasticity cut has the important effect of significantly reducing the variation in the $K$ factors. Since the MUonE experiment proposes to measure ratios of cross sections of different kinematic regions to cancel systematic uncertainties as opposed to absolute values, the flatness of the corrections is highly advantageous.

We further observe that the NNLO mixed corrections are larger than naively expected based on the counting of the collinear logarithms $L_m$. A possible explanation for this is that the soft enhancements are more relevant than the collinear effects. This is supported by Figure~\ref{fig:thetae_nobandcut} where the electronic contribution clearly dominates in the bulk of the distribution. Only in the region of strong soft enhancement, i.e. for small scattering angles or equivalently large electron energies, the two contributions are similar in size.

\begin{figure}
    \centering
    \subfloat[\texttt{S1}]{
        \includegraphics[width=.8\textwidth]{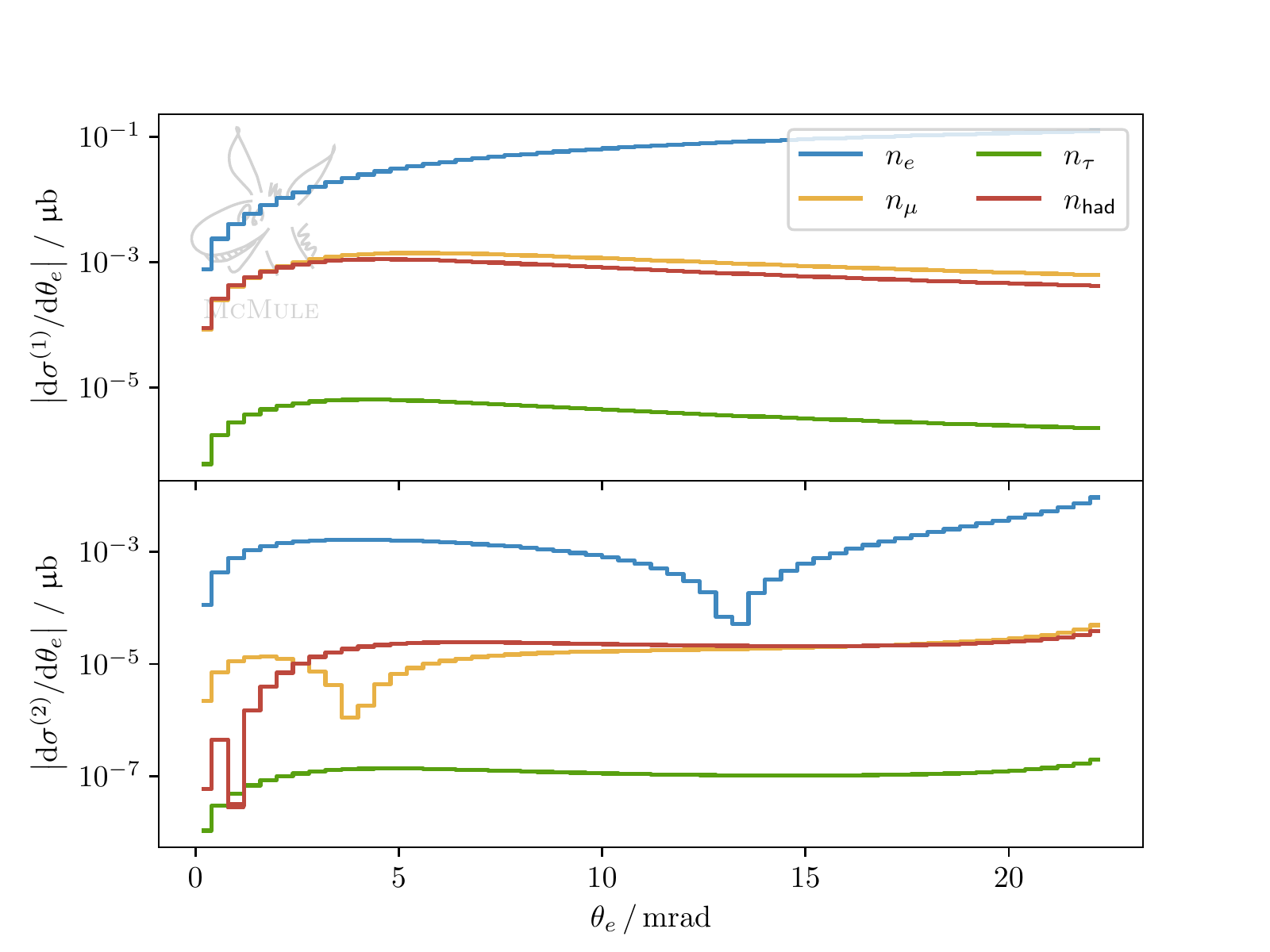}
    \label{fig:thetae_vp_nobandcut}
    } \\
    \subfloat[\texttt{S2}]{
        \includegraphics[width=.8\textwidth]{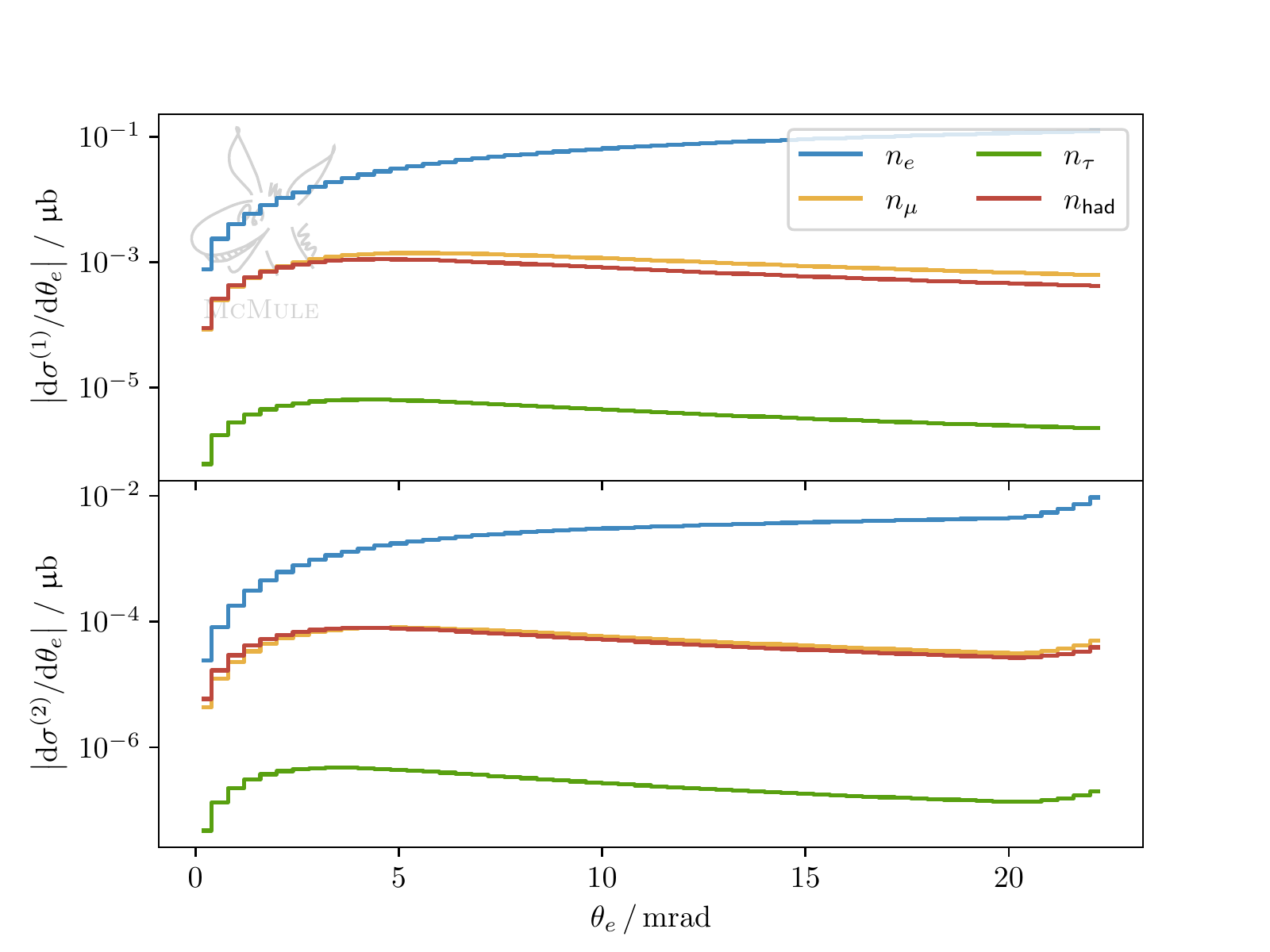}
    \label{fig:thetae_vp_bandcut}
    }\caption{The absolute value of the NLO (upper panels) and NNLO (lower panels) VP corrections to the differential cross section w.r.t. $\theta_e$ for \texttt{S1} and \texttt{S2}. The electron ($n_e$), muon ($n_\mu$), tau ($n_\tau$), and hadronic ($n_\text{had}$) contributions are shown separately.}
\label{fig:thetae_vp}
\end{figure}

\begin{figure}
    \centering
    \subfloat[\texttt{S1}]{
        \includegraphics[width=.8\textwidth]{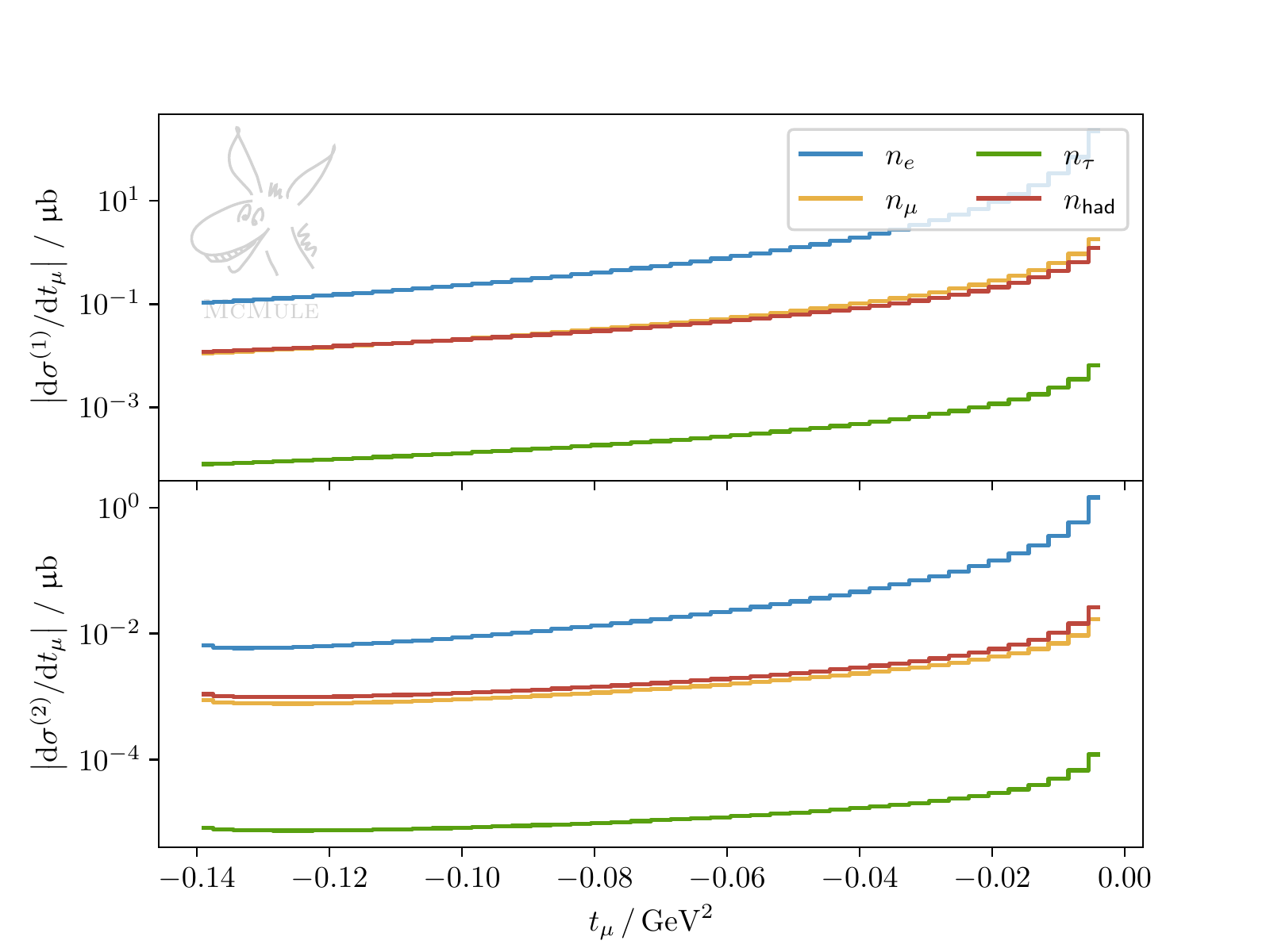}
    \label{fig:tm_vp_nobandcut}
    } \\
    \subfloat[\texttt{S2}]{
        \includegraphics[width=.8\textwidth]{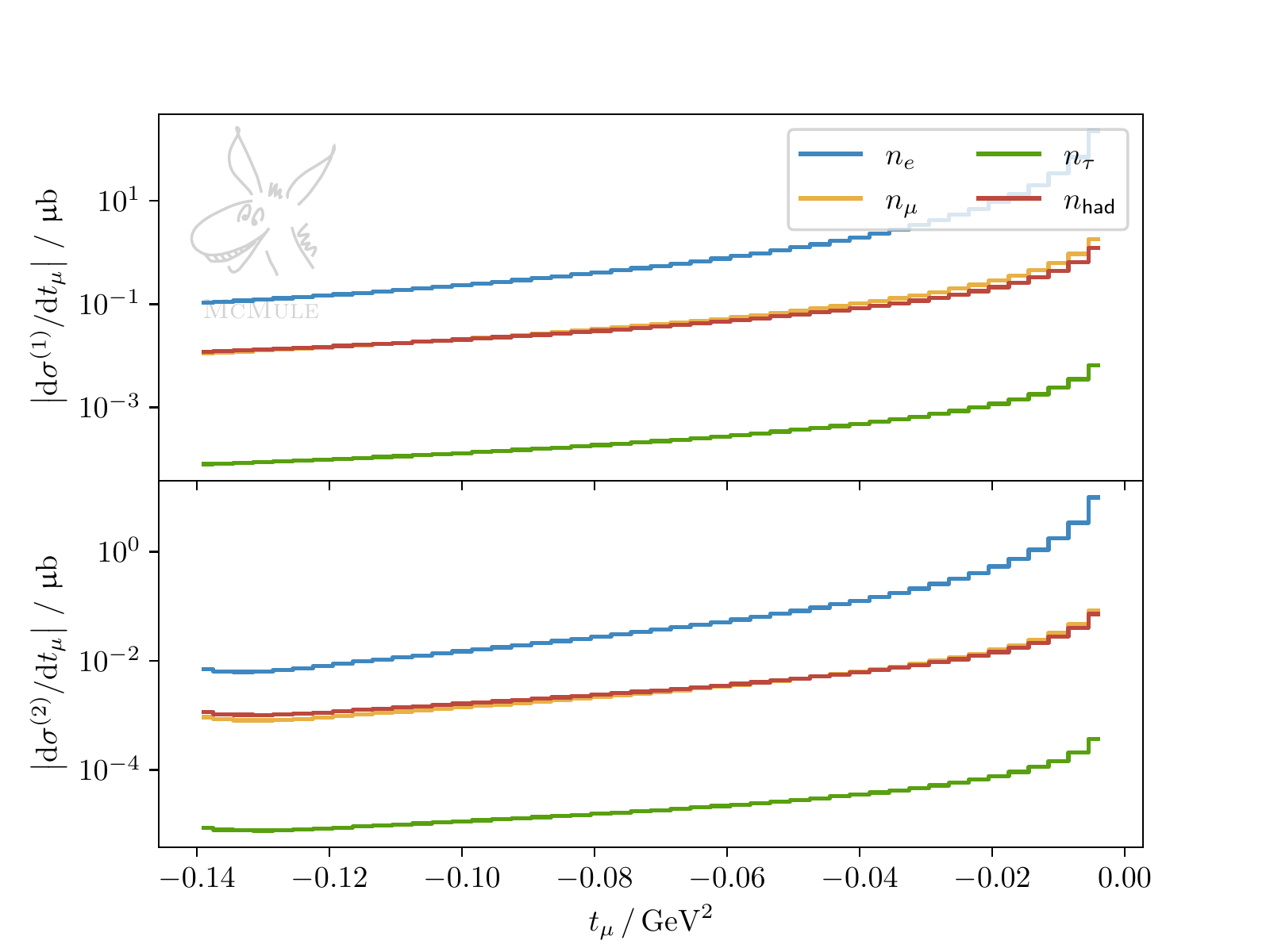}
    \label{fig:tm_vp_bandcut}
    }\caption{The absolute value of NLO (upper panels) and NNLO (lower panels) VP corrections to the differential cross section w.r.t. $t_\mu$ for \texttt{S1} and \texttt{S2}. The electron ($n_e$), muon ($n_\mu$), tau ($n_\tau$), and hadronic ($n_\text{had}$) contributions are shown separately.}
\label{fig:tmm_vp}
\end{figure}

Finally, we provide a comparison of the individual fermion contributions in Figure~\ref{fig:thetae_vp} and Figure~\ref{fig:tmm_vp}. The upper and lower panels show the absolute value of the NLO and NNLO corrections, respectively. Separate results are shown for the electron ($n_e$), muon ($n_\mu$), tau ($n_\tau$), and hadronic ($n_\text{had}$) contributions. The importance of the corrections clearly obey the expectation based on the mass hierarchy
\begin{align}
	m_e^2 \ll m_\mu^2 \sim m_\pi^2 \sim s \ll m_\tau^2\, .
\end{align}
The similarity of the hadronic and muonic contribution is particularly striking. The dips that can be observed in Figure~\ref{fig:thetae_vp_nobandcut} are due to zero-crossings.

In summary, the calculation of the complete set of NNLO QED corrections to $\mu$-$e$ scattering presented here is a major step towards the ambitious $10\, \text{ppm}$ goal of the MUonE experiment. For the integrated cross section, this calculation already ensures this level of precision. At the differential level, however, the corrections are strongly enhanced due to soft photon emission. As a consequence, a parton shower approach is needed to resum the corresponding large logarithms at LL accuracy. The relative simplicity of QED might also allow to fully exploit QCD efforts towards NLL showers. Furthermore, a fixed order calculation at N$^3$LO of the electronic corrections could be envisaged in order to quantify the missing NLL corrections.

\chapter{Outlook towards N$^3$LO}\label{chap:outlook}

\begin{figure}
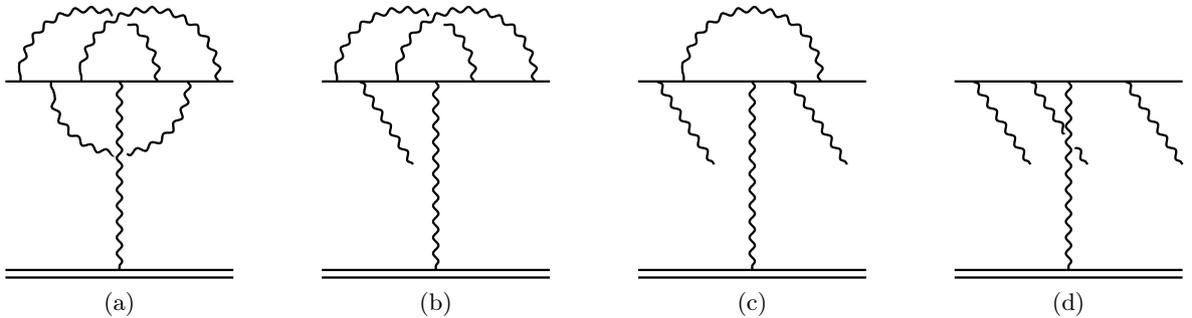

    \centering
    \subfloat[]{
        \begin{tikzpicture}[scale=1,baseline={(0,0)}]
        \input{tikz/muone_virtvirtvirt}
    \end{tikzpicture}
    \label{fig:muone_virtvirtvirt}
    }
    \hspace{.75cm}
    \subfloat[]{
        \begin{tikzpicture}[scale=1,baseline={(0,0)}]
        \input{tikz/muone_realvirtvirt}
    \end{tikzpicture}
    \label{fig:muone_realvirtvirt}
    }
    \hspace{.75cm}
    \subfloat[]{
        \begin{tikzpicture}[scale=1,baseline={(0,0)}]
        \input{tikz/muone_realrealvirt}
    \end{tikzpicture}
    \label{fig:muone_realrealvirt}
    }
    \hspace{.75cm}
    \subfloat[]{
        \begin{tikzpicture}[scale=1,baseline={(0,0)}]
        \input{tikz/muone_realrealreal}
    \end{tikzpicture}
    \label{fig:muone_realrealreal}
    }
\caption{Examples of electron-line corrections at N$^3$LO. This includes virtual-virtual-virtual, real-virtual-virtual, real-real-virtual, and real-real-real contributions.
}
\label{fig:muone_n3lo}
\end{figure}

The results presented in the previous chapter show that a N$^3$LO computation of the electron-line corrections is needed to reach the $10\, \text{ppm}$ target precision of the MUonE experiment. At this perturbative order the cross section has the four contributions
\begin{align}
	\sigma^{(3)}
	=\int \D\Phi_n \mathcal{M}_n^{(3)}
	+\int \D\Phi_{n+1} \mathcal{M}_{n+1}^{(2)}
	+\int \D\Phi_{n+2} \mathcal{M}_{n+2}^{(1)}
	+\int \D\Phi_{n+3} \mathcal{M}_{n+3}^{(0)}
\end{align}
given by virtual-virtual-virtual, real-virtual-virtual, real-real-virtual, and real-real-real corrections. Some sample diagrams are shown in Figure~\ref{fig:muone_n3lo}. The all-order FKS$^\ell$ subtraction scheme, presented in Section~\ref{sec:fks}, can be used to subtract the soft divergences arising in the phase-space integration. In complete analogy to the NNLO master formula~\eqref{eq:fks_nnlo}, we find at N$^3$LO the four separately finite contributions
\begin{subequations}
\label{eq:fks_nnnlo}
\begin{align}
\sigma^{(3)} 
&= \sigma^{(3)}_n(\xc)  + \sigma^{(3)}_{n+1}(\xc)
    + \sigma^{(3)}_{n+2}(\xc)+ \sigma^{(3)}_{n+3}(\xc)\, , \\
\sigma^{(3)}_n(\xc) &= 
\int\! \D\Phi_n^{d=4}\, \fM n3(\xc)
\,
\label{eq:fks_nnnlo:n}
,\\
\sigma^{(3)}_{n+1}(\xc) &= \int\!
 \D\Phi^{d=4}_{n+1}
  \cdis{\xi_1} 
  \, \Big(\xi_1\, \fM{n+1}2(\xc)\Big)\label{eq:fks_nnnlo:n1}
\, ,\\
\sigma^{(3)}_{n+2}(\xc) &= \int\!
 \D\Phi^{d=4}_{n+2}
  \cdis{\xi_1}\,
  \cdis{\xi_2}\,
   \Big(\xi_1\xi_2 \, \fM{n+2}1(\xc)\Big)\label{eq:fks_nnnlo:n2}
\, ,\\
\sigma^{(3)}_{n+3}(\xc) &= \int\!
  \D\Phi_{n+3}^{d=4}
   \cdis{\xi_1}\,
   \cdis{\xi_2}\,
   \cdis{\xi_3}\,
     \Big(\xi_1\xi_2\xi_3\, \fM{n+3}0\Big) \label{eq:fks_nnnlo:n3}\, .
\end{align}
\end{subequations}
The eikonal-subtracted finite $\ell$-loop squared amplitude, $\fM n\ell$, is defined in~\eqref{eq:soft_exp}. In the case of the three-loop contribution ($\ell=3$) it reads
\begin{align}
	\fM n3(\xc)
	= \M n3
   	+\ieik(\xc)\,\M n2
   	+\frac1{2!}\M n1 \ieik(\xc)^2
   	+\frac1{3!}\M n0 \ieik(\xc)^3\, .
\end{align}
The major challenge of this fully differential N$^3$LO calculation is thus the computation of the corresponding amplitudes.

In the case of the real-real-real and the real-real-virtual corrections, $\mathcal{M}_{n+3}^{(0)}$ and $\mathcal{M}_{n+2}^{(1)}$, automated one-loop tools such as OpenLoops~\cite{Buccioni:2017yxi,Buccioni:2019sur} can be used. Nevertheless, the numerical stability of the highly complex one-loop $6$-particle amplitude, $\mathcal{M}_{n+2}^{(1)}$, will be challenging. Based on the discussion of Chapter~\ref{chap:nts}, we can expect soft photon emission to be particularly problematic. We can rely on the one-loop LBK theorem~\eqref{eq:lbk_oneloop} to stabilise the amplitude in the limit where only one photon becomes soft. In fact, the analogous amplitude for Bhabha scattering was considered in Section~\ref{sec:lbk_validation} as a non-trivial validation of the LBK formula. The theorem in its current version is, however, not applicable to the case where both photons become soft. A corresponding extension of the LBK formalism could therefore turn out to be very useful in this context.

The virtual-virtual-virtual correction, $\mathcal{M}_{n}^{(3)}$, amounts to the calculation of the heavy quark form factor at three loop. In the case of massless quarks (light quark form factor) this has been computed more than 10 years ago~\cite{Baikov:2009bg}. Today even the four-loop correction is known~\cite{Lee:2021uqq,Lee:2022nhh}. The analytic expression for the massive form factor, on the other hand, has only been calculated at two loop~\cite{Mastrolia:2003yz,Bonciani:2003ai,Bernreuther:2004ih,Gluza:2009yy}. Very recently, however, the fully massive calculation was performed in~\cite{Fael:2022rgm,Fael:2022miw} with a semi-numerical approach combining the method of differential equations with expansions around regular and singular points. With these impressive results available, one of the main obstacles towards N$^3$LO is overcome. Nevertheless, an independent check of the calculation is desirable. One possible strategy for this is the extension of the method of massification, presented in Section~\ref{sec:massification}, to three loop. Based on the corresponding massless result one could then obtain the massive amplitude up to polynomially suppressed mass effects.

\begin{figure}
    \centering
   \begin{tikzpicture}[scale=1.2,baseline={(0,0)}]
        \input{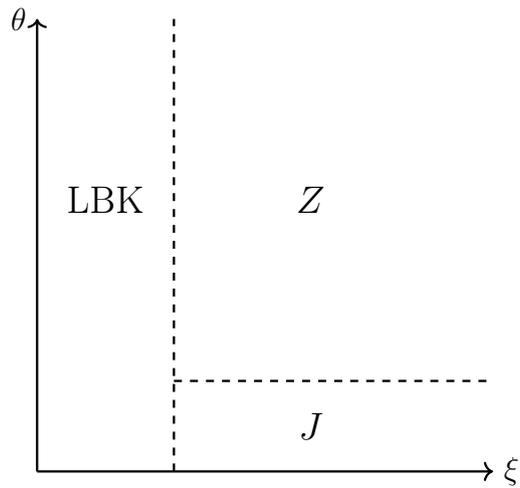}
    \end{tikzpicture}
    \caption{Schematic illustration to approximate radiative amplitudes using a combination of massification, the LBK theorem, and the massive splitting function. For the bulk of the phase space the massification constant $Z$ can be used to obtain the small-mass expansion. This approximation breaks down in the soft and in the collinear limit where $\xi \to 0$ and $\theta \to 0$, respectively. In the soft limit the LBK theorem can be applied to compute the soft expansion at NLP. In the collinear limit one can rely on the universality of the massive splitting function $J$ to compute the corresponding approximation at LP.}
\label{fig:real_approx}
\end{figure}

This leaves the real-virtual-virtual amplitude, $\mathcal{M}_{n+1}^{(2)}$, as the main bottleneck. Also in this case the massless result has been known for some time now~\cite{Garland:2001tf,Garland:2002ak}. A corresponding analytic computation for massive fermions is not currently feasible. For an exact calculation one therefore has to resort to numerical techniques. On the other hand, with the massless amplitude available the question arises whether massification could be used in this case. In principle, the results of Section~\ref{sec:massification_twoloop} can also be applied to radiative amplitudes. Contrary to purely virtual corrections, however, the massification scale hierarchy~\eqref{eq:massification_scaling} is only valid for part of the phase space. In the soft as well as in the collinear limit additional parameters become small and the small-mass expansion breaks down. Nevertheless, we have seen in Chapter~\ref{chap:nts} and Chapter~\ref{chap:coll} that the amplitude also reduces to universal quantities in these limits (at least at one loop). One could therefore switch to the corresponding next-to-soft and collinear approximation accordingly. A schematic illustration of this idea is shown in Figure~\ref{fig:real_approx}. This requires the generalisation of the LBK theorem beyond one loop as well as the calculation of the massive splitting function at two loop. Combined with the two-loop massification constant of Section~\ref{sec:massification_twoloop} this would allow us to obtain an approximation of the massive real-virtual-virtual amplitude in all relevant kinematic regions.

To summarise, many ingredients that enter the fully differential calculation of the electronic N$^3$LO corrections to $\mu$-$e$ scattering are already available. A successful completion of this ambitious project in the near-term future is therefore indeed conceivable. At the time of writing, the main bottleneck is the calculation of the real-virtual-virtual amplitude. One promising approach in this regard is based on massification, the LBK theorem, and the massive splitting function. The further development of these concepts is therefore planned for the near future and represents an important building block of the MUonE endeavour towards N$^3$LO.

\appendix
\chapter{Conventions}\label{sec:notation}

We denote the amplitude for a process with $n$ final state particles by $\mathcal{A}_n$ and the corresponding QED $l$-loop correction by $\mathcal{A}_n^{(l)}$. Analogously, we use $\mathcal{M}_n$ and $\mathcal{M}_n^{(l)}$ for the unpolarised squared amplitude and $\sigma_n$ and $\sigma_n^{(l)}$ for the cross section. In particular, $\mathcal{M}_n^{(l)}$ includes in addition to the genuine $l$-loop correction also all other interference terms. At two-loop, for example, we have
\begin{align}
	\mathcal{M}_n^{(2)}
	= 2 \text{Re} \big\{ \mathcal{A}_n^{(2)} {\mathcal{A}_n^{(0)}}^\dagger \big\} 
	+ \mathcal{A}_n^{(1)} {\mathcal{A}_n^{(1)}}^\dagger \, .
\end{align}
The corresponding quantities for the radiative process with $r$ additional photons in the final state are given by $n \to n+r$. The symbol $\Gamma$ is used in various places to denote part of an amplitude. The perturbative expansion of the amplitudes is done in terms of the fermion charge $Q$ where $Q=-e$ for an incoming particle or an outgoing antiparticle and $Q=+e$ otherwise. All amplitudes are regularised in $d=4-2\epsilon$ dimensions and renormalised in the on-shell scheme. We conveniently define the $d$-dimensional loop measure as
\begin{align}
	[\D\ell] = C(\epsilon) \frac{\D^d\ell}{i \pi^{d/2}}\, ,\quad\quad C(\epsilon) = \mu^{2\epsilon} \Gamma(1-\epsilon)\, ,
\end{align}
with $\mu$ the scale of dimensional regularisation. This ensures that no spurious terms of Euler's constant $\gamma_E$ and $\log(4\pi)$ occur in the result. In order to convert to the standard loop measure $ \mu^{2\epsilon}\D^d\ell/(2\pi)^d$ the final result then has to be multiplied with
\begin{align}\label{eq:loopfactor}
	K(\epsilon) = \frac{(4\pi)^\epsilon}{16\pi^2\Gamma(1-\epsilon)}\, .
\end{align}

Throughout this thesis we denote loop momenta by $\ell$. Furthermore, we consistently take $k$ to represent the momentum of the emitted photon in a radiative process and $\xi = 2 E_k / \sqrt{s}$ as its dimensionless energy. The corresponding CMS energy is given by $\sqrt{s}$. Other on-shell momenta are denoted by $p_i$, i.e. $p_i^2=m_i^2$ with $m_i$ the particle mass. The corresponding velocity is then given by $\beta_i=(1-m_i^2/E_i^2)^{1/2}$ where $E_i$ is the energy. Since we mainly study processes with electrons and muons as external states, the simplifying notation
\begin{align}
	m_e \to m, \quad\quad\quad m_\mu \to M
\end{align}
is used for the corresponding masses. In many results combinations of kinematic invariants occur that can be compactly written in terms of the Källén function
\begin{align}
	\lambda(x,y,z) = x^2+y^2+z^2-2xy-2yz-2zx\, .
\end{align}
The Lorentz-invariant $d$-dimensional phase space for $n$ final state particles is given by
\begin{align}
	\D\Phi_n(P;p_1,…,p_n)
	=\prod_{i=1}^{n} \frac{\D^{d-1} p_i}{(2\pi)^{d-1}2E_i} (2\pi)^d \delta^d\Big(P-\sum_{i=1}^n p_i \Big)
\end{align}
where $P$ is the total initial-state momentum.

In the course of the thesis we discuss three types of scale hierarchies: the small-mass expansion (massification) as well as the soft and collinear limit of radiative amplitudes. In order to make the corresponding power counting transparent we introduce the book-keeping parameters $\lambda$, $\lambda_s$, and $\lambda_c$, respectively. The three hierarchies can then be defined in the following way with $S$ representing all other relevant scales in the process:
\begin{subequations}
\begin{alignat}{3}
    &\mbox{massification:}& 	\quad\quad\quad &p_i^2=m_i^2\sim\lambda^2 \ll S \\
    &\mbox{soft limit:}&         	\quad\quad\quad &k\sim\lambda_s \ll m_i^2, S \\
    &\mbox{collinear limit:}&        \quad\quad\quad &k\cdot p_i\sim m_i^2 \sim \lambda_c^2 \ll S
\end{alignat}
\end{subequations}
All three limits are governed by universal quantities. Massification and the collinear limit are described in terms of the massification constant $Z$ and the splitting function $J$. The LP soft limit is given by the well-known eikonal factor that we denote by $\eik$. The integrated eikonal, $\ieik$, is obtained by integration over the $d$-dimensional photon phase space.

\chapter{Vacuum polarisation}\label{sec:vacuum_polarisation}

We define the QED VP as in~\eqref{eq:vp_definition}. Note that this definition varies in the literature up to an overall sign. In the following, we provide the expression for the leptonic VP
\begin{align}
	\Pi_\ell(q^2) = 1 + \frac{\alpha}{2\pi} \Pi_\ell^{(1)}(q^2)
		 + \Big(\frac{\alpha}{2\pi}\Big)^2 \Pi_\ell^{(2)}(q^2)
		 + \mathcal{O}(\alpha^3)
\end{align}
up to two loops. We do not provide contributions of $\mathcal{O}(\epsilon)$ here since they are not needed in the calculation of physical observables due to the cancellation of IR poles.

The one-loop contribution, $\Pi_\ell^{(1)}$, can be calculated straightforwardly and reads
\begin{align}
	\Pi_\ell^{(1)}(q^2) 
	= \frac{2}{9}\Bigg(
	-5+12\lambda-3(1-2\lambda)\sqrt{1+4\lambda}H_{0}(x)\Bigg)
\end{align}
with
\begin{align}
	x = \frac{2\lambda}{1+2\lambda+\sqrt{1+4\lambda}}\, , \quad\quad
	\lambda = - \frac{m_\ell^2}{q^2}
\end{align}
and $m_\ell$ the mass of the lepton. 

The unsubtracted two-loop result, $\bar{\Pi}_\ell^{(2)}(q^2)=\Pi^{(2)}_\ell(q^2)+\bar{\Pi}^{(2)}_\ell(0)$, is given in~\cite{Djouadi:1993ss}. The limit $\bar{\Pi}^{(2)}_\ell(0)$ corresponds to the two-loop charge renormalisation and can be taken from (6.62) of~\cite{Grozin:2005yg}. We then find for the real part of the fully renormalised result
\begin{align}
	\Pi_\ell^{(2)}(q^2) 
	&= \frac{1}{6} \Bigg(
	-5+52\lambda+4(1-4\lambda^2)\big(6\zeta_3-2F(x)+F(x^2)\big)
	-8(1-2\lambda)\sqrt{1+4\lambda}\big(G(x) \\ \nonumber
	&\quad-G(x^2)\big)
	-6(1-6\lambda)\sqrt{1+4\lambda}H_{0}(x)+8\lambda(4+\lambda)H_{0,0}(x)
	\Bigg)
\end{align}
with
\begin{subequations}
\begin{align}
	F(x) &= -2 H_{2,0}(x)+2H_{1,0,0}(x) \, , \\
	G(x) &= \frac{2x}{1-x}H_{0,0}(x)-2H_{1,0}(x)\, .
\end{align}
\end{subequations}
Above the threshold, i.e. for $q^2>4m_l^2$, the VP develops an imaginary part. The corresponding expression is given in (5.2) of~\cite{Djouadi:1993ss}.\footnote{There is a typo in this result that has been corrected in~\cite{PhysRevD.53.4111}.}

\chapter{Hyperspherical master kernels}\label{sec:hyperspherical_masterintegrals}

We give here the general expressions for the hyperspherical kernels in the case of one, two, and three angular dependent propagators. These are all the integrals needed for non-factorisable VP diagrams for $2 \to 2$ processes at two loop . As discussed in Section~\ref{sec:int_hyperspherical} the one and two propagator case can be calculated using Gegenbauer polynomials. In the case of three angular dependent propagators (box integral), on the other hand, the integration over the hyperspherical angles has to be done explicitely. The corresponding formulas are taken from~\cite{Laporta:1994mb} where this calculation was performed. All results that are given below are defined in the Euclidean region. The analytic continuation to the physical region has to be done case by case as discussed in Section~\ref{sec:cont_physical}. The results below are written in terms of the logarithm
\begin{align}
	L(x) \equiv \frac{1}{2}\log(\frac{1+x}{1-x})\, .
\end{align}

The simple one propagator case reads
\begin{align}
	\int \frac{\D\Omega_Q}{2\pi^2}
	\frac{1}{[(Q-P)^2+m^2]}
	= \frac{Z}{|Q|\,|P|}
\end{align}
with
\begin{align}
	Z = \frac{Q^2+P^2+m^2-\lambda^{1/2}(Q^2,P^2,-m^2)}{2|Q|\,|P|}\, .
\end{align}

For two angular dependent propagators we find
\begin{align}
	\int \frac{\D\Omega_Q}{2\pi^2}
	\frac{1}{[(Q-P_1)^2+m_1^2][(Q-P_2)^2-m_2^2]}
	= \frac{L(x_1) + L(x_2) + L(x_3)}
	  {Q^2\lambda^{1/2}\big(P_1^2,P_2^2,(P_1-P_2)^2\big)}
\end{align}
with
\begin{subequations}
\begin{align}
	x_1 &= \lambda^{1/2}\big(P_1^2,P_2^2,(P_1-P_2)^2\big) \lambda^{1/2}\big(Q^2,P_1^2,-m_1^2\big)/C_1\, , \\
	x_2 &= \lambda^{1/2}\big(P_1^2,P_2^2,(P_1-P_2)^2\big) \lambda^{1/2}\big(Q^2,P_2^2,-m_2^2\big)/C_2\,, \\
	x_3 &= \lambda^{1/2}\big(P_1^2,P_2^2,(P_1-P_2)^2\big)/C_3\, .
\end{align}
\end{subequations}
and
\begin{subequations}
\begin{align}
	C_1 &= \big(P_2^2-P_1^2-(P_1-P_2)^2 \big)Q^2
		+(P_1^2)^2+m_1^2\big(P_1^2+P_2^2-(P_1-P_2)^2\big) \\ \nonumber
		&\quad-P_1^2\big(P_2^2+(P_1-P_2)^2+2m_2^2\big)\, , \\
	C_2 &= \big(P_1^2-P_2^2-(P_1-P_2)^2\big)Q^2
		+(P_2^2)^2+m_2^2\big(P_1^2+P_2^2-(P_1-P_2)^2\big) \\ \nonumber
		&\quad-P_2^2\big(P_1^2+(P_1-P_2)^2+2m_1^2\big)\, , \\
	C_3 &= (P_1-P_2)^2-P_1^2-P_2^2\, .
\end{align}
\end{subequations}

Finally, the result for three propagators is given by
\begin{align}
	\int \frac{\D\Omega_Q}{2\pi^2}
	\frac{1}{[(Q-P_1)^2+m_1^2][(Q-P_2)^2-m_2^2][(Q-P_3)^2-m_3^2]}
	= \frac{L(x_4) + L(x_5) + L(x_6)}
	  {Q^2 \sqrt{\Delta}}
\end{align}
with
\begin{subequations}
\begin{align}
	x_4 &= \lambda^{1/2}(Q^2,P_1^2,-m_1^2)\sqrt{\Delta}/C_4\, , \\
	x_5 &= \lambda^{1/2}(Q^2,P_2^2,-m_2^2)\sqrt{\Delta}/C_5\, , \\
	x_6 &= \lambda^{1/2}(Q^2,P_3^2,-m_3^2)\sqrt{\Delta}/C_6
\end{align}
\end{subequations}
where $\Delta$ is the polynomial of degree 2 in $Q^2$
\begin{align}
	\Delta = \alpha_2 (Q^2)^2+\alpha_1 Q^2+\alpha_0\, .
\end{align}
The corresponding coefficients read
\begin{subequations}
\begin{align}
	\alpha_2 &= \lambda(a_1,a_2,a_3)\, , \\
	\alpha_1 &= 2u_1(2a_1b_1-a_1b_2-a_1b_3+a_2b_3+a_3b_2-a_2b_2-a_3b_3) \\ \nonumber
		 &\quad + 2(u_1-b_1)a_1(a_1-a_2-a_3)-\frac{4}{3}a_1a_2a_3
		 +(\text{cyclic permutations})\, , \\
	\alpha_0 &= u_1^2(a_1^2-2a_1b_2-2a_1b_3-2b_2b_3+b_2^2+b_3^2) \\ \nonumber
		 &\quad + 2u_1u_2(-a_1a_2+a_1b_1+a_1b_3+a_2b_2+a_2b_3
		 -2a_3b_3-b_1b_2+b_1b_3+b_2b_3-b_3^2) \\ \nonumber
		 &\quad +2u_1(a_1^2b_1-a_1a_2b_2-a_1a_3b_3-a_1b_1b_2
		 -a_1b_1b_3+2a_1b_2b_3+a_2b_2^2-a_2b_2b_3 \\ \nonumber
		 &\quad -a_3b_2b_3+a_3b_3^2)+a_1^2b_1^2-2a_1a_2b_1b_2 
		 + (\text{cyclic permutations})
\end{align}
\end{subequations}
where $a_1=(P_2-P_3)^2$, $a_2=(P_1-P_3)^2$, $a_3=(P_1-P_2)^2$, $b_i=P_i^2$, and $u_i=m_i^2$. The complete expression is obtained by summing up also the two cyclic permutations of the indices $\{1,2,3\}$. Furthermore, the arguments of the logarithm contain
\begin{subequations}
\begin{align}
	C_4 &= \beta_2 (Q^2)^2+\beta_1 Q^2 + \beta_0\, , \\
	C_5 &= C_4(\text{with cyclic permutation $1\to2$, $2\to3$,$3\to1$})\, , \\
	C_6 &= C_4(\text{with cyclic permutation $1\to3$, $2\to1$,$3\to2$})
\end{align}
\end{subequations}
where
\begin{subequations}
\begin{align}
	\beta_2 &= a_1-a_2-a_3\, , \\
	\beta_1 &= u_1(2a_1-a_2-a_3+2b_1-b_2-b_3)
		   +u_2(-a_2-b_1+b_3)+u_3(-a_3-b_1+b_2) \\ \nonumber
		   &\quad -2a_1b_1-2a_2a_3+a_2b_1+a_2b_2+a_3b_1+a_3b_3\, , \\
	\beta_3 &= u_1^2(a_1-b_2-b_3)+u_1u_2(-a_2+b_1+b_3)
		   +u_1u_3(-a_3+b_1+b_2) \\ \nonumber
		   &\quad +u_2b_1(-a_2+b_1-b_3)+u_3b_1(-a_3+b_1-b_2)
		   +u_2u_3(-2b_1) \\ \nonumber
		   &\quad +u_1(2a_1b_1-a_2b_2-a_3b_3-b_1b_2-b_1b_3+2b_2b_3)
		   +a_1b_1^2-a_2b_1b_2-a_3b_1b_3\, .
\end{align}
\end{subequations}

\chapter{UV counterterms for dispersive method}\label{sec:counterterms_dispersive}

The UV renormalisation in the dispersive approach to VP proceeds in complete analogy to the hyperspherical case discussed in Section~\ref{sec:uv_renormalisation}. The renormalisation is performed at the level of the kernel~\eqref{eq:dispersive_kernel} of the dispersive integral~\eqref{eq:disperive_method_master}. This corresponds to the amplitude with the VP replaced by a photon of mass $\sqrt{z}$. The corresponding on-shell wave function and mass counterterms are then related through~\eqref{eq:onshell_const} to the fermion self-energy diagram
 \begin{align}
    -i \Sigma_z =
    \begin{tikzpicture}[scale=.8,baseline={(1,0)}]

	\draw[line width=.3mm]  (-1.5,0) node[left] {$p$} -- (1.5,0);
	\centerarc [line width=0.3mm,photon](0,0)(0:180:.9);
	\draw[line width=.3mm]  [fill=math3] (0,1) circle (0.3) node[] {$z$};

	\centerarc [line width=0.3mm,->](0,0)(125:165:1.2);
	\node at (-1.2,.8) {$\ell$};

    \end{tikzpicture}
    = (-ie)^2 \mu^{2\epsilon}\int \frac{\D \ell^d}{(2\pi)^d}
	\frac{\gamma_\mu(\slashed{\ell}+\slashed{p}+m)\gamma^\mu}
	{[\ell^2-z][(\ell+p)^2-m^2]}\, ,
\end{align}
where the green blob indicates the massive photon. The counterterms then read
\begin{subequations}
\begin{align}
	\big(m \tilde{K}(\epsilon)\big)^{-1}\langle \delta m \rangle_\ell (z)
	=& -\frac{3}{\epsilon}
	-6 - \frac{1}{x}-x
	+(-3-4x-x^2)H_0(x)
	+\frac{(1+x)^4}{x^2} H_{-1}(x)\, , \\
	\tilde{K}(\epsilon)^{-1} \langle \delta Z \rangle_\ell (z)
	=&-\frac{1}{\epsilon}
	-10-\frac{3}{x}-3x
	-\frac{1-x-9x^2-3x^3}{1-x}H_0(x)\\ \nonumber
	&+(14+\frac{3}{x^2}+\frac{12}{x}+12x+3x^2) H_{-1}(x)
\end{align}
\end{subequations}
with
\begin{align}
	\tilde{K}(\epsilon) = 4\pi\alpha \Big( \frac{\mu^2}{m^2} \Big) K(\epsilon)
\end{align}
and $K(\epsilon)$ defined in~\eqref{eq:loopfactor}. We have expressed the result in terms of HPLs with the letter
\begin{align}
	x = \frac{\sqrt{z}-\sqrt{z-4m^2}}{\sqrt{z}+\sqrt{z-4m^2}}\, .
\end{align}
Since $0<z$ in the dispersive integral~\eqref{eq:disperive_method_master}, the $i\delta$ prescription for $x$ is irrelevant in this case.

\chapter{Soft integrals}\label{sec:softints}

In~\eqref{eq:softinta} and~\eqref{eq:softintb} we have defined the two integrals necessary to
construct the soft contribution. These integrals
are universal and are given here in $d=4-2\epsilon$ dimensions with $\mu$ denoting the scale of dimensional regularisation. With
$s_{ij} = 2p_i\cdot p_j$ and $s_{i\gamma} = 2 k\cdot p_i$ we have
\begin{subequations}
\begin{align}
I_1(p_i,k) &= i\mu^{2\epsilon}\int \frac{\text{d}^d \ell}{(2\pi)^d} 
        \frac{1}{[\ell^2+i\delta][\ell\cdot p_i-k\cdot p_i+i\delta]} \\
 &= -2K(\epsilon)\bigg(\frac{m_i^2}{s_{i\gamma}}+i\delta\bigg)^{2\epsilon-1}
    \bigg(\frac{\mu^2}{m_i^2}\bigg)^\epsilon
    \Gamma(1-\epsilon)^2\Gamma(2\epsilon-1)\, ,
\\
I_2(p_i,p_j,k) &
    = i\mu^{2\epsilon}\int \frac{\text{d}^d \ell}{(2\pi)^d} 
        \frac{1}{[\ell^2+i\delta][-\ell\cdot p_j+i\delta][\ell\cdot p_i-k\cdot p_i+i\delta]} \\
   \begin{split}
   &= 
  \frac{8K(\epsilon)}{s_{ij}}
  \bigg(\frac{m_i^2}{s_{i\gamma}}+i\delta\bigg)^{2\epsilon}
  \bigg(\frac{\mu^2}{m_i^2}\bigg)^\epsilon
  \bigg(\frac{|s_{ij}|}{2m_im_j}\bigg)^{2\epsilon}
  \Bigg\{
    \Gamma(1-\epsilon)^2\Gamma(2\epsilon-1)
    \\& \quad \ \,
    \times \pFq{2}{1}{\tfrac12-\epsilon,1-\epsilon}{\tfrac32-\epsilon}{v_{ij}^2}
   + i\pi
    \Big(-\frac14+i\delta\Big)^{-\epsilon}
    v_{ij}^{-1+2\epsilon}
    \Gamma(1-2\epsilon)\Gamma(2\epsilon)
    \Theta(s_{ij})
  \Bigg\}
  \end{split}
  \\
  \begin{split}
  &= \frac{4K(\epsilon)}{s_{ij}}\frac{1+\chi}{1-\chi}
  \bigg(\frac{m_i^2}{s_{i\gamma}}+i\delta\bigg)^{2\epsilon}
  \bigg(\frac{\mu^2}{m_i^2}\bigg)^\epsilon
  \Bigg\{
    \frac{H_0(\chi)}{2\epsilon}-\zeta_2-\frac{1}{2}H_{0,0}(\chi)
    \\&\quad \
   -H_{1,0}(\chi) + \Theta(s_{ij}) 6\zeta_2
    +i \pi \Theta(s_{ij}) \Big(\frac{1}{\epsilon}-H_{0}(\chi)-2H_{1}(\chi) \Big)
    +\mathcal{O}(\epsilon)
  \Bigg\}\, ,
  \end{split}
\end{align}
\end{subequations}
where $v_{ij}=\sqrt{1-4m_i^2m_j^2/s_{ij}^2}=(1-\chi)/(1+\chi)$ and $K(\epsilon)$ is defined in~\eqref{eq:loopfactor}. Since $0<v_{ij}<1$ and thus $0<\chi<1$ all HPLs are
manifestly real.

\chapter{Splitting functions}\label{sec:splitfunc}

In the following we give the explicit expressions for all the
quantities that enter the collinear factorisation
formulas~\eqref{eq:collfac_isr} (ISR) and~\eqref{eq:collfac_fsr}
(FSR). The results are presented in a form that can be used in three
major flavours of dimensional regularisation: the four-dimensional
helicity scheme (\FDH), 't Hooft-Veltman scheme (\HV), and
conventional dimensional regularisation (\CDR)
(see~\cite{Gnendiger:2017pys} and references therein for the
definitions of these schemes). To this end, we keep the dimensionality
of $\epsilon$ scalars, $n_\epsilon$, explicit in the poles but set it to
zero in the finite parts. The regularisation-scheme dependence is
therefore manifest as terms $\propto n_\epsilon$.  The corresponding
results in \HV and \CDR can be obtained by setting $n_\epsilon=0$.
Inserting $n_\epsilon=2\epsilon$, on the other hand, retrieves the
expressions in \FDH. Furthermore, in the case of \HV and \FDH
$\epsilon$ has to be set to zero in the tree-level splitting function.

In agreement with~\cite{Engel:2018fsb} the massification
constant reads
\begin{align}\label{eq:jetfunction_general}
    Z^{(1)} =
    4\pi\alpha K(\epsilon)
    \Big(\frac{\mu^2}{m^2}\Big)^\epsilon
    \Big(
    \frac{2}{\epsilon^2}
    +\frac{1}{\epsilon}(1-\frac{1}{2}n_\epsilon)
    +4+2\zeta_2
    \Big)
    +\mathcal{O}(\epsilon)\, ,
\end{align}
with $\alpha = Q^2/(4\pi) = e^2/(4\pi)$ and $K(\epsilon)$ defined in~\eqref{eq:loopfactor}.

We then define the invariant $s_{kp}=2k\cdot p$ where the photon
momentum $k$ is collinear to an initial- or final-state fermion $p$.
The initial-state collinear splitting function
\begin{align}
    J_\text{ISR} 
    = J_\text{ISR}^{(0)}+J_\text{ISR}^{(1)}+\mathcal{O}(\alpha^2)
\end{align}
can conveniently be written in terms of
\begin{align}
    x=\frac{E_p-E_k}{E_p},\qquad u=\frac{s_{kp}-m^2}{s_{kp}}\, ,
\end{align}
as
\begin{subequations}
\begin{align}
J_\text{ISR}^{(0)}
=& \frac{8\pi\alpha(1-u)}{m^2(1-x)x}
\big(1-2x+3x^2+2xu-2x^2u-\epsilon(1-x)^2\big)\, , \\
J_\text{ISR}^{(1)}
=& 8\pi\alpha J_\text{ISR}^{(0)} K(\epsilon)\Big(\frac{\mu^2}{m^2}\Big)^\epsilon
\Big(
\frac{1}{\epsilon^2} 
+\frac{1}{4\epsilon} (2-8 H_{0}(x)-n_\epsilon)
\Big)
+ \alpha^2 \frac{4(1-u)}{x m^2} \tilde{J}_\text{ISR}
+\mathcal{O}(\epsilon)\, ,
\end{align}
\end{subequations}
with
\begin{align}
\begin{split}
\tilde{J}_\text{ISR}
=& \frac{1}{u(1-x)}(-2 u^2 x^2+2 u^2+2 u x^2+2 u x+x^2-x) \\
+& \frac{\zeta_2}{1-x}(2 u^2 x^2-4 u^2 x+2 u^2-12 u x^2+16 u x-4 u+13 x^2-12 x+5) \\
+& \frac{H_{1}(u)}{u^2}(2 u^3 x-5 u^2 x-2 u^3+3 u^2+2 u x-u+x) \\
+& \frac{H_{1}(u)H_{0}(x)+H_{1,0}(x)}{1-x}(-8 u x^2+8 u x+10 x^2-8 x+2) \\
+& \big(H_{2}(u)+H_{1,1}(u)\big)(-2 u^2 x+2 u^2+2 u x-4 u+2)\, .
\end{split}
\end{align}
Because $0<x<1$ and $u<1$ the above expression is always real. 

The result for the final-state splitting function
\begin{align}
    J_\text{FSR} 
    = J_\text{FSR}^{(0)}+J_\text{FSR}^{(1)}+\mathcal{O}(\alpha^3)
\end{align}
can be obtained from $J_\text{ISR}$ via the crossing relation
$p\to-p$. In particular, this implies $x \to z^{-1}$ and $u \to
v^{-1}$ with
\begin{align}
    z=\frac{E_p}{E_p+E_k},\qquad v=\frac{s_{kp}}{m^2+s_{kp}}\, .
\end{align}
The corresponding analytic continuation is unambiguously defined via 
$s_{kp} \to s_{kp} + i \delta$ or equivalently $u \to
u - i \delta$. We then find
\begin{subequations}
\begin{align}
J_\text{ISR}^{(0)}
=& \frac{8\pi\alpha(1-v)}{m^2v^2(1-z)}\big(v z^2-2 v z+3 v+2 z-2-\epsilon v(1-z)^2\big)\, , \\
J_\text{FSR}^{(1)}
=& 8\pi\alpha J_\text{FSR}^{(0)} K(\epsilon)\Big(\frac{\mu^2}{m^2}\Big)^\epsilon
\Big(
\frac{1}{\epsilon^2} 
+\frac{1}{4\epsilon} (2+8 H_{0}(z)-n_\epsilon)
\Big)
+ \alpha^2 \frac{4(1-v)}{m^2v^2} \Big( \tilde{J}_\text{FSR}^\text{Re}+i\pi \tilde{J}_\text{FSR}^\text{Im} \ \Big)
+\mathcal{O}(\epsilon)\, ,
\end{align}
\end{subequations}
with
\begin{subequations}
\begin{align}
\begin{split}
\tilde{J}_\text{FSR}^\text{Re}
=& \frac{1}{1-z}(-v^2z+v^2+2 v z+2 v+2 z^2-2) \\
+& \frac{\zeta_2}{1-z}(-v z^2+6 v z-7 v-6 z+6) \\
+&\big(H_{0}(v)+H_{1}(v)\big)(v^2 z-v^3-2 v^2-3 v z+5 v+2 z-2) \\
+& \frac{H_{0}(v)H_{0}(z)+H_{1}(v)H_{0}(z)+H_{0,0}(z)+H_{1,0}(z)}{1-z}
(-2 v z^2+8 v z-10 v-8 z+8) \\
+& \frac{H_{1,0}(v)+H_{1,1}(v)}{v}(-2 v^2 z+4 v z-2 v-2 z+2)\, ,
\end{split}
\\ \vspace{.1cm}
\begin{split}
\tilde{J}_\text{FSR}^\text{Im}
=& -v^2 z+v^3+2 v^2+3 v z-5 v-2 z+2 \\
+& \frac{H_{0}(z)}{1-z}(2 v z^2-8 v z+10 v+8 z-8) \\
+& \frac{H_{1}(v)}{v}(2 v^2 z-4 v z+2 v+2 z-2)\, .
\end{split}
\end{align}
\end{subequations}
The imaginary part is given explicitly leaving all of the HPLs real for the physical region where $0<z,u<1$. The massless version of the FSR splitting function entering~\eqref{eq:massless_splittings} can be extracted from the spin-summed result of equations (II.10) and (II.11) in~\cite{Bern:1994zx} by taking the QED limit. The corresponding expressions in the \FDH scheme read
\begin{subequations}
\begin{align}
    &\bar{J}_\text{FSR}^{(0)} =
    \frac{8\pi\alpha}{s_{kp}} \frac{1+z^2}{1-z}\, , \\
    & \bar{J}_\text{FSR}^{(1)} =
    16\pi\alpha K(\epsilon) \Big\{ \bar{J}_\text{FSR}^{(0)} 
    \Big(
    -\frac{1}{\epsilon^2}\Big(-\frac{\mu^2}{z s_{kp}}\Big)^\epsilon
    +\frac{1}{\epsilon^2}\Big(-\frac{\mu^2}{s_{kp}}\Big)^\epsilon
    -H_{1,0}(z)-\zeta_2
    \Big)
    -\frac{4\pi\alpha}{s_{kp}} \Big\}\, .
\end{align}
\end{subequations}

%!TEX root=thesis

\chapter*{List of Abbbreviations}
\addcontentsline{toc}{chapter}{Index}
\markboth{INDEX}{INDEX}

We have used the following acronyms and abbreviations
\makeatletter
\write\@auxout{%
\unexpanded{\global\@namedef{notfirstrun}{1}}%
}
\makeatother
\ifcsname notfirstrun\endcsname {
\begin{multicols}{2}
\begin{acronym}[MMCT]
\renewcommand*{\aclabelfont}[1]{
    \def\textsc{}
    \textbf{\MakeUppercase{\acsfont{#1}}}
}

\acro{1PI}{one-particle irreducible}
\acro{BSM}{beyond the Standard Model}
\acro{CMS}{centre-of-mass system}
\acro{CPS}{collinear pseudo-singularity}
\acro{EFT}{effective field theory}
\acro{EW}{electroweak}
\acro{FDH}{four-dimensional helicity scheme}
\acro{FKS}{Frixione-Kunszt-Signer}
\acro{FSR}{final-state radiation}
\acro{GPL}{Goncharov polylogarithm}
\acro{HLbL}{hadronic light-by-light}
\acro{HPL}{harmonic polylogarithm}
\acro{HQET}{heavy-quark effective theory}
\acro{HVP}{hadronic vacuum polarisation}
\acro{IR}{infrared}
\acro{ISR}{initial-state radiation}
\acro{KLN}{Kinoshita-Lee-Nauenberg}
\acro{LAB}{laboratory}
\acro{LBK}{Low-Burnett-Kroll}
\acro{LbL}{light-by-light}
\acro{LHC}{Large Hadron Collider}
\acro{LL}{leading logarithm}
\acro{LO}{leading order}
\acro{LP}{leading power}
\acro{LSZ}{Lehmann-Symanzik-Zimmermann}
\acro{MoR}{method of regions}
\acro{NLL}{next-to-leading logarithm}
\acro{NLO}{next-to-leading order}
\acro{NLP}{next-to-leading power}
\acro{NNLO}{next-to-next-to leading order}
\acro{N$^3$LO}{next-to-next-to-next-to-leading order}
\acro{QCD}{quantum chromodynamics}
\acro{QED}{quantum electrodynamics}
\acro{QFT}{quantum field theory}
\acro{SCET}{soft-collinear effective theory}
\acro{SM}{Standard Model}
\acro{UV}{ultraviolet}
\acro{VP}{vacuum polarisation}
\acro{YFS}{Yennie-Frautschi-Suura}
\end{acronym}
\end{multicols}
}
\else yy \fi

\bibliographystyle{JHEP}
\bibliography{thesis}{}

\providecommand{\href}[2]{#2}\begingroup\raggedright\begin{thebibliography}{100}

\bibitem{Engel:2018}
T.~Engel, \emph{{Two-loop corrections to the muon decay}},  master's thesis,
  ETH Zurich, 5, 2018.

\bibitem{Engel:2018fsb}
T.~Engel, C.~Gnendiger, A.~Signer and Y.~Ulrich, \emph{{Small-mass effects in
  heavy-to-light form factors}},
  \href{https://doi.org/10.1007/JHEP02(2019)118}{\emph{JHEP} {\bfseries 02}
  (2019) 118} [\href{https://arxiv.org/abs/1811.06461}{{\ttfamily
  1811.06461}}].

\bibitem{Engel:2019nfw}
T.~Engel, A.~Signer and Y.~Ulrich, \emph{{A subtraction scheme for massive
  QED}}, \href{https://doi.org/10.1007/JHEP01(2020)085}{\emph{JHEP} {\bfseries
  01} (2020) 085} [\href{https://arxiv.org/abs/1909.10244}{{\ttfamily
  1909.10244}}].

\bibitem{Banerjee:2020tdt}
P.~Banerjee et~al., \emph{{Theory for muon-electron scattering @ 10 ppm: A
  report of the MUonE theory initiative}},
  \href{https://doi.org/10.1140/epjc/s10052-020-8138-9}{\emph{Eur. Phys. J. C}
  {\bfseries 80} (2020) 591}
  [\href{https://arxiv.org/abs/2004.13663}{{\ttfamily 2004.13663}}].

\bibitem{Banerjee:2020rww}
P.~Banerjee, T.~Engel, A.~Signer and Y.~Ulrich, \emph{{QED at NNLO with
  McMule}}, \href{https://doi.org/10.21468/SciPostPhys.9.2.027}{\emph{SciPost
  Phys.} {\bfseries 9} (2020) 027}
  [\href{https://arxiv.org/abs/2007.01654}{{\ttfamily 2007.01654}}].

\bibitem{Banerjee:2021mty}
P.~Banerjee, T.~Engel, N.~Schalch, A.~Signer and Y.~Ulrich, \emph{{Bhabha
  scattering at NNLO with next-to-soft stabilisation}},
  \href{https://doi.org/10.1016/j.physletb.2021.136547}{\emph{Phys. Lett. B}
  {\bfseries 820} (2021) 136547}
  [\href{https://arxiv.org/abs/2106.07469}{{\ttfamily 2106.07469}}].

\bibitem{Banerjee:2021qvi}
P.~Banerjee, T.~Engel, N.~Schalch, A.~Signer and Y.~Ulrich, \emph{{M\o{}ller
  scattering at NNLO}},
  \href{https://doi.org/10.1103/PhysRevD.105.L031904}{\emph{Phys. Rev. D}
  {\bfseries 105} (2022) L031904}
  [\href{https://arxiv.org/abs/2107.12311}{{\ttfamily 2107.12311}}].

\bibitem{Engel:2021ccn}
T.~Engel, A.~Signer and Y.~Ulrich, \emph{{Universal structure of radiative QED
  amplitudes at one loop}},
  \href{https://doi.org/10.1007/JHEP04(2022)097}{\emph{JHEP} {\bfseries 04}
  (2022) 097} [\href{https://arxiv.org/abs/2112.07570}{{\ttfamily
  2112.07570}}].

\bibitem{Frixione:2022ofv}
S.~Frixione et~al., \emph{{Initial state QED radiation aspects for future
  $e^+e^-$ colliders}},  in \emph{{2022 Snowmass Summer Study}}, 3, 2022,
  \href{https://arxiv.org/abs/2203.12557}{{\ttfamily 2203.12557}}.

\bibitem{Ulrich:2020frs}
Y.~Ulrich, \emph{{McMule -- QED Corrections for Low-Energy Experiments}},  phd
  thesis, University of Zurich, 8, 2020.

\bibitem{Bloch:1937pw}
F.~Bloch and A.~Nordsieck, \emph{{Note on the Radiation Field of the
  electron}}, \href{https://doi.org/10.1103/PhysRev.52.54}{\emph{Phys. Rev.}
  {\bfseries 52} (1937) 54}.

\bibitem{Kunszt:1992tn}
Z.~Kunszt and D.~E. Soper, \emph{{Calculation of jet cross-sections in hadron
  collisions at order $\alpha_s^3$}},
  \href{https://doi.org/10.1103/PhysRevD.46.192}{\emph{Phys. Rev. D} {\bfseries
  46} (1992) 192}.

\bibitem{Kinoshita:1962ur}
T.~Kinoshita, \emph{{Mass singularities of Feynman amplitudes}},
  \href{https://doi.org/10.1063/1.1724268}{\emph{J. Math. Phys.} {\bfseries 3}
  (1962) 650}.

\bibitem{Lee:1964is}
T.~D. Lee and M.~Nauenberg, \emph{{Degenerate Systems and Mass Singularities}},
  \href{https://doi.org/10.1103/PhysRev.133.B1549}{\emph{Phys. Rev.} {\bfseries
  133} (1964) B1549}.

\bibitem{Muong-2:2006rrc}
{\scshape Muon g-2} collaboration, G.~W. Bennett et~al., \emph{{Final Report of
  the Muon E821 Anomalous Magnetic Moment Measurement at BNL}},
  \href{https://doi.org/10.1103/PhysRevD.73.072003}{\emph{Phys. Rev. D}
  {\bfseries 73} (2006) 072003}
  [\href{https://arxiv.org/abs/hep-ex/0602035}{{\ttfamily hep-ex/0602035}}].

\bibitem{Muong-2:2021ojo}
{\scshape Muon g-2} collaboration, B.~Abi et~al., \emph{{Measurement of the
  Positive Muon Anomalous Magnetic Moment to 0.46 ppm}},
  \href{https://doi.org/10.1103/PhysRevLett.126.141801}{\emph{Phys. Rev. Lett.}
  {\bfseries 126} (2021) 141801}
  [\href{https://arxiv.org/abs/2104.03281}{{\ttfamily 2104.03281}}].

\bibitem{Aoyama:2020ynm}
T.~Aoyama et~al., \emph{{The anomalous magnetic moment of the muon in the
  Standard Model}},
  \href{https://doi.org/10.1016/j.physrep.2020.07.006}{\emph{Phys. Rept.}
  {\bfseries 887} (2020) 1} [\href{https://arxiv.org/abs/2006.04822}{{\ttfamily
  2006.04822}}].

\bibitem{Saito:2012zz}
{\scshape J-PARC $g$-$2$/EDM} collaboration, N.~Saito, \emph{{A novel precision
  measurement of muon g-2 and EDM at J-PARC}},
  \href{https://doi.org/10.1063/1.4742078}{\emph{AIP Conf. Proc.} {\bfseries
  1467} (2012) 45}.

\bibitem{Aoyama:2012wk}
T.~Aoyama, M.~Hayakawa, T.~Kinoshita and M.~Nio, \emph{{Complete Tenth-Order
  QED Contribution to the Muon g-2}},
  \href{https://doi.org/10.1103/PhysRevLett.109.111808}{\emph{Phys. Rev. Lett.}
  {\bfseries 109} (2012) 111808}
  [\href{https://arxiv.org/abs/1205.5370}{{\ttfamily 1205.5370}}].

\bibitem{Aoyama:2019ryr}
T.~Aoyama, T.~Kinoshita and M.~Nio, \emph{{Theory of the Anomalous Magnetic
  Moment of the Electron}},
  \href{https://doi.org/10.3390/atoms7010028}{\emph{Atoms} {\bfseries 7} (2019)
  28}.

\bibitem{Czarnecki:2002nt}
A.~Czarnecki, W.~J. Marciano and A.~Vainshtein, \emph{{Refinements in
  electroweak contributions to the muon anomalous magnetic moment}},
  \href{https://doi.org/10.1103/PhysRevD.67.073006}{\emph{Phys. Rev. D}
  {\bfseries 67} (2003) 073006}
  [\href{https://arxiv.org/abs/hep-ph/0212229}{{\ttfamily hep-ph/0212229}}].

\bibitem{Gnendiger:2013pva}
C.~Gnendiger, D.~St\"ockinger and H.~St\"ockinger-Kim, \emph{{The electroweak
  contributions to $(g-2)_\mu$ after the Higgs boson mass measurement}},
  \href{https://doi.org/10.1103/PhysRevD.88.053005}{\emph{Phys. Rev. D}
  {\bfseries 88} (2013) 053005}
  [\href{https://arxiv.org/abs/1306.5546}{{\ttfamily 1306.5546}}].

\bibitem{Aoki:2021kgd}
Y.~Aoki et~al., \emph{{FLAG Review 2021}},
  \href{https://arxiv.org/abs/2111.09849}{{\ttfamily 2111.09849}}.

\bibitem{Davier:2017zfy}
M.~Davier, A.~Hoecker, B.~Malaescu and Z.~Zhang, \emph{{Reevaluation of the
  hadronic vacuum polarisation contributions to the Standard Model predictions
  of the muon $g-2$ and ${\alpha (m_Z^2)}$ using newest hadronic cross-section
  data}}, \href{https://doi.org/10.1140/epjc/s10052-017-5161-6}{\emph{Eur.
  Phys. J. C} {\bfseries 77} (2017) 827}
  [\href{https://arxiv.org/abs/1706.09436}{{\ttfamily 1706.09436}}].

\bibitem{Keshavarzi:2018mgv}
A.~Keshavarzi, D.~Nomura and T.~Teubner, \emph{{Muon $g-2$ and $\alpha(M_Z^2)$:
  a new data-based analysis}},
  \href{https://doi.org/10.1103/PhysRevD.97.114025}{\emph{Phys. Rev. D}
  {\bfseries 97} (2018) 114025}
  [\href{https://arxiv.org/abs/1802.02995}{{\ttfamily 1802.02995}}].

\bibitem{Colangelo:2018mtw}
G.~Colangelo, M.~Hoferichter and P.~Stoffer, \emph{{Two-pion contribution to
  hadronic vacuum polarization}},
  \href{https://doi.org/10.1007/JHEP02(2019)006}{\emph{JHEP} {\bfseries 02}
  (2019) 006} [\href{https://arxiv.org/abs/1810.00007}{{\ttfamily
  1810.00007}}].

\bibitem{Hoferichter:2019mqg}
M.~Hoferichter, B.-L. Hoid and B.~Kubis, \emph{{Three-pion contribution to
  hadronic vacuum polarization}},
  \href{https://doi.org/10.1007/JHEP08(2019)137}{\emph{JHEP} {\bfseries 08}
  (2019) 137} [\href{https://arxiv.org/abs/1907.01556}{{\ttfamily
  1907.01556}}].

\bibitem{Davier:2019can}
M.~Davier, A.~Hoecker, B.~Malaescu and Z.~Zhang, \emph{{A new evaluation of the
  hadronic vacuum polarisation contributions to the muon anomalous magnetic
  moment and to $\alpha(m_Z^2)$}},
  \href{https://doi.org/10.1140/epjc/s10052-020-7792-2}{\emph{Eur. Phys. J. C}
  {\bfseries 80} (2020) 241}
  [\href{https://arxiv.org/abs/1908.00921}{{\ttfamily 1908.00921}}].

\bibitem{Keshavarzi:2019abf}
A.~Keshavarzi, D.~Nomura and T.~Teubner, \emph{{$g-2$ of charged leptons,
  $\alpha (M^2_Z)$ , and the hyperfine splitting of muonium}},
  \href{https://doi.org/10.1103/PhysRevD.101.014029}{\emph{Phys. Rev. D}
  {\bfseries 101} (2020) 014029}
  [\href{https://arxiv.org/abs/1911.00367}{{\ttfamily 1911.00367}}].

\bibitem{Borsanyi:2020mff}
S.~Borsanyi et~al., \emph{{Leading hadronic contribution to the muon magnetic
  moment from lattice QCD}},
  \href{https://doi.org/10.1038/s41586-021-03418-1}{\emph{Nature} {\bfseries
  593} (2021) 51} [\href{https://arxiv.org/abs/2002.12347}{{\ttfamily
  2002.12347}}].

\bibitem{Jegerlehner:2017gek}
F.~Jegerlehner, \emph{{The Anomalous Magnetic Moment of the Muon}}, vol.~274.
  Springer, Cham, 2017,
  \href{https://doi.org/10.1007/978-3-319-63577-4}{10.1007/978-3-319-63577-4}.

\bibitem{Abbiendi:2022liz}
G.~Abbiendi et~al., \emph{{Mini-Proceedings of the STRONG2020 Virtual Workshop
  on ''Space-like and Time-like determination of the Hadronic Leading Order
  contribution to the Muon $g-2$''}},  in \emph{{STRONG2020 Virtual Workshop
  \textquotedblleft{}Space-like and Time-like determination of the Hadronic
  Leading Order contribution to the Muon g \ensuremath{-}
  2\textquotedblright{}}}, 1, 2022,
  \href{https://arxiv.org/abs/2201.12102}{{\ttfamily 2201.12102}}.

\bibitem{refId0}
{Jegerlehner, Fred}, \emph{Leading-order hadronic contribution to the electron
  and muon g - 2},
  \href{https://doi.org/10.1051/epjconf/201611801016}{\emph{EPJ Web of
  Conferences} {\bfseries 118} (2016) 01016}.

\bibitem{Abbiendi:2016xup}
G.~Abbiendi et~al., \emph{{Measuring the leading hadronic contribution to the
  muon g-2 via $\mu e$ scattering}},
  \href{https://doi.org/10.1140/epjc/s10052-017-4633-z}{\emph{Eur. Phys. J. C}
  {\bfseries 77} (2017) 139}
  [\href{https://arxiv.org/abs/1609.08987}{{\ttfamily 1609.08987}}].

\bibitem{CarloniCalame:2015obs}
C.~M. Carloni~Calame, M.~Passera, L.~Trentadue and G.~Venanzoni, \emph{{A new
  approach to evaluate the leading hadronic corrections to the muon $g$-2}},
  \href{https://doi.org/10.1016/j.physletb.2015.05.020}{\emph{Phys. Lett. B}
  {\bfseries 746} (2015) 325}
  [\href{https://arxiv.org/abs/1504.02228}{{\ttfamily 1504.02228}}].

\bibitem{Abbiendi:2022oks}
G.~Abbiendi, \emph{{Status of the MUonE experiment}},
  \href{https://doi.org/10.1088/1402-4896/ac6297}{\emph{Phys. Scripta}
  {\bfseries 97} (2022) 054007}
  [\href{https://arxiv.org/abs/2201.13177}{{\ttfamily 2201.13177}}].

\bibitem{Masiero:2020vxk}
A.~Masiero, P.~Paradisi and M.~Passera, \emph{{New physics at the MUonE
  experiment at CERN}},
  \href{https://doi.org/10.1103/PhysRevD.102.075013}{\emph{Phys. Rev. D}
  {\bfseries 102} (2020) 075013}
  [\href{https://arxiv.org/abs/2002.05418}{{\ttfamily 2002.05418}}].

\bibitem{Dev:2020drf}
P.~S.~B. Dev, W.~Rodejohann, X.-J. Xu and Y.~Zhang, \emph{{MUonE sensitivity to
  new physics explanations of the muon anomalous magnetic moment}},
  \href{https://doi.org/10.1007/JHEP05(2020)053}{\emph{JHEP} {\bfseries 05}
  (2020) 053} [\href{https://arxiv.org/abs/2002.04822}{{\ttfamily
  2002.04822}}].

\bibitem{Schubert:2019nwm}
U.~Schubert and C.~Williams, \emph{{Interplay between SM precision, BSM
  physics, and the measurements of $\alpha_{\textrm{had}}$ in $\mu$-$e$
  scattering}}, \href{https://doi.org/10.1103/PhysRevD.100.035030}{\emph{Phys.
  Rev. D} {\bfseries 100} (2019) 035030}
  [\href{https://arxiv.org/abs/1907.01574}{{\ttfamily 1907.01574}}].

\bibitem{diCortona:2022zjy}
G.~G. di~Cortona and E.~Nardi, \emph{{Probing light mediators at the MUonE
  experiment}},  \href{https://arxiv.org/abs/2204.04227}{{\ttfamily
  2204.04227}}.

\bibitem{Galon:2022xcl}
I.~Galon, D.~Shih and I.~R. Wang, \emph{{Dark Photons and Displaced Vertices at
  the MUonE Experiment}},  \href{https://arxiv.org/abs/2202.08843}{{\ttfamily
  2202.08843}}.

\bibitem{Budassi:2022kqs}
E.~Budassi, C.~M.~C. Calame, C.~L. Del~Pio and F.~Piccinini, \emph{{Single
  $\pi^0$ production in $\mu e$ scattering at MUonE}},
  \href{https://arxiv.org/abs/2203.01639}{{\ttfamily 2203.01639}}.

\bibitem{Budassi:2021twh}
E.~Budassi, C.~M. Carloni~Calame, M.~Chiesa, C.~L. Del~Pio, S.~M. Hasan,
  G.~Montagna et~al., \emph{{NNLO virtual and real leptonic corrections to
  muon-electron scattering}},
  \href{https://doi.org/10.1007/JHEP11(2021)098}{\emph{JHEP} {\bfseries 11}
  (2021) 098} [\href{https://arxiv.org/abs/2109.14606}{{\ttfamily
  2109.14606}}].

\bibitem{Fael:2019nsf}
M.~Fael and M.~Passera, \emph{{Muon-Electron Scattering at
  Next-To-Next-To-Leading Order: The Hadronic Corrections}},
  \href{https://doi.org/10.1103/PhysRevLett.122.192001}{\emph{Phys. Rev. Lett.}
  {\bfseries 122} (2019) 192001}
  [\href{https://arxiv.org/abs/1901.03106}{{\ttfamily 1901.03106}}].

\bibitem{Fael:2018dmz}
M.~Fael, \emph{{Hadronic corrections to $\mu$-$e$ scattering at NNLO with
  space-like data}}, \href{https://doi.org/10.1007/JHEP02(2019)027}{\emph{JHEP}
  {\bfseries 02} (2019) 027}
  [\href{https://arxiv.org/abs/1808.08233}{{\ttfamily 1808.08233}}].

\bibitem{Levine:1974xh}
M.~J. Levine and R.~Roskies, \emph{{Hyperspherical approach to quantum
  electrodynamics - sixth-order magnetic moment}},
  \href{https://doi.org/10.1103/PhysRevD.9.421}{\emph{Phys. Rev. D} {\bfseries
  9} (1974) 421}.

\bibitem{Levine:1975jz}
M.~J. Levine, R.~C. Perisho and R.~Roskies, \emph{{Analytic Contributions to
  the G Factor of the electron}},
  \href{https://doi.org/10.1103/PhysRevD.13.997}{\emph{Phys. Rev. D} {\bfseries
  13} (1976) 997}.

\bibitem{Alacevich:2018vez}
M.~Alacevich, C.~M. Carloni~Calame, M.~Chiesa, G.~Montagna, O.~Nicrosini and
  F.~Piccinini, \emph{{Muon-electron scattering at NLO}},
  \href{https://doi.org/10.1007/JHEP02(2019)155}{\emph{JHEP} {\bfseries 02}
  (2019) 155} [\href{https://arxiv.org/abs/1811.06743}{{\ttfamily
  1811.06743}}].

\bibitem{Melnikov:1996na}
K.~Melnikov and V.~G. Serbo, \emph{{New type of beam size effect and the W
  boson production at $\mu^+\mu^-$ colliders}},
  \href{https://doi.org/10.1103/PhysRevLett.76.3263}{\emph{Phys. Rev. Lett.}
  {\bfseries 76} (1996) 3263}
  [\href{https://arxiv.org/abs/hep-ph/9601221}{{\ttfamily hep-ph/9601221}}].

\bibitem{Mastrolia:2003yz}
P.~Mastrolia and E.~Remiddi, \emph{{Two loop form-factors in QED}},
  \href{https://doi.org/10.1016/S0550-3213(03)00405-X}{\emph{Nucl. Phys. B}
  {\bfseries 664} (2003) 341}
  [\href{https://arxiv.org/abs/hep-ph/0302162}{{\ttfamily hep-ph/0302162}}].

\bibitem{Bonciani:2003ai}
R.~Bonciani, P.~Mastrolia and E.~Remiddi, \emph{{QED vertex form-factors at two
  loops}}, \href{https://doi.org/10.1016/j.nuclphysb.2003.10.031}{\emph{Nucl.
  Phys. B} {\bfseries 676} (2004) 399}
  [\href{https://arxiv.org/abs/hep-ph/0307295}{{\ttfamily hep-ph/0307295}}].

\bibitem{Bernreuther:2004ih}
W.~Bernreuther, R.~Bonciani, T.~Gehrmann, R.~Heinesch, T.~Leineweber,
  P.~Mastrolia et~al., \emph{{Two-loop QCD corrections to the heavy quark
  form-factors: The Vector contributions}},
  \href{https://doi.org/10.1016/j.nuclphysb.2004.10.059}{\emph{Nucl. Phys. B}
  {\bfseries 706} (2005) 245}
  [\href{https://arxiv.org/abs/hep-ph/0406046}{{\ttfamily hep-ph/0406046}}].

\bibitem{Gluza:2009yy}
J.~Gluza, A.~Mitov, S.~Moch and T.~Riemann, \emph{{The QCD form factor of heavy
  quarks at NNLO}},
  \href{https://doi.org/10.1088/1126-6708/2009/07/001}{\emph{JHEP} {\bfseries
  07} (2009) 001} [\href{https://arxiv.org/abs/0905.1137}{{\ttfamily
  0905.1137}}].

\bibitem{CarloniCalame:2020yoz}
C.~M. Carloni~Calame, M.~Chiesa, S.~M. Hasan, G.~Montagna, O.~Nicrosini and
  F.~Piccinini, \emph{{Towards muon-electron scattering at NNLO}},
  \href{https://doi.org/10.1007/JHEP11(2020)028}{\emph{JHEP} {\bfseries 11}
  (2020) 028} [\href{https://arxiv.org/abs/2007.01586}{{\ttfamily
  2007.01586}}].

\bibitem{Mastrolia:2017pfy}
P.~Mastrolia, M.~Passera, A.~Primo and U.~Schubert, \emph{{Master integrals for
  the NNLO virtual corrections to $\mu e$ scattering in QED: the planar
  graphs}}, \href{https://doi.org/10.1007/JHEP11(2017)198}{\emph{JHEP}
  {\bfseries 11} (2017) 198}
  [\href{https://arxiv.org/abs/1709.07435}{{\ttfamily 1709.07435}}].

\bibitem{DiVita:2018nnh}
S.~Di~Vita, S.~Laporta, P.~Mastrolia, A.~Primo and U.~Schubert, \emph{{Master
  integrals for the NNLO virtual corrections to $\mu e$ scattering in QED: the
  non-planar graphs}},
  \href{https://doi.org/10.1007/JHEP09(2018)016}{\emph{JHEP} {\bfseries 09}
  (2018) 016} [\href{https://arxiv.org/abs/1806.08241}{{\ttfamily
  1806.08241}}].

\bibitem{Bonciani:2021okt}
R.~Bonciani et~al., \emph{{Two-Loop Four-Fermion Scattering Amplitude in QED}},
  \href{https://doi.org/10.1103/PhysRevLett.128.022002}{\emph{Phys. Rev. Lett.}
  {\bfseries 128} (2022) 022002}
  [\href{https://arxiv.org/abs/2106.13179}{{\ttfamily 2106.13179}}].

\bibitem{Becher:2007cu}
T.~Becher and K.~Melnikov, \emph{{Two-loop QED corrections to Bhabha
  scattering}},
  \href{https://doi.org/10.1088/1126-6708/2007/06/084}{\emph{JHEP} {\bfseries
  06} (2007) 084} [\href{https://arxiv.org/abs/0704.3582}{{\ttfamily
  0704.3582}}].

\bibitem{Penin:2005eh}
A.~A. Penin, \emph{{Two-loop photonic corrections to massive Bhabha
  scattering}},
  \href{https://doi.org/10.1016/j.nuclphysb.2005.11.016}{\emph{Nucl. Phys. B}
  {\bfseries 734} (2006) 185}
  [\href{https://arxiv.org/abs/hep-ph/0508127}{{\ttfamily hep-ph/0508127}}].

\bibitem{Weinzierl:2006qs}
S.~Weinzierl, \emph{{The Art of computing loop integrals}}, {\emph{Fields Inst.
  Commun.} {\bfseries 50} (2007) 345}
  [\href{https://arxiv.org/abs/hep-ph/0604068}{{\ttfamily hep-ph/0604068}}].

\bibitem{Weinzierl:2022eaz}
S.~Weinzierl, \emph{{Feynman Integrals}},
  \href{https://arxiv.org/abs/2201.03593}{{\ttfamily 2201.03593}}.

\bibitem{Smirnov:2012gma}
V.~A. Smirnov, \emph{{Analytic tools for Feynman integrals}}, vol.~250. 2012,
  \href{https://doi.org/10.1007/978-3-642-34886-0}{10.1007/978-3-642-34886-0}.

\bibitem{Smirnov:1999gc}
V.~A. Smirnov, \emph{{Analytical result for dimensionally regularized massless
  on shell double box}},
  \href{https://doi.org/10.1016/S0370-2693(99)00777-7}{\emph{Phys. Lett. B}
  {\bfseries 460} (1999) 397}
  [\href{https://arxiv.org/abs/hep-ph/9905323}{{\ttfamily hep-ph/9905323}}].

\bibitem{Tausk:1999vh}
J.~B. Tausk, \emph{{Nonplanar massless two loop Feynman diagrams with four
  on-shell legs}},
  \href{https://doi.org/10.1016/S0370-2693(99)01277-0}{\emph{Phys. Lett. B}
  {\bfseries 469} (1999) 225}
  [\href{https://arxiv.org/abs/hep-ph/9909506}{{\ttfamily hep-ph/9909506}}].

\bibitem{Huber:2005yg}
T.~Huber and D.~Maitre, \emph{{HypExp: A Mathematica package for expanding
  hypergeometric functions around integer-valued parameters}},
  \href{https://doi.org/10.1016/j.cpc.2006.01.007}{\emph{Comput. Phys. Commun.}
  {\bfseries 175} (2006) 122}
  [\href{https://arxiv.org/abs/hep-ph/0507094}{{\ttfamily hep-ph/0507094}}].

\bibitem{Huber:2007dx}
T.~Huber and D.~Maitre, \emph{{HypExp 2, Expanding Hypergeometric Functions
  about Half-Integer Parameters}},
  \href{https://doi.org/10.1016/j.cpc.2007.12.008}{\emph{Comput. Phys. Commun.}
  {\bfseries 178} (2008) 755}
  [\href{https://arxiv.org/abs/0708.2443}{{\ttfamily 0708.2443}}].

\bibitem{Remiddi:1999ew}
E.~Remiddi and J.~A.~M. Vermaseren, \emph{{Harmonic polylogarithms}},
  \href{https://doi.org/10.1142/S0217751X00000367}{\emph{Int. J. Mod. Phys. A}
  {\bfseries 15} (2000) 725}
  [\href{https://arxiv.org/abs/hep-ph/9905237}{{\ttfamily hep-ph/9905237}}].

\bibitem{Maitre:2005uu}
D.~Maitre, \emph{{HPL, a mathematica implementation of the harmonic
  polylogarithms}},
  \href{https://doi.org/10.1016/j.cpc.2005.10.008}{\emph{Comput. Phys. Commun.}
  {\bfseries 174} (2006) 222}
  [\href{https://arxiv.org/abs/hep-ph/0507152}{{\ttfamily hep-ph/0507152}}].

\bibitem{Maitre:2007kp}
D.~Maitre, \emph{{Extension of HPL to complex arguments}},
  \href{https://doi.org/10.1016/j.cpc.2011.11.015}{\emph{Comput. Phys. Commun.}
  {\bfseries 183} (2012) 846}
  [\href{https://arxiv.org/abs/hep-ph/0703052}{{\ttfamily hep-ph/0703052}}].

\bibitem{Patel:2015tea}
H.~H. Patel, \emph{{Package-X: A Mathematica package for the analytic
  calculation of one-loop integrals}},
  \href{https://doi.org/10.1016/j.cpc.2015.08.017}{\emph{Comput. Phys. Commun.}
  {\bfseries 197} (2015) 276}
  [\href{https://arxiv.org/abs/1503.01469}{{\ttfamily 1503.01469}}].

\bibitem{Urwyler:2019}
D.~Urwyler, \emph{{Mellin-Barnes for two-loop integrals}},  bachelor's thesis,
  University of Zurich, 7, 2019.

\bibitem{Henn:2014qga}
J.~M. Henn, \emph{{Lectures on differential equations for Feynman integrals}},
  \href{https://doi.org/10.1088/1751-8113/48/15/153001}{\emph{J. Phys. A}
  {\bfseries 48} (2015) 153001}
  [\href{https://arxiv.org/abs/1412.2296}{{\ttfamily 1412.2296}}].

\bibitem{Beneke:1997zp}
M.~Beneke and V.~A. Smirnov, \emph{{Asymptotic expansion of Feynman integrals
  near threshold}},
  \href{https://doi.org/10.1016/S0550-3213(98)00138-2}{\emph{Nucl. Phys. B}
  {\bfseries 522} (1998) 321}
  [\href{https://arxiv.org/abs/hep-ph/9711391}{{\ttfamily hep-ph/9711391}}].

\bibitem{Smirnov:1999bza}
V.~A. Smirnov, \emph{{Problems of the strategy of regions}},
  \href{https://doi.org/10.1016/S0370-2693(99)01061-8}{\emph{Phys. Lett. B}
  {\bfseries 465} (1999) 226}
  [\href{https://arxiv.org/abs/hep-ph/9907471}{{\ttfamily hep-ph/9907471}}].

\bibitem{Becher:2014oda}
T.~Becher, A.~Broggio and A.~Ferroglia, \emph{{Introduction to Soft-Collinear
  Effective Theory}}, vol.~896. Springer, 2015,
  \href{https://doi.org/10.1007/978-3-319-14848-9}{10.1007/978-3-319-14848-9},
  [\href{https://arxiv.org/abs/1410.1892}{{\ttfamily 1410.1892}}].

\bibitem{Jantzen:2012mw}
B.~Jantzen, A.~V. Smirnov and V.~A. Smirnov, \emph{{Expansion by regions:
  revealing potential and Glauber regions automatically}},
  \href{https://doi.org/10.1140/epjc/s10052-012-2139-2}{\emph{Eur. Phys. J. C}
  {\bfseries 72} (2012) 2139}
  [\href{https://arxiv.org/abs/1206.0546}{{\ttfamily 1206.0546}}].

\bibitem{Heinrich:2021dbf}
G.~Heinrich, S.~Jahn, S.~P. Jones, M.~Kerner, F.~Langer, V.~Magerya et~al.,
  \emph{{Expansion by regions with pySecDec}},
  \href{https://doi.org/10.1016/j.cpc.2021.108267}{\emph{Comput. Phys. Commun.}
  {\bfseries 273} (2022) 108267}
  [\href{https://arxiv.org/abs/2108.10807}{{\ttfamily 2108.10807}}].

\bibitem{Frenkel:1976bj}
J.~Frenkel and J.~C. Taylor, \emph{{Exponentiation of Leading Infrared
  Divergences in Massless Yang-Mills Theories}},
  \href{https://doi.org/10.1016/0550-3213(76)90320-5}{\emph{Nucl. Phys. B}
  {\bfseries 116} (1976) 185}.

\bibitem{Bauer:2001yt}
C.~W. Bauer, D.~Pirjol and I.~W. Stewart, \emph{{Soft collinear factorization
  in effective field theory}},
  \href{https://doi.org/10.1103/PhysRevD.65.054022}{\emph{Phys. Rev. D}
  {\bfseries 65} (2002) 054022}
  [\href{https://arxiv.org/abs/hep-ph/0109045}{{\ttfamily hep-ph/0109045}}].

\bibitem{Bauer:2000yr}
C.~W. Bauer, S.~Fleming, D.~Pirjol and I.~W. Stewart, \emph{{An Effective field
  theory for collinear and soft gluons: Heavy to light decays}},
  \href{https://doi.org/10.1103/PhysRevD.63.114020}{\emph{Phys. Rev. D}
  {\bfseries 63} (2001) 114020}
  [\href{https://arxiv.org/abs/hep-ph/0011336}{{\ttfamily hep-ph/0011336}}].

\bibitem{Beneke:2002ph}
M.~Beneke, A.~P. Chapovsky, M.~Diehl and T.~Feldmann, \emph{{Soft collinear
  effective theory and heavy to light currents beyond leading power}},
  \href{https://doi.org/10.1016/S0550-3213(02)00687-9}{\emph{Nucl. Phys. B}
  {\bfseries 643} (2002) 431}
  [\href{https://arxiv.org/abs/hep-ph/0206152}{{\ttfamily hep-ph/0206152}}].

\bibitem{Bern:2000ie}
Z.~Bern, L.~J. Dixon and A.~Ghinculov, \emph{{Two loop correction to Bhabha
  scattering}}, \href{https://doi.org/10.1103/PhysRevD.63.053007}{\emph{Phys.
  Rev. D} {\bfseries 63} (2001) 053007}
  [\href{https://arxiv.org/abs/hep-ph/0010075}{{\ttfamily hep-ph/0010075}}].

\bibitem{Anastasiou:2002zn}
C.~Anastasiou, E.~W.~N. Glover and M.~E. Tejeda-Yeomans, \emph{{Two loop QED
  and QCD corrections to massless fermion boson scattering}},
  \href{https://doi.org/10.1016/S0550-3213(02)00140-2}{\emph{Nucl. Phys. B}
  {\bfseries 629} (2002) 255}
  [\href{https://arxiv.org/abs/hep-ph/0201274}{{\ttfamily hep-ph/0201274}}].

\bibitem{Gnendiger:2017pys}
C.~Gnendiger et~al., \emph{{To ${d}$, or not to ${d}$: recent developments and
  comparisons of regularization schemes}},
  \href{https://doi.org/10.1140/epjc/s10052-017-5023-2}{\emph{Eur. Phys. J. C}
  {\bfseries 77} (2017) 471}
  [\href{https://arxiv.org/abs/1705.01827}{{\ttfamily 1705.01827}}].

\bibitem{Smirnov:1997gx}
V.~A. Smirnov, \emph{{Asymptotic expansions of two loop Feynman diagrams in the
  Sudakov limit}},
  \href{https://doi.org/10.1016/S0370-2693(97)00545-5}{\emph{Phys. Lett. B}
  {\bfseries 404} (1997) 101}
  [\href{https://arxiv.org/abs/hep-ph/9703357}{{\ttfamily hep-ph/9703357}}].

\bibitem{Becher:2010tm}
T.~Becher and M.~Neubert, \emph{{Drell-Yan Production at Small $q_T$,
  Transverse Parton Distributions and the Collinear Anomaly}},
  \href{https://doi.org/10.1140/epjc/s10052-011-1665-7}{\emph{Eur. Phys. J. C}
  {\bfseries 71} (2011) 1665}
  [\href{https://arxiv.org/abs/1007.4005}{{\ttfamily 1007.4005}}].

\bibitem{Chiu:2011qc}
J.-y. Chiu, A.~Jain, D.~Neill and I.~Z. Rothstein, \emph{{The Rapidity
  Renormalization Group}},
  \href{https://doi.org/10.1103/PhysRevLett.108.151601}{\emph{Phys. Rev. Lett.}
  {\bfseries 108} (2012) 151601}
  [\href{https://arxiv.org/abs/1104.0881}{{\ttfamily 1104.0881}}].

\bibitem{Anastasiou:2005pn}
C.~Anastasiou, K.~Melnikov and F.~Petriello, \emph{{The electron energy
  spectrum in muon decay through $\mathcal{O}(\alpha^2)$}},
  \href{https://doi.org/10.1088/1126-6708/2007/09/014}{\emph{JHEP} {\bfseries
  09} (2007) 014} [\href{https://arxiv.org/abs/hep-ph/0505069}{{\ttfamily
  hep-ph/0505069}}].

\bibitem{Chen:2018dpt}
L.-B. Chen, \emph{{Two-Loop master integrals for heavy-to-light form factors of
  two different massive fermions}},
  \href{https://doi.org/10.1007/JHEP02(2018)066}{\emph{JHEP} {\bfseries 02}
  (2018) 066} [\href{https://arxiv.org/abs/1801.01033}{{\ttfamily
  1801.01033}}].

\bibitem{Arbuzov:2002pp}
A.~Arbuzov, A.~Czarnecki and A.~Gaponenko, \emph{{Muon decay spectrum: Leading
  logarithmic approximation}},
  \href{https://doi.org/10.1103/PhysRevD.65.113006}{\emph{Phys. Rev. D}
  {\bfseries 65} (2002) 113006}
  [\href{https://arxiv.org/abs/hep-ph/0202102}{{\ttfamily hep-ph/0202102}}].

\bibitem{Arbuzov:2002cn}
A.~Arbuzov and K.~Melnikov, \emph{{$\mathcal{O}(\alpha^2 \ln( m_\mu/m_e))$
  corrections to electron energy spectrum in muon decay}},
  \href{https://doi.org/10.1103/PhysRevD.66.093003}{\emph{Phys. Rev. D}
  {\bfseries 66} (2002) 093003}
  [\href{https://arxiv.org/abs/hep-ph/0205172}{{\ttfamily hep-ph/0205172}}].

\bibitem{Yennie:1961ad}
D.~R. Yennie, S.~C. Frautschi and H.~Suura, \emph{{The infrared divergence
  phenomena and high-energy processes}},
  \href{https://doi.org/10.1016/0003-4916(61)90151-8}{\emph{Annals Phys.}
  {\bfseries 13} (1961) 379}.

\bibitem{Frederix:2009yq}
R.~Frederix, S.~Frixione, F.~Maltoni and T.~Stelzer, \emph{{Automation of
  next-to-leading order computations in QCD: The FKS subtraction}},
  \href{https://doi.org/10.1088/1126-6708/2009/10/003}{\emph{JHEP} {\bfseries
  10} (2009) 003} [\href{https://arxiv.org/abs/0908.4272}{{\ttfamily
  0908.4272}}].

\bibitem{Frixione:1995ms}
S.~Frixione, Z.~Kunszt and A.~Signer, \emph{{Three jet cross-sections to
  next-to-leading order}},
  \href{https://doi.org/10.1016/0550-3213(96)00110-1}{\emph{Nucl. Phys. B}
  {\bfseries 467} (1996) 399}
  [\href{https://arxiv.org/abs/hep-ph/9512328}{{\ttfamily hep-ph/9512328}}].

\bibitem{Djouadi:1993ss}
A.~Djouadi and P.~Gambino, \emph{{Electroweak gauge bosons selfenergies:
  Complete QCD corrections}},
  \href{https://doi.org/10.1103/PhysRevD.49.3499}{\emph{Phys. Rev. D}
  {\bfseries 49} (1994) 3499}
  [\href{https://arxiv.org/abs/hep-ph/9309298}{{\ttfamily hep-ph/9309298}}].

\bibitem{Jegerlehner:2001ca}
F.~Jegerlehner, \emph{{The Effective fine structure constant at TESLA
  energies}},  \href{https://arxiv.org/abs/hep-ph/0105283}{{\ttfamily
  hep-ph/0105283}}.

\bibitem{Jegerlehner:2006ju}
F.~Jegerlehner, \emph{{Precision measurements of $\sigma_\text{hadronic}$ for
  $\alpha_\text{eff}(E)$ at ILC energies and $(g-2)_\mu$}},
  \href{https://doi.org/10.1016/j.nuclphysbps.2006.09.060}{\emph{Nucl. Phys. B
  Proc. Suppl.} {\bfseries 162} (2006) 22}
  [\href{https://arxiv.org/abs/hep-ph/0608329}{{\ttfamily hep-ph/0608329}}].

\bibitem{Jegerlehner:2011mw}
F.~Jegerlehner, \emph{{Electroweak effective couplings for future precision
  experiments}}, \href{https://doi.org/10.1393/ncc/i2011-11011-0}{\emph{Nuovo
  Cim. C} {\bfseries 034S1} (2011) 31}
  [\href{https://arxiv.org/abs/1107.4683}{{\ttfamily 1107.4683}}].

\bibitem{Davier:2010rnx}
M.~Davier, A.~Hoecker, B.~Malaescu, C.~Z. Yuan and Z.~Zhang,
  \emph{{Reevaluation of the hadronic contribution to the muon magnetic anomaly
  using new $e^+e^-\to\pi^+\pi^-$ cross section data from BABAR}},
  \href{https://doi.org/10.1140/epjc/s10052-010-1246-1}{\emph{Eur. Phys. J. C}
  {\bfseries 66} (2010) 1} [\href{https://arxiv.org/abs/0908.4300}{{\ttfamily
  0908.4300}}].

\bibitem{PhysRev.124.1577}
N.~Cabibbo and R.~Gatto, \emph{Electron-positron colliding beam experiments},
  \href{https://doi.org/10.1103/PhysRev.124.1577}{\emph{Phys. Rev.} {\bfseries
  124} (1961) 1577}.

\bibitem{vanRitbergen:1998hn}
T.~van Ritbergen and R.~G. Stuart, \emph{{Hadronic contributions to the muon
  lifetime}}, \href{https://doi.org/10.1016/S0370-2693(98)00895-8}{\emph{Phys.
  Lett. B} {\bfseries 437} (1998) 201}
  [\href{https://arxiv.org/abs/hep-ph/9802341}{{\ttfamily hep-ph/9802341}}].

\bibitem{Davydychev:2000ee}
A.~I. Davydychev, K.~Schilcher and H.~Spiesberger, \emph{{Hadronic corrections
  at $\mathcal{O}(\alpha^2)$ to the energy spectrum of muon decay}},
  \href{https://doi.org/10.1007/s100520100577}{\emph{Eur. Phys. J. C}
  {\bfseries 19} (2001) 99}
  [\href{https://arxiv.org/abs/hep-ph/0011221}{{\ttfamily hep-ph/0011221}}].

\bibitem{Actis:2007fs}
S.~Actis, M.~Czakon, J.~Gluza and T.~Riemann, \emph{{Virtual hadronic and
  leptonic contributions to Bhabha scattering}},
  \href{https://doi.org/10.1103/PhysRevLett.100.131602}{\emph{Phys. Rev. Lett.}
  {\bfseries 100} (2008) 131602}
  [\href{https://arxiv.org/abs/0711.3847}{{\ttfamily 0711.3847}}].

\bibitem{Kuhn:2008zs}
J.~H. Kuhn and S.~Uccirati, \emph{{Two-loop QED hadronic corrections to Bhabha
  scattering}},
  \href{https://doi.org/10.1016/j.nuclphysb.2008.08.002}{\emph{Nucl. Phys. B}
  {\bfseries 806} (2009) 300}
  [\href{https://arxiv.org/abs/0807.1284}{{\ttfamily 0807.1284}}].

\bibitem{CarloniCalame:2011zq}
C.~Carloni~Calame, H.~Czyz, J.~Gluza, M.~Gunia, G.~Montagna, O.~Nicrosini
  et~al., \emph{{NNLO leptonic and hadronic corrections to Bhabha scattering
  and luminosity monitoring at meson factories}},
  \href{https://doi.org/10.1007/JHEP07(2011)126}{\emph{JHEP} {\bfseries 07}
  (2011) 126} [\href{https://arxiv.org/abs/1106.3178}{{\ttfamily 1106.3178}}].

\bibitem{Laporta:1994mb}
S.~Laporta, \emph{{Hyperspherical integration and the triple cross vertex
  graphs}}, \href{https://doi.org/10.1007/BF02780705}{\emph{Nuovo Cim. A}
  {\bfseries 107} (1994) 1729}
  [\href{https://arxiv.org/abs/hep-ph/9404203}{{\ttfamily hep-ph/9404203}}].

\bibitem{Bonciani:2004gi}
R.~Bonciani, A.~Ferroglia, P.~Mastrolia, E.~Remiddi and J.~J. van~der Bij,
  \emph{{Two-loop $N_F=1$ QED Bhabha scattering differential cross section}},
  \href{https://doi.org/10.1016/j.nuclphysb.2004.09.015}{\emph{Nucl. Phys. B}
  {\bfseries 701} (2004) 121}
  [\href{https://arxiv.org/abs/hep-ph/0405275}{{\ttfamily hep-ph/0405275}}].

\bibitem{Denner:2016kdg}
A.~Denner, S.~Dittmaier and L.~Hofer, \emph{{Collier: a fortran-based Complex
  One-Loop LIbrary in Extended Regularizations}},
  \href{https://doi.org/10.1016/j.cpc.2016.10.013}{\emph{Comput. Phys. Commun.}
  {\bfseries 212} (2017) 220}
  [\href{https://arxiv.org/abs/1604.06792}{{\ttfamily 1604.06792}}].

\bibitem{Lepage:1980jk}
G.~P. Lepage, \emph{{VEGAS}: An adaptive multidimensional integration program}.

\bibitem{Weinzierl:2000wd}
S.~Weinzierl, \emph{{Introduction to Monte Carlo methods}},
  \href{https://arxiv.org/abs/hep-ph/0006269}{{\ttfamily hep-ph/0006269}}.

\bibitem{Buccioni:2017yxi}
F.~Buccioni, S.~Pozzorini and M.~Zoller, \emph{{On-the-fly reduction of open
  loops}}, \href{https://doi.org/10.1140/epjc/s10052-018-5562-1}{\emph{Eur.
  Phys. J. C} {\bfseries 78} (2018) 70}
  [\href{https://arxiv.org/abs/1710.11452}{{\ttfamily 1710.11452}}].

\bibitem{Buccioni:2019sur}
F.~Buccioni, J.-N. Lang, J.~M. Lindert, P.~Maierh\"ofer, S.~Pozzorini, H.~Zhang
  et~al., \emph{{OpenLoops 2}},
  \href{https://doi.org/10.1140/epjc/s10052-019-7306-2}{\emph{Eur. Phys. J. C}
  {\bfseries 79} (2019) 866}
  [\href{https://arxiv.org/abs/1907.13071}{{\ttfamily 1907.13071}}].

\bibitem{max}
M.~Zoller, \emph{{private communication}}.

\bibitem{Low:1958sn}
F.~E. Low, \emph{{Bremsstrahlung of very low-energy quanta in elementary
  particle collisions}},
  \href{https://doi.org/10.1103/PhysRev.110.974}{\emph{Phys. Rev.} {\bfseries
  110} (1958) 974}.

\bibitem{Burnett:1967km}
T.~H. Burnett and N.~M. Kroll, \emph{{Extension of the low soft photon
  theorem}}, \href{https://doi.org/10.1103/PhysRevLett.20.86}{\emph{Phys. Rev.
  Lett.} {\bfseries 20} (1968) 86}.

\bibitem{Cachazo:2014fwa}
F.~Cachazo and A.~Strominger, \emph{{Evidence for a New Soft Graviton
  Theorem}},  \href{https://arxiv.org/abs/1404.4091}{{\ttfamily 1404.4091}}.

\bibitem{Bern:2014vva}
Z.~Bern, S.~Davies, P.~Di~Vecchia and J.~Nohle, \emph{{Low-Energy Behavior of
  Gluons and Gravitons from Gauge Invariance}},
  \href{https://doi.org/10.1103/PhysRevD.90.084035}{\emph{Phys. Rev. D}
  {\bfseries 90} (2014) 084035}
  [\href{https://arxiv.org/abs/1406.6987}{{\ttfamily 1406.6987}}].

\bibitem{Beneke:2021umj}
M.~Beneke, P.~Hager and R.~Szafron, \emph{{Gravitational soft theorem from
  emergent soft gauge symmetries}},
  \href{https://doi.org/10.1007/JHEP03(2022)199}{\emph{JHEP} {\bfseries 03}
  (2022) 199} [\href{https://arxiv.org/abs/2110.02969}{{\ttfamily
  2110.02969}}].

\bibitem{DelDuca:1990gz}
V.~Del~Duca, \emph{{High-energy Bremsstrahlung Theorems for Soft Photons}},
  \href{https://doi.org/10.1016/0550-3213(90)90392-Q}{\emph{Nucl. Phys. B}
  {\bfseries 345} (1990) 369}.

\bibitem{Bonocore:2015esa}
D.~Bonocore, E.~Laenen, L.~Magnea, S.~Melville, L.~Vernazza and C.~D. White,
  \emph{{A factorization approach to next-to-leading-power threshold
  logarithms}}, \href{https://doi.org/10.1007/JHEP06(2015)008}{\emph{JHEP}
  {\bfseries 06} (2015) 008}
  [\href{https://arxiv.org/abs/1503.05156}{{\ttfamily 1503.05156}}].

\bibitem{Bonocore:2016awd}
D.~Bonocore, E.~Laenen, L.~Magnea, L.~Vernazza and C.~D. White,
  \emph{{Non-abelian factorisation for next-to-leading-power threshold
  logarithms}}, \href{https://doi.org/10.1007/JHEP12(2016)121}{\emph{JHEP}
  {\bfseries 12} (2016) 121}
  [\href{https://arxiv.org/abs/1610.06842}{{\ttfamily 1610.06842}}].

\bibitem{Laenen:2020nrt}
E.~Laenen, J.~Sinninghe~Damst\'e, L.~Vernazza, W.~Waalewijn and L.~Zoppi,
  \emph{{Towards all-order factorization of QED amplitudes at next-to-leading
  power}}, \href{https://doi.org/10.1103/PhysRevD.103.034022}{\emph{Phys. Rev.
  D} {\bfseries 103} (2021) 034022}
  [\href{https://arxiv.org/abs/2008.01736}{{\ttfamily 2008.01736}}].

\bibitem{Larkoski:2014bxa}
A.~J. Larkoski, D.~Neill and I.~W. Stewart, \emph{{Soft Theorems from Effective
  Field Theory}}, \href{https://doi.org/10.1007/JHEP06(2015)077}{\emph{JHEP}
  {\bfseries 06} (2015) 077} [\href{https://arxiv.org/abs/1412.3108}{{\ttfamily
  1412.3108}}].

\bibitem{Beneke:2019oqx}
M.~Beneke, A.~Broggio, S.~Jaskiewicz and L.~Vernazza, \emph{{Threshold
  factorization of the Drell-Yan process at next-to-leading power}},
  \href{https://doi.org/10.1007/JHEP07(2020)078}{\emph{JHEP} {\bfseries 07}
  (2020) 078} [\href{https://arxiv.org/abs/1912.01585}{{\ttfamily
  1912.01585}}].

\bibitem{Liu:2021mac}
Z.~L. Liu, M.~Neubert, M.~Schnubel and X.~Wang, \emph{{Radiative quark jet
  function with an external gluon}},
  \href{https://doi.org/10.1007/JHEP02(2022)075}{\emph{JHEP} {\bfseries 02}
  (2022) 075} [\href{https://arxiv.org/abs/2112.00018}{{\ttfamily
  2112.00018}}].

\bibitem{Adler:1966gc}
S.~L. Adler and Y.~Dothan, \emph{{Low-energy theorem for the weak axial-vector
  vertex}}, \href{https://doi.org/10.1103/PhysRev.151.1267}{\emph{Phys. Rev.}
  {\bfseries 151} (1966) 1267}.

\bibitem{Bonocore:2021cbv}
D.~Bonocore and A.~Kulesza, \emph{{Soft photon bremsstrahlung at
  next-to-leading power}},
  \href{https://doi.org/10.1016/j.physletb.2022.137325}{\emph{Phys. Lett. B}
  {\bfseries 833} (2022) 137325}
  [\href{https://arxiv.org/abs/2112.08329}{{\ttfamily 2112.08329}}].

\bibitem{Belle-II:2018jsg}
{\scshape Belle-II} collaboration, W.~Altmannshofer et~al., \emph{{The Belle II
  Physics Book}}, \href{https://doi.org/10.1093/ptep/ptz106}{\emph{PTEP}
  {\bfseries 2019} (2019) 123C01}
  [\href{https://arxiv.org/abs/1808.10567}{{\ttfamily 1808.10567}}].

\bibitem{Dittmaier:1999mb}
S.~Dittmaier, \emph{{A General approach to photon radiation off fermions}},
  \href{https://doi.org/10.1016/S0550-3213(99)00563-5}{\emph{Nucl. Phys. B}
  {\bfseries 565} (2000) 69}
  [\href{https://arxiv.org/abs/hep-ph/9904440}{{\ttfamily hep-ph/9904440}}].

\bibitem{Catani:1996jh}
S.~Catani and M.~H. Seymour, \emph{{The Dipole formalism for the calculation of
  QCD jet cross-sections at next-to-leading order}},
  \href{https://doi.org/10.1016/0370-2693(96)00425-X}{\emph{Phys. Lett. B}
  {\bfseries 378} (1996) 287}
  [\href{https://arxiv.org/abs/hep-ph/9602277}{{\ttfamily hep-ph/9602277}}].

\bibitem{Catani:1996vz}
S.~Catani and M.~H. Seymour, \emph{{A General algorithm for calculating jet
  cross-sections in NLO QCD}},
  \href{https://doi.org/10.1016/S0550-3213(96)00589-5}{\emph{Nucl. Phys. B}
  {\bfseries 485} (1997) 291}
  [\href{https://arxiv.org/abs/hep-ph/9605323}{{\ttfamily hep-ph/9605323}}].

\bibitem{Dittmaier:2008md}
S.~Dittmaier, A.~Kabelschacht and T.~Kasprzik, \emph{{Polarized QED splittings
  of massive fermions and dipole subtraction for non-collinear-safe
  observables}},
  \href{https://doi.org/10.1016/j.nuclphysb.2008.03.010}{\emph{Nucl. Phys. B}
  {\bfseries 800} (2008) 146}
  [\href{https://arxiv.org/abs/0802.1405}{{\ttfamily 0802.1405}}].

\bibitem{Bern:1994zx}
Z.~Bern, L.~J. Dixon, D.~C. Dunbar and D.~A. Kosower, \emph{{One loop $n$ point
  gauge theory amplitudes, unitarity and collinear limits}},
  \href{https://doi.org/10.1016/0550-3213(94)90179-1}{\emph{Nucl. Phys. B}
  {\bfseries 425} (1994) 217}
  [\href{https://arxiv.org/abs/hep-ph/9403226}{{\ttfamily hep-ph/9403226}}].

\bibitem{Kosower:1999rx}
D.~A. Kosower and P.~Uwer, \emph{{One loop splitting amplitudes in gauge
  theory}}, \href{https://doi.org/10.1016/S0550-3213(99)00583-0}{\emph{Nucl.
  Phys. B} {\bfseries 563} (1999) 477}
  [\href{https://arxiv.org/abs/hep-ph/9903515}{{\ttfamily hep-ph/9903515}}].

\bibitem{Bern:2004cz}
Z.~Bern, L.~J. Dixon and D.~A. Kosower, \emph{{Two-loop $g \to gg$ splitting
  amplitudes in QCD}},
  \href{https://doi.org/10.1088/1126-6708/2004/08/012}{\emph{JHEP} {\bfseries
  08} (2004) 012} [\href{https://arxiv.org/abs/hep-ph/0404293}{{\ttfamily
  hep-ph/0404293}}].

\bibitem{Badger:2004uk}
S.~D. Badger and E.~W.~N. Glover, \emph{{Two loop splitting functions in QCD}},
  \href{https://doi.org/10.1088/1126-6708/2004/07/040}{\emph{JHEP} {\bfseries
  07} (2004) 040} [\href{https://arxiv.org/abs/hep-ph/0405236}{{\ttfamily
  hep-ph/0405236}}].

\bibitem{Baier:1973ms}
V.~N. Baier, V.~S. Fadin and V.~A. Khoze, \emph{{Quasireal electron method in
  high-energy quantum electrodynamics}},
  \href{https://doi.org/10.1016/0550-3213(73)90291-5}{\emph{Nucl. Phys. B}
  {\bfseries 65} (1973) 381}.

\bibitem{Berends:1981uq}
F.~A. Berends, R.~Kleiss, P.~De~Causmaecker, R.~Gastmans, W.~Troost and T.~T.
  Wu, \emph{{Multiple Bremsstrahlung in Gauge Theories at High-Energies. 2.
  Single Bremsstrahlung}},
  \href{https://doi.org/10.1016/0550-3213(82)90489-8}{\emph{Nucl. Phys. B}
  {\bfseries 206} (1982) 61}.

\bibitem{Kleiss:1986ct}
R.~Kleiss, \emph{{Hard Bremsstrahlung Amplitudes for $e^+ e^-$ Collisions With
  Polarized Beams at {LEP} / {SLC} Energies}},
  \href{https://doi.org/10.1007/BF01552550}{\emph{Z. Phys. C} {\bfseries 33}
  (1987) 433}.

\bibitem{muonennlo}
A.~Broggio, T.~Engel, A.~Ferroglia, M.~K. Mandal, P.~Mastrolia, M.~Passera
  et~al., \emph{{Muon-electron scattering at NNLO}}, {\emph{in preparation}
  (2022)}.

\bibitem{NOGUEIRA1993279}
P.~Nogueira, \emph{Automatic feynman graph generation},
  \href{https://doi.org/https://doi.org/10.1006/jcph.1993.1074}{\emph{Journal
  of Computational Physics} {\bfseries 105} (1993) 279}.

\bibitem{Gehrmann:2001pz}
T.~Gehrmann and E.~Remiddi, \emph{{Numerical evaluation of harmonic
  polylogarithms}},
  \href{https://doi.org/10.1016/S0010-4655(01)00411-8}{\emph{Comput. Phys.
  Commun.} {\bfseries 141} (2001) 296}
  [\href{https://arxiv.org/abs/hep-ph/0107173}{{\ttfamily hep-ph/0107173}}].

\bibitem{Goncharov:1998kja}
A.~B. Goncharov, \emph{{Multiple polylogarithms, cyclotomy and modular
  complexes}}, \href{https://doi.org/10.4310/MRL.1998.v5.n4.a7}{\emph{Math.
  Res. Lett.} {\bfseries 5} (1998) 497}
  [\href{https://arxiv.org/abs/1105.2076}{{\ttfamily 1105.2076}}].

\bibitem{Naterop:2019xaf}
L.~Naterop, A.~Signer and Y.~Ulrich, \emph{{handyG \textemdash{}Rapid numerical
  evaluation of generalised polylogarithms in Fortran}},
  \href{https://doi.org/10.1016/j.cpc.2020.107165}{\emph{Comput. Phys. Commun.}
  {\bfseries 253} (2020) 107165}
  [\href{https://arxiv.org/abs/1909.01656}{{\ttfamily 1909.01656}}].

\bibitem{Vollinga:2004sn}
J.~Vollinga and S.~Weinzierl, \emph{{Numerical evaluation of multiple
  polylogarithms}},
  \href{https://doi.org/10.1016/j.cpc.2004.12.009}{\emph{Comput. Phys. Commun.}
  {\bfseries 167} (2005) 177}
  [\href{https://arxiv.org/abs/hep-ph/0410259}{{\ttfamily hep-ph/0410259}}].

\bibitem{carlo}
C.~C. Calame, \emph{{private communication}}.

\bibitem{Abbiendi:2677471}
G.~Abbiendi, \emph{{Letter of Intent: the MUonE project}},  tech. rep., CERN,
  Geneva, Jun, 2019.

\bibitem{PhysRevD.98.030001}
{\scshape Particle Data Group} collaboration, M.~Tanabashi, K.~Hagiwara,
  K.~Hikasa, K.~Nakamura, Y.~Sumino, F.~Takahashi et~al., \emph{Review of
  particle physics},
  \href{https://doi.org/10.1103/PhysRevD.98.030001}{\emph{Phys. Rev. D}
  {\bfseries 98} (2018) 030001}.

\bibitem{Baikov:2009bg}
P.~A. Baikov, K.~G. Chetyrkin, A.~V. Smirnov, V.~A. Smirnov and M.~Steinhauser,
  \emph{{Quark and gluon form factors to three loops}},
  \href{https://doi.org/10.1103/PhysRevLett.102.212002}{\emph{Phys. Rev. Lett.}
  {\bfseries 102} (2009) 212002}
  [\href{https://arxiv.org/abs/0902.3519}{{\ttfamily 0902.3519}}].

\bibitem{Lee:2021uqq}
R.~N. Lee, A.~von Manteuffel, R.~M. Schabinger, A.~V. Smirnov, V.~A. Smirnov
  and M.~Steinhauser, \emph{{Fermionic corrections to quark and gluon form
  factors in four-loop QCD}},
  \href{https://doi.org/10.1103/PhysRevD.104.074008}{\emph{Phys. Rev. D}
  {\bfseries 104} (2021) 074008}
  [\href{https://arxiv.org/abs/2105.11504}{{\ttfamily 2105.11504}}].

\bibitem{Lee:2022nhh}
R.~N. Lee, A.~von Manteuffel, R.~M. Schabinger, A.~V. Smirnov, V.~A. Smirnov
  and M.~Steinhauser, \emph{{Quark and Gluon Form Factors in Four-Loop QCD}},
  \href{https://doi.org/10.1103/PhysRevLett.128.212002}{\emph{Phys. Rev. Lett.}
  {\bfseries 128} (2022) 212002}
  [\href{https://arxiv.org/abs/2202.04660}{{\ttfamily 2202.04660}}].

\bibitem{Fael:2022rgm}
M.~Fael, F.~Lange, K.~Sch\"onwald and M.~Steinhauser, \emph{{Massive Vector
  Form Factors to Three Loops}},
  \href{https://doi.org/10.1103/PhysRevLett.128.172003}{\emph{Phys. Rev. Lett.}
  {\bfseries 128} (2022) 172003}
  [\href{https://arxiv.org/abs/2202.05276}{{\ttfamily 2202.05276}}].

\bibitem{Fael:2022miw}
M.~Fael, F.~Lange, K.~Sch\"onwald and M.~Steinhauser, \emph{{Singlet and
  nonsinglet three-loop massive form factors}},
  \href{https://doi.org/10.1103/PhysRevD.106.034029}{\emph{Phys. Rev. D}
  {\bfseries 106} (2022) 034029}
  [\href{https://arxiv.org/abs/2207.00027}{{\ttfamily 2207.00027}}].

\bibitem{Garland:2001tf}
L.~W. Garland, T.~Gehrmann, E.~W.~N. Glover, A.~Koukoutsakis and E.~Remiddi,
  \emph{{The Two loop QCD matrix element for $e^+e^- \to 3~\text{Jets}$}},
  \href{https://doi.org/10.1016/S0550-3213(02)00057-3}{\emph{Nucl. Phys. B}
  {\bfseries 627} (2002) 107}
  [\href{https://arxiv.org/abs/hep-ph/0112081}{{\ttfamily hep-ph/0112081}}].

\bibitem{Garland:2002ak}
L.~W. Garland, T.~Gehrmann, E.~W.~N. Glover, A.~Koukoutsakis and E.~Remiddi,
  \emph{{Two loop QCD helicity amplitudes for $e^+e^- \to 3~\text{Jets}$}},
  \href{https://doi.org/10.1016/S0550-3213(02)00627-2}{\emph{Nucl. Phys. B}
  {\bfseries 642} (2002) 227}
  [\href{https://arxiv.org/abs/hep-ph/0206067}{{\ttfamily hep-ph/0206067}}].

\bibitem{Grozin:2005yg}
A.~Grozin, \emph{{Lectures on QED and QCD}},  in \emph{{3rd Dubna International
  Advanced School of Theoretical Physics}}, 8, 2005,
  \href{https://arxiv.org/abs/hep-ph/0508242}{{\ttfamily hep-ph/0508242}}.

\bibitem{PhysRevD.53.4111}
A.~Djouadi and P.~Gambino, \emph{Electroweak gauge boson self-energies:
  Complete qcd corrections},
  \href{https://doi.org/10.1103/PhysRevD.53.4111}{\emph{Phys. Rev. D}
  {\bfseries 53} (1996) 4111}.

\end{thebibliography}\endgroup



\providecommand{\href}[2]{#2}\begingroup\raggedright\endgroup

\end{document}